\documentclass[fleqn,usenatbib]{mnras}

\usepackage{newtxtext,newtxmath}
\usepackage{longtable}

\usepackage[T1]{fontenc}
\usepackage{ae,aecompl}

\usepackage{graphicx}	
\usepackage{amsmath}	
\usepackage{enumitem}
\usepackage{times}
\usepackage{url}
\usepackage[usenames]{color}
\usepackage{xcolor}
\usepackage{longtable}
\usepackage{threeparttable}
\usepackage{bigints}
\usepackage{mathtools}
 \usepackage{orcidlink}
\defcitealias{Ferrara_2019}{F19}
\defcitealias{Kennicutt_1998}{K98}
\defcitealias{De_Looze_2011}{L11}
\defcitealias{De_Looze_2014}{L14}
\defcitealias{Herrera_Camus_2015}{H15}

\def \simless {\mathbin{\lower 3pt\hbox{$\rlap{\raise 4pt
              \hbox{$\char'074$}}\mathchar"7218$}}}
\def \simgreat {\mathbin{\lower 3pt\hbox{$\rlap{\raise 4pt
              \hbox{$\char'076$}}\mathchar"7218$}}}
\def\ie {{\it i.e.}}
\def\cf {{\it c.f.}}
\def\eg {{\it e.g.}}

\usepackage{pifont}
\title[$\rm C_{II}$ emission as an indicator of galaxy SFR]{$\rm [C_{II}]$ 158 $\rm \mu m$ emission as an indicator of galaxy star formation rate} 
\author[Liang et al.]{Lichen Liang$^{1,2}$\thanks{Email: lliang@cita.utoronto.ca}\orcidlink{0000-0001-9422-0095}, Robert Feldmann$^{2}$\orcidlink{0000-0002-1109-1919}, Norman Murray$^{1,3}$\orcidlink{0000-0002-8659-3729}, Desika Narayanan$^{4,5,6}$\orcidlink{0000-0002-7064-4309}, \newauthor  
Christopher C. Hayward$^{7}$\orcidlink{0000-0003-4073-3236}, Daniel Angl\'{e}s-Alc\'{a}zar$^{8, 7}$\orcidlink{0000-0001-5769-4945}, Luigi Bassini$^{2}$\orcidlink{0000-0002-6864-7762}, Alexander J. Richings$^{9, 10}$\orcidlink{0000-0003-0502-9235} \newauthor 
Claude-Andr\'{e} Faucher-Gigu\`{e}re$^{11}$\orcidlink{0000-0002-4900-6628}, Dongwoo T. Chung$^{1,12}$\orcidlink{0000-0003-2618-6504}, Jennifer Y. H. Chan$^{1,12}$\orcidlink{0000-0003-0314-7027},  \newauthor
Do\v{g}a Tolgay$^{1,13}$\orcidlink{0000-0002-3155-946X}, Onur \c{C}atmabacak$^{2}$\orcidlink{0000-0003-4067-1434}, Du\v{s}an Kere\v{s}$^{14}$\orcidlink{0000-0002-1666-7067}, Philip F. Hopkins$^{15}$\orcidlink{0000-0003-3729-1684} \\
$^{1}$Canadian Institute for Theoretical Astrophysics, University of Toronto, 60 St. George Street, Toronto, ON M5S 3H8, Canada\\
$^{2}$Institute for Computational Science, University of Zurich, Winterthurerstrasse 190, Zurich CH-8057, Switzerland\\
$^{3}$Canada Research Chair in Theoretical Astrophysics\\
$^{4}$Department of Astronomy, University of Florida, 211 Bryant Space Sciences Center, Gainesville, FL 32611, USA\\
$^{5}$University of Florida Informatics Institute, 432 Newell Drive, CISE Bldg E251, Gainesville, FL 32611, USA\\
$^{6}$Cosmic Dawn Center at the Niels Bohr Institute, University of Copenhagen and DTU-Space, Technical University of Denmark, Denmark\\
$^{7}$Center for Computational Astrophysics, Flatiron Institute, 162 Fifth Avenue, New York, NY 10010, USA\\
$^{8}$Department of Physics, University of Connecticut, 196 Auditorium Road, U-3046, Storrs, CT 06269-3046, USA \\
$^{9}$E. A. Milne Centre for Astrophysics, University of Hull, Cottingham Road, Hull, HU6 7RX, UK\\
$^{10}$DAIM, University of Hull, Cottingham Road, Hull, HU6 7RX, UK\\
$^{11}$Department of Physics and Astronomy and CIERA, Northwestern University, Evanston, IL 60208, USA\\
$^{12}$Dunlap Institute for Astronomy and Astrophysics, University of Toronto, 50 St. George Street, Toronto, ON M5S 3H4, Canada\\
$^{13}$Department of Physics, University of Toronto, 60 St. George Street, Toronto, Ontario M5S 1A7, Canada\\
$^{14}$Department of Physics, Center for Astrophysics and Space Sciences, University of California at San Diego, La Jolla, CA 92093, USA\\
$^{15}$TAPIR, Mailcode 350-17, California Institute of Technology, Pasadena, CA 91125, USA}
\date{Accepted 2022. Received 2022; in original form 2022}

\pubyear{2022}

\begin{document}
\label{firstpage}
\pagerange{\pageref{firstpage}--\pageref{lastpage}}
\maketitle

\begin{abstract}
\noindent Observations of local star-forming galaxies (SFGs) show a tight correlation between their singly ionized carbon line luminosity ($L_{\rm [C_{II}]}$) and star formation rate (SFR), suggesting that $L_{\rm [C_{II}]}$ may be a useful SFR tracer for galaxies. Some other galaxy populations, however, are found to have lower $L_{\rm [C_{II}]}{}/{}\rm SFR$ than local SFGs, including the infrared-luminous, starburst galaxies at low and high redshifts as well as some moderately star-forming galaxies at the epoch of re-ionization (EoR). The {origins} of this `$\rm [C_{II}]$ deficit' is unclear. In this work, we study the $L_{\rm [C_{II}]}$-SFR relation of galaxies using a sample of $z=0-8$ galaxies with $M_*\approx10^7-5\times10^{11}\,M_\odot$ extracted from cosmological volume and zoom-in simulations from the Feedback in Realistic Environments ({\sc\small FIRE}) project. We find a simple analytic expression for $L_{\rm [C_{II}]}$/SFR of galaxies in terms of the following parameters: mass fraction of $\rm [C_{II}]$-emitting gas ($f_{\rm [C_{II}]}$), gas metallicity ($Z_{\rm gas}$), gas density ($n_{\rm gas}$) and gas depletion time ($t_{\rm dep}{}={}M_{\rm gas}{}/{}\rm SFR$). We find two distinct physical regimes: $\rm H_2$-rich galaxies where $t_{\rm dep}$ is the main driver of the $\rm [C_{II}]$ deficit and $\rm H_2$-poor galaxies where $Z_{\rm gas}$ is the main driver. The observed $\rm [C_{II}]$ deficit of IR-luminous galaxies and early EoR galaxies, corresponding to the two different regimes, is due to short gas depletion time and low gas metallicity, respectively. Our result indicates that {the} $\rm [C_{II}]$ deficit is a common phenomenon of galaxies, and caution needs to be taken when applying a constant $L_{\rm [C_{II}]}$-to-SFR conversion factor derived from local SFGs to estimate cosmic SFR density at high redshifts and interpret data from upcoming $\rm [C_{II}]$ line intensity mapping experiments. 

\end{abstract}

\section{Introduction}
\label{Sec:1}

The census of cosmic star formation from the present day to the highest redshifts imposes a key constraint on galaxy evolution theory and physical cosmology \citep[see \eg,][and references therein]{Madau_2014, Dayal_2018}. The rest-frame ultra-violet (UV) luminosity ($L_{\rm UV}$) of galaxies, tracing the young, massive stars, is a common star formation rate (SFR) indicator of galaxies \citep[\eg][]{Hao_2011, Kennicutt_2012}. However, a large fraction of the UV light from galaxies in the Universe is absorbed by interstellar dust and gets re-emitted as thermal radiation at far-infrared (far-IR) wavelength \citep[\eg][]{Fixsen_1998, Takeuchi_2005, Dole_2006, Magnelli_2009, Gruppioni_2013, Burgarella_2013, Whitaker_2017, Salim_2020}. Therefore, an accurate estimate of the cosmic SF history depends on a multi-wavelength, UV-to-millimetre (mm) analysis that accounts for both the direct, unobscured stellar light and the dust thermal emission of galaxies over cosmic time. 

In practice, however, our capability of constraining the two components of stellar radiation is largely imbalanced \citep[\eg][]{Casey_2018a}. While the rest-frame UV-based, unobscured component has been constrained to {up to redshifts} $z{}\sim{}15$ \citep[\eg][]{Bouwens_2007, Bouwens_2011, Bouwens_2015, Bouwens_2019, Oesch_2012, Oesch_2016, Oesch_2018, Ellis_2013, McLure_2013, Finkelstein_2015, McLeod_2015, Bowler_2020, Naidu_2022, Leethochawalit_2023a, Leethochawalit_2023b, Donnan_2023, Harikane_2023} through deep imaging with the {\it Hubble} and the {\it James Webb Space Telescopes}, the obscured component is still not well constrained beyond $z{}\sim{}3$ due to the lack of statistically-representative, unbiased galaxy samples in that regime \citep{Casey_2014, Casey_2018a, Dayal_2018, Zavala_2021}. It is therefore important to have other SFR diagnostics in addition to UV+IR for early galaxies \citep[see \eg][and references therein]{Khusanova_2021}. 

The $\rm 158\,\mu m$ (1900.5 GHz) fine structure transition (${}^{2}P_{3/2}\rightarrow {}^{2}P_{1/2}$) of singly ionized carbon ($\rm [C_{II}]$) has been {proposed} as a promising alternative SFR indicator, particularly for high-$z$ galaxies \citep{Hodge_2020}. It is a major coolant of the neutral atomic gas of the interstellar medium (ISM) and {often} the strongest emission line of star-forming galaxies at rest-frame far-IR wavelength \citep{Carilli_2013}. The $\rm [C_{II}]$ line of galaxies is usually not much affected by dust extinction \citep[\eg][]{Abel_2007}. 

To first order, a correlation between $L_{\rm [C_{II}]}$ and global SFR of galaxies is expected. Much of the $\rm [C_{II}]$ emission of galaxy originates from the neutral atomic gas regions \citep{Hollenbach_1999, Wolfire_2003, Ferrara_2019}, where the far-UV (FUV) photons produced by the young O and B-type stars heat the gas via the photoelectric (PE) effect on small dust grains and polycyclic aromatic hydrocarbon (PAH) molecules \citep{Tielens_1985, Hollenbach_1991, Weingartner_2001a, Helou_2001}. The photo-electrons ejected from the dust grains/PAH molecules collisionally couple to and heat the gas. Since the PE heating rate ($\dot{E}_{\rm PE}$) traces galaxy SFR, and $L_{\rm [C_{II}]}$ balances $\dot{E}_{\rm PE}$ given that $\rm [C_{II}]$ line is the dominant coolant in those regions (assuming a thermal equilibrium), $L_{\rm [C_{II}]}$ should therefore be correlated to SFR. Observations of local star-forming galaxies (SFGs) have indeed found a linear correlation between $L_{\rm [C_{II}]}$ and SFR over the broad SFR range of $\approx{}10^{-4}-10\,M_\odot\,\rm yr^{-1}$ \citep[\eg][]{Stacey_1991, Leech_1999, Boselli_2002, De_Looze_2011, De_Looze_2014, Herrera_Camus_2015}. These observations suggest that the $\rm [C_{II}]$ line can be a useful SFR indicator for galaxies.

There is evidence, however, showing that this scaling relationship does not hold {in all environments}. For instance, observations find that local ultra-luminous infrared galaxies (ULIRGs, galaxies having $L_{\rm IR}{}\simgreat{}10^{12}\,L_\odot$) show a significant lower $L_{\rm [C_{II}]}{}/{}L_{\rm IR}$ ($\sim L_{\rm [C_{II}]}{}/{}\rm SFR$) ratio than normal SFGs by up to an order of magnitude \citep{Malhotra_1997, Malhotra_2001, Luhman_1998, Luhman_2003, Brauher_2008, Farrah_2013, Magdis_2014}, the so-called `$\rm [C_{II}]$ deficit' problem. This result was at first revealed with the {\it Infrared Space Observatory} \citep[{\it ISO};][]{Kessler_1996} and later confirmed by observations with the \textit{Herschel Space Observatory} \citep[hereafter {\it Herschel};][]{Pilbratt_2010} that has improved far-IR observing capabilities. Subsequent observations with {\it Herschel} also show that the $\rm [C_{II}]$ deficit extends to lower $L_{\rm IR}$ and that the $L_{\rm [C_{II}]}{}/{}L_{\rm IR}$ ratio of galaxies exhibits a continuous decrease with increasing $L_{\rm IR}$ at $L_{\rm IR}{}\simgreat {} 10^{11}\,L_\odot$ \citep[\eg][]{Gracia_Carpio_2011, Sargsyan_2012, Diaz_Santos_2013, Cormier_2015, Herrera_Camus_2015, Herrera_Camus_2018, Contursi_2017, Diaz_Santos_2017, Hughes_2017, Smith_2017}.

Studies have investigated the $L_{\rm [C_{II}]}$-SFR relation of galaxies at higher redshifts \citep[\eg][]{Stacey_2010, Gullberg_2015, Gullberg_2018, Brisbin_2015, Spilker_2016, Zanella_2018, Cooke_2018, Rybak_2019, Mckinney_2020}. At $z{}\approx{}1-5$, the selected galaxies are mostly uncovered by sub-mm surveys, which are traditionally classified as `sub-millimetre-bright galaxies (SMGs\footnote{In the literature, `SMGs' typically refer to the galaxies detectable by single-dish sub-mm telescopes, of which the observed sub-mm flux density is above $\sim{}1$ mJy \citep{Casey_2014, Hodge_2020}.})'. These are heavily dust-obscured systems having $L_{\rm IR}{}\simgreat{}10^{12}\,L_\odot$ (corresponding to {SFR${}\simgreat{}100\,M_\odot\,\rm yr^{-1}$}; \citealt{Kennicutt_1998}). In general, it is found that $\rm [C_{II}]$ deficit persists at high $L_{\rm IR}$ at high redshifts, although the high-$z$ populations appear to show larger scatter of $L_{\rm [C_{II}]}{}/{}\rm SFR$ at given $L_{\rm IR}$ than the local ones. 

The advent of the Atacama Large Millimetre/submillimetre Array ({\sc\small ALMA}) Telescope \citep[\eg][]{Wootten_2009} has triggered particular interest in searching for $\rm [C_{II}]$ emitters at $z{}\simgreat{}5$, and accumulating efforts have been made to constrain the $L_{\rm [C_{II}]}$-SFR relation of galaxies at this epoch \citep[\eg][]{Ouchi_2013, Ota_2014, Maiolino_2015, Capak_2015, Willott_2015b, Pentericci_2016, Matthee_2017, Matthee_2019, Carniani_2018a, Smit_2018, Schaerer_2020, Fujimoto_2021, Ferrara_2022, Schouws_2023}. The {\sc\small ALMA} observational programs are often designed to target the Lyman-$\alpha$ emitters (LAEs), Lyman-break galaxies (LBGs) and the quasar host galaxies (hereafter quasar hosts for simplicity) having pre-determined redshift \citep{Hodge_2020}. Though the earliest attempts targeting the bright LAEs were mostly unsuccessful \citep[\eg][]{Maiolino_2005, Ouchi_2013, Ota_2014, Inoue_2016}, follow-up programs targeting the LBGs and quasar hosts generally have had much higher success rate of $\rm [C_{II}]$ line detection. Overall, there have been $>{}200$ galaxies at $z{}\simgreat{}5$ that have confirmed detection of $\rm [C_{II}]$ line to date. While the quasar hosts {are typically very luminous and have substantial SFR} \citep[\eg][]{Banados_2016, Decarli_2018, Venemans_2020}, many of the selected LBGs/LAEs at $z{}\simgreat{}5$ are normal SFGs having moderate SFR ($\approx10\,M_\odot\,\rm yr^{-1}$). In particular, the {\sc\small ALMA} Large Program to INvestigate $\rm [C_{II}]$ at Early times ({\sc\small ALPINE}) survey \citep{Le_Fevre_2020, Bethermin_2020, Faisst_2020a} in Cycle-5, targeting a sample of 118 star-forming galaxies at $z{}\approx{}5-6$, has contributed more than a third ($\sim75/200$) of the total number of successful detections at $z{}\simgreat{}5$ \citep{Schaerer_2020}. More recently, the {\sc\small ALMA} Reionization Era Bright Emission Line Survey \citep[{\sc\small REBELS};][]{Bouwens_2022} in Cycle-7 has targeted a sample of 40 UV-bright, star-forming galaxies at $z{}\approx{}7$ and confirmed $\rm [C_{II}]$ line detection for 18 galaxies in their sample \citep{Ferrara_2022}. 

Observations have drawn divergent conclusions on the $L_{\rm [C_{II}]}$-SFR relation at $z{}\simgreat{}5$. While some have argued a clear $\rm [C_{II}]$ deficit of galaxies at $z{}\simgreat{}5$ with respect to the local normal SFGs \citep[\eg][]{Ouchi_2013, Ota_2014, Maiolino_2015, Inoue_2016, Knudsen_2016, Pentericci_2016, Bradac_2017, Ferrara_2019, Laporte_2019, Carniani_2020, Fujimoto_2022, Fudamoto_2023b}, others have argued that they follow the same linear scaling relation \citep[\eg][]{Matthee_2017, Carniani_2018a, Schaerer_2020, Fujimoto_2021, Ferrara_2022, Schouws_2023, Fudamoto_2023a, Fujimoto_2023}. It should be noted, however, that the SFR estimates at such high redshifts can be highly uncertain. Galaxies at $z{}\simgreat{}5$ typically have very few reliable photometric data points in the dust thermal continuum that are measured with {\sc\small ALMA} (at band 6 or 7). A number of recent studies, both observational \citep{Capak_2015, Bouwens_2016, Casey_2018a, Faisst_2020b} and theoretical \citep{Liang_2019, Liang_2021, Ma_2019, Sommovigo_2020, Sommovigo_2021}, have pointed out that based on the {\sc\small ALMA} broad-band flux(es) alone, $L_{\rm IR}$ (and hence the obscured SFR) of galaxies at $z{}\simgreat{}5$ is likely to be poorly constrained due to the large variation in the shape of the spectral energy distribution (SED) of their dust emission. The reported (in)consistencies of the $L_{\rm [C_{II}]}$-SFR relation at $z{}\simgreat{}5$ with the local SFGs by the observations therefore need to be more carefully assessed. 

Much effort has been made to model $\rm [C_{II}]$ emission of galaxies and explain the origins of the observed $\rm [C_{II}]$ deficit over the last two decades. A broad variety of different methods are used by different studies, including pure analytic approaches \citep[\eg][]{Munoz_2016, Ferrara_2019}, numerical models of idealized gas clouds \citep[\eg][]{Abel_2009, Narayanan_2017}, semi-analytic galaxy models \citep[SAMs, \eg][]{Popping_2014, Popping_2016, Popping_2019, Lagache_2018, Yang_2021,Yang_2022} and hydrodynamic galaxy simulations \citep[\eg][]{Vallini_2013, Vallini_2015, Olsen_2015, Olsen_2017, Pallottini_2017, Pallottini_2019, Katz_2019, Leung_2020, Lupi_2020a, Lupi_2020b, kannan_2022b, Richings_2022, Bisbas_2022}.  A pure analytic approach and/or a simplified cloud model can capture the key physical mechanisms that determine $L_{\rm [C_{II}]}$ of galaxies and provide useful insights at low computational cost, but does not provide the necessary galaxy statistics. SAMs can produce statistically significant galaxy samples probing a very wide dynamic range (in stellar mass, SFR, redshift and etc.) and are computationally efficient \citep{Somerville_2015a}, but they do not provide any information of structures on the sub-galactic scales. Hydrodynamic simulations, in contrast, can calculate the detailed sub-galactic structures and thus provide more accurate prediction for the $\rm [C_{II}]$ emission properties of galaxies, at the cost of more computational expense.

Different explanations for the $\rm [C_{II}]$ deficit {\em in the high $L_{\rm IR}$ regime} have been proposed by the theory groups \citep[see also \eg][for a summary]{Casey_2014, Narayanan_2017}. For instance, some studies argue that the deficit is due to a strong UV radiative intensity ($G$) in the IR luminous galaxies \citep[\eg][]{Malhotra_1997, Malhotra_2001, Luhman_1998, Genzel_2000, Helou_2001, Luhman_2003, Abel_2009, Stacey_2010, Gracia_Carpio_2011, Lagache_2018}. This can have two important effects on the thermal balance of $\rm [C_{II}]$-emitting gas. First of all, a high $G$ leads to large positive grain charges, thereby reducing the kinetic energy of the ejected photo-electrons and hence the rate of PE heating ($\dot{E}_{\rm PE}$) of gas \citep{Tielens_1985, Kaufman_1999}. As a result, $\rm [C_{II}]$ cooling rate drops. Besides, $\rm H^+$ regions in those galaxies may become `dust bounded' rather than `ionization bounded' (\eg~\citealt{Bottorff_1998, Abel_2009}; see also \citealt{Ferrara_2019}). In this scenario, most of the UV radiation from young stars is absorbed by dust in the $\rm H^+$ regions, leading to both an excess of IR emission in the $\rm H^+$ regions and a reduced $\dot{E}_{\rm PE}$ (and hence $L_{\rm [C_{II}]}$) in gas outside the $\rm H^+$ regions due to a starvation of UV photons there.

Alternatively, \citet{Narayanan_2017} suggest that a high gas density can lead to a $\rm [C_{II}]$ deficit of galaxy in addition to having a high $G$. Using a stratified gas cloud model, the authors demonstrate that with increasing gas density, a larger fraction of carbon in gas turns into neutral (\ie~in CO and $\rm C_I$) and $L_{\rm [C_{II}]}$ decreases due to a reduced mass fraction of $\rm [C_{II}]$-emitting gas.

Apart from these studies, \citet{Munoz_2016} posit an analytic model where $\rm [C_{II}]$ deficit is due to thermal saturation of the upper fine structure transition state (${}^{2}P_{3/2}$) of {$\rm C^+$} ions\footnote{Throughout this paper, we use `$\rm [C_{II}]$' when referring to the observable emission line, and {`$\rm C^+$'} when discussing ionized carbon under the context of chemical abundances of gas.}. At above 91.8 K (note: $T_*{}={}91.8$ K is the equivalent temperature of the $\rm [C_{II}]$ transition), $L_{\rm [C_{II}]}$ does not increase much with gas kinetic temperature and this has been suggested to be the reason for $L_{\rm [C_{II}]}$ not increasing much with SFR at high $L_{\rm IR}$ ($\rm \sim{}SFR$) {\citep[see also discussions in the observational studies by][]{Diaz_Santos_2017, Croxall_2017}}. Note, however, that the \citet{Munoz_2016} model assumes that the bulk of the $\rm [C_{II}]$ emission of galaxies originates from the gas having density in excess of the critical density for the $\rm [C_{II}]$ transition \citep{Goldsmith_2012}.

With the recent success of the {\sc\small ALMA} programs in searching for $\rm [C_{II}]$-emitters, there has been an increasing amount of effort to predict $\rm [C_{II}]$ emission properties of galaxies at $z{}\simgreat{}5$ by coupling cosmological hydrodynamic simulations or SAMs with photo-ionization codes (\eg~{\sc\small CLOUDY}, \citealt{Ferland_1998, Ferland_2013}; {\sc\small DESPOTIC}, \citealt{Krumholz_2014}; {\sc\small RADMC-3D}, \citealt{Dullemond_2012}). The predicted $L_{\rm [C_{II}]}$-SFR relation for galaxies, however, shows non-trivial discrepancy between different groups in both normalization and slope \citep[\eg][]{Katz_2019, Leung_2020}, which can be ascribed to the differences in the simulation methodology and $\rm [C_{II}]$ modelling techniques adopted by the different groups. Despite the discrepancy, many have predicted a $\rm [C_{II}]$ deficit of galaxies at $z{}\simgreat{}5$ with respect to the local normal SFGs. For instance, \citet{Lagache_2018} couple a sample of $\sim{}20$ K SAM galaxies at $4{}\le{}z{}\le{}8$ with {\sc\small CLOUDY} and report a $\rm [C_{II}]$ deficit of $>{}0.5$ dex and a trend of decreasing normalization of the relation with redshift. \citet{Olsen_2017} post-process 30 star-forming galaxies at $z{}={}6$ extracted from the {\sc\small MUFASA} `zoom-in' simulations \citep{Dave_2016} using {\sc\small CLOUDY} and predict a $\rm [C_{II}]$ deficit of about one {dex}. A similarly strong $\rm [C_{II}]$ deficit is reported by \citet{Pallottini_2017, Pallottini_2019} using the {\sc\small SERRA} `zoom-in' simulations that include more sophisticated chemical networks. More recently, \citet{kannan_2022b} predict an even more prominent $\rm [C_{II}]$ deficit at $z{}\ge{}5$ than the above-mentioned earlier studies, especially at low SFR, using a galaxy sample produced by the {\sc\small THESAN} `zoom-in' suite \citep{kannan_2022a, Garaldi_2022}, which includes the {\sc\small Illustris-TNG} galaxy formation model \citep{Pillepich_2018a, Pillepich_2018b}.
 
It has been generally thought that gas metallicity ($Z_{\rm gas}$) is the key factor in determining the $\rm [C_{II}]$ luminosity of the early galaxies \citep[\eg][]{Vallini_2015, Olsen_2017, Ferrara_2019, Heintz_2021, Heintz_2022} since $\rm [C_{II}]$ emissivity is linearly scaled with $Z_{\rm gas}$. The early work by \citet{Vallini_2015} shows that the $L_{\rm [C_{II}]}$-SFR relation of EoR galaxies {is sensitive to} $Z_{\rm gas}$, and the significant $\rm [C_{II}]$ deficit found with the LAEs at $z{}\approx{}5-7$, such as Himiko \citep{Ouchi_2013, Ota_2014} and  IOK-1 \citep{Ota_2014}, can be well accounted for by assigning a very low gas metallicity ($Z_{\rm gas}{}<{}0.05\,Z_\odot$) to the simulated galaxy in an ad hoc manner. The $\rm [C_{II}]$ deficit of galaxies at $z{}\simgreat{}5$ commonly found in the recent simulations, as mentioned above, is likely due to the much lower $Z_{\rm gas}$ of the early galaxies than the $z{}={}0$ ones predicted by these simulations. Observationally, however, direct measurement of $Z_{\rm gas}$ at $z{}\simgreat{}5$ is still very challenging, though some preliminary attempts have been made recently \citep[\eg][]{Rigopoulou_2018, Schaerer_2022, Curti_2023, Heintz_2023a, Heintz_2023b, Rhoads_2023, Trump_2023}.

A few recent studies have predicted $\rm [C_{II}]$ emission of galaxies at lower redshifts using simulations. For instance, \citet{Popping_2019} and \citet{Yang_2021} predict the $L_{\rm [C_{II}]}$-SFR relation for the catalog derived from the `Santa Cruz' semi-analytic models \citep{Somerville_1999, Somerville_2015b} using {\sc\small DESPOTIC}. Their result is in good agreement with the observational data at $z{}\approx{}2$, except that at high SFR (\ie~${\rm SFR}{}\simgreat{}10\,M_\odot\,\rm yr^{-1}$), they produce a noticeably weaker $\rm [C_{II}]$ deficit than is observed. More recently, \citet{Richings_2022} ran a set of hydrodynamic simulations of isolated (dwarf and Milky Way-mass) galaxies implemented with the {\sc\small CHIMES} non-equilibrium chemistry module \citep{Richings_2014a, Richings_2014b} (including a dust-depletion model) and predict the $\rm [C_{II}]$ emission of their galaxy sample using {\sc\small RADMC-3D}. Despite having a small sample size, the predicted $L_{\rm [C_{II}]}$ of their galaxies appears to be in agreement with the observational result of local galaxies \citep[\eg][]{De_Looze_2011, De_Looze_2014, Herrera_Camus_2015} at similar SFR (see also another recent work by \citealt{Bisbas_2022} using isolated dwarf simulations).

Apart from these studies, there has been limited effort to predict the $L_{\rm [C_{II}]}$-SFR relation of galaxies at $z{}={}0-5$ using statistically representative galaxy samples and compare the result to the fruitful observational data in this regime. In particular, the origin of the $\rm [C_{II}]$ deficit of the IR-luminous galaxies has not yet been studied in detail using cosmological hydrodynamic simulations. This is largely because producing a statistically representative sample in this regime with well-resolved ISM is computationally demanding, which is possible only for a few large simulation consortiums. It is, however, of critical importance that a robust $\rm [C_{II}]$ model should be able to simultaneously reproduce the data of different galaxy populations over the entire SFR and redshift ranges.

In this study, we conduct a comprehensive analysis of the galaxy $L_{\rm [C_{II}]}$-SFR relation using a simulated sample spanning an unprecedentedly broad redshift range of $z{}={}0-8$ extracted from the {\sc\small MassiveFIRE} \citep{Feldmann_2016,Feldmann_2017b, Angles_Alcazar_2017} and {\sc\small FIREbox} \citep{Feldmann_2023} cosmological hydrodynamic simulations from the {\it Feedback in Realistic Environments} ({\sc\small FIRE}) project\footnote{{\sc\small FIRE} project website: \url{http://fire.northwestern.edu}} \citep{Hopkins_2014, Hopkins_2018, Hopkins_2023}. The sample covers a very broad range of galaxy stellar mass and SFR, allowing us to make direct comparison with the observational data in different regimes. In particular, the sample includes local normal SFGs (having SFR${}\approx{}0.1-10\,M_\odot\,\rm yr^{-1}$) that can be compared with the observations where a linear $L_{\rm [C_{II}]}$-SFR correlation has been found by the observations. It also includes IR-luminous ($L_{\rm IR}>10^{11}\,L_\odot$) galaxies at $z{}={}0-5$ that are candidates for (U)LIRGs and SMGs, where observations have shown to have $\rm [C_{II}]$ deficit. Moreover, the sample includes early galaxies at above $z{}={}5$ spanning a broad SFR range. Many of these galaxies have similar mass and SFR to the samples of the {\sc\small ALPINE} and {\sc\small REBELS} projects and therefore can be used to provide useful interpretations for a variety of their recent observational results \citep[\eg][]{Fujimoto_2020, Ginolfi_2020, Schaerer_2020, Fudamoto_2021, Fudamoto_2022, Ferrara_2022, Sommovigo_2022}.

The main goal of this work is to predict the $L_{\rm [C_{II}]}$-SFR relation for the {\sc\small FIRE} galaxy sample (spanning $z{}={}0-8$ and ${\rm SFR}{}\approx{}0.1-10^3\,M_\odot\,\rm yr^{-1}$) and to understand what physical parameters of galaxies determine their overall $L_{\rm [C_{II}]}$-to-SFR ratio. This will then help us find the origin of the observed $\rm [C_{II}]$ deficit of galaxies at both high $L_{\rm IR}$ and high redshifts.

Note that the results from this work will be useful for interpreting the data of several upcoming $\rm [C_{II}]$ line intensity mapping (LIM) experiments \citep[see \eg~][and references therein]{Kovetz_2017, Kovetz_2019, Bernal_2022, Horlaville_2023}, such as {\sc\small TIME}\footnote{\url{https://cosmology.caltech.edu/projects/TIME}} \citep{Sun_2021}, {\sc\small CCAT-prime}\footnote{\url{http://www.ccatobservatory.org}} \citep{CCAT_2021}, {\sc\small CONCERTO}\footnote{\url{https://www.apex-telescope.org/ns/concerto/}} \citep{CONCERTO_2020, Gkogkou_2023} and {\sc\small EXCLAIM} \citep{Switzer_2021, Pullen_2023}. The LIM experiments have been designed to measure the emission from spectral lines originating from galaxies at all luminosities, including the ones that cannot be resolved by the current surveys (\eg~with {\sc\small ALMA}). The experiments that will target $\rm [C_{II}]$ emission, in particular, will be useful for constraining the cosmic star-formation history \citep[see \eg][]{Gong_2012, Silva_2015, Serra_2016, Fonseca_2017, Padmanabhan_2019, Yue_2019, Chung_2020, Padmanabhan_2022, Karoumpis_2022, Sun_2023}. It is, however, not yet certain whether the $\rm [C_{II}]$ line always acts as a reliable SFR tracer for galaxies of all types and at all redshifts. 

This paper is structured as follows. We describe in Section~\ref{Sec:2} the simulation methodology and in Section~\ref{Sec:3}, the method used to simulate $\rm [C_{II}]$ emission. In Section~\ref{Sec:4}, we compare the predicted $L_{\rm [C_{II}]}$-SFR relation of the {\sc\small FIRE} galaxy sample with the observational data at different redshifts. In Section~\ref{Sec:5}, we investigate the origin of the tight $L_{\rm [C_{II}]}$-SFR linear scaling relation of normal SFGs at $z{}={}0$ and the causes of the $\rm [C_{II}]$ deficit of galaxies. We discuss our results in Section~\ref{Sec:6} and finally summarize and conclude this study in Section~\ref{Sec:7}. Throughout this paper, we adopt the cosmological parameters of the Planck 2015 Cosmology \citep{Planck_2015}, specifically $\Omega_{\rm m}{}={}0.309$, $\Omega_\Lambda{}={}0.691$, $\Omega_{\rm b}{}={}0.049$, $\sigma_8{}={}0.816$, and $H_0{}={}67.74\,\rm km\,s^{-1}\,Mpc^{-1}$.

\section{Simulation methodology}
\label{Sec:2}

In this section, we introduce the simulation suites ({\sc\small FIREbox} and {\sc\small MassiveFIRE}) from which we extract the galaxy sample used for this study. 

\subsection{Simulation set-up and galaxy catalogue}
\label{Sec:2a}

We adopt a sample that spans the wide redshift range $z{}={}0-8$, stellar mass ($M_*$) range $M_*{}\approx{}10^7-5\times10^{11}\,M_\odot$ and SFR range ${\rm SFR}{} \approx{}0.1-10^3\,M_\odot\,\rm yr^{-1}$. The sample consists primarily of galaxies at $z{}={}0-8$ produced by {\sc \small FIREbox}  \citep{Feldmann_2023}, the new-generation simulation suite of {\sc\small FIRE} run with full cosmological volume boxes. It is supplemented by a number of high-$z$ ($z{}={}1-8$) massive galaxies ($M_*{}\simgreat{}10^{10}\,M_\odot$) extracted from the `zoom-in' suite, {\sc MassiveFIRE} \citep{Feldmann_2016, Feldmann_2017b}, rerun with {\sc\small FIRE}-2 physics \citep{Angles_Alcazar_2017, Catmabacak_2022, Bassini_2023}. Many of the {\sc MassiveFIRE} galaxies have the $L_{\rm IR}$ close to that of the SMGs \citep{Liang_2018, Cochrane_2019} that are used by the observational studies on the $L_{\rm [C_{II}]}$-SFR relation at high redshifts. All simulations used for this study are run with the same {\sc\small FIRE}-2 physics and numerics \citep{Hopkins_2018}.  
 
\paragraph*{FIREbox simulations\\}

\begin{figure}
 \includegraphics[height=135mm]{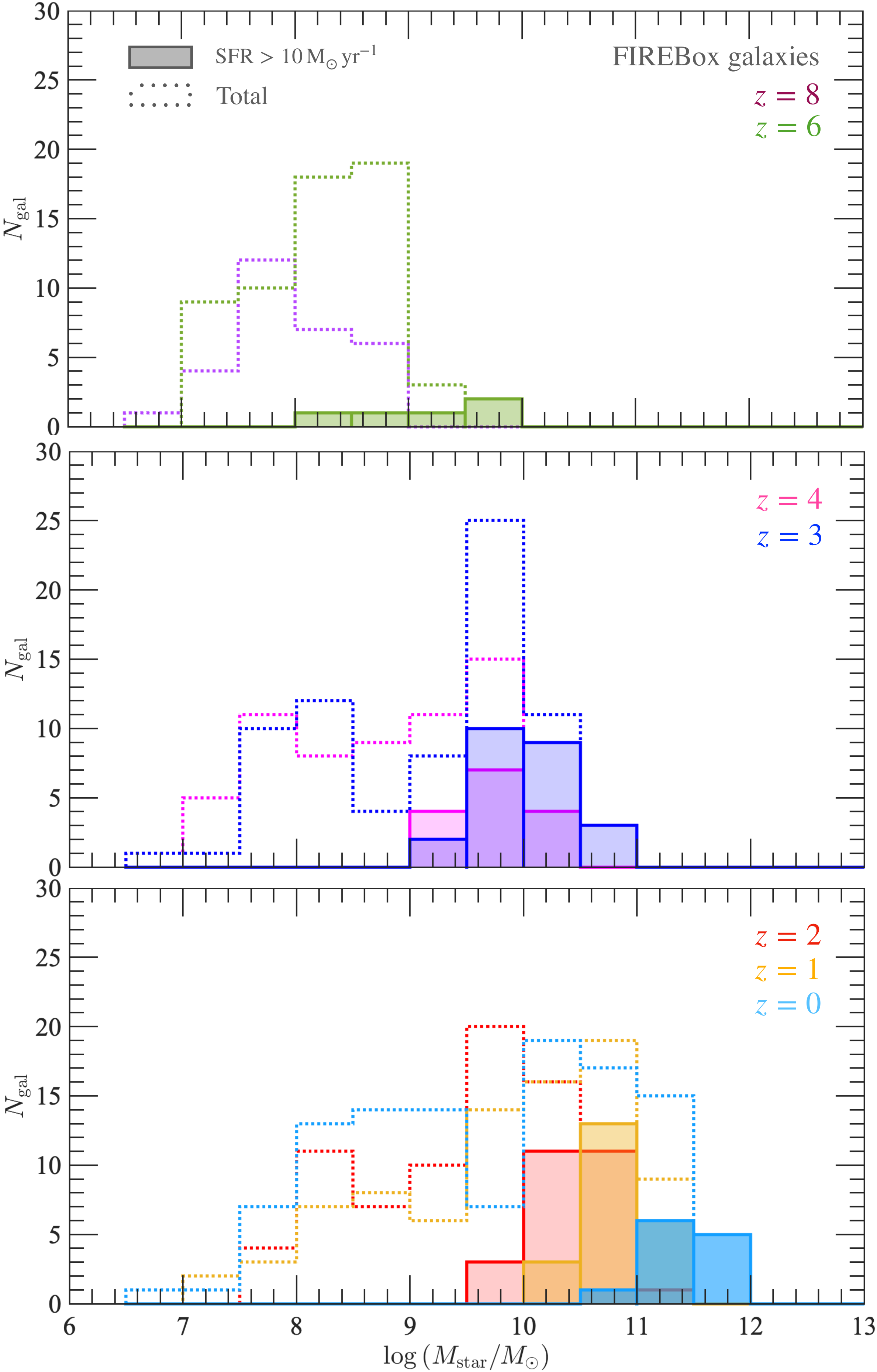}
 \caption{Histograms of the stellar mass distribution of the {\sc\small FIREbox} sample at different redshifts. Violet, green, magenta, blue, red, yellow and cyan histograms correspond to $z{}={}8$, $z{}={}6$, $z{}={}4$, $z{}={}3$, $z{}={}2$, $z{}={}1$ and $z{}={}0$, respectively. For each redshift, the unfilled histograms indicate the result of the entire galaxy sample, whereas the filled histograms indicate specifically the result of the galaxies having SFR${}\ge{}10\,M_\odot\,\rm yr^{-1}$ (corresponding to $L_{\rm IR}{}\simgreat{}10^{11}\,L_\odot$ based on the \citealt{Kennicutt_1998} relation. Note: $\rm [C_{II}]$ deficit is observed at $L_{\rm IR}{}\simgreat{}10^{11}\,L_\odot$.). For clarity of presentation, we separately show the result of the 7 snapshots in 3 separate panels ({\it top} panel for $z{}={}8$ and $z{}={}6$, {\it middle} panel for $z{}={}3$ and $z{}={}4$ and {\it bottom} panel for $z{}={}0$, $z{}={}1$ and $z{}={}2$).}
\label{fig.1}
\end{figure}

{\sc\small FIREbox} \citep{Feldmann_2023} is a new-generation simulation suite using {\sc\small FIRE} physics. Different from all previous simulations of {\sc\small FIRE}, {\sc\small FIREbox} simulates full cosmological volumes instead of using `zoom-in' set-up to study galaxy evolution. {\sc\small FIREbox} simulations are run in cubic boxes with periodic boundary conditions, and with initial conditions at redshift $z{}={}120$ generated using the {\sc \small MUSIC} (Multi-Scale Initial Conditions) code \citep{Hahn_2011}. The simulations use the Planck 2015 Cosmology \citep{Planck_2015}.

All {\sc\small FIREbox} simulations use the same initial conditions and cosmology but differ in numerical resolution. For this study, we extract galaxies from the fiducial {\sc\small FIREbox} hydrodynamic simulation, which is run with a box length of $15\,h^{-1}$ cMpc and with the following number of dark matter (DM) and baryonic particles: $N_{\rm DM}{}={}1024^3$ and $N_{\rm b}{}={}1024^3$. The mass resolution of DM and baryon particles are $m_{\rm DM}{}={}3.3\times10^5$ and $m_{\rm b}{}={}6.3\times10^4\,M_\odot$. The gravitational softening lengths are kept fixed in proper (comoving) coordinates at $z{}\le{}9$ ($z{}\ge{}9$) and are set to $h_{\rm DM}{}={}80$ pc for DM particles and $h_*{}={}12$ pc for star particles. The softening length for gas particles ($h_{\rm gas}$) is fully adaptive and is set equal to their kernel smoothing length down to a minimum of 1.5 proper pc, which is reached in the densest parts of the ISM. {\sc\small FIREbox} is evolved down to $z{}={}0$. 

We identify galaxies in different snapshots of the {\sc\small FIREbox} simulation using the {\sc \small AMIGA Halo Finder}\footnote{Code available at: \url{popia.ft.uam.es/AHF/Download.html}} \citep[{\sc\small AHF};][]{Gill_2004, Knollmann_2009}. We use the galaxies extracted from 7 snapshots corresponding to redshift $z{}={}0$, 1, 2, 3, 4, 6 and 8. For each snapshot, we include the central galaxy of the 30 most massive halos identified by {\sc\small AHF}. To enlarge our sample, we also include the central galaxy of a number of additional, randomly chosen halos having ${\rm log}\,(M_{\rm vir}/M_\odot){}>{}10$. We show in Fig.~\ref{fig.1} the histograms of the $M_*$ distribution of the selected {\sc\small FIREbox} galaxies at different redshifts. 

The number of galaxies selected at $z{}={}0$, 1, 2, 3, 4, 6, and 8 are 113, 84, 80, 75, 64, 61 and 30, respectively. As is shown in Fig.~\ref{fig.1}, all but a few selected galaxies have stellar mass greater than $10^7\,M_\odot$ (corresponding to $\sim{}160$ times of the mass resolution). The most massive galaxy of the {\sc\small FIREbox} sample has $M_*{}={}4.8\times 10^{11}\,M_\odot$ (at $z{}={}0$). 

In the same figure, we also show the $M_*$ distribution of the galaxies having ${\rm SFR}{}\simgreat{}10\,M_\odot\,\rm yr^{-1}$ (filled histograms). These galaxies have $L_{\rm IR}{}\ge{}10^{11}\,L_\odot$, the regime where a $\rm [C_{II}]$ deficit is observed (see Section~\ref{Sec:3}). They apparently are more massive than the galaxies having SFR${}<{}10\,M_\odot\,\rm yr^{-1}$. In our catalogue, we find most galaxies with ${\rm SFR}{}\ge{}10\,M_\odot\,\rm yr^{-1}$ at $z{}={}2$ (red histogram, $N{}={}29$) and $z{}={}3$ (blue histogram, $N{}={}28$). These redshifts are at the `cosmic noon', where massive galaxies start to form and they are more gas-rich and actively star-forming than galaxies at lower redshifts. 

Since the {\sc\small FIREbox} simulation is run with a volume of $(15\,h^{-1}\rm cMpc)^3$, it does not produce enough galaxies at high redshifts that are as massive and luminous as the galaxy samples selected by the observational studies. We therefore supplement our sample with a handful of more massive galaxies ($M_*{}\approx{}10^9-5\times10^{11}\,M_\odot$) extracted from the {\sc MassiveFIRE} `zoom-in' simulations (see below). 

\paragraph*{MassiveFIRE simulations\\}

\begin{table*}
\caption{List of {\sc \small MassiveFIRE} simulations used for this work.}
\begin{threeparttable}
\begin{tabular}{ p{1 cm}  p{1 cm}  p{0.8 cm}  p{1 cm} p{1.3 cm} p{1.3 cm}  p{1.3 cm}  p{1.3 cm}  p{1.3 cm}  p{1.3 cm} }    
 \hline                                     
\multicolumn{1}{c|}{Sim ID}$\dagger$  & \multicolumn{1}{c|}{Box Size } & \multicolumn{1}{c|}{$z_{\rm final}$} & \multicolumn{1}{c|}{$M_{\rm vir}$ $\ddagger$ } & \multicolumn{6}{c|}{$M_*$ ($M_\odot$)} \\
&  \multicolumn{1}{c|}{($h^{-1}$ Mpc)} &  &  \multicolumn{1}{c|}{($10^{12}\,M_\odot$)} &  \multicolumn{1}{c|}{$z=1$} &  \multicolumn{1}{c|}{$z=2$}  &  \multicolumn{1}{c|}{$z=3$} &   \multicolumn{1}{c|}{$z=4$} &  \multicolumn{1}{c|}{$z=6$} & \multicolumn{1}{c|}{$z=8$}    \\
 \hline
A1 &  \multicolumn{1}{c|}{100} & \multicolumn{1}{c|}{1} & \multicolumn{1}{c|}{2.4}  & $5.4\times10^{11}$ & $5.1\times10^{10}$ & $9.6\times10^9$ & $1.2\times10^9$ &  \multicolumn{1}{c|}{/}  &  \multicolumn{1}{c|}{/} \\
A2 &  \multicolumn{1}{c|}{100}  & \multicolumn{1}{c|}{1} &  \multicolumn{1}{c|}{3.0}  & $4.1\times10^{11}$ & $2.9\times10^{11}$ & $1.3\times10^{11}$ & $2.7\times10^{10}$ &  \multicolumn{1}{c|}{/}  &  \multicolumn{1}{c|}{/}  \\
A4 &  \multicolumn{1}{c|}{100} & \multicolumn{1}{c|}{1} &  \multicolumn{1}{c|}{2.9}  & $2.3\times10^{11}$ & $1.3\times10^{11}$ & $2.2\times10^{10}$ & $6.5\times10^9$ &  \multicolumn{1}{c|}{/}   &  \multicolumn{1}{c|}{/}  \\
A8 &  \multicolumn{1}{c|}{100} & \multicolumn{1}{c|}{1} &  \multicolumn{1}{c|}{3.6}  & $2.8\times10^{11}$ & $1.8\times10^{11}$ & $9.8\times10^{10}$ & $5.1\times10^{10}$ &  \multicolumn{1}{c|}{/}  &   \multicolumn{1}{c|}{/}  \\
D3 &  \multicolumn{1}{c|}{400} & \multicolumn{1}{c|}{6} &   \multicolumn{1}{c|}{4.5} & \multicolumn{1}{c|}{/}  &  \multicolumn{1}{c|}{/}  &  \multicolumn{1}{c|}{/}  &   \multicolumn{1}{c|}{/} & $3.9\times10^{11}$ & $7.0\times10^{10}$ \\
D7$\mathsection$ &  \multicolumn{1}{c|}{400} & \multicolumn{1}{c|}{6} &   \multicolumn{1}{c|}{2.5} & \multicolumn{1}{c|}{/}  &  \multicolumn{1}{c|}{/}  &  \multicolumn{1}{c|}{/} &  \multicolumn{1}{c|}{/} & \multicolumn{1}{c|}{/}  & $5.8\times10^{10}$ \\
D9&  \multicolumn{1}{c|}{400} & \multicolumn{1}{c|}{6} &   \multicolumn{1}{c|}{1.0} & \multicolumn{1}{c|}{/}  &  \multicolumn{1}{c|}{/}  &  \multicolumn{1}{c|}{/}  & \multicolumn{1}{c|}{/}  & $3.9\times10^{10}$  & $1.3\times10^9$ \\
E1 & \multicolumn{1}{c|}{762} & \multicolumn{1}{c|}{6} &   \multicolumn{1}{c|}{6.8} & \multicolumn{1}{c|}{/}  &  \multicolumn{1}{c|}{/}  & \multicolumn{1}{c|}{/}  &   \multicolumn{1}{c|}{/} & $1.6\times10^{10}$  & $3.2\times10^9$ \\
E2 &  \multicolumn{1}{c|}{762} & \multicolumn{1}{c|}{6} &   \multicolumn{1}{c|}{6.5} & \multicolumn{1}{c|}{/}  & \multicolumn{1}{c|}{/}   &  \multicolumn{1}{c|}{/}  &  \multicolumn{1}{c|}{/}  & $7.2\times10^9$  & $5.3\times10^9$ \\
E3 &  \multicolumn{1}{c|}{762} & \multicolumn{1}{c|}{6} &   \multicolumn{1}{c|}{6.1} & \multicolumn{1}{c|}{/}  &  \multicolumn{1}{c|}{/}  & \multicolumn{1}{c|}{/}  &  \multicolumn{1}{c|}{/}  & $8.6\times10^9$ & $2.7\times10^9$ \\
 \hline
\end{tabular}
  \begin{tablenotes}
    \item[$\dagger$] The A (D and E) Series of {\sc\small MassiveFIRE} were published in \citet{Angles_Alcazar_2017} (\citealt{Catmabacak_2022}) for the first time.
    \item[$\ddagger$] Virial mass at $z_{\rm final}$. 
    \item[$\mathsection$] The HR simulation of D7 has been run only down to $z=7.2$.
  \end{tablenotes}
  \end{threeparttable}
\label{T1}
\end{table*}

{\sc\small MassiveFIRE} \citep{Feldmann_2016, Feldmann_2017b} is a set of simulations of massive galaxies at high redshifts using the `zoom-in' method. A number of low-resolution (LR) DM-only simulations were run with the initial conditions generated using the {\sc\small MUSIC} code within periodic boxes. From the outputs of these LR DM-runs, we then select a number of model haloes to re-simulate at much higher resolution and with baryons included (HR runs). The selected haloes have a variety of masses, accretion history, and environmental over-densities. 

For this study, we use the galaxies produced by 10 {\sc\small MassiveFIRE} simulations, which are from the A \citep{Angles_Alcazar_2017}, D and E Series \citep{Catmabacak_2022, Bassini_2023}. The A, D and E Series were run in the periodic boxes with size of $(100\,h^{-1} \,\rm Mpc)^3$, $(400\,h^{-1} \,\rm Mpc)^3$ and $(762\,h^{-1} \,\rm Mpc)^3$, respectively. The model haloes of the A Series are selected from the snapshot of $z_{\rm final}{}={}1$, those of the D and E Series are selected from the snapshot of $z_{\rm final}{}={}6$. All the HR runs were run down to $z_{\rm final}$ except D7, where the HR run is evolved to only $z{}={}7.2$. This is because part of the ISM in D7 became too compact so that the gas particles with the highest densities were evolved at extremely small time-steps and it became infeasible to run the simulation down to the target redshift.

Initial conditions for the HR runs are set up using a convex hull surrounding all particles within $3R_{\rm vir}$ at $z_{\rm final}$ of the chosen halo defining the Lagrangian HR region following the method of \citet{Hahn_2011}. The mass resolutions and force softening lengths of the HR runs are similar to those of the {\sc\small FIREbox} simulation. Specifically, $m_{\rm DM}$ and $m_{\rm b}$ are set to $1.9 \times 10^5\,M_\odot$, $3.6\times10^4\,M_\odot$, respectively. Both $h_{\rm DM}$ and $h_*$ are fixed in proper (comoving) coordinates at $z{}\le{}9$ ($z{}\ge{}9$) and are set equal to 57 pc and 7 pc, respectively. $h_{\rm gas}$ is set equal to the smoothing length of the gas particles down to a minimum of 0.7 proper pc.

We include the central galaxy of the chosen haloes at $z_{\rm final}$ except for that of the D7 run. In addition, we also include the most massive progenitors (MMPs) of the central galaxies at higher redshifts. Specifically, for the 4 A Series runs, we include the MMPs at $z{}={}2$, $z{}={}3$ and $z{}={}4$, while for the D and E Series, we include the MMPs at $z{}={}8$. The galaxies are identified in the simulation snapshots using {\sc\small AHF} \citep{Gill_2004, Knollmann_2009}. In Table~\ref{T1}, we summarize the information\footnote{Physical properties, including \eg~$M_*$ , SFR, $L_{\rm IR}$ and $L_{\rm [C_{II}]}$, of the {\sc\small FIRE} galaxies reported in this paper are estimated using a radial kernel of $0.1R_{\rm vir}$ around the DM halo centre, \ie~{the maximum density centre provided by {\sc\small AHF}}.} of the 10 {\sc\small MassiveFIRE} simulations used for this study. \\

Both the {\sc\small MassiveFIRE} and {\sc\small FIREbox} simulations used in this work are run using the N-body+hydrodynamics code {\sc \small GIZMO} ({\sc\small FIRE}-2 version) in the Meshless-Finite-Mass (MFM) mode \citep{Hopkins_2018}. The simulations incorporate various gas cooling processes (free-free, photoionization/recombination, Compton, photoelectric, metal-line, molecular and fine structure processes) and a uniform UV background following the FG09 prescription \citep{Faucher_Giguere_2009}, Star formation occurs in dense, self-gravitating and self-shielding molecular gas based on a sink-particle prescription. The simulations explicitly incorporate several different stellar feedback channels (but not feedback from supermassive black holes) including 1) local and long-range momentum flux from radiative pressure, 2) energy, momentum, mass and metal injection from supernovae (Types Ia and II), 3) stellar mass loss (both OB and AGB stars) and 4) photo-ionization and photo-electric heating processes. We refer the reader to \citet{Hopkins_2014, Hopkins_2018} for details of the star formation and feedback prescriptions of {\sc\small FIRE}. 

{\sc\small FIRE} has demonstrated success at reproducing a variety of key galaxy properties that are relevant to this work, such as the stellar-to-halo mass relation \citep{Hopkins_2014, Feldmann_2017b}, the specific SFR (sSFR) of galaxies at the cosmic noon ($z{}\sim{}2$) \citep{Hopkins_2014, Feldmann_2016, Sparre_2017, Feldmann_2023}, the galaxy molecular (atomic) hydrogen gas mass and stellar mass relations at $z{}={}0$ \citep{Feldmann_2023}, the gas-phase and stellar mass-metallicity relation at $z{}={}0-2$ \citep{Ma_2016, Feldmann_2023}, the observational effective dust temperatures at $z{}={}2-4$ \citep{Liang_2019} as well as the UV luminosity functions and UV-based cosmic star formation rate density (CSFRD) at $z{}>{}5$ \citep{Ma_2019}.
 
\section{Simulating observational properties}
\label{Sec:3} 

In this section, we describe the method used to predict the observational properties for the {\sc\small FIRE} galaxy sample, which we compare to the observational data. In Section~\ref{Sec:3a}, we describe our $\rm [C_{II}]$ emission model. In Section~\ref{Sec:3b}, we describe the prescription for the dust RT modelling of the {\sc\small FIRE} galaxies using {\sc\small SKIRT} code, based on which we derive the multi-wavelength SED and {the distribution of the interstellar radiation field (ISRF)} for the galaxies. The ISRF distribution is essential for predicting the $\rm [C_{II}]$ emission properties of the galaxies.

\subsection{Predicting $\rm [C_{II}]$ emission using CLOUDY}
\label{Sec:3a}

We predict the $\rm [C_{\rm II}]$ line luminosity for the {\sc\small FIRE} sample using the spectral synthesis code {\sc \small CLOUDY} version 17.01 \citep{Ferland_2017}. {\sc \small CLOUDY} is a plasma simulation code designed to simulate the ionization, level populations, molecular state and thermal state of gas over a wide range of density and temperature in different astrophysical environments (\eg~black hole accretion disks, PDRs, molecular clouds, etc). It solves for the ionization structure for all stages of ionization for the lightest 30 elements \citep{Abel_2008}. 

We treat each gas particle of the galaxies as an idealized spherical uniform `gas cloud'. The $\rm [C_{II}]$ luminosity of each `cloud' is calculated based on its physical conditions, including `cloud' (or gas particle) mass ($M_{\rm cl}$), gas density\footnote{In this paper, `gas density' consistently refers to the number density of hydrogen nuclei ($n_{\rm H}$) in the gas, rather than mass density. {\sc CLOUDY} takes $n_{\rm H}$ as an input.} ($n_{\rm H}$), gas metallicity ($Z_{\rm gas}$), gas turbulent velocity dispersion ($\sigma$) and local UV ISRF strength ($G$\footnote{Conventionally, $G$ is used to denote the mean ISRF in the Habing band ($6.0-13.6$ eV). It is indicated in {units} of $G_0=1.6\times10^{-3}\,\rm erg\,s^{-1}\,cm^{-2}$, the observed value in the solar neighbourhood \citep{Habing_1968}.}). $M_{\rm cl}$, $n_{\rm H}$, $Z_{\rm gas}$ of each `cloud' are known directly from the {\sc\small FIRE} simulations. {$\sigma$ is the mass-weighted standard deviation of the velocities in gas at the location of the `cloud', which is calculated in post-processing.} Finally, $G$ at the location of each `cloud' in the galaxy is calculated using the dust RT code {\sc \small SKIRT} \citep{Baes_2011, Baes_2015, Camps_2015} in post-processing (see Section~\ref{Sec:3b} for the details). 

We calculate the $\rm [C_{II}]$ luminosity for each `cloud' ($L_{\rm [C_{II}],\,cl}$) by integrating the $\rm [C_{II}]$ line cooling rate, $\Lambda_{\rm [C_{II}]}$ ($\rm erg\,s^{-1}\,cm^{-3}$; see Appendix~\ref{Sec:Ap1} for its analytic expression), obtained from the output of the {\sc\small CLOUDY} simulations, over the volume of the `cloud'\footnote{{Note that we do not derive $L_{\rm [C_{II}],\,cl}$ using the `{\em emergent intensity}' ($I_{\rm em}$, with physical unit $\rm erg\,s^{-1}\,cm^{-2}$) output by {\sc\small CLOUDY} because $I_{\rm em}$ is calculated for a plane-parallel geometry instead of a spherical geometry. The conversion factor between the two geometries is not simply a constant but depends on the profile of $\rm [C_{II}]$ emissivity \citep{Olsen_2017, Olsen_2018}.}}: 
\begin{equation}
L_{\rm [C_{II}],\,cl} = 4\pi \int^{R_{\rm cl}}_0 \Lambda_{\rm [C_{II}]} (x)\;x^2 {\rm d}x,
\label{eq.1}
\end{equation}
\noindent where $R_{\rm cl}$ indicates the `radius' of the `cloud', {approximated by a Sobolev-like length scale ($L_{\rm sob}$) defined using local density gradients \citep{Sobolev_1957, Gnedin_2009}, \ie}
\begin{equation}
R_{\rm cl} \sim L_{\rm sob} \equiv \frac{\rho}{2 |\nabla \rho|}.
\label{eq.2}
\end{equation}
{This length scale was introduced by \citet{Gnedin_2011} to derive the {\em effective} column densities of the `clouds' for determining their $\rm H_2$ abundances, knowing that small-scale star-forming molecular clumps are typically unresolved by galaxy-scale simulations.} We then calculate the $\rm [C_{II}]$ luminosity of the galaxy ($L_{\rm [C_{II}]}$) by summing over $L_{\rm [C_{II}],\,cl}$ of all gas `clouds' calculated using equation~(\ref{eq.1}). We treat the $\rm [C_{II}]$ emission of our galaxy sample as being optically thin.

In practice, to run {\sc \small CLOUDY} simulations for every gas particle for the whole {\sc \small FIRE} sample ($>{}400$ galaxies in total) is computationally formidable: a {\sc \small CLOUDY} simulation is typically completed (\ie~when iterative convergence is reached) in $0.1{}-{}0.5$ CPU hour, depending on the gas column density, and hence to analyze one single galaxy snapshot that contains $\sim{}1$ million gas particles would cost $100{}-{}500$ K CPU hours in total. We therefore use a lookup-table method similar to the previous studies \citep[\eg][]{Vallini_2015, Vallini_2018, Vallini_2021, Katz_2017, Katz_2019, Olsen_2017, Lagache_2018, Li_2018, Pallottini_2019, Keating_2020, Leung_2020, Lupi_2020a, Yang_2021, Lupi_2020b}. Specifically, for each of the 7 snapshots, we build a grid of {\sc\small CLOUDY} models that covers a gas density range $-1{}<{}{\rm log}\,(n_{\rm H}/{\rm cm^{-3}}){}<{}5$, a gas metallicity range $-2{}<{}{\rm log}\,(Z_{\rm gas}/Z_\odot){}<{}0.8$, a turbulent velocity dispersion range $0{}<{}{\rm log}\,(\sigma/\rm km\,s^{-1}){}<{}2.4$ and a UV ISRF range $-1{}<{}{\rm log}\,(G/G_0){}<{}4$. The grid spacing is set 0.5 dex for $n_{\rm H}$ and $G$, and 0.4 dex for $Z_{\rm gas}$ and $\sigma$. In total, the default look-up table that we use for calculating the $\rm [C_{II}]$ luminosity of our galaxy sample consists of 8,008 ($13{}\times8{}\times{}7{}\times{}11$) models for each redshift. We include the CMB background in the {\sc\small CLOUDY} simulations for each redshift and the predicted $\rm [C_{II}]$ luminosity is corrected for the CMB attenuation effect \citep{daCunha_2013}. Cosmic-ray (CR) hydrogen ionization rate in these models is fixed to the fiducial value of $2{}\times{}10^{-16}\,\rm s^{-1}$, the observed value in the Milky Way \citep{Indriolo_2007, Indriolo_2012, Neufeld_2017}. {We assume a constant dust-to-metal mass ratio $\delta_{\rm dzr}{}={}0.4$ \citep{Dwek_1998, Draine_2007, Watson_2011, Li_2019} and adopt the default interstellar medium metal abundances ({\sc\small abundance ISM}) stored in {\sc\small CLOUDY}.} The simulations are run till sufficiently large distance from the surface of the slab is reached\footnote{{The {\sc thickness} of the slab is specified as a stopping criterion and is set at 400 pc in all our models, which is typically much larger than $R_{\rm cl}$ of the gas `clouds' (defined using equation~\ref{eq.2}).}}. Given $n_{\rm H}$\footnote{{We calculate $n_{\rm H}$ for each individual `cloud' using the values of $M_{\rm cl}$ (cloud mass) and $R_{\rm cl}$ as defined in equation~(\ref{eq.2}): $n_{\rm H}{}={}(3M_{\rm cl}){}/{}(4\pi R^3_{\rm cl} \mu_{\rm H} m_{\rm H})$, where $m_{\rm H}$ represents the proton mass, and $\mu_{\rm H}$ represents the mean molecular weight of the gas.}}, $Z_{\rm gas}$, $G$ and $N_{\rm H}$ of each `cloud', we interpolate $\rm [C_{II}]$ luminosity of the `cloud' from the values found in the computed grid.

The treatment of the ISM as an aggregate of spherical gas `clouds' in our model (and in the models of the previous theoretical studies mentioned above) is undoubtedly an idealization, since the ISM in real galaxies is a continuous medium with complex spatial configurations at and below the scale of these idealized `clouds'. Nevertheless, this approach allows us to crudely sample the surface densities of gas within the ISM, thereby enabling us to capture the essential physics responsible for the observed $\rm [C_{II}]$ deficit in galaxies.

\paragraph*{CLOUDY simulation: an example\\}

\begin{figure*}
 \includegraphics[height=168mm]{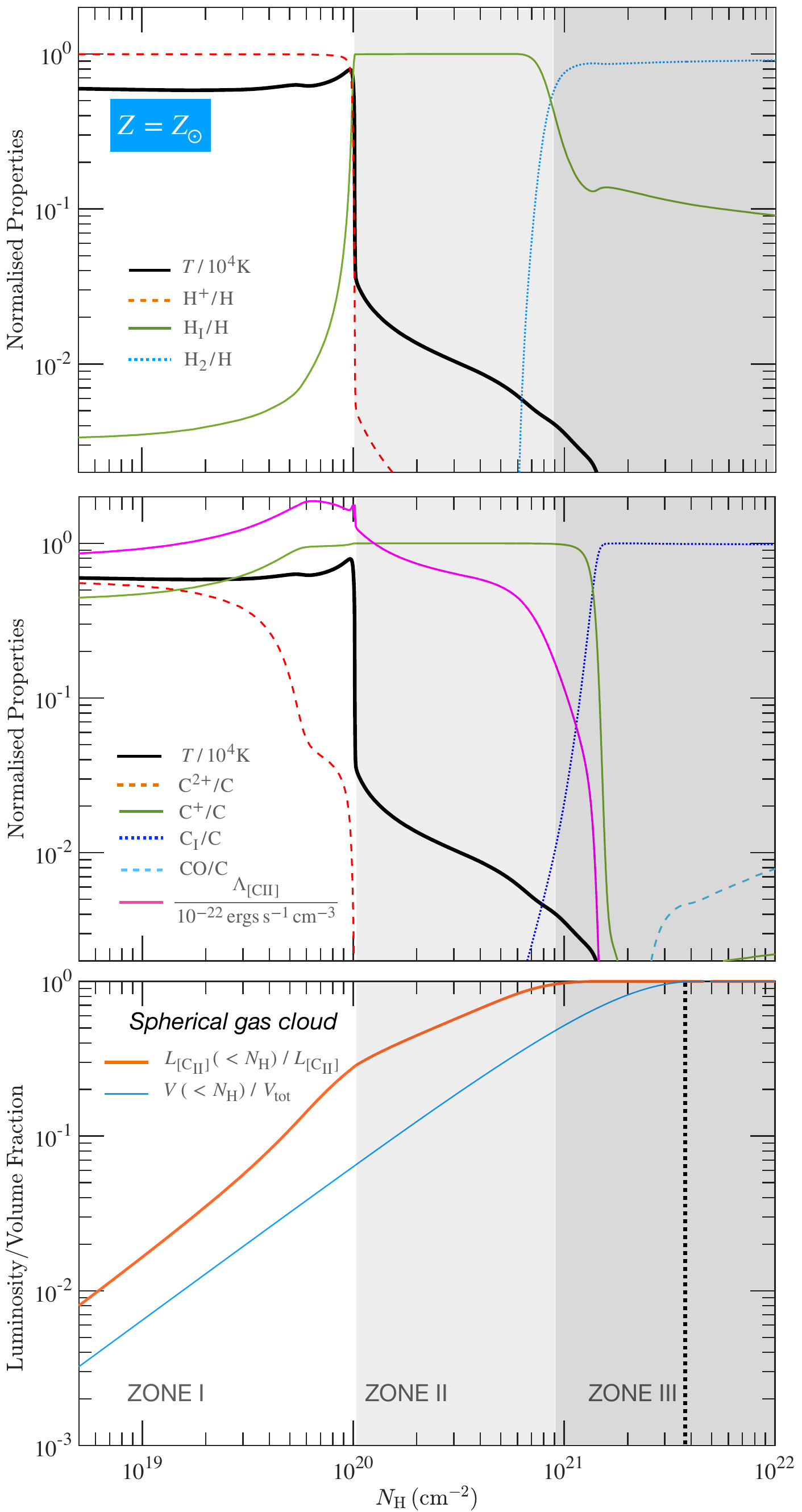}
 \includegraphics[height=168mm]{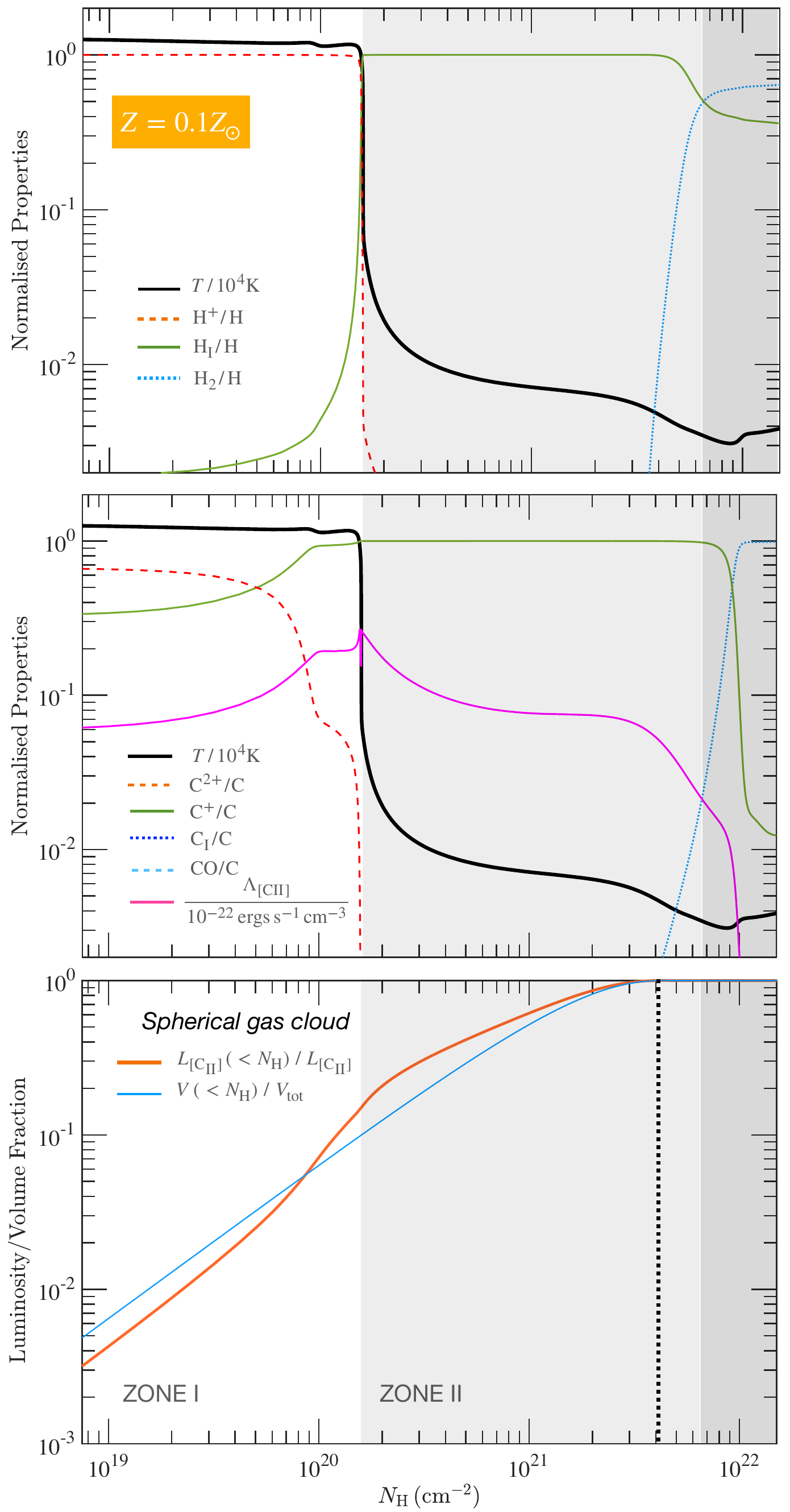}
 \caption{{\it Top} and {\it middle} panels: Ionization structures of a plane-parallel gas slab ($n_{\rm H}=50\,\rm cm^{-3}$) irradiated by an external radiation field ($G=200\,G_0$) incident from the left in the figure predicted by the {\sc\small CLOUDY} code. Dashed red, solid green and dotted blue lines in the {\it top} ({\it middle}) panels represent the abundance profiles for $\rm H^+$ ({$\rm C^{2+}$}), $\rm H_I$ ({$\rm C^+$}) and $\rm H_2$ ($\rm C_I$), respectively. 
 {Dashed cyan line in the {\it middle} panels represents the abundance profile for $\rm CO$}. Solid black line in the {\it top} and {\it middle} panels shows the profile of gas kinetic temperature (normalized by $10^4$ K). Solid magenta line in the \textit{middle} panels indicates the profile of $\rm [C_{II}]$ cooling rate (normalized by $10^{-22}\,\rm erg\,s^{-1}\,cm^{-3}$). {\it Bottom} panels: Cumulative fraction of $\rm [C_{II}]$ luminosity (thick orange line) and volume (thin blue) as a function of gas column density (from the surface) of a spherical gas cloud ($M_{\rm cl}=10^5\,M_\odot$, $n_{\rm H}=50\,\rm cm^{-3}$) irradiated by an external radiation field ($G{}={}200\,G_0$). {Black} dotted line marks the surface-to-centre column density of the cloud ($N_{\rm H}=4\times10^{21}\,\rm cm^{-2}$). The {\it left} and {\it right} columns correspond to the metal-rich and metal-poor models where gas metallicity of the slab (cloud) is set to $Z_\odot$ and $1/10\,Z_\odot$. For the metal-poor model, the dust-to-gas mass ratio ($\delta_{\rm dgr}$) becomes lower and therefore Lyman-Werner photons can penetrate deeper into the slab (cloud), resulting in larger $\rm [C_{II}]$-emitting region (Zone I + Zone II).}
\label{fig.2}
\end{figure*}

Here we show the conditions of a plane-parallel gas slab calculated by {\sc\small CLOUDY} (Fig.~\ref{fig.2}). The slab has a uniform gas density $n_{\rm H}{}={}50\,\rm cm^{-3}$ and is illuminated by an external radiation field having $G{}={}200\,G_0$. We present {\sc\small CLOUDY} simulations for two different models, where $Z_{\rm gas}$ is set to $Z_\odot$ and $1/10\,Z_\odot$. {We include the $z{}={}0$ CMB background and the CR hydrogen ionization rate is set to the default value.} We show the results of the dust-rich and dust-poor models in the {\it left} and {\it right} panels of Fig.~\ref{fig.2}, respectively. 

The slab is characterized by three distinct zones based on the ionization state of hydrogen gas. In the {\it upper} panels, we show the abundance profiles for ionized hydrogen ($\rm H^+$; dashed red line), atomic hydrogen ($\rm H_I$; solid green line) and molecular hydrogen ($\rm H_2$; dotted blue line), as well as the profile for gas temperature (solid black line). We can see that a $\rm H^+$ region (Zone I) is created near the surface of the slab by the ionizing photons ($E_\gamma{}>{}13.6$ eV) of the incident radiation field. Gas in this region is heated to high temperature ($T{}\approx{}10^4$ K). The slab then transits to a $\rm H_I$-dominated region (Zone II) at a distance where ionizing radiation gets fully absorbed. The photons in the Lyman-Werner (LW) band ($11.2{}<{}E_\gamma{}<{}13.6$ eV) dissociate $\rm H_2$ in this region, while maintaining gas temperature at about $10^2$ K. Finally, the slab transits to a $\rm H_2$-dominated region (Zone III) at some larger distance, beyond which the LW radiation becomes sufficiently absorbed and the majority of hydrogen turns into $\rm H_2$. 

Like hydrogen, carbon has a very different ionization state in the three zones. This can be seen from the {\it middle} panels of Fig.~\ref{fig.2}, where we explicitly show the abundance profiles for atomic carbon ($\rm C_I$; dotted blue line), singly ionized carbon ({$\rm C^+$}; solid green line) and doubly ionized carbon ({$\rm C^{2+}$}; dashed red line) for the two models. Carbon is mostly ionized in Zone I and II. Specifically, in Zone I, it gets excited to {$\rm C^+$} level as well as higher ionization levels (\eg~{$\rm C^{2+}$}). In Zone II, on the contrary, carbon is singly ionized by LW photons but not excited to higher levels since ionizing photons are shielded from the region\footnote{The ionization energy of {$\rm C^{2+}$} is 24.39 eV, which is greater than the ionization energy of hydrogen atom (13.6 eV).}. Finally, in Zone III, carbon turns into $\rm C_I$ and CO since the region is UV-dark\footnote{{The first ionization energy of carbon is 11.26 eV, which coincides with the lower frequency limit of the LW band (11.2 eV). Consequently, the transition from the $\rm H_I$ to the $\rm H_2$ regions should align with the shift from $\rm C^+$-rich to $\rm C_I$-rich regions when neglecting self-shielding of $\rm H_2$ from LW radiation \citep{Stecher_1967, Black_1977, Federman_1979, van_Dishoeck_1986}. It is worth noting that $\rm H_2$ self-shielding can be significant in high column density and low-metallicity environments \citep{Draine_1996, Madden_1997, Madden_2020, Wolfire_2010, Gnedin_2014}. In these environments, a substantial amount of {$\rm C^+$} can be found within the envelope of the $\rm H_2$ regions (Zone III). This has motivated some studies that suggest using $\rm [C_{II}]$ as a tracer for CO-dark $\rm H_2$ gas \citep[\eg][]{Madden_1997, Madden_2020, Langer_2010, Langer_2014, Velusamy_2010, Pineda_2013, Pineda_2014, Requena_Torres_2016, Li_2018, Dessauges_Zavadsky_2020, Vizgan_2022}. However, in our simulations, we find that only a small fraction ($<10\%$) of the $\rm [C_{II}]$ emission from our galaxies originates from the $\rm H_2$-dominated regions (see Section~\ref{Sec:5b}). Therefore, we have not explicitly incorporated an additional zone in our model that is both $\rm H_2$ and $\rm C^+$-rich, which would be situated between the current Zone II and Zone III.}}. 

$\rm [C_{II}]$ emission originates mostly from the ionized (Zone I) and atomic hydrogen (Zone II) phases {in our models}. We show in the {\it middle} panels the profile for $\rm [C_{II}]$ cooling rate ($\rm erg\,s^{-1}\,cm^{-3}$), $\Lambda_{\rm [C_{II}]}$, for the two models (solid magenta line). It is clear that $\Lambda_{\rm [C_{II}]}$ drops sharply in Zone III, which is due to the low abundance of {$\rm C^+$} ions (solid green line) in this region (note: most carbon is in neutral state in Zone III). For the chosen models, $\Lambda_{\rm [C_{II}]}$ appears to be similar in the ionized and atomic hydrogen phases, varying by less than a factor of few. Comparing the metal-rich ({\it left} panel) and metal-poor ({\it right} panel) models, it can be seen that $\Lambda_{\rm [C_{II}]}$ of the metal-rich model is about a factor of ten higher. This is due to the fact that $\Lambda_{\rm [C_{II}]}$ is linearly scaled to $Z_{\rm gas}$ and $Z_{\rm gas}$ of the metal-rich model is set as ten times that of the metal-poor model.

Using the $\Lambda_{\rm [C_{II}]}$ profile output by {\sc\small CLOUDY}, we subsequently derive the $\rm [C_{II}]$ luminosity profile (cumulative $\rm [C_{II}]$ luminosity as a function of column depth from the surface) for a uniform spherical cloud having $n_{\rm H}{}={}50\,\rm cm^{-3}$ (same as the gas slab) and $M_{\rm cl}{}={}10^5\,M_\odot$ that is irradiated by an external field having $G{}={}200\,G_0$ (same as the gas slab) following equation~(\ref{eq.1}). We calculate the result for the metal-rich ($Z_{\rm gas}{}={}Z_\odot$) and metal-poor ($Z_{\rm gas}{}={}0.1Z_\odot$) models, which are shown in the {\it lower left} and {\it lower right} panels of the figure, respectively. It can be seen that about $30\%$ ($10\%$) of the total $\rm [C_{II}]$ luminosity of the cloud is produced by the $\rm H^+$ region for the metal-rich (poor) model, while the remainder originates almost totally from the $\rm H_I$ region. The $\rm H_2$ region contributes very limited fraction of the $\rm [C_{II}]$ luminosity. Note that the $\Lambda_{\rm [C_{II}]}$ profile, the size of the different zones, and their relatively contribution to the total $\rm [C_{II}]$ luminosity of the cloud depends on $G$, $n_{\rm H}$ and $Z_{\rm gas}$ (see Section~\ref{Sec:5a} for a detailed discussion).

One major difference between the two models (metal-rich vs. metal-poor) is whether or not the gas cloud has an $\rm H_2$ region in the core, as can be seen from the {\it bottom} panels. For the metal-poor model ({\it bottom right} panel), because dust column density is small, LW photons are able to penetrate the entire cloud, making it $\rm H_2$-free. The metal-rich model ({\it bottom left} panel), in contrast, has an $\rm H_2$ core owing to the high dust column density, which accounts for roughly half of $M_{\rm cl}$. The two cloud models correspond to the two distinct regimes where $L_{\rm [C_{II}]\,,cl}$ has different scaling with $Z_{\rm gas}$. When the cloud has no $\rm H_2$ core, $L_{\rm [C_{II}]\,,cl}$ scales linearly with $Z_{\rm gas}$. As $Z_{\rm gas}$ (and hence the dust-to-gas mass ratio, $\delta_{\rm dgr}$) increases, the depth of Zone I+Zone II decreases \citep{Ferrara_2019}. When $Z_{\rm gas}$ is high enough that $\rm H_2$ becomes abundant (\ie~, Zone III forms) in the core, $L_{\rm [C_{II}]\,,cl}$ saturates and no longer depends sensitively on $Z_{\rm gas}$. In Section~\ref{Sec:5}, we will discuss in detail how the $L_{\rm [C_{II}]}{}/{}\rm SFR$ ratio of the {\sc\small FIRE} galaxies depends on gas metallicity, and interpret the results using the insights obtained from the toy models presented here.

\begin{figure*}
 \includegraphics[width=176mm]{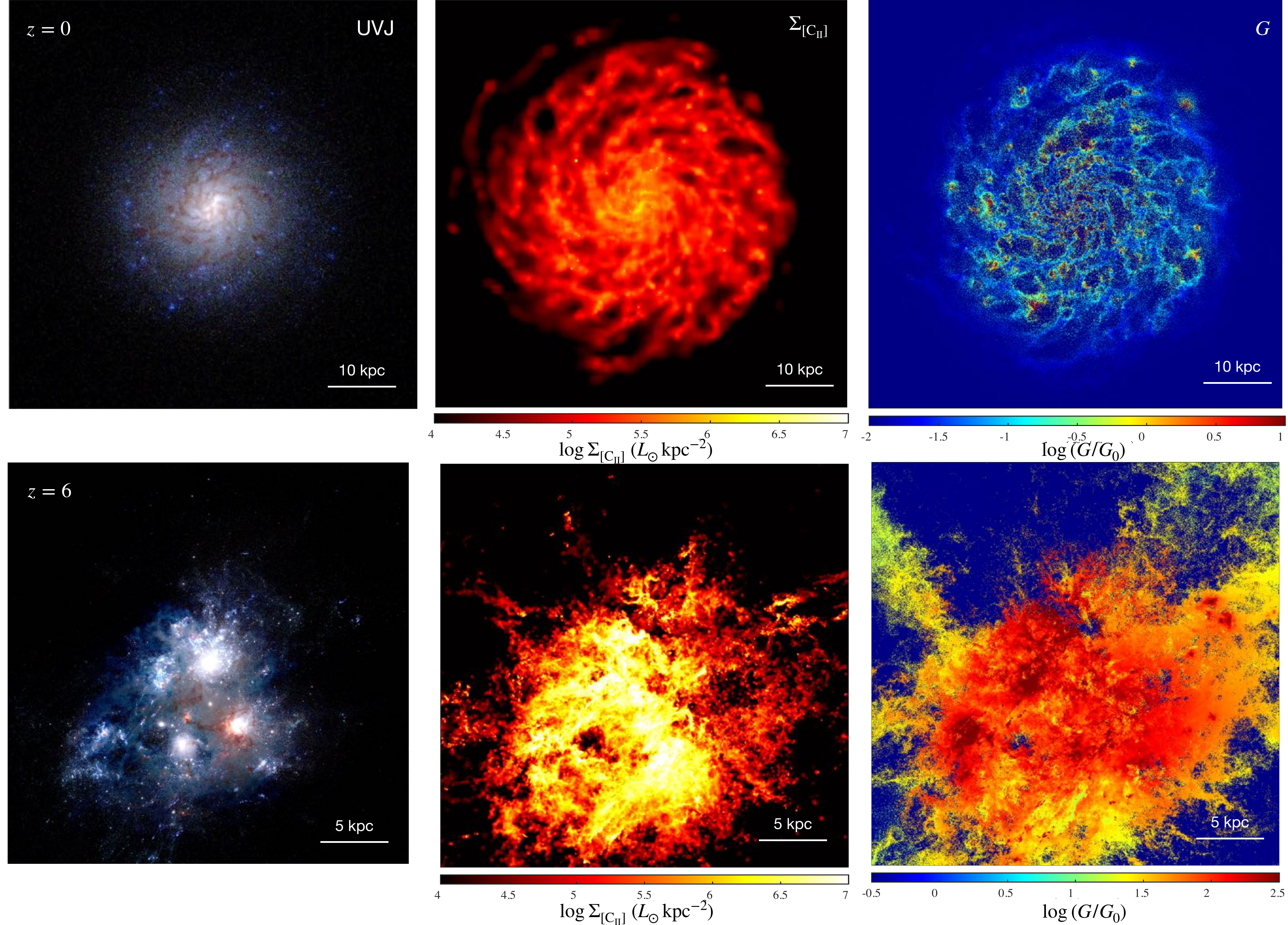}
 \caption{The UVJ false-colour image ({\it left}), ${\rm [C_{II}]}$ surface brightness ({\it middle}) and the distribution of UV ISRF strength ($G$) ({\it right}) of selected {\sc\small FIRE} galaxies. The {\it upper} panels show the results of a $z{}={}0$ disc galaxy from {\sc\small FIREbox} {(\cf~Fig.3 of \citealt{Feldmann_2023})}, while the {\it lower} panels correspond to {a galaxy undergoing multiple mergers at $z{}={}6$} extracted from the {\sc\small MassiveFIRE} `zoom-in' suite.}
    \label{fig.3}
\end{figure*}

\subsection{Calculating ISRF distribution and multi-wavelength SEDs of galaxies using SKIRT}
\label{Sec:3b}

To predict the $\rm [C_{II}]$ luminosity of the ISM, it is essential to know the local UV ISRF strength. We calculate the ISRF distribution for the {\sc\small FIRE} galaxies using the open-source\footnote{Code repository: \url{https://skirt.ugent.be/version8/}} 3D Monte Carlo dust RT code {\sc\small SKIRT} \citep{Baes_2011, Baes_2015, Camps_2015} (version 8).  {\sc\small  SKIRT} provides full treatment of absorption and anisotropic scattering by dust, and self-consistently computes dust thermal re-emission and dust temperature distribution for various astrophysical systems. 

To prepare the galaxy snapshots as RT input models for {\sc\small SKIRT}, we follow the prescription of \citet{Camps_2016} (see also \citealt{Trayford_2017, Camps_2018}). We summarize the key points of the prescription here, and refer interested readers to the above-mentioned papers for the details.

For the analysis, each star particle of the galaxy is treated as a `single stellar population' (SSP), and a spectrum of stellar emission is assigned to each particle using the {\sc\small STARBURST99} \citep{Leitherer_1999, Vazquez_2005} SED libraries according to the age, metallicity and initial mass of the particle. The RT calculations are performed on an equally spaced logarithmic wavelength grid consisting of 250 wavelength points spanning the wavelength range $\lambda{}={}0.01-1000 \,\rm \mu m$. We launch $10^6$ photon packages for each of the 250 point in the wavelength grid and for each of the stellar emission and following dust emission stages. The calculation iterates until convergence. To produce mock images and SEDs for the galaxies, we place mock detectors at an arbitrary ‘local’ distance of 10 Mpc from galaxy along multiple viewing angles to accumulate both spatially resolved as well as integrated fluxes at each wavelength grid point.

We assume that dust mass traces metal mass in galaxies \citep{Hayward_2011, Narayanan_2015, Camps_2016, Trayford_2017, Liang_2018, Liang_2019, Liang_2021, Ma_2019, Cochrane_2019, Cochrane_2022, Vogelsberger_2020, Shen_2022} and adopt a constant dust-to-metal mass ratio $\delta_{\rm dzr}=0.4$ in gas cooler than $10^6$ K. Hotter gas is assumed to be dust-free due to thermal sputtering \citep{Draine_1979, Tielens_1994}. We adopt the \citet{Weingartner_2001b} dust model with Milky-Way size distribution for the case of $R_{\rm V}=3.1$. We discretize the spatial domain using an octree grid and keep subdividing grid cells until the cell contains less than $f{}={}3\times10^{-6}$ of the total dust mass and the $V$-band (0.55 $\rm \mu m$) optical depth in each cell is less than unity. The highest grid level corresponds to a cell width of $\sim20$ pc, \ie~about twice the minimal SPH smoothing length. We self-consistently calculate the self-absorption of dust emission and include the transient heating function to calculate non-local thermal equilibrium dust emission by transiently heated small grains and PAH molecules \citep{Baes_2011, Camps_2015}. To account for the heating of dust by the cosmic microwave background, we adopt a correction to the dust temperature using equation (12) of \citet{daCunha_2013}. 

The final output of the {\sc \small SKIRT} simulations includes the ISRF, $J_\lambda$ ($\rm W\,cm^{-3}\,sr^{-1}$), of each adaptive grid cell. We calculate the UV ISRF strength ($G$) for each cell by integrating $J_\lambda$ over the Habing band ($6-13.6$ eV) and solid angle ($\Omega$). $G$ is assigned to every {gas particle (`cloud')} inside the cell for predicting its $\rm [C_{II}]$ luminosity. 

In Fig.~\ref{fig.3}, we show the UVJ composite image ({\it left} panels), $\rm [C_{II}]$ surface brightness ({\it middle} panels), and $G$ distribution ({\it right} panels) for the {two} selected {\sc\small FIRE} galaxies calculated using {\sc\small CLOUDY} and {\sc\small SKIRT}. The {\it upper} panels show the results of a disc galaxy at $z{}={}0$ extracted from {\sc\small FIREbox}, whilst the {\it lower} panels show the results of {a galaxy undergoing multiple mergers at $z{}={}6$} extracted from the {\sc\small MassiveFIRE} simulation (Sim ID: {D9}). The $z{}={}6$ {galaxy system} has much stronger strength of ISRF ({\it right} panels) due to its higher SFR ($220\,M_\odot\,\rm yr^{-1}$ vs. $4.5\,M_\odot\rm \,yr^{-1}$) and shows higher $\rm [C_{II}]$ surface brightness. $L_{\rm [C_{II}]}$ of the $z{}={}6$ {system} and the $z{}={}0$ galaxy are {$5.5\times10^8\,L_\odot$ and $1.0\times10^8\,L_\odot$}, respectively. 

\section{Comparison with observations}
\label{Sec:4}

In this section, we compare the $L_{\rm [C_{II}]}$-SFR relation of the {\sc\small FIRE} galaxies predicted by our model with the observational data at various redshifts. We separately discuss the results for three redshift regimes, $z{}={}0$ (Section~\ref{Sec:4a}), $1{}\simless{}z{}\simless{}5$ (Section~\ref{Sec:4b}) and $z{}\simgreat{}5$ (Section~\ref{Sec:4c}). We make this distinction because observations use different sample selection methods and the SFR of galaxies is estimated by different means of calibration in the three different regimes. 

\subsection{Local Universe (redshift $z{}={}0$)}
\label{Sec:4a}

Observations of the $L_{\rm [C_{II}]}$-SFR relation at $z{}={}0$ probe a very wide SFR range across several orders of magnitude. The selected samples include low-SFR systems such as dwarf galaxies as well as the extreme IR-luminous starbursts.

Three primary samples of nearby galaxies have been employed to calibrate the relation between $L_{\rm [C_{II}]}$ and the SFR in normal star-forming galaxies (${\rm SFR}{}\approx{}10^{-5}-10\,M_\odot \rm yr^{-1}$): \citet[][hereafter referred to as L11]{De_Looze_2011}, \citet[][hereafter referred to as L14]{De_Looze_2014}, and \citet[][hereafter referred to as H15]{Herrera_Camus_2015}. These studies have consistently found a linear correlation between $L_{\rm [C_{II}]}$ and SFR, and their calibrations are often used as benchmarks for high-redshift observations (galaxies below their $L_{\rm [C_{II}]}$-SFR relation are considered to have a `$\rm [C_{II}]$ deficit'). However, it's important to note that other evidence suggests this linear correlation can break down at high SFR at $z{}={}0$ \citep[\eg][]{Diaz_Santos_2013, Diaz_Santos_2017, Herrera_Camus_2018}, and whether we should use these relations as a ‘standard ruler’ is highly doubtful.

The \citetalias{De_Looze_2011} sample consists of 24 star-forming galaxies selected from the early compilation by \citet{Brauher_2008} that have measurements at both the \textit{Galaxy Evolution Explorer} ({\sc \small GALEX}) FUV and the \textit{Multiband Imaging Photometer for Spitzer} ({\sc\small MIPS}) $\rm 24\,\mu m$ bands. The sample of  \citetalias{De_Looze_2014} includes 48 nearby low-metallicity ($Z_{\rm gas}\approx0.03-0.55\,Z_\odot$) dwarf galaxies extracted from the Dwarf Galaxy Survey \citep[DGS,][]{Madden_2013} catalogue. Lastly, \citetalias{Herrera_Camus_2015} study a sample consisting of 46 local star-forming galaxies chosen from the {\sc\small KINGFISH} catalogue \citep{Kennicutt_2011}, having very diverse integrated galaxy properties and ISM environments. All these studies have excluded the sources with AGN features. 

\begin{figure*}
 \includegraphics[width=175mm]{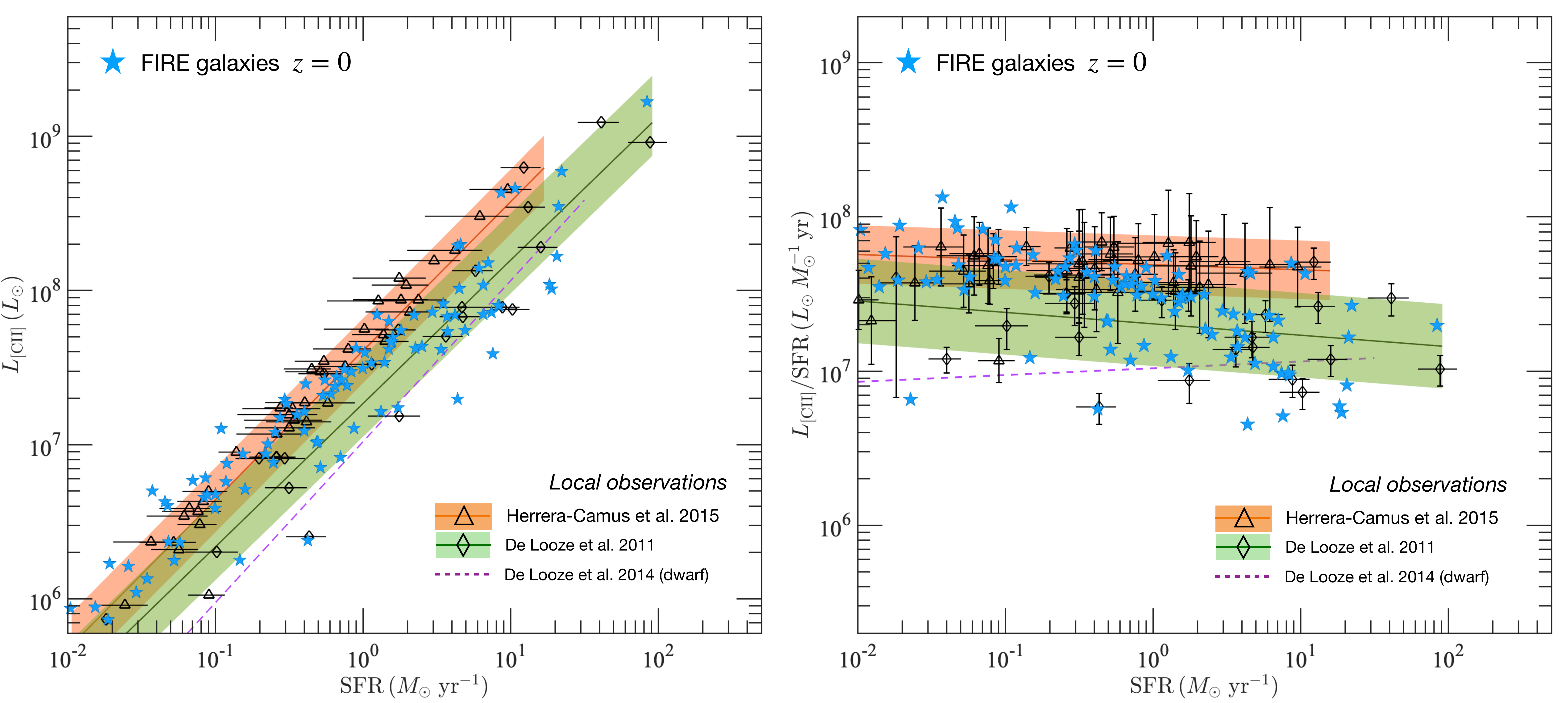}
 \caption{The {$L_{\rm [C_{II}]}$ vs. SFR} (\textit{left} panel) and {$L_{\rm [C_{II}]}{}/{}\rm SFR$ vs. SFR} (\textit{right} panel) relations of the $z{}={}0$ galaxies. The filled cyan stars in the two panels show the result of the {\sc\small FIRE} galaxies. Black triangles and diamonds show the observational data of \citet{Herrera_Camus_2015} (\citetalias{Herrera_Camus_2015})  and \citet{De_Looze_2011} (\citetalias{De_Looze_2011}), and the orange and green lines indicate the best-fit linear relation of the \citetalias{Herrera_Camus_2015} and \citetalias{De_Looze_2011} samples, respectively. The coloured shaded regions indicate the $1\sigma$ scatter of the data around the best-fit linear relation of the observed samples. Purple dashed line in the two panels represents the best-fit linear relation to the low-metallicity dwarf galaxy sample of \citet{De_Looze_2014} (\citetalias{De_Looze_2014}). The result of the {\sc\small FIRE} galaxies at $z{}={}0$ is in good agreement {with the observational data}. }
    \label{fig.4}
\end{figure*}

\begin{table*}
\caption{Observed {and simulated} scaling relations between SFR and $L_{\rm [C_{II}]}$ of local galaxies, \ie~$L_{\rm [C_{II}]} {}/{} L_\odot{}={}A{}({\rm SFR}{}/{}M_\odot\,{\rm yr}^{-1})^B$.}
\begin{threeparttable}
\begin{tabular}{ p{3.1 cm} p{2.7 cm}   p{2.7 cm}  p{2 cm} p{2 cm}  p{2 cm} }
 \hline
\multicolumn{1} {}{} \;\;\; Galaxy sample\;\;\; &  SFR range ($M_\odot\,\rm yr^{-1}$) & Median SFR ($M_\odot\,\rm yr^{-1}$) & \multicolumn{1}{c|}{$A$} & \multicolumn{1}{c|}{$B$}  & \multicolumn{1}{c|}{$1\sigma$ scatter} \\
 \hline
  \citet{De_Looze_2011} & \multicolumn{1}{c|}{$0.02-88$} & \multicolumn{1}{c|}{1.75} & $7.31\pm0.06$ & $0.93\pm0.06$ & \multicolumn{1}{c|}{0.26  dex} \\
  \hline
  \citet{De_Looze_2014} &  \multicolumn{1}{c|}{$6\times10^{-4}-56$} & \multicolumn{1}{c|}{0.12} & $7.10\pm0.11$ & $1.05\pm0.07$ & \multicolumn{1}{c|}{0.43  dex} \\
 \hline
  \citet{Herrera_Camus_2015} &  \multicolumn{1}{c|}{$10^{-3}-9.6$} & \multicolumn{1}{c|}{0.34} & $7.63\pm0.03$ & $0.97\pm0.03$ & \multicolumn{1}{c|}{0.21 dex} \\
  \hline
 \multicolumn{1}{c|}{{{\sc\small FIRE} (this work)}} & \multicolumn{1}{c|}{$0.01-1$$\dagger$} & \multicolumn{1}{c|}{0.19} & $7.48\pm0.06$  & $0.87\pm0.06$ &  \multicolumn{1}{c|}{0.27 dex} \\
  \hline
\end{tabular}
  \begin{tablenotes}
  \item[$\dagger$] Here we do not include the galaxies in the sample having ${\rm SFR}{}>{}1\,M_\odot\,\rm yr^{-1}$ for the fitting because they exhibit a reduced $L_{\rm [C_{II}]}{}/{}\rm SFR$ ratio (a $\rm [C_{II}]$ deficit).
  \end{tablenotes}
\end{threeparttable}
\label{T2}
\end{table*}

Both \citetalias{De_Looze_2011} and \citetalias{De_Looze_2014} derive the SFR of their sample using {\sc\small GALEX} FUV and {\sc\small MIPS} $24\,\rm \mu m$ fluxes (\ie~${\rm SFR}= \beta\,(L_{\rm FUV,\,obs}+\alpha\times L_{\rm 24\,\mu m})$) but with different calibration.  Specifically, \citetalias{De_Looze_2011} and \citetalias{De_Looze_2014} use the calibration by \citet{Zhu_2008} ($\alpha{}={}6.31$) and \citet{Hao_2011} ($\alpha{}={}3.89$), respectively. \citetalias{Herrera_Camus_2015}, on the other hand, derive the SFR of their sample using a hybrid of different methods: for 27 galaxies in their sample, SFR is derived using the $\rm H_\alpha$+$24\,\rm \mu m$ calibration by \citet[]{Calzetti_2007} (equation 7). For the other 8 galaxies, they use the FUV+$24\,\rm \mu m$ calibration by \citet{Leroy_2008} (equations D10 and D11). And lastly, for the remaining 11 galaxies having no measurement of either $\rm H_\alpha$ nor FUV flux, SFR is derived based solely on their $24\,\rm \mu m$ flux using the calibration by \citet{Calzetti_2007} (equation 6). In Table~\ref{T2}, we show the SFR range as well as the median SFR of the three samples (\citetalias{De_Looze_2011}, \citetalias{De_Looze_2014} and \citetalias{Herrera_Camus_2015}). We also show in the table the best-fit parameter values for the scaling relation
\begin{equation}
{\rm log}{}(L_{\rm [C_{II}]} {}/{} L_\odot){}={}A+B \;{\rm log}{}({\rm SFR}{}/{}M_\odot\,\rm yr^{-1})
\label{eq.3}
\end{equation}
 for the three samples as well as the $1\sigma$ scatter (in dex) of the data around the best-fit relation. Note that for the galaxies of the \citetalias{De_Looze_2011} and \citetalias{Herrera_Camus_2015} samples whose SFR is derived using the FUV+$24\,\rm \mu m$ fluxes, we have re-calibrated their SFR following \citet{Hao_2011} as has been done by \citetalias{De_Looze_2014} for a fair comparison. All the SFR calibrations are based on the \citet{Kroupa_2002} initial mass function (IMF). 
 
From Table~\ref{T2}, we can see that the three samples all exhibit an almost linear correlation between $L_{\rm [C_{II}]}$ and SFR, though having noticeable difference in the normalization. The \citetalias{Herrera_Camus_2015} sample has the highest normalization among the three samples. It is higher than that of the \citetalias{De_Looze_2011} sample by 0.32 dex. This offset may partly be due to the difference in sample selection. Another potential cause is that \citetalias{Herrera_Camus_2015} adopt different SFR indicators and calibration methods compared with \citetalias{De_Looze_2011} for a large fraction of the galaxies in their sample. The offset between the \citetalias{De_Looze_2011} and \citetalias{De_Looze_2014} samples (0.21 dex), on the other hand, is mainly due to the difference in {\em sample selection} since \citetalias{De_Looze_2011} and \citetalias{De_Looze_2014} adopt the same SFR indicators (FUV+$\rm 24\,\mu m$ fluxes) for their entire samples and we have re-calibrated their results following the same method of \citet{Hao_2011}. The lower normalization of the \citetalias{De_Looze_2014} relation is very likely due to the relatively lower $Z_{\rm gas}$ of the dwarf galaxies they use for the study, as has been explicitly stated in L14. 

In Fig.~\ref{fig.4}, we show the $L_{\rm [C_{II}]}$-SFR relation of the three samples (\citetalias{De_Looze_2011},  \citetalias{De_Looze_2014} and \citetalias{Herrera_Camus_2015})  in the {\it left} panel. To more clearly show the difference in the normalization of these scaling relations, we present the {$L_{\rm [C_{II}]}{}/{}\rm SFR$ vs. SFR} relation of the same samples in the {\it right} panel. In both panels, we also present the results for the {\sc\small FIRE} sample\footnote{We calculate the SFR of the {\sc\small FIRE} galaxies by averaging over a timescale of the last 100 Myrs.} $z{}={}0$ (filled cyan stars) for comparison with the observational data. Note that for the \citetalias{De_Looze_2011} and \citetalias{Herrera_Camus_2015} samples, we show both the data of the individual sources as well as the best-fit scaling relation for each sample, whereas for the \citetalias{De_Looze_2014} sample, we only present the best-fit scaling relation (purple dashed line) for reference. The \citetalias{De_Looze_2014} sample has systematically lower gas metallicity than the other two observational samples as well as the {\sc\small FIRE} galaxy sample at $z{}={}0$. 

The {\sc\small FIRE} simulations, combined with our line model, produce the $L_{\rm [C_{II}]}$-SFR relation at $z{}={}0$ (cyan stars) that is in good agreement with the local star-forming samples of \citetalias{De_Looze_2011} (black diamonds) and \citetalias{Herrera_Camus_2015} (black triangles). The best-fit parameter values for the {\sc\small FIRE} galaxies over the SFR range of $0.01-1\,M_\odot \,\rm yr^{-1}$ are $A{}={}7.48\pm 0.06$ and $B{}={}0.87\pm0.06$, and the $1\sigma$ scatter of the data points around the best-fit relation is 0.27 dex, similar to the \citetalias{De_Looze_2011} and \citetalias{Herrera_Camus_2015} samples (see Table~\ref{T2}). When including galaxies with ${\rm SFR}{}>{}1\,M_\odot \,\rm yr^{-1}$, the best-fit parameters become $A{}={}7.42\pm 0.03$ and $B{}={}0.78\pm0.03$. Note that we have excluded galaxies with SFR${}<{}0.01\,M_\odot \,\rm yr^{-1}$ from the fitting to avoid the regime where galaxy statistics can be contaminated by shot noise due to the resolution limit of the simulation \citep{Feldmann_2017a}.

The reduced linearity in the $L_{\rm [C_{II}]}$-SFR relation at high SFR is driven by galaxies with ${\rm SFR}{}\simgreat{}1\,M_\odot \,\rm yr^{-1}$, showing a reduced $L_{\rm [C_{II}]}{}/{}\rm SFR$ ratio compared to those with lower SFR (see the {\it right} panel of Fig.~\ref{fig.4}). Such a trend is not clearly present in any of the three (\citetalias{De_Looze_2011}, \citetalias{De_Looze_2014}, and \citetalias{Herrera_Camus_2015}) observational samples. However, it's important to note that these samples do not contain a statistically large number of galaxies at SFR${}\simgreat{}1\,M_\odot \,\rm yr^{-1}$. Other studies examining local LIRGs and ULIRGs have found clear evidence of a $\rm [C_{II}]$ deficit at high $L_{\rm IR}$ ($\rm \sim SFR$) (see below).

\begin{figure}
 \includegraphics[width=85mm]{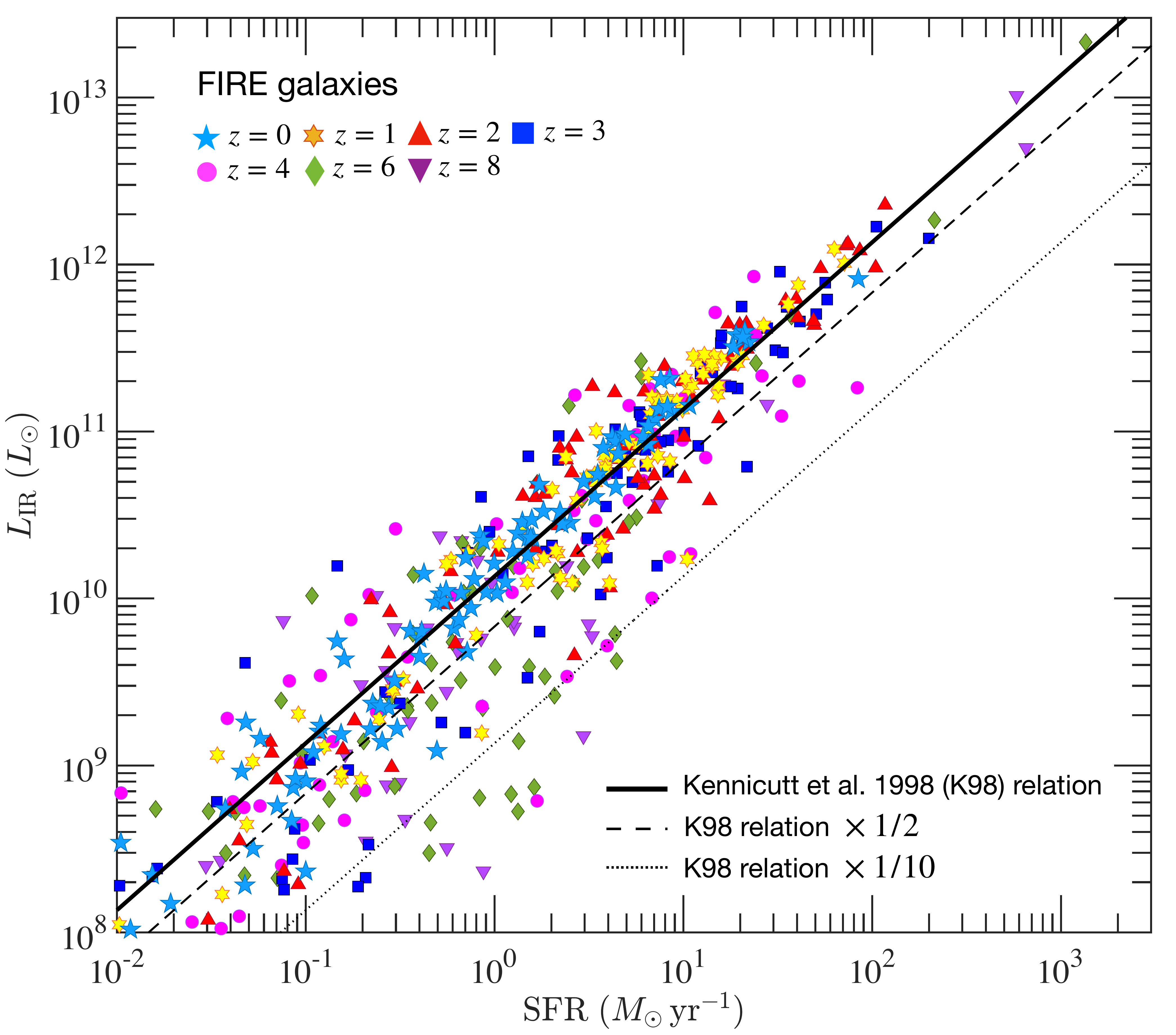}
 \caption{The {$L_{\rm IR}$ vs. SFR} relation of {\sc \small FIRE} galaxies at different redshifts (cyan stars for $z{}={}0$, yellow hexagons for $z{}={}1$, red triangles for $z{}={}2$, blue squares for $z{}={}3$, magenta circles for $z{}={}4$, green diamonds for $z{}={}6$ and purple downward diamonds for $z{}={}8$). The diagonal solid black line indicates the \citetalias{Kennicutt_1998} relation, \ie~$L_{\rm IR}\, (L_\odot) {}={}1.36\times10^{10}\,{\rm SFR} \,(M_\odot\,{\rm yr^{-1}})$.  The dashed and dotted lines indicate the modified \citetalias{Kennicutt_1998} relations where the normalization is lower than the solid black line by a factor of 2 and 10, respectively. {The \citetalias{Kennicutt_1998} relation (solid black line) fits well to the galaxies at high SFR}.}
    \label{fig.5}
\end{figure}

\begin{figure*}
 \includegraphics[width=178mm]{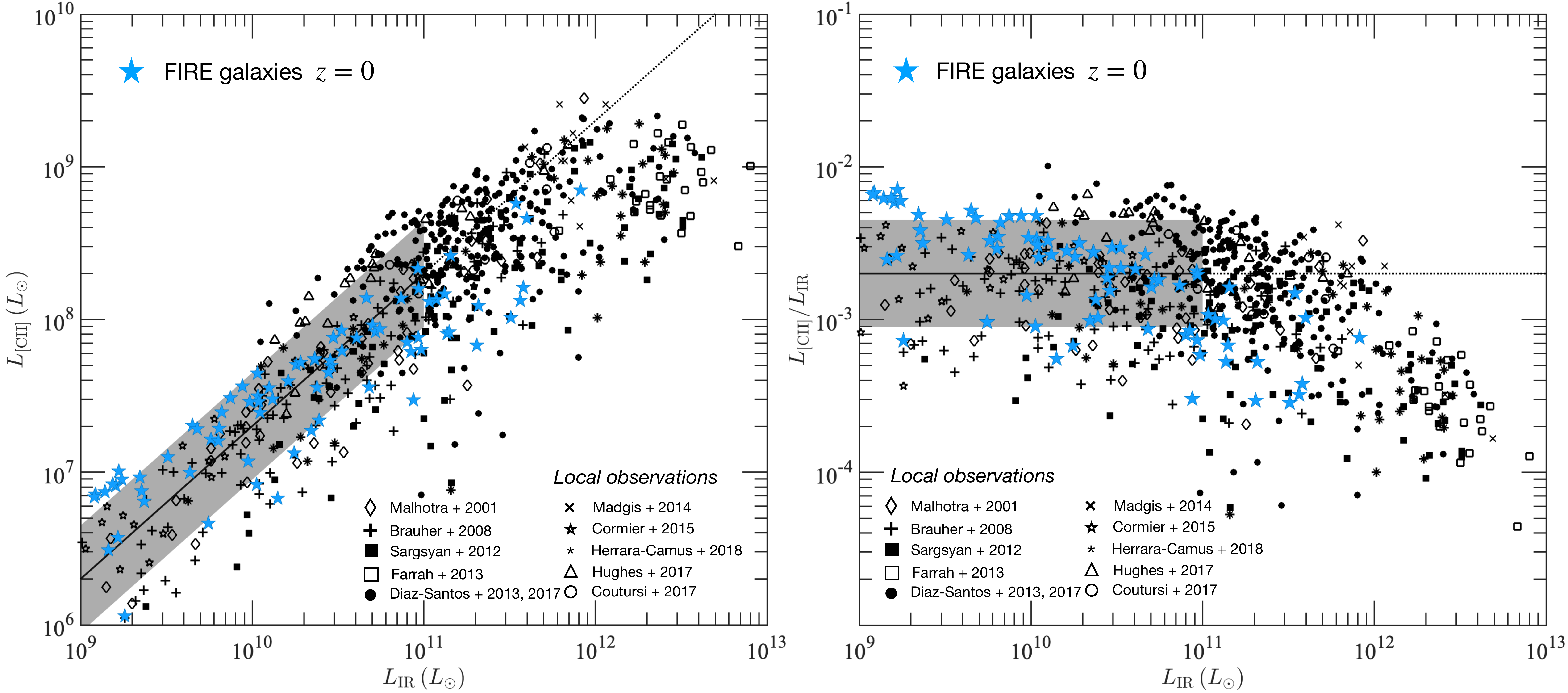}
 \caption{The {$L_{\rm [C_{II}]}$ vs. $L_{\rm IR}$} (\textit{left} panel) and the {$L_{\rm [C_{II}]}{}/{}L_{\rm IR}$ vs. $L_{\rm IR}$} (\textit{right} panel) relations of $z=0$ galaxies. In the two panels, cyan stars show the result of the {\sc\small FIRE} galaxies, whereas black symbols indicate the observational data from different studies, including \citet{Malhotra_2001} (diamond), \citet{Brauher_2008} ({crosses}), \citet{Sargsyan_2012} (filled squares), \citet{Farrah_2013} (empty squares), \citet{Diaz_Santos_2013, Diaz_Santos_2017} (filled circles), \citet{Magdis_2014} ({`X's}), \citet{Cormier_2015} (empty stars), \citet{Herrera_Camus_2018} (asterisks), \citet{Hughes_2017} (triangles) and \citet{Contursi_2017} (empty circles). Observations show that $L_{\rm [C_{II}]}/L_{\rm IR}$ ratio of galaxies is nearly a constant at $10^9{}\simless{}L_{\rm IR}{}\simless{}10^{11}\,L_\odot$, but declines with $L_{\rm IR}$ at $L_{\rm IR}{}\simgreat{}10^{11}\,L_\odot$. In the two panels, black line (solid at $L_{\rm IR}{}<{}10^{11}\,L_\odot$ and dotted at $L_{\rm IR}{}\ge{}10^{11}\,L_\odot$) indicates the median $L_{\rm [C_{II}]}{}/{}L_{\rm IR}$ ratio ($\approx{}2\times10^{-3}$) of the galaxies having $L_{\rm IR}{}<{}10^{11}\,L_\odot$ and grey shaded bar indicates the $1\sigma$ scatter of the $L_{\rm [C_{II}]}{}/{}L_{\rm IR}$ ratio of these galaxies. {\sc\small FIREbox} successfully reproduces the observed {$L_{\rm [C_{II}]}$ vs. $L_{\rm IR}$ (and the $L_{\rm [C_{II}]}{}/{}L_{\rm IR}$ vs. $L_{\rm IR}$)} relation at $z{}={}0$.}
    \label{fig.6}
\end{figure*}

\paragraph*{The $L_{\rm [C_{II}]}$ - $L_{\rm IR}$ relation of $z{}=0{}$ galaxies\\}

A number of observational studies have probed the relation between $L_{\rm [C_{II}]}$ and $L_{\rm IR}$ (or $L_{\rm FIR}$\footnote{In the literature, `$L_{\rm IR}$' is used to denote the bolometric IR luminosity of galaxy that is integrated over the wavelength range $\rm 8-1000\,\mu m$, whereas `$L_{\rm FIR}$' represents the FIR luminosity of galaxy ($\rm 42.5-122.5\,\mu m$). Both $L_{\rm IR}$ and $L_{\rm FIR}$ are commonly adopted as SFR indicators for heavily dust-obscured galaxies.}) of local galaxies.

$L_{\rm IR}$ (or $L_{\rm FIR}$) can be a good proxy for galaxy SFR when the stellar light of a galaxy is heavily absorbed by dust \citep[\eg][]{Kennicutt_1998, Salim_2020}. Galaxies having higher SFR tend to be more gas/dust-rich and have higher gas density. Therefore, they tend to have higher dust opacity \citep[\eg][]{Whitaker_2017}. We show in {Fig.~\ref{fig.5}} the $L_{\rm IR}$ vs. SFR relation of the {\sc\small FIRE} galaxies at different redshifts, {where $L_{\rm IR}$ is calculated using their SEDs produced by {\sc\small SKIRT}}. It can be seen that at $z{}={}0$, the {\sc\small FIRE} galaxies (cyan stars) well follow the \citet[][hereafter K98]{Kennicutt_1998} relation\footnote{We adopt the \citetalias{Kennicutt_1998} relation for the \citet{Kroupa_2002} IMF using the stellar population synthesis (SPS) model {\sc\small STARBURST99}, assuming a constant star formation history lasting for 1 Gyr (see \citealt{Hao_2011} for the details). The original relation (\ie,~$L_{\rm IR}{}/{}L_\odot=5.8\times10^9\;{\rm SFR} {}/{}(M_\odot\,\rm yr^{-1}$)) was derived for the Salpeter IMF based on the older SPS model of \citet{Leitherer_1995}, and for a shorter starburst period ($t_{\rm age}=10-100\,\rm Myrs$). }, \ie
\begin{equation}
L_{\rm IR}\;(L_\odot)=1.36\times10^{10}\;{\rm SFR} \;(M_\odot\,\rm yr^{-1})
\label{eq.4}
\end{equation}
at $\rm SFR{}\simgreat{}1\,M_\odot\,\rm yr^{-1}$ (or $L_{\rm IR}{}\simgreat{}10^{10}\,L_\odot$). The \citetalias{Kennicutt_1998} relation is derived assuming that all radiative energy of the young stars is absorbed and re-emitted by dust {and AGN radiation does not contribute to dust heating}. At $\rm SFR{}<{}1\,M_\odot\,\rm yr^{-1}$, however, the $z{}={}0$ {\sc\small FIRE} galaxies show larger scatter. Some of these galaxies are below the \citetalias{Kennicutt_1998} relation by over 0.3 dex (indicating that less than half of the radiative energy of the young stars gets re-emitted at FIR by dust). These are the galaxies having relatively low dust opacity\footnote{It can be seen from Fig.~\ref{fig.5} that some of the simulated galaxies (particularly those having low SFR) lie above the \citetalias{Kennicutt_1998} relation, which seem to break the energy conservation law. These are in fact the galaxies that are recently quenched after a strong starburst whose dust is heated mainly by the stars older than $100\,\rm Myrs$ \citep[see \eg][]{Hayward_2014}.}. Nonetheless, $L_{\rm IR}$ appears to be a good SFR tracer for the $z{}={}0$ galaxies at $\rm SFR{}\simgreat{}1\,M_\odot\,\rm yr^{-1}$ in the {\sc\small FIRE} simulations. 

In Fig.~\ref{fig.6}, we present the observed {$L_{\rm [C_{II}]}$ vs. $L_{\rm IR}$} ({\it left} panel) and the {$L_{\rm [C_{II}]}{}/{}L_{\rm IR}$ vs. $L_{\rm IR}$} ({\it right} panel) relations for local galaxy samples sourced from various studies. These samples include the {\sc\small GOALS} \citep[``Great Observatories All-sky LIRG Survey";][]{Armus_2009} sample,  consisting of 241 galaxies studied by \citet{Diaz_Santos_2013, Diaz_Santos_2017}, the SHINING (``Survey with Herschel of the ISM in Nearby INfrared Galaxies"; PI: Sturm) sample of 52 galaxies analyzed by \citet{Herrera_Camus_2018}, as well as those studied by \citet{Malhotra_2001}, \citet{Brauher_2008}, \citet{Sargsyan_2012}, \citet{Farrah_2013}, \citet{Magdis_2014}, \citet{Cormier_2015} (note: the same DGS sample as in \citetalias{De_Looze_2014}), \citet{Hughes_2017} and \citet{Contursi_2017}. For those studies that use $L_{\rm FIR}$ as an SFR indicator, we convert the reported $L_{\rm FIR}$ of the galaxies to $L_{\rm IR}$ by multiplying it by 1.6, following \citet{Sanders_2003}. Additionally, in the same figure, we include the data of the $z{}={}0$ {\sc\small FIRE} galaxies, where $L_{\rm IR}$ is determined by integrating the SED produced by {\sc\small SKIRT} over the wavelength range of $\rm 8-1000\,\mu m$.

The observed samples contain a large number of galaxies that are IR-luminous ($L_{\rm IR}\simgreat10^{11}\,L_\odot$, corresponding to ${\rm SFR}{}\simgreat{}10\,M_\odot\,{\rm yr}^{-1}$ following equation~\ref{eq.4}). With these statistically large samples, the $L_{\rm [C_{II}]}{}/{}L_{\rm IR}$ ($\sim L_{\rm [C_{II}]}{}/{}{\rm SFR}$) ratio of the $z{}={}0$ galaxies appear to show a clear decline with $L_{\rm IR}$ at $L_{\rm IR}{}\simgreat{}10^{11}\,L_\odot$ ($\rm [C_{II}]$ deficit), albeit with a large scatter ($1\sigma{}={}0.3$ dex) at given $L_{\rm IR}$. From $L_{\rm IR}{}={}10^{11}$ to $10^{13}\,L_\odot$,  $L_{\rm [C_{II}]}{}/{}L_{\rm IR}$ decreases from $2{}\times{}10^{-3}$ to $10^{-4}$, over a factor of ten. At $L_{\rm IR}{}\simless{}10^{11}\,L_\odot$, on the other hand, $L_{\rm [C_{II}]}{}/{}L_{\rm IR}$ of the observed galaxies is a constant. Overall, the observational and the simulated data agree well with each other (on both the mean value and level of scatter). In particular, the {\sc\small FIRE} sample exhibits a mild $\rm [C_{II}]$ deficit at $L_{\rm IR}{}\simgreat{}10^{11}\,L_\odot$ at $z{}={}0$, which is in agreement with the observational data. Note, however, that our {\sc\small FIRE} sample at $z{}={}0$ does not include any ULIRGs (\ie~$L_{\rm IR}{}\simgreat{}10^{12}\,L_\odot$) at $z{}={}0$. 

\subsection{High redshifts ($1{}\simless{}z{}\simless{}5$)}
\label{Sec:4b}

\begin{figure*}
 \includegraphics[height=77mm]{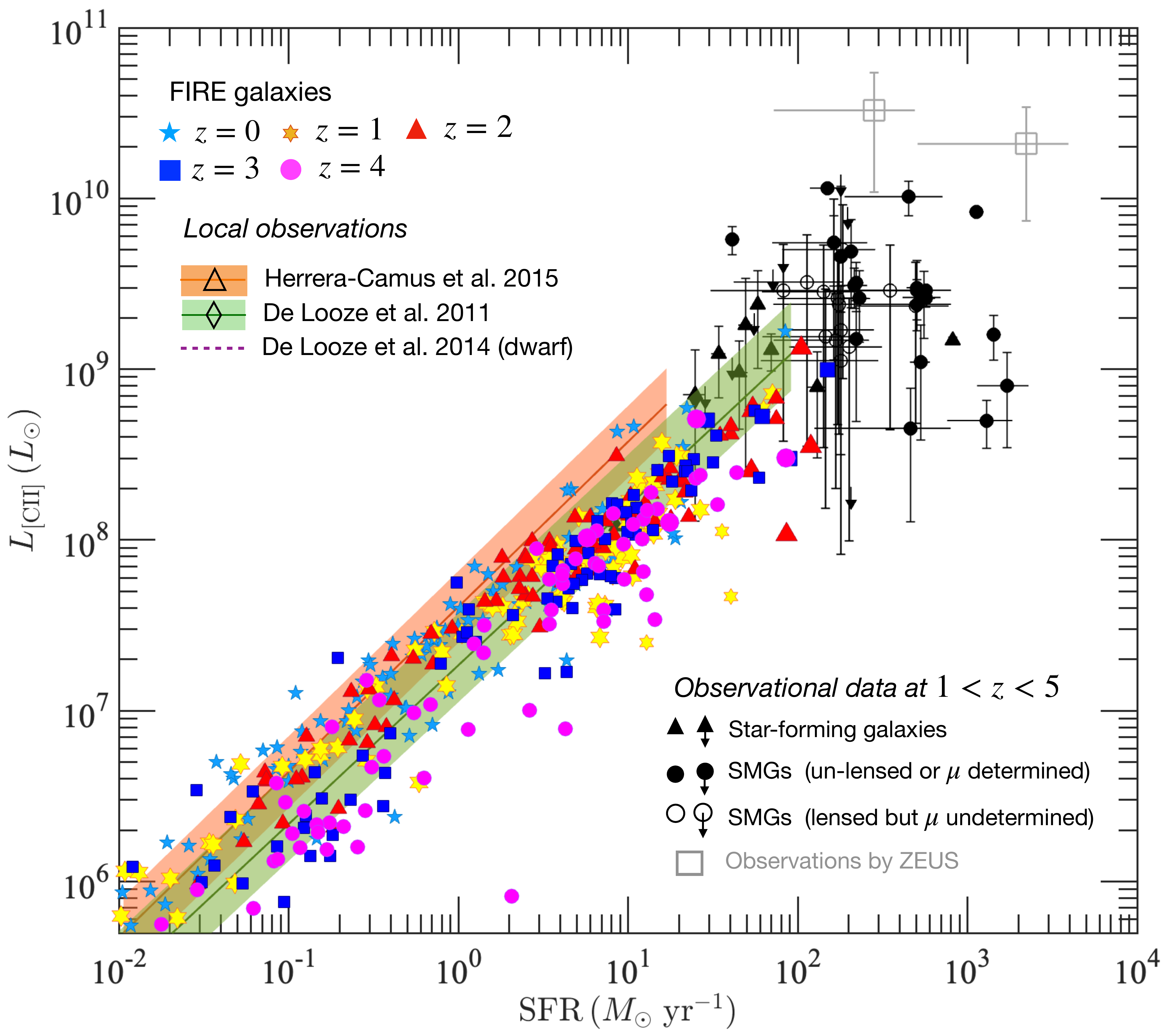}
  \includegraphics[height=77mm]{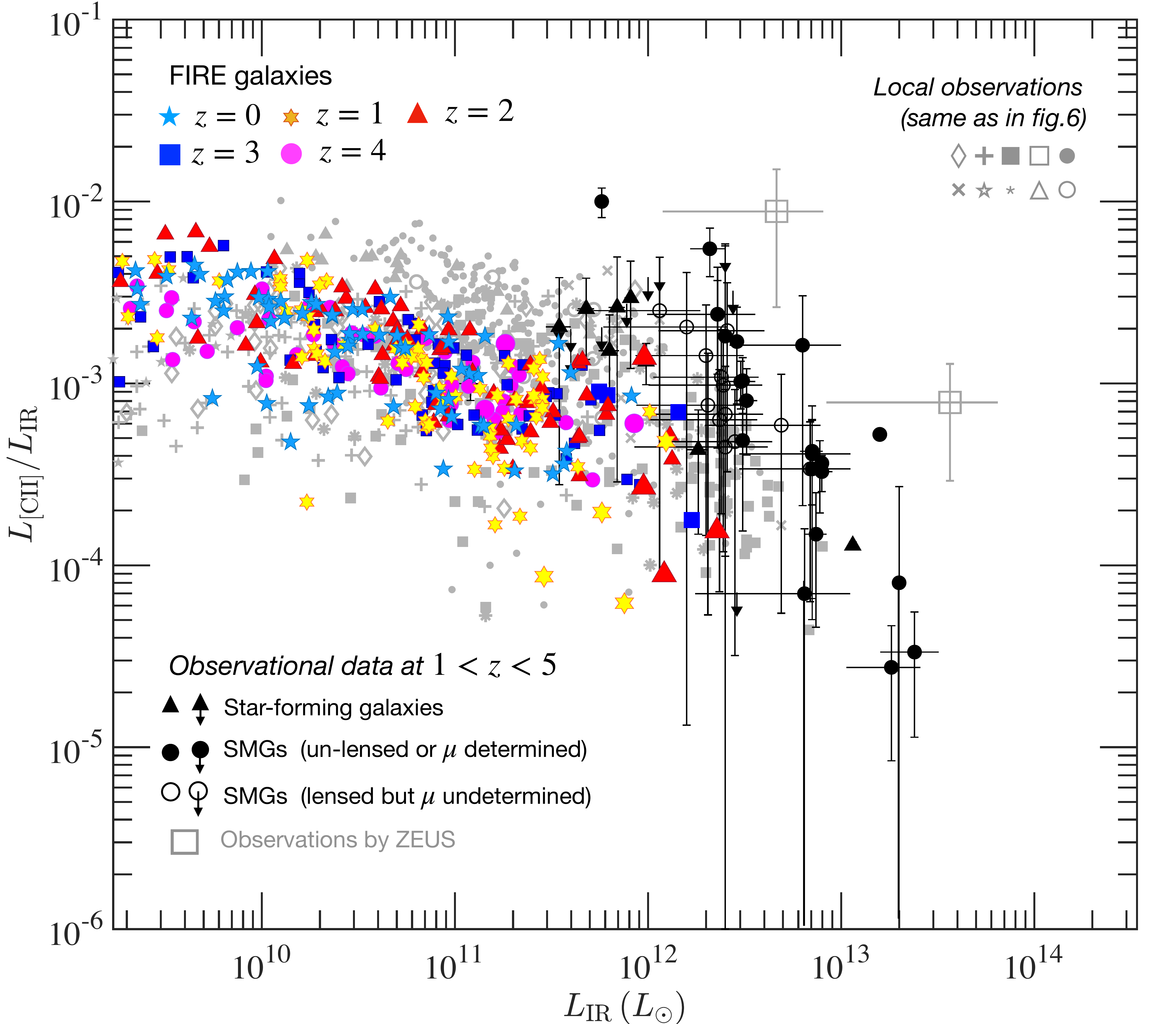}
 \caption{The  {$L_{\rm [C_{II}]}$ vs. SFR} (\textit{left} panel) and the {$L_{\rm [C_{II}]}/L_{\rm IR}$ vs. $L_{\rm IR}$} (\textit{right} panel) relations of galaxies at $z{}={}0$ and high redshifts. In both panels, filled coloured symbols represent the {\sc\small FIRE} galaxies (cyan stars for $z{}={}0$, yellow hexagons for $z{}={}1$, red triangles for $z{}={}2$, blue squares for $z{}={}3$ and magenta circles for $z{}={}4$). Black symbols (filled and empty) show the observational data of galaxies at $1{}\simless{}z{}\simless{}5$ (see Table~\ref{T3} for the details). Specifically, black circles and black triangles correspond to SMGs and other star-forming galaxies, respectively. For the gravitationally-lensed galaxies, their $\rm [C_{II}]$ and IR luminosities have been corrected by the lensing magnification factor $\mu$ reported in the literature. Those having direct measurement of $\mu$ as well as the un-lensed galaxies are marked by filled symbols (triangles and circles), whereas the 16 lensed {\sc\small SPT} galaxies whose $\mu$ is extrapolated ($\mu$ has been assumed to be 14.1 by \citealt{Gullberg_2015}) are shown by empty circles. The two grey empty squares represent the stacked result of the galaxy samples of \citet{Stacey_2010} and \citet{Brisbin_2015}. The $\rm [C_{II}]$ line of the two samples is measured with the redshift and Early Universe Spectrometer ({\sc\small ZEUS}) and their data systematically offsets from that of the other galaxy samples. For reference, we also show in the {\it left} ({\it right}) panel the observational results of the local galaxy samples as shown in Fig.~\ref{fig.4} (Fig.~\ref{fig.6}).  Both observations and {\sc\small FIRE} simulations show that high-$z$ ($1{}\simless{}z{}\simless{}5$) galaxies exhibit a $\rm [C_{II}]$ deficit at high $L_{\rm IR}$ similar to local galaxies.}
    \label{fig.7}
\end{figure*}

Observational studies have investigated the $L_{\rm [C_{II}]}$-SFR relation of galaxies at $1{}\simless{}z{}\simless{}5$, including \eg~\citet{Ivison_2010, Stacey_2010, Valtchanov_2011, Brisbin_2015, Gullberg_2015, Gullberg_2018, Schaerer_2015a, Umehata_2017, Zanella_2018, Hashimoto_2019b, Mckinney_2020}. Their samples consist of roughly 80 galaxies in total (see Table~\ref{T3} for the details). Most of these galaxies have substantial SFR (${\rm SFR}{}\simgreat{}100\,M_\odot\,\rm yr^{-1}$) and are IR-luminous ($L_{\rm IR}{}\simgreat{}10^{12}\,L_\odot$). This is in stark contrast with the local observations (see Section~\ref{Sec:4a}), which probe the galaxies having much lower SFR (see Table~\ref{T2}).  Note that a large fraction of the selected galaxies in this redshift regime are uncovered by wide-field sub-mm galaxy surveys, \eg~the South Pole Telescope \citep[{\sc\small SPT};][]{Vieira_2010, Carlstrom_2011} survey \citep{Weiss_2013, Gullberg_2015}. 

We derive the SFR of the selected galaxies from their measured $L_{\rm IR}$ (see Table~\ref{T3}) using the \citetalias{Kennicutt_1998} relation (equation~\ref{eq.4}) assuming that the galaxies are heavily dust-obscured. Note that at high redshifts, the \citetalias{Kennicutt_1998} relation may only apply to {the more massive and starburst galaxies}. High-$z$ galaxies are metal and dust-poorer than the $z{}={}0$ galaxies {at given mass (or SFR)}, and therefore only {the more massive and gas-rich systems} have high enough dust opacity leading to total obscuration of stellar light. We can see from Fig.~\ref{fig.5} that the \citetalias{Kennicutt_1998} relation (solid black line) fits well the high-$z$ {\sc\small FIRE} galaxies at ${\rm SFR}{}\simgreat{}100\,M_\odot\,\rm yr^{-1}$ (or $L_{\rm IR}{}\simgreat{}10^{12}\,L_\odot$. Note: For the $z{}={}1$ galaxies, the \citetalias{Kennicutt_1998} relation {fits well} to the data down to $L_{\rm IR}{}\approx{}10^{11}\,L_\odot$). At lower SFR, the high-$z$ galaxies exhibit larger scatter and they, on the average, have lower $L_{\rm IR}$ at given SFR than the $z{}={}0$ galaxies due to their reduced dust opacity.

The galaxies selected at $1{}\simless{}z{}\simless{}5$ typically have a good sampling of photometric data points in the dust continuum, which are obtained by observations with multiple IR and millimetre instruments (\textit{Spitzer}, \textit{Herschel}, {\sc\small ALMA} and etc). The shape of the dust SED of these galaxies is therefore well constrained. This results in relatively small uncertainty in the estimate of their $L_{\rm IR}$.

The $\rm [C_{II}]$ line of these galaxies is measured with different instruments (see Table~\ref{T3}). For instance, \citet{Stacey_2010} and \citet{Brisbin_2015} measure the $\rm [C_{II}]$ line of the 20 galaxies at $z{}\approx{}1-2$ of their samples using the redshift ($z$) and Early Universe Spectrometer \citep[{\sc\small ZEUS};][]{Stacey_2007, Hailey_Dunsheath_2009} on the 10.4 m Caltech Submillimeter Observatory ({\sc\small CSO}). \citet{Gullberg_2015} measure the $\rm [C_{II}]$ line of the 16 SMGs selected from the {\sc\small SPT} catalogue \citep{Weiss_2013} using the {\sc\small SPIRE} Fourier Transform Spectrometer \citep[{\sc\small FTS};][]{Griffin_2010} onboard {\it Herschel} (for the galaxies at $z{}<{}3$) and the First Light APEX Sub-millimetre Heterodyne receiver \citep[{\sc\small FLASH};][]{Heyminck_2006} (for the galaxies at $z{}>{}3$). For the remaining galaxies ($\sim40$), their $\rm [C_{II}]$ line is measured with {\sc\small ALMA} (at Band 7, 8 and 9 for the galaxies at $z\sim4$, $z\sim3$ and $z\sim2$, respectively). {\sc\small ALMA} observations often marginally resolve a galaxy spatially in $\rm [C_{II}]$, whereas observations with {\sc\small ZEUS}, {\sc\small APEX/FLASH} and {\sc\small SPIRE FTS} do not. 

\begin{table*}
\centering
\caption{The observed $L_{\rm [C_{II}]}$-SFR relation of galaxies at high redshifts. }
\begin{threeparttable}
\begin{tabular}{ p{2.4 cm} p{1.1 cm}  p{1.4 cm}  p{3.3 cm} p{1.8 cm} p{0.8cm} p{1.4 cm}  p{1.7 cm}}
 \hline
  \hline
 \multicolumn{1}{c|}{Name\tnote{$\dagger$}} &  \multicolumn{1}{c|}{$z$}  &  \multicolumn{1}{c|}{${\rm log}\,(L_{\rm IR}{}/{}L_\odot)$\tnote{$\mathsection$}} & ${\rm log}\,(L_{\rm [C_{II}]}{}/{}L_\odot)$\tnote{$\ddagger$, $\mathsection$, $\parallel$} & \multicolumn{1}{c|}{Galaxy type\tnote{$\#$}} & \multicolumn{1}{c|}{AGN}  & \multicolumn{1}{c|}{$\mu$} & References\tnote{$\ast$} \\
 \hline
ID 7118 & 1.7290  & $12.06\pm0.01$ & $<9.70$ ({\sc\small ALMA} 9) & \multicolumn{1}{c|}{MS} & \multicolumn{1}{c|}{No} & \multicolumn{1}{c|}{$-$} &   [1, 2]  \\
GS IRS61 & 1.759 & $12.46\pm0.13$ & $<8.31$ ({\sc\small ALMA} 9) & \multicolumn{1}{c|}{SB} & \multicolumn{1}{c|}{No} & \multicolumn{1}{c|}{$-$} &   [3, 4] \\
ID 9834 & 1.7644  & $11.99\pm0.02$ & $9.11\pm0.07$ ({\sc\small ALMA} 9) & \multicolumn{1}{c|}{MS} & \multicolumn{1}{c|}{No} & \multicolumn{1}{c|}{$-$} &   [1, 2]  \\
ID 2910 & 1.7686  & $11.76\pm0.08$ & $<9.08$ ({\sc\small ALMA} 9) & \multicolumn{1}{c|}{MS} & \multicolumn{1}{c|}{No} & \multicolumn{1}{c|}{$-$} &   [1, 2]  \\
ID 2861 & 1.8102  & $12.00\pm0.03$ & $<9.58$ ({\sc\small ALMA} 9) & \multicolumn{1}{c|}{MS} & \multicolumn{1}{c|}{No} & \multicolumn{1}{c|}{$-$} &   [1, 2]  \\
ID 6515 & 1.8438  & $11.68\pm0.04$ & $9.09\pm0.12$ ({\sc\small ALMA} 9) & \multicolumn{1}{c|}{MS} & \multicolumn{1}{c|}{No} & \multicolumn{1}{c|}{$-$} &  [1, 2]  \\
ID 9347 & 1.8505  & $11.80\pm0.05$ & $8.98\pm0.14$ ({\sc\small ALMA} 9) & \multicolumn{1}{c|}{MS} & \multicolumn{1}{c|}{No} & \multicolumn{1}{c|}{$-$} &  [1, 2]  \\
ID 9681 & 1.8852  & $11.84\pm0.04$ & $9.26\pm0.20$ ({\sc\small ALMA} 9) & \multicolumn{1}{c|}{MS} & \multicolumn{1}{c|}{No} & \multicolumn{1}{c|}{$-$}  & [1, 2]  \\  
ID 8490 & 1.9056 & $11.54\pm0.06$ & $8.85\pm0.20$ ({\sc\small ALMA} 9) & \multicolumn{1}{c|}{MS} & \multicolumn{1}{c|}{No} & \multicolumn{1}{c|}{$-$} &   [1, 2]  \\  
ID 10049 & 1.9200 & $11.60\pm0.06$ & $<8.78$ ({\sc\small ALMA} 9) & \multicolumn{1}{c|}{MS} & \multicolumn{1}{c|}{Yes} & \multicolumn{1}{c|}{$-$} &  [1, 2]  \\  
GS IRS20 & 1.923 & $13.06\pm0.12$ & $9.17\pm0.01$ ({\sc\small ALMA} 9) & \multicolumn{1}{c|}{SB} & \multicolumn{1}{c|}{Yes} & \multicolumn{1}{c|}{$-$} &   [3, 4] \\
ID 10076 & 1.9462 & $11.91\pm0.03$ & $9.38\pm0.14$ ({\sc\small ALMA} 9) & \multicolumn{1}{c|}{MS} & \multicolumn{1}{c|}{No} & \multicolumn{1}{c|}{$-$} &   [1, 2]  \\  
MACS J0451+0006 & 2.013 & $11.08\pm0.04$ & $8.08\pm0.04$ ({\sc\small ALMA} 9) & \multicolumn{1}{c|}{MS} & \multicolumn{1}{c|}{No} & \multicolumn{1}{c|}{$49\pm5$} &   [5, 6, 7]  \\  
GRB 080207 & 2.0865 & $12.26\pm0.05$ & $8.89\pm0.12$ ({\sc\small ALMA} 9) & \multicolumn{1}{c|}{MS} & \multicolumn{1}{c|}{No} & \multicolumn{1}{c|}{$-$} &  [8]  \\  
SPT 0551-50 & 2.123 & $11.89\pm0.05$ & $<9.33$ (SPIRE FTS) & \multicolumn{1}{c|}{SMG} & \multicolumn{1}{c|}{No} & ($14.1\pm7.8$) &  [9, 10] \\  
SPT 0512-59 & 2.234 & $12.29\pm0.04$ & $9.45\pm0.09$ (SPIRE FTS) & \multicolumn{1}{c|}{SMG} & \multicolumn{1}{c|}{No} & ($14.1\pm7.8$) &   [9, 10] \\  
SMM J2135 & 2.3259 & $12.08\pm0.07$ & $8.25\pm0.11$ (SPIRE FTS) & \multicolumn{1}{c|}{SMG} & \multicolumn{1}{c|}{No} & $32.5\pm4.5$ &   [12, 13] \\  
SDP.130 & 2.625 & $12.40\pm0.02$ & $<10.14$ (SPIRE FTS) & \multicolumn{1}{c|}{SMG} & \multicolumn{1}{c|}{No} & \multicolumn{1}{c|}{$6\pm1$} & [14, 15] \\  
SPT 0538-50 & 2.782 & $12.44\pm0.03$ & $<9.95$ (SPIRE FTS) & \multicolumn{1}{c|}{SMG} & \multicolumn{1}{c|}{No} & $20.9\pm4.2$ & [9, 10] \\
ALESS 49.1 & 2.943 & $12.85\pm0.06$ & $9.48\pm0.12$ ({\sc\small ALMA} 8) & \multicolumn{1}{c|}{SMG} & \multicolumn{1}{c|}{No} & \multicolumn{1}{c|}{$-$} & [16, 17, 18] \\  
ALESS 57.1 & 2.943 & $12.87\pm0.06$ & $9.04\pm0.17$ ({\sc\small ALMA} 8) & \multicolumn{1}{c|}{SMG} & \multicolumn{1}{c|}{No} & \multicolumn{1}{c|}{$-$} & [16, 17, 18] \\  
SDP.81 & 3.042 & $12.32\pm0.08$ & $10.06\pm0.01$ (SPIRE FTS) & \multicolumn{1}{c|}{SMG} & \multicolumn{1}{c|}{No} & \multicolumn{1}{c|}{$25\pm7$} & [14, 15] \\  
SPT 0103-45 & 3.090 & $12.38\pm0.02$ & $9.41\pm0.06$ ({\sc\small APEX/FLASH}) & \multicolumn{1}{c|}{SMG} & \multicolumn{1}{c|}{No} & ($14.1\pm7.8$)  &   [9, 10] \\  
LAB1-{\sc\small ALMA}3 & 3.0993 & 11.76 & $9.41\pm0.06$ ({\sc\small ALMA} 8) & \multicolumn{1}{c|}{MS} & \multicolumn{1}{c|}{No} & \multicolumn{1}{c|}{$-$} & [19, 20]\\  
LAB1-{\sc\small ALMA}1 & 3.1 & 11.54 & $<8.9$ ({\sc\small ALMA} 8) & \multicolumn{1}{c|}{MS} & \multicolumn{1}{c|}{No} & \multicolumn{1}{c|}{$-$} & [19, 20] \\  
LAB1-{\sc\small ALMA}2 & 3.1 & 11.60 & $<8.9$ ({\sc\small ALMA} 8) & \multicolumn{1}{c|}{MS} & \multicolumn{1}{c|}{No} & \multicolumn{1}{c|}{$-$} & [19, 20] \\  
SPT 0550-53 & 3.129 & $12.08\pm0.09$ & $9.46\pm0.09$ ({\sc\small APEX/FLASH}) & \multicolumn{1}{c|}{SMG} & \multicolumn{1}{c|}{No} & ($14.1\pm7.8$) &   [9, 10] \\  
SPT 0529-54 & 3.369 & $12.36\pm0.04$ & $9.74\pm0.04$ ({\sc\small APEX/FLASH}) & \multicolumn{1}{c|}{SMG} & \multicolumn{1}{c|}{No} & \multicolumn{1}{c|}{$9.4\pm1.0$} &   [9, 10] \\  
SPT 0532-50 & 3.399 & $12.69\pm0.07$ & $9.46\pm0.08$ ({\sc\small APEX/FLASH}) & \multicolumn{1}{c|}{SMG} & \multicolumn{1}{c|}{No} & ($14.1\pm7.8$) &   [9, 10] \\  
SPT 0300-46 & 3.596 & $12.40\pm0.11$ & $9.05\pm0.11$ ({\sc\small APEX/FLASH}) & \multicolumn{1}{c|}{SMG} & \multicolumn{1}{c|}{No} & ($14.1\pm7.8$) &    [9, 10] \\  
SPT 2147-50 & 3.761 & $12.39\pm0.06$ & $9.38\pm0.06$ ({\sc\small APEX/FLASH}) & \multicolumn{1}{c|}{SMG} & \multicolumn{1}{c|}{No} & ($14.1\pm7.8$) &   [9, 10] \\  
SPT 0418-47 & 4.224 & $12.48\pm0.03$ & $9.49\pm0.03$ ({\sc\small APEX/FLASH}) & \multicolumn{1}{c|}{SMG} & \multicolumn{1}{c|}{No} & \multicolumn{1}{c|}{$21.0\pm3.5$} &  [9, 10] \\  
SPT 0113-46 & 4.232 & $12.20\pm0.09$ & $9.51\pm0.10$ ({\sc\small APEX/FLASH}) & \multicolumn{1}{c|}{SMG} & \multicolumn{1}{c|}{No} & ($14.1\pm7.8$) &   [9, 10] \\  
SDP.141 & 4.24 & $12.52\pm0.12$ & $9.48\pm0.07$ ({\sc\small APEX/FLASH}) & \multicolumn{1}{c|}{SMG} & \multicolumn{1}{c|}{No} & \multicolumn{1}{c|}{$10-30$} &   [11] \\  
SPT 2311-54 & 4.281 & $12.40\pm0.04$ & $9.23\pm0.06$ ({\sc\small APEX/FLASH}) & \multicolumn{1}{c|}{SMG} & \multicolumn{1}{c|}{No} & ($14.1\pm7.8$)  &   [9, 10] \\  
SPT 0345-47 & 4.296 & $12.84\pm0.04$ & $9.37\pm0.04$ ({\sc\small APEX/FLASH}) & \multicolumn{1}{c|}{SMG} & \multicolumn{1}{c|}{No} & ($14.1\pm7.8$) &   [9, 10] \\  
COSMOS-AzTEC-1 & 4.342 & $13.21\pm0.09$ & $9.80\pm0.04$ ({\sc\small ALMA} 7) & \multicolumn{1}{c|}{SMG} & \multicolumn{1}{c|}{No} & \multicolumn{1}{c|}{$-$} &   [21, 22] \\  
AS2UDS.0568.0 & 4.404 & $13.30\pm0.08$ & $9.20\pm0.08$ ({\sc\small ALMA} 7) & \multicolumn{1}{c|}{SMG} & \multicolumn{1}{c|}{No} & \multicolumn{1}{c|}{$-$} & [23, 24] \\  
ALESS 61.1 & 4.4189 & $12.49\pm0.03$ & $9.18\pm0.17$ ({\sc\small ALMA} 7) & \multicolumn{1}{c|}{SMG} & \multicolumn{1}{c|}{No} & \multicolumn{1}{c|}{$-$} & [24, 25, 26] \\  
UDS 47.0 & 4.4201 & $12.50\pm0.06$ & $9.42\pm0.12$ ({\sc\small ALMA} 7) & \multicolumn{1}{c|}{SMG} & \multicolumn{1}{c|}{No} & \multicolumn{1}{c|}{$-$} & [24, 26] \\  
AS2UDS.0051.0 & 4.421 & $12.85\pm0.20$ & $9.38\pm0.05$ ({\sc\small ALMA} 7) & \multicolumn{1}{c|}{SMG} & \multicolumn{1}{c|}{No} & \multicolumn{1}{c|}{$-$} & [23, 24] \\  
AS2UDS.0104.0  & 4.423 & $12.85\pm0.20$ & $9.46\pm0.05$ ({\sc\small ALMA} 7) & \multicolumn{1}{c|}{SMG} & \multicolumn{1}{c|}{No} & \multicolumn{1}{c|}{$-$} & [23, 24] \\ 
SGP 38326 (SMG1) & 4.4237 & $13.20\pm0.09$ & $9.92\pm0.05$ ({\sc\small ALMA} 7) & \multicolumn{1}{c|}{SMG (SB)} & \multicolumn{1}{c|}{No} & \multicolumn{1}{c|}{$-$} & [27] \\ 
SGP 38326 (SMG2) & 4.4289 & $12.90\pm0.09$ & $9.46\pm0.05$ ({\sc\small ALMA} 7) & \multicolumn{1}{c|}{SMG (SB)}  & \multicolumn{1}{c|}{No} & \multicolumn{1}{c|}{$-$} & [27] \\ 
BRI 0952-0115 & 4.4337 & $12.40\pm0.25$ & $9.66\pm0.25$ ({\sc\small APEX/FLASH}) & \multicolumn{1}{c|}{SMG (SB)} & \multicolumn{1}{c|}{No} & \multicolumn{1}{c|}{$4.5\pm2.8$} & [28, 29, 30] \\ 
SPT 2103-60 & 4.435 & $12.41\pm0.03$ & $9.70\pm0.06$ ({\sc\small APEX/FLASH}) & \multicolumn{1}{c|}{SMG} & \multicolumn{1}{c|}{No} & ($14.1\pm7.8$) &    [9, 10] \\  
AS2UDS.0232.0  & 4.443 & $13.26\pm0.15$ & $8.70\pm0.09$ ({\sc\small ALMA} 7) & \multicolumn{1}{c|}{SMG} & \multicolumn{1}{c|}{No} & \multicolumn{1}{c|}{$-$} & [23, 24] \\ 
ALESS 65.1 & 4.4445 & $12.49\pm0.03$ & $9.51\pm0.09$ ({\sc\small ALMA} 7) & \multicolumn{1}{c|}{SMG} & \multicolumn{1}{c|}{No} & \multicolumn{1}{c|}{$-$} & [24, 25, 26] \\  
AS2UDS.0109.0  & 4.450 & $12.90\pm0.06$ & $9.42\pm0.03$ ({\sc\small ALMA} 7) & \multicolumn{1}{c|}{SMG} & \multicolumn{1}{c|}{No} & \multicolumn{1}{c|}{$-$} & [23, 24] \\ 
AS2UDS.0002.1  & 4.4611 & $13.38\pm0.08$ & $8.90\pm0.11$ ({\sc\small ALMA} 7) & \multicolumn{1}{c|}{SMG} & \multicolumn{1}{c|}{No} & \multicolumn{1}{c|}{$\simless1.5-2$} & [23, 24] \\ 
AS2UDS.0643.0  & 4.4614 & $13.11\pm0.22$ & $8.95\pm0.15$ ({\sc\small ALMA} 7) & \multicolumn{1}{c|}{SMG} & \multicolumn{1}{c|}{No} & \multicolumn{1}{c|}{$\simless1.5-2$}  & [23, 24] \\ 
AS2UDS.0208.0  & 4.4615 & $12.89\pm0.01$ & $9.42\pm0.06$ ({\sc\small ALMA} 7) & \multicolumn{1}{c|}{SMG} & \multicolumn{1}{c|}{No} & \multicolumn{1}{c|}{$\simless1.5-2$}  & [23, 24] \\ 
SPT 0441-46 & 4.477 & $12.45\pm0.02$ & $9.13\pm0.11$ ({\sc\small APEX/FLASH}) & \multicolumn{1}{c|}{SMG} & \multicolumn{1}{c|}{No} & ($14.1\pm7.8$) &   [9, 10] \\  
SPT 2146-55 & 4.567 & $12.31\pm0.05$ & $9.19\pm0.10$ ({\sc\small APEX/FLASH}) & \multicolumn{1}{c|}{SMG} & \multicolumn{1}{c|}{No} & ($14.1\pm7.8$) &   [9, 10] \\  
W2246–0526 & 4.601 & $14.34\pm0.08$ & $9.79\pm0.03$ ({\sc\small ALMA} 7) & \multicolumn{1}{c|}{DOG} & \multicolumn{1}{c|}{Yes} &  \multicolumn{1}{c|}{$-$}  &   [31] \\  
ALESS 73.1 & 4.7555 & $12.46\pm0.03$ & $9.69\pm0.14$ ({\sc\small ALMA} 7) & \multicolumn{1}{c|}{SMG (SB)} & \multicolumn{1}{c|}{Yes} & \multicolumn{1}{c|}{$-$} & [24, 26, 32] \\  
SPT 2132-58 & 4.768 & $12.37\pm0.04$ & $9.17\pm0.08$ ({\sc\small APEX/FLASH}) & \multicolumn{1}{c|}{SMG} & \multicolumn{1}{c|}{No} & ($14.1\pm7.8$) &   [9, 10] \\  
HDF850.1 & 5.185 & $12.58\pm0.07$ & $9.38\pm0.05$ (IRAM/PdBI) & \multicolumn{1}{c|}{SMG} & \multicolumn{1}{c|}{No} & \multicolumn{1}{c|}{$1.5-1.7$} &   [33, 34] \\  
HLSJ091828.6+514223 & 5.24 & $13.04\pm0.10$ & $9.98\pm0.01$ (SMA) & \multicolumn{1}{c|}{SMG} & \multicolumn{1}{c|}{No} & \multicolumn{1}{c|}{$8.9\pm1.9$} &   [35] \\  
SPT 2319-55 & 5.293 & $12.28\pm0.03$ & $9.00\pm0.06$ ({\sc\small APEX/FLASH}) & \multicolumn{1}{c|}{SMG} & \multicolumn{1}{c|}{No} & ($14.1\pm7.8$) &   [9, 10] \\   
SPT 0346-52 & 5.656 & $13.39\pm0.02$ & $9.97\pm0.06$ ({\sc\small APEX/FLASH}) & \multicolumn{1}{c|}{SMG} & \multicolumn{1}{c|}{No} & \multicolumn{1}{c|}{$5.4\pm0.2$} &  [9, 10] \\ 
SPT 0243-49 & 5.699 & $12.40\pm0.04$ & $<9.40$ ({\sc\small APEX/FLASH}) & \multicolumn{1}{c|}{SMG} & \multicolumn{1}{c|}{No} & ($14.1\pm7.8$) & [9, 10] \\ 
 \hline
\end{tabular}
\;\;\;\;\;(Continue on next page)
  \end{threeparttable}
\label{T3}
\end{table*}
\begin{table*}
 \contcaption{The observed $L_{\rm [C_{II}]}$-SFR relation of galaxies at high redshifts.}
 \label{T3x}
\begin{threeparttable}
\begin{tabular}{ p{2.6 cm} p{1.1 cm}  p{1.4 cm}  p{3.3 cm} p{1.8 cm} p{0.8cm} p{1.4 cm}  p{1.5 cm}}
 \hline
  \hline
 \multicolumn{1}{c|}{Name\tnote{$\dagger$}}  &  \multicolumn{1}{c|}{$z$}  &   \multicolumn{1}{c|}{${\rm log}\,(L_{\rm IR}{}/{}L_\odot)$\tnote{$\mathsection$}} & ${\rm log}\,(L_{\rm [C_{II}]}{}/{}L_\odot)$\tnote{$\ddagger$, $\mathsection$, $\parallel$} &  \multicolumn{1}{c|}{Galaxy type\tnote{$\#$}} &  \multicolumn{1}{c|}{AGN}  & \multicolumn{1}{c|}{$\mu$} & References\tnote{$\ast$} \\
 \hline
HerMESFLS3 & 6.3369 & $13.34\pm0.05$ & $9.83\pm0.10$ (CARMA) & \multicolumn{1}{c|}{SMG} & \multicolumn{1}{c|}{No} & \multicolumn{1}{c|}{$2.2\pm0.3$} &  [36, 37] \\  
SPT 0311-58-E & 6.900 & $12.66\pm0.12$ & $9.62\pm0.06$ ({\sc\small ALMA} 6) & \multicolumn{1}{c|}{SMG} & \multicolumn{1}{c|}{No} & \multicolumn{1}{c|}{$1.3$} &  [38] \\  
SPT 0311-58-W & 6.900 & $13.52\pm0.09$ & $9.66\pm0.06$ ({\sc\small ALMA} 6) & \multicolumn{1}{c|}{SMG} & \multicolumn{1}{c|}{No} & \multicolumn{1}{c|}{$2.2$} &  [38] \\  
  \hline
 \end{tabular}
  \begin{tablenotes}
  \item[$\dagger$] The table does not include the 20 galaxies ($z\approx2$) in the samples of \citet{Stacey_2010} and \citet{Brisbin_2015}, of which the $\rm [C_{II}]$ line is measured by {\sc\small ZEUS}. The {$L_{\rm [C_{II}]}{}/{}L_{\rm IR}$ vs $L_{\rm IR}$} relation of these two samples systematically offsets from the others that use different instrument to measure $\rm [C_{II}]$ line (see Fig.~\ref{fig.7}). 
  \item[$\mathsection$] For the gravitationally-lensed galaxies, $L_{\rm [C_{II}]}$ and $L_{\rm IR}$ have been de-magnified by the reported lensing magnification factor $\mu$. For those {\sc\small SPT} galaxies having no direct measurement of $\mu$ (galaxies are not spatially resolved by any observation), we adopt a constant $\mu{}={}14.1$ as is done by \citet{Gullberg_2015}, which is the mean of the four galaxies (SPT 0538-50, SPT 0529-54, SPT 0418-47 and SPT 0346-52) in the same sample that is observationally determined via lensing modelling.
  \item[$\ddagger$] For the $\rm [C_{II}]$-undetected galaxies, we show the $3\sigma$ upper confidence limit. 
    \item[$\parallel$] IRAM/PdBI: the IRAM Plateau de Bure Interferometer \citep{Guilloteau_1992}; SMA: the Submillimeter Array \citep{Ho_2004}; CARMA: the Combined Array for Research in Millimeter-wave Astronomy \citep{Woody_2004}. Note that the three telescopes have produced spatially resolved line emission maps of $\rm [C_{II}]$ for high-$z$ SMGs (HDF850.1, HLSJ091828.6+514223 and HerMESFLS3) as {\sc\small ALMA} does. 
 \item[$\#$] SMG: sub-mm galaxies; MS: `main-sequence' galaxies; SB: starburst galaxies; DOG: hot dust-obscured galaxies \citep[galaxies uncovered by surveys at near-infrared wavelengths, which have strong IR emission from warm dust, \eg][]{Dey_2008, Eisenhardt_2012}. 
 \item[$\ast$]  References: (1): \citet{Zanella_2018}, [2]: \citet{Elbaz_2011}, [3]: \citet{Mckinney_2020}, [4]: \citet{Kirkpatrick_2015}, [5]:\citet{Schaerer_2015a}, [6]: \citet{Sklias_2014}, [7]: \citet{Jones_2010}, [8]: \citet{Hashimoto_2019b}, [9]: \citet{Gullberg_2015}, [10]: \citet{Weiss_2013}, [11]: \citet{Cox_2011}, [12]: \citet{Ivison_2010}, [13]: \citet{Swinbank_2010}, [14]: \citet{Valtchanov_2011}, [15]: \citet{Hopwood_2011}, [16]: \citet{Rybak_2019}, [17]: \citet{Wardlow_2018}, [18]: \citet{daCunha_2021}, [19]: \citet{Umehata_2017}, [20]: \citet{Geach_2016}, [21]: \citet{Tadaki_2018}, [22]: \citet{Tadaki_2020}, [23]: \citet{Cooke_2018}, [24]: \citet{Swinbank_2014}, [25]: \citet{Swinbank_2012}, [26]: \citet{Gullberg_2018}, [27]: \citet{Oteo_2016}, [28]: \citet{Maiolino_2009}, [29]: \citet{Priddey_2001}, [30]: \citet{Lehar_2000}, [31]: \citet{Diaz_Santos_2015}, [32]: \citet{De_Breuck_2014}, [33]: \citet{Neri_2014}, [34]: \citet{Walter_2012}, [35]: \citet{Rawle_2014}, [36]: \citet{Riechers_2013}, [37]: \citet{Cooray_2014}, [38]: \citet{Marrone_2018}.
  \end{tablenotes}
  \end{threeparttable}
\end{table*}

It should be particularly noted that a large number (26) of the selected galaxies (mostly SMGs) in this regime are gravitationally-lensed systems (see Table~\ref{T3}). Hence, one important source of uncertainty in the estimates of their {\em intrinsic} $L_{\rm [C_{II}]}$ and $L_{\rm IR}$ ($\sim$SFR) is the lensing magnification factor $\mu$. To observationally determine $\mu$ of a lensed source requires spatially resolved imaging. Note that 16 of the selected {\sc\small SPT} galaxies in this regime, however, are not spatially resolved by the observations and their $\mu$ is unknown. \citet{Gullberg_2015} adopt a constant $\mu{}={}14.1$ to de-magnify the luminosities of all the 16 galaxies. This is the mean of the $\mu$ of the only 4 galaxies in their selected {\sc\small SPT} sample, which is determined using the spatially resolved {\sc\small ALMA} $\rm 860\,\mu m$ broadband imaging of dust continuum by \citet{Hezaveh_2013}. 

In Fig.~\ref{fig.7}, we show the $L_{\rm [C_{II}]}$-SFR relation (\textit{left} panel) of the observed samples at $1{}\simless{}z{}\simless{}5$, where we have converted the SFR of all galaxies from their $L_{\rm IR}$ using the \citetalias{Kennicutt_1998} relation following the observational studies. We show the stacked result for the samples of \citet{Stacey_2010} and \citet{Brisbin_2015} by grey empty squares. Both studies measure $\rm [C_{II}]$ line with {\sc\small ZEUS}, and both obtain systematically higher $L_{\rm [C_{II}]}{}/{}\rm SFR$ ratio of galaxies than the other studies using different instruments (by about one dex) at similar SFR. For the other studies, we explicitly show the data of each individual source in their samples. Specifically, we show the result of the SMGs by black circles (empty and filled), whilst the other star-forming galaxies are denoted by black triangles. For all the lensed galaxies, both $L_{\rm [C_{II}]}$ and $L_{\rm IR}$ are de-magnified by the observationally determined $\mu$ when available. For the 16 {\sc\small SPT} galaxies having no determined $\mu$ (indicated by empty black circles in Fig.~\ref{fig.7}), we correct their luminosities by an assumed $\mu{}={}14.1$ following \citet{Gullberg_2015}. For reference, we also show the $L_{\rm [C_{II}]}$-SFR relation of local galaxies by \citetalias{De_Looze_2011},  \citetalias{De_Looze_2014} and \citetalias{Herrera_Camus_2015} in the same ({\it left}) panel. 

The bulk of the selected samples at $1{}\simless{}z{}\simless{}5$ have higher SFR than the local samples of \citetalias{De_Looze_2011},  \citetalias{De_Looze_2014} and  \citetalias{Herrera_Camus_2015}. Only the few galaxies at $z{}\approx{}1-2$ of the \citet{Zanella_2018} sample overlap with the SFR range of the {most actively star-forming galaxies} of the \citetalias{De_Looze_2011} sample, and they appear to follow the same $L_{\rm [C_{II}]}$-SFR relation. At higher SFR (\ie~${\rm SFR}{}\simgreat{}100\,M_\odot\,\rm yr^{-1}$), the high-$z$ galaxy samples show a larger scatter in the $L_{\rm [C_{II}]}$-SFR relation compared to the local samples (\citetalias{De_Looze_2011},  \citetalias{De_Looze_2014} and \citetalias{Herrera_Camus_2015}) . Apart from that, the high-$z$ samples show a decline of $L_{\rm [C_{II}]}{}/{}\rm SFR$ ratio with increasing SFR at above $100\,M_\odot\,\rm yr^{-1}$ (corresponding to $L_{\rm IR}\simgreat10^{12}\,L_\odot$). This trend can be more clearly seen in the {\it right} panel, where we show the $L_{\rm [C_{II}]}{}/{}L_{\rm IR}$ ($\approx{}L_{\rm [C_{II}]}{}/{}\rm SFR$ at high SFR) ratio of the same high-$z$ galaxy samples as a function of their $L_{\rm IR}$ ($\sim$SFR). From $L_{\rm IR}=10^{12}\,L_\odot$ to $10^{13}\,L_\odot$, the $L_{\rm [C_{II}]}{}/{}L_{\rm IR}$ (or $L_{\rm [C_{II}]}{}/{}\rm SFR$) ratio of the high-$z$ samples decreases by roughly a factor of 50 (excluding the \citealt{Stacey_2010} and \citealt{Brisbin_2015} samples). This $\rm [C_{II}]$ deficit at high $L_{\rm IR}$ is similar to what has been found with the local galaxy samples (indicated by the filled grey symbols in Fig.~\ref{fig.7}). 

In the same figure, we also show the results of the {\sc\small FIRE} galaxies at high redshifts. Specifically, we show the $L_{\rm [C_{II}]}$-SFR ({\it left} panel) and the $L_{\rm [C_{II}]}{}/{}L_{\rm IR}$-$L_{\rm IR}$ ({\it right} panel) relations of the {\sc\small FIRE} galaxies at $z{}={}1$ (yellow hexagons), $z{}={}2$ (red triangles), $z{}={}3$ (blue squares) and $z{}={}4$ (magenta circles). For reference, we also show in the two panels the results of the {\sc\small FIRE} sample at $z{}={}0$ (cyan stars). 

The {\sc\small FIRE} galaxies follow a roughly linear $L_{\rm [C_{II}]}$-SFR scaling relation over the SFR range of $\approx{}0.01-100\,M_\odot\,\rm yr^{-1}$ at each redshift ({\it left} panel), though having {considerable} scatter ($1\sigma{}\approx{}0.2-0.35$ dex). The normalization of the relation, however, shows clear redshift evolution. From $z{}={}0$ to $z{}={}4$, the mean $L_{\rm [C_{II}]}{}/{}\rm SFR$ ratio of the {\sc\small FIRE} sample declines by about one dex (see the {\it left} panel of Fig.~\ref{fig.7}). This indicates that using the $L_{\rm [C_{II}]}$-SFR relation derived by \citetalias{De_Looze_2011} or \citetalias{Herrera_Camus_2015} will lead to a systematic underestimate of SFR of galaxies at high redshifts. 

On the other hand, the $L_{\rm [C_{II}]}{}/{}L_{\rm IR}$ ratio of the {\sc\small FIRE} galaxies does not evolve as much with redshift between $z{}={}0-4$ ({\it right} panel). From $z{}={}0$ to $z{}={}4$, the mean $L_{\rm [C_{II}]}{}/{}L_{\rm IR}$ ratio of the {\sc\small FIRE} galaxies decreases by 0.5 dex, which is less than the decrease of the $L_{\rm [C_{II}]}{}/{}\rm SFR$ ratio ($\sim1$ dex). Obviously, the reason for the discrepancy in the redshift evolution of the two ratios ($L_{\rm [C_{II}]}{}/{}\rm SFR$ and $L_{\rm [C_{II}]}{}/{}L_{\rm IR}$) is the redshift evolution of the $L_{\rm IR}$-SFR relation of the galaxies (see Fig.~\ref{fig.5} for the result of the {\sc\small FIRE} galaxies, and also the observational data of \eg~\citealt{Whitaker_2017}) --- at fixed SFR, galaxies at higher redshift have on average lower dust opacity and thus a smaller fraction of stellar radiation is absorbed and re-emitted at far-IR. The mean $L_{\rm IR}{}/{}\rm SFR$ ratio of galaxies therefore decreases with redshift. 

Apart from that, it is clear from the {\it right} panel that the {\sc\small FIRE} galaxies at $z{}={}1-4$ show a similar decrease of $L_{\rm [C_{II}]}{}/{}L_{\rm IR}$ ratio with $L_{\rm IR}$ like the local $z{}={}0$ {\sc\small FIRE} galaxies (cyan stars), and the decrease appears to be more significant at $L_{\rm IR}{}\simgreat{} 5\times10^{11}\,L_\odot$. The sharp decrease of $L_{\rm [C_{II}]}{}/{}L_{\rm IR}$ at the high $L_{\rm IR}$ end is in line with the trend in the observational data at similar redshifts. In Section~\ref{Sec:5}, we will examine in detail the origin of this `$[\rm C_{II}]$ deficit' at high $L_{\rm IR}$ and we will show that it is mainly driven by the decrease of gas mass per unit SFR, or {\em depletion timescale} ($t_{\rm dep}{}\equiv{}M_{\rm gas}{}/{}\rm SFR$), of galaxies with SFR.

Note that at $L_{\rm IR}{}\approx{}10^{12}\,L_\odot$, the observed $L_{\rm [C_{II}]}{}/{}L_{\rm IR}$ ratio of the galaxies at high redshifts (black symbols) appears to be higher than that of the observed $z{}={}0$ galaxy samples (grey symbols) as well as the {\sc\small FIRE} galaxies (coloured symbols). The mean $L_{\rm [C_{II}]}{}/{}L_{\rm IR}$ ratio is roughly in agreement with the upper bound of the {\sc\small FIRE} galaxies at similar $L_{\rm IR}$. This can possibly be due to selection effect. Those galaxies at $L_{\rm IR}{}\approx{}10^{12}\,L_\odot$ are mostly the `main-sequence' (MS) galaxies at $z{}\approx{}1.5-2$ selected by \citet{Zanella_2018}, which are expected to have {longer} $t_{\rm dep}$ (\ie~gas mass per unit SFR) than starburst galaxies at the same redshift \citep[\eg][]{Genzel_2015, Aravena_2016, Miettinen_2017, Tacconi_2018, Feldmann_2020} and hence higher $L_{\rm [C_{II}]}{}/{}L_{\rm IR}$ (note: $L_{\rm [C_{II}]}{}/{}{\rm SFR}{}\propto{} t^{0.7}_{\rm dep}$, equation~\ref{eq.30}). The {\sc\small FIRE} sample as well as the local observed galaxy samples, on the contrary, consist of galaxies across the star-forming MS as well as starburst galaxies, exhibiting a wide range of $t_{\rm dep}$. 

Finally, we note that the observational data in this redshift regime has large {uncertainties} due to the large fraction of gravitationally-lensed galaxies included in the samples (see Table~\ref{T3}). First of all, as mentioned above, many of the lensed galaxies do not have determined magnification factor $\mu$ (marked by empty circles in Fig.~\ref{fig.7}). Even for those whose $\mu$ is derived from either the rest-UV (with {\it Hubble Space Telescope}) or dust continuum imaging (with {\sc\small ALMA}), it is not yet certain whether their $\rm [C_{II}]$ luminosity is magnified by the same level, given that the $\rm [C_{II}]$ and stellar/dust emission of galaxies may have different spatial configuration \citep[\eg][]{Cochrane_2019, Fujimoto_2019, Matthee_2019, Novak_2020, Fudamoto_2022} and thus the different emission components may have different $\mu$ due to the effect of differential lensing \citep[\eg][]{Blain_1999, Hezaveh_2012, Serjeant_2012, Canameras_2018, Yang_2019, Harrington_2021}. Hence, it is important to obtain spatially resolved imaging of both $\rm [C_{II}]$ and dust emission for lensed galaxies and re-examine the {\em intrinsic} $L_{\rm [C_{II}]}{}/{}L_{\rm IR}$ ratio of these galaxies (note: most of the lensed SMGs do not have spatially resolved $\rm [C_{II}]$ imaging, see Table~\ref{T3}).  

\subsection{Early galaxies (redshift $z{}\simgreat{}5$)}
\label{Sec:4c}

\begin{figure*}
 \includegraphics[width=178mm]{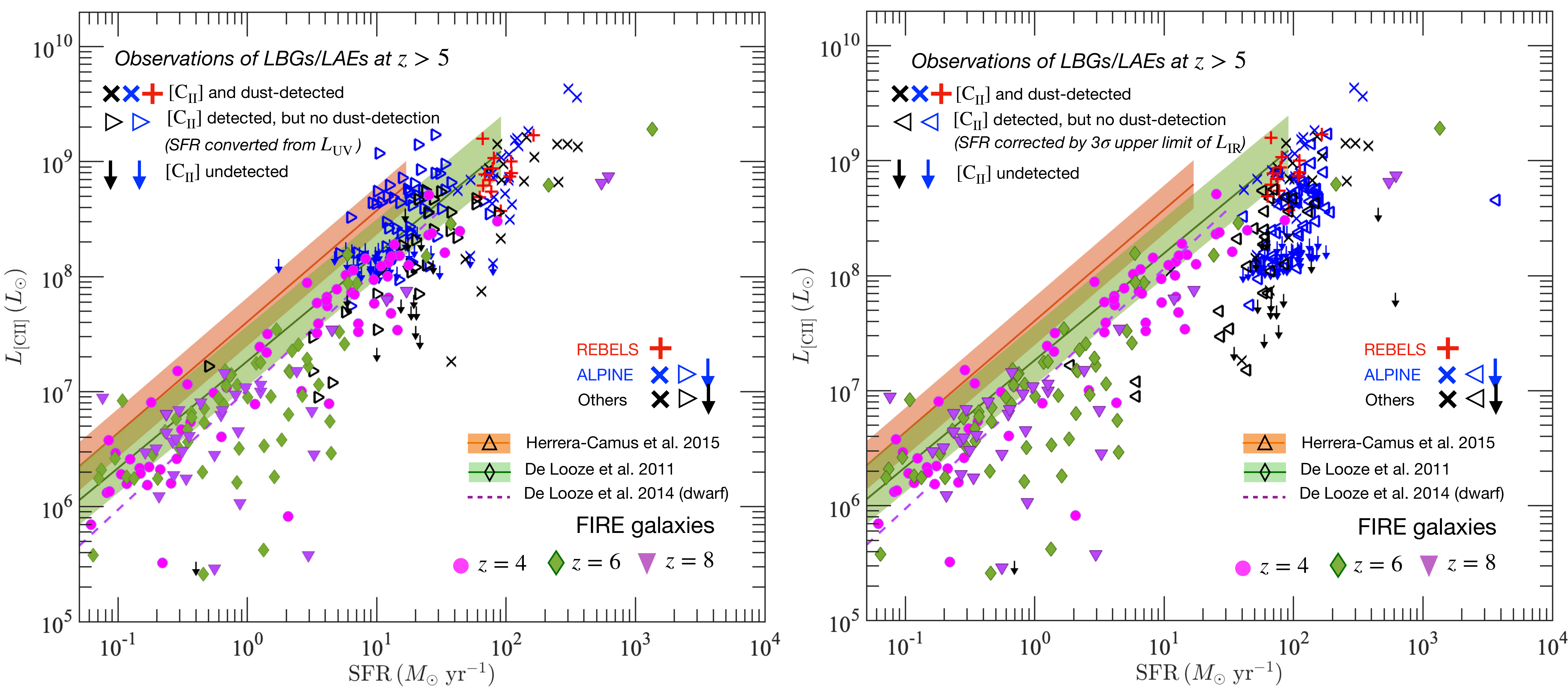}
 \caption{Comparison of the $L_{\rm [C_{II}]}$-SFR relation of the {\sc\small FIRE} galaxies with the observational data at high redshifts. In the two panels, we show the result of the {\sc\small FIRE} galaxies at $z{}={}4$, $z{}={}6$ and $z{}={}8$ by magenta circles, green diamonds and purple downward triangles, respectively. We also show in the two panels the observational data of the rest-UV-selected star-forming galaxies at $z{}\simgreat{}5$, including the ones targeted by the {\sc\small ALPINE} (blue symbols) and {\sc\small REBELS} (red symbols) {\sc\small ALMA} surveys as well as the others targeted by the other observations (black symbols) (see Table~\ref{T4} for the details). The galaxies having both confirmed $\rm [C_{II}]$ and dust continuum detection are indicated by {crosses} ({\sc\small REBELS}) and {`X's} (red for {\sc\small ALPINE} and black for others). The galaxies having no $\rm [C_{II}]$ detection are shown by downward arrows in both panels. The location of the arrows indicate the $3\sigma$ upper limit of their $L_{\rm [C_{II}]}$. For the ones having $\rm [C_{II}]$ but without dust detection (meaning that their $\rm SFR_{IR}$ is uncertain), we show the relation between their $L_{\rm [C_{II}]}$ and the lower (upper) SFR limit in the {\it left} ({\it right}) panel by rightward (leftward) triangles. For reference, we also show the result of local ($z{}={}0$) observations of normal star-forming galaxies (SFGs) by \citetalias{De_Looze_2011}, \citetalias{De_Looze_2014} and \citetalias{Herrera_Camus_2015} in the two panels. The {\sc\small FIRE} sample at $z{}={}4-8$ shows systematically lower $L_{\rm [C_{II}]}{}/{}\rm SFR$ ratio than the local SFGs, in particular at low SFR. The observed galaxy samples at $z{}\simgreat{}5$ show similar $\rm [C_{II}]$ deficit if $T_{\rm eqv}$ follows equation~(\ref{eq.7}) (assuming $\delta_{\rm dzr}{}={}0.4$). }
    \label{fig.8}
\end{figure*}

Observational studies on the $L_{\rm [C_{II}]}$-SFR relation at $z{}\simgreat{}5$ {depend} mainly on the rest-frame UV-selected galaxies whose redshift has previously been confirmed either spectroscopically or via the Lyman break `drop-out' technique \citep{Hodge_2020}. Their $\rm [C_{II}]$ and dust emission are constrained in follow-up observational campaigns with {\sc\small ALMA}, which has the power to spatially resolve the distant galaxies down to the scale of $\sim1$ physical kpc. The majority of the UV-selected galaxies at this epoch are unlensed.

There have been two major observational campaigns for searching for $\rm [C_{II}]$ line of galaxies at $z{}\simgreat{}5$. The {\sc\small ALPINE} {\sc\small ALMA} Large Program \citep{Le_Fevre_2020, Bethermin_2020} in cycle-5 targeted a sample of 118 UV-selected star-forming galaxies at $4.5{}<{}z{}<{}6$ with $M_{\rm UV,\,AB}{}<{}-20.2$ and identified $\rm [C_{II}]$ emission (at $>{}3.5\sigma$ level) in 75 galaxies of them \citep{Schaerer_2020}. More recently, the {\sc\small REBELS} Large Program \citep{Bouwens_2022} in Cycle-7 studied a sample of 40 UV-bright ($M_{\rm UV,\,AB}<-21.4$) galaxies at $6.5<{}z{}<{}7.7$ and confirmed $\rm [C_{II}]$ detection (at $>{}7\sigma$) in 18 galaxies in their sample \citep{Ferrara_2022}. Other observations targeting the LBGs/LAEs at $z{}\simgreat{}5$ have identified $\rm [C_{II}]$ emission in another $>{}35$ sources in total. The most distant galaxy that has a $\rm [C_{II}]$ detection to date is MACS1149-JD1 \citep{Hashimoto_2018}, a gravitationally-lensed ($\mu{}={}10$) galaxy at $z{}={}9.11$ (\citealt{Carniani_2020}; see also \citealt{Inoue_2016} and \citealt{Laporte_2019}). We provide a summary of the star-forming galaxies at $z{}\simgreat{}5$ having confirmed $\rm [C_{II}]$ detection in Table~\ref{T4} (excluding quasar host galaxies). 

The SFR of these UV-selected galaxies has been derived based on their $L_{\rm UV}$ and $L_{\rm IR}$. Because the galaxies at $z{}\simgreat{}5$ typically do not have good photometric sampling of the dust continuum \citep[\eg][]{Casey_2018b, Liang_2019, Faisst_2020b},  $L_{\rm IR}$ has frequently been converted from the {\sc\small ALMA} broad-band flux density (measured at band 6 or 7 for galaxies at $z{}\simgreat{}5$) using the standard modified-blackbody (MBB) function of the form \citep[\eg][]{Hildebrand_1983, Hayward_2011}
\begin{equation}
S_{\nu_0} = \frac{(1+z)}{d^2_{\rm L}}\kappa_\nu M_{\rm dust} B_\nu (T), 
\label{eq.5}
\end{equation}
where $\nu_0$ is the observing frequency (note: $\nu_0{}={}345$ GHz for {\sc\small ALMA} band 7 and $\nu_0{}={}230$ GHz for {\sc\small ALMA} band 6), $S_{\nu_0}$ is the broad-band flux density at $\nu_0$, $\nu{}={}(1+z)\nu_0$ is the rest-frame frequency, $\kappa_\nu$ is the dust opacity (per unit dust mass) at $\nu$, $M_{\rm dust}$ is the dust mass of galaxy, $T$ is the `dust temperature', $B_\nu (T)$ is the Planck function and $d_{\rm L}$ is the luminosity distance. $L_{\rm IR}$ is then converted from $S_{\nu_0}$ using (see Section 3.1.3 of \citealt{Liang_2019} for the details)
\begin{equation}
L_{\rm IR}  = \frac{\mathcal{D}d^2_{\rm L}T^{4+\beta_{\rm dust}} }{(1+z)\kappa_\nu B_\nu (T)} S_{\nu_0} , 
\label{eq.6}
\end{equation}
where $\beta_{\rm dust}{}\approx{}2.0$ is the dust emissivity spectral index \citep[\eg][]{Dunne_2000, Draine_2007} and $\mathcal{D}$ is a parameter that depends on the shape of the dust opacity curve. The derived $L_{\rm IR}$ (and hence the obscured SFR) therefore depends mainly on the assumed `dust temperature'. It should be noted that recent cosmological simulations show that the true SED of high-$z$ galaxies may significantly differ from the standard MBB function (\eg~\citealt{Liang_2019, Ma_2019}, and also \citealt{Casey_2012}, \citealt{Casey_2018b}) and $T$ does not faithfully reflect the {\em physical} temperature of dust in galaxies \citep[\eg][]{Behrens_2018, Liang_2019}. \citet{Liang_2019} defines the `dust temperature' that one would need to obtain the correct $L_{\rm IR}$ and match the observed $S_{\nu_0}$ under the assumption that the SED has the shape of a standard MBB function (equation~\ref{eq.5}) to be the `equivalent dust temperature' ($T_{\rm eqv}$).

\begin{table*}
\centering
\caption{Properties of the star-forming galaxies at $z\simgreat5$ targeted for search for $\rm [C_{II}]$ emission. }
\begin{threeparttable}
\begin{tabular}{ p{2.1 cm} p{0.8 cm}  p{1 cm}  p{2.2 cm} p{1.3 cm} p{1.3 cm}  p{2.1 cm} p{1.3 cm} p{1.3 cm}}
 \hline
  \hline
 \multicolumn{1}{c|}{Name\tnote{$\dagger$}} &  \multicolumn{1}{c|}{$z$}  &  $\rm SFR_{\rm UV}$\tnote{$\S$, $\#$} ($M_\odot\,\rm yr^{-1}$) & $S$ ($\rm \mu Jy$)\tnote{$\ddagger$, $\P$, $\#$} & ${\rm log}\,(L_{\rm IR}/L_\odot)$\tnote{$\parallel$}  &  \;\;\;\;$\rm SFR$\tnote{$\dagger\dagger$} ($M_\odot\, \rm yr^{-1}$) &  \multicolumn{1}{c|}{${\rm log}\,(L_{\rm [C_{II}]}/L_\odot)$\tnote{$\P$, $\#$}} & \multicolumn{1}{c|}{$\mu$} & References \tnote{$\ast$} \\
  \hline
HZ7 & 5.253 & 31.2 & $<108$ ({\sc\small ALMA} 7) & $<11.6$ & $<62.7$ & 8.74 ({\sc\small ALMA} 7) & \multicolumn{1}{c|}{$-$} & [1, 2, 3]\\
HZ9 & 5.541 & 22.1 & 516 ({\sc\small ALMA} 7) & 11.9 & 174.5 & 9.21 ({\sc\small ALMA} 7) & \multicolumn{1}{c|}{$-$} & [1, 2, 3]\\
HZ10 & 5.657 & 58.2 & 1261 ({\sc\small ALMA} 7) & 12.7 & 432.8 & 9.13 ({\sc\small ALMA} 7) & \multicolumn{1}{c|}{$-$} & [1, 2, 3]\\
NB816-S-61269 & 5.684 & 19.9 & $<66$ ({\sc\small ALMA} 7) & $<11.4$ & $<39.6$ & 8.32 ({\sc\small ALMA} 7) & \multicolumn{1}{c|}{$-$} & [4, 5]\\
WMH13 & 5.985 & 87.1 & $<48$ ({\sc\small ALMA} 6) & $<11.7$ & $<131.3$ & 8.56 ({\sc\small ALMA} 6) & \multicolumn{1}{c|}{$-$} & [4, 5]\\
A383-5.1 & 6.029 & 3.5 & $<2.9$ ({\sc\small ALMA} 6) & $<10.5$ & $<6.2$ & 6.95 ({\sc\small ALMA} 6) & $11.4\pm1.9$ & [6]\\
J1211-0118 & 6.029 & 55.2 & 220 ({\sc\small ALMA} 6) & 12.4 & 257.3 & 9.15 ({\sc\small ALMA} 6) & \multicolumn{1}{c|}{$-$} & [7]\\
NTTDF2313 & 6.07 & 18.4 & $<54$ ({\sc\small ALMA} 6) & $<11.8$ & $<68.0$ & $<7.7$ ({\sc\small ALMA} 6) & \multicolumn{1}{c|}{$-$} & [8]\\
WMH5  & 6.07 & 63.2 & 218 ({\sc\small ALMA} 6) & 12.4 & 263.5 & 8.82 ({\sc\small ALMA} 6) & \multicolumn{1}{c|}{$-$} & [9, 10]\\
RXCJ0600-z6 & 6.0719 & 2.8 & $9.5$ ({\sc\small ALMA} 6) & 11.0 & 11.5 & 8.04 ({\sc\small ALMA} 6) & \multicolumn{1}{c|}{$21\pm10$} & [11]\\
J0235-0532 & 6.089 & 58.4 & $<101$ ({\sc\small ALMA} 6) & $<12.1$ & $<150.5$ & 8.63 ({\sc\small ALMA} 6) & \multicolumn{1}{c|}{$-$} & [7]\\
BDF2203 & 6.12 & 24.2 & $<69$ ({\sc\small ALMA} 6) & $<11.9$ & $<87.6$ & 8.1 ({\sc\small ALMA} 6) & \multicolumn{1}{c|}{$-$} & [8]\\
CLM1  & 6.166 & 56.0 & 40 ({\sc\small ALMA} 6) & 11.7 & 92.9 & 8.33 ({\sc\small ALMA} 6) & \multicolumn{1}{c|}{$1.13$} & [4, 9]\\
J0217-0208 & 6.203 & 86.6 & 239 ({\sc\small ALMA} 6) & 12.4 & 307.3 & 9.15 ({\sc\small ALMA} 6) & \multicolumn{1}{c|}{$-$} & [7]\\
MACS0308- zD1 & 6.2078 & 3.2 & $<27$ ({\sc\small ALMA} 6) & $<11.5$ & $<33.7$ & 7.47 ({\sc\small ALMA} 6) & \multicolumn{1}{c|}{22} & [12,13] \\
GOODS3203 & 6.27 & 27.2 & $<123$ ({\sc\small ALMA} 6) & $<12.2$ & $<140.4$ & $<8.1$ ({\sc\small ALMA} 6) & \multicolumn{1}{c|}{$-$} & [8]\\
NIRCam 12053 & 6.3254 & 34 & 66.0 ({\sc\small ALMA} 6) & 11.9 & 92.0 & 8.84 ({\sc\small ALMA} 6) & \multicolumn{1}{c|}{1.97} & [14] \\
COSMOS20521 & 6.36 & 20.2 & $<60$ ({\sc\small ALMA} 6) & $<11.8$ & $<75.5$ & $<7.7$ ({\sc\small ALMA} 6) & \multicolumn{1}{c|}{$-$} & [8]\\
VR7 & 6.529 & 58.2 & $<31.8$ ({\sc\small ALMA} 6) & $<11.6$ & $<87.5$ & 8.68 ({\sc\small ALMA} 6) & \multicolumn{1}{c|}{$-$} & [15]\\
MASOSA & 6.543 & 13.0 & $<27.6$ ({\sc\small ALMA} 6) & $<11.5$ & $<35.5$ & $<7.34$ ({\sc\small ALMA} 6) & \multicolumn{1}{c|}{$-$} & [15]\\
HCM6A & 6.56 & 5.9 & $<680$ (PdBI) & $<12.9$ & $<631.1$ & $<7.81$ (PdBI) & \multicolumn{1}{c|}{$4.5$} & [16, 17]\\
UDS4812 & 6.561 & 19.3 & $<72$ ({\sc\small ALMA} 6) & $<11.9$ & $<85.7$ & $<7.8$ ({\sc\small ALMA} 6) & \multicolumn{1}{c|}{$-$} & [8] \\
Himiko & 6.591 & 19.8 & $<27$ ({\sc\small ALMA} 6) & $<11.5$ & $<44.8$ & 8.08 ({\sc\small ALMA} 6) & \multicolumn{1}{c|}{$-$} & [18, 19]\\
CR7 & 6.600 & 41.7 & $<21$ ({\sc\small ALMA} 6) & $<11.4$ & $<61.1$ & 8.34 ({\sc\small ALMA} 6) & \multicolumn{1}{c|}{$-$} & [20, 21]\\
COSMOS24108  & 6.629 & 25.6 & $<54$ ({\sc\small ALMA} 6) & $<11.8$ & $<68.2$ & 8.04 ({\sc\small ALMA} 6) & \multicolumn{1}{c|}{$-$} & [22]\\
UDS16291  & 6.638 & 13.4 & $<60$ ({\sc\small ALMA} 6) & $<11.8$ & $<65.4$ & 7.85 ({\sc\small ALMA} 6) & \multicolumn{1}{c|}{$-$} & [22]\\
NTTDF6345  & 6.701 & 21.2 & $<48$ ({\sc\small ALMA} 6) & $<11.7$ & $<60.2$ & 8.26 ({\sc\small ALMA} 6) & \multicolumn{1}{c|}{$-$} & [22]\\
MS0451-H & 6.703 & 0.4 & $<0.33$ ({\sc\small ALMA} 6) & $<9.6$ & $<0.7$ & $<5.48$ ({\sc\small ALMA} 6) & \multicolumn{1}{c|}{$100\pm20$} & [6]\\
UVISTA-Z-007 & 6.7496 & 23.7 & $<52.2$ ({\sc\small ALMA} 6) & $<11.8$ & $<72.0$ & 8.75 ({\sc\small ALMA} 6) & \multicolumn{1}{c|}{$-$} & [23, 24]\\
UVISTA-Z-019 & 6.7534 & 15.8 & 66 ({\sc\small ALMA} 6) & 11.9 & 74.1 & 8.94 ({\sc\small ALMA} 6) & \multicolumn{1}{c|}{$-$} & [23, 24]\\
RXJ1347-1216 & 6.766 & 2.4 & $<45$ ({\sc\small ALMA} 6) & $<11.7$ & $<44.8$ & 7.18 ({\sc\small ALMA} 6) & \multicolumn{1}{c|}{$5.0\pm0.3$} & [25]\\
COS-2987030247  & 6.808 & 24.6 & $<75$ ({\sc\small ALMA} 6) & $<11.9$ & $<94.3$ & 8.56 ({\sc\small ALMA} 6) & \multicolumn{1}{c|}{$-$} & [26]\\
A1703-zD1 & 6.827 & 10.1 & $<24.5$ ({\sc\small NOEMA}) & $<11.5$ & $<32.8$ & 7.54 ({\sc\small NOEMA}) & \multicolumn{1}{c|}{$9.0\pm2.7$} & [27]\\
SDF-46975  & 6.844 & 15.4 & $<57.6$ ({\sc\small ALMA} 6) & $<11.8$ & $<68.7$ & $<7.75$ ({\sc\small ALMA} 6) & \multicolumn{1}{c|}{$-$} & [28]\\
COS-3018555981  & 6.854 & 20.8 & $<87$ ({\sc\small ALMA} 6) & $<12.0$ & $<101.3$ & 8.67 ({\sc\small ALMA} 6) & \multicolumn{1}{c|}{$-$} & [26]\\
UVISTA-Z-009 & 6.86 & 16.9 & $<38.0$ ({\sc\small ALMA} 6) & $<11.6$ & $<52.1$ & $<8.12$ ({\sc\small ALMA} 6) & \multicolumn{1}{c|}{$\simless{}1.5$} & [23, 24]\\	
IOK-1 & 6.965 & 20.0 & $<63$ ({\sc\small ALMA} 6) & $<11.9$ & $<78.4$ & $<7.53$ ({\sc\small ALMA} 6) & \multicolumn{1}{c|}{$-$} & [29] \\
BDF-512 & 7.008 & 6.0 & $<55.2$ ({\sc\small ALMA} 6) & $<11.8$ & $<54.2$ & $<7.78$ ({\sc\small ALMA} 6) & \multicolumn{1}{c|}{$-$} & [28] \\
UVISTA-Z-013 & 7.02 & 22.1 & $<45.0$ ({\sc\small ALMA} 6) & $<11.7$ & $<63.8$ & $<8.30$ ({\sc\small ALMA} 6) & \multicolumn{1}{c|}{$-$} & [23, 24]\\
UVISTA-Z-001 & 7.0599 & 45.8 & 104 ({\sc\small ALMA} 6) & 12.1 & 137.8 & 8.83 ({\sc\small ALMA} 6) & \multicolumn{1}{c|}{$-$} & [23, 24]\\
UVISTA-Z-010 & 7.06 & 17.4 & $<44.1$ ({\sc\small ALMA} 6) & $<11.7$ & $<58.3$ & $<8.30$ ({\sc\small ALMA} 6) & \multicolumn{1}{c|}{$-$} & [23, 24]\\
BDF-3299 & 7.109 & 5.7 & $<23.4$ ({\sc\small ALMA} 6) & $<11.4$ & $<27.4$ & 7.83 ({\sc\small ALMA} 6) & \multicolumn{1}{c|}{$-$} & [28, 30, 31]\\
A1689-zD1 & 7.137 & 4.7 & 60.2 ({\sc\small ALMA} 6) & 11.9 & 67.5 & 7.87 ({\sc\small ALMA} 6) & \multicolumn{1}{c|}{$9.3$} & [32, 33, 34] \\
COSMOS13679 & 7.145 & 21.1 & $<42$ ({\sc\small ALMA} 6) & $<11.7$ & $<60.1$ & 7.85 ({\sc\small ALMA} 6) & \multicolumn{1}{c|}{$-$} & [22]\\
B14-65666   & 7.152 & 50.2 & 130 ({\sc\small ALMA} 6) & 12.2 & 170.2 & 9.12 ({\sc\small ALMA} 6) & \multicolumn{1}{c|}{$-$} &[35, 36]\\
SXDF-NB1006-2  & 7.212 & 21.6 & $<42$ ({\sc\small ALMA} 6) & $<11.7$ & $<60.6$ & $<7.45$ ({\sc\small ALMA} 6) & \multicolumn{1}{c|}{$-$} & [37] \\
z8-GND-5296 & 7.508 & 16.6 & $<480$ (PdBI) & $<12.7$ & $<464.1$ & $<8.55$ (PdBI) & \multicolumn{1}{c|}{$-$} & [38, 39] \\
MACS0416-Y1 & 8.311 & 11.7 & 137 ({\sc\small ALMA} 7) & 11.8 & 56.8 & 8.15 ({\sc\small ALMA} 5) & $1.43\pm0.04$ & [40, 41, 42] \\
A2744-YD4 & 8.380 & 11.2 & 99 ({\sc\small ALMA} 7) & 11.6 & 43.8 & 7.26 ({\sc\small ALMA} 5) & \multicolumn{1}{c|}{$1.8\pm0.3$} & [31, 43, 44] \\
S04590 & 8.4931 &  0.5 &  $<{}4.81$ ({\sc\small ALMA} 7) & $<10.3$ & $<2.0$ & 7.22 ({\sc\small ALMA} 5) & \multicolumn{1}{c|}{$8.69\pm2.5$} &  [45, 46] \\
MACS1149-JD1 & 9.110 & 4.5 & $<5.3$ ({\sc\small ALMA} 7) & $<10.4$ & $<6.5$ & 7.08 ({\sc\small ALMA} 5) & \multicolumn{1}{c|}{10} & [31, 44, 47] \\
\hline
\multicolumn{9}{c|}{{\sc\small REBELS}\tnote{$\ddagger\ddagger$}}  \\
 \hline
REBELS-05 & 6.496 & 15.1 & 67.2 ({\sc\small ALMA} 6) & 11.9 & 77.2 & 8.84 ({\sc\small ALMA} 6) & \multicolumn{1}{c|}{$-$} & [48, 49, 50]\\
REBELS-38 & 6.577  & 19.5 & 163.0 ({\sc\small ALMA} 6) & 12.3 & 170.2 & 9.23 ({\sc\small ALMA} 6) & \multicolumn{1}{c|}{$-$} & [48, 49, 50]\\
REBELS-29 & 6.685 & 27.0 & 56.1 ({\sc\small ALMA} 6) & 11.8 & 78.9 & 8.74 ({\sc\small ALMA} 6) & \multicolumn{1}{c|}{$-$} & [48, 49, 50]\\
REBELS-32 & 6.729 & 15.1 & 60.4 ({\sc\small ALMA} 6) & 11.8 & 71.0 & 8.89 ({\sc\small ALMA} 6) & \multicolumn{1}{c|}{$-$} &[48, 49, 50]\\
REBELS-08 & 6.749 & 17.3 & 101.4 ({\sc\small ALMA} 6) & 12.1 & 111.2 & 8.87 ({\sc\small ALMA} 6) &  \multicolumn{1}{c|}{$-$} & [48, 49, 50]\\
REBELS-39 & 6.847 & 40.0 & 79.6 ({\sc\small ALMA} 6) & 12.0 & 113.7 & 8.90 ({\sc\small ALMA} 6) &  \multicolumn{1}{c|}{$-$} & [48, 49, 50]\\
REBELS-14 & 7.084 & 37.9 & 60.0 ({\sc\small ALMA} 6) & 11.8 & 93.6 & 8.57 ({\sc\small ALMA} 6) &  \multicolumn{1}{c|}{$-$} &[48, 49, 50]\\ 
REBELS-27 & 7.090 & 21.6 & 50.6 ({\sc\small ALMA} 6) & 11.8 & 68.5 & 8.79 ({\sc\small ALMA} 6) &  \multicolumn{1}{c|}{$-$} & [48, 49, 50]\\
 \hline
\end{tabular}
\;\;\;\;(Continue on next page)  
  \end{threeparttable}
\label{T4}
\end{table*}
 
\begin{table*}
\contcaption{}
\begin{threeparttable}
\begin{tabular}{ p{2.1 cm} p{0.8 cm}  p{1 cm}  p{2.2 cm} p{1.3 cm} p{1.3 cm}  p{2 cm} p{1.3 cm} p{1.3 cm}}
 \hline
  \hline
 \multicolumn{1}{c|}{Name\tnote{$\dagger$}} &  \multicolumn{1}{c|}{$z$}  &  $\rm SFR_{\rm UV}$\tnote{$\S$, $\#$} ($M_\odot\,\rm yr^{-1}$) & $S$ ($\rm \mu Jy$)\tnote{$\ddagger$, $\P$, $\#$} & ${\rm log}\,(L_{\rm IR}/L_\odot)$\tnote{$\parallel$}  &  \;\;\;\;$\rm SFR$\tnote{$\dagger\dagger$} ($M_\odot\, \rm yr^{-1}$) &  \multicolumn{1}{c|}{${\rm log}\,(L_{\rm [C_{II}]}/L_\odot)$\tnote{$\P$, $\#$}} & \multicolumn{1}{c|}{$\mu$} & References \tnote{$\ast$} \\
  \hline
 REBELS-25 & 7.306 & 16.2 & 56.1 ({\sc\small ALMA} 6) & 11.8 & 68.3 & 9.20 ({\sc\small ALMA} 6) &  \multicolumn{1}{c|}{$-$} & [48, 49, 50]\\
REBELS-12 & 7.349 & 32.5 & 86.8 ({\sc\small ALMA} 6) & 12.0 & 113.2 & 9.00 ({\sc\small ALMA} 6) &  \multicolumn{1}{c|}{$-$} & [48, 49, 50]\\
REBELS-40 & 7.365 & 18.4 & 48.3 ({\sc\small ALMA} 6) & 11.8 & 64.5 & 8.69 ({\sc\small ALMA} 6) &  \multicolumn{1}{c|}{$-$} & [48, 49, 50]\\
REBELS-19 & 7.369 & 15.1 & 71.2 ({\sc\small ALMA} 6) & 11.9 & 81.3 & 8.94 ({\sc\small ALMA} 6) &  \multicolumn{1}{c|}{$-$} & [48, 49, 50]\\
REBELS-18 & 7.675 & 33.5 & 52.9 ({\sc\small ALMA} 6) & 11.8 & 82.8 & 9.03 ({\sc\small ALMA} 6) &  \multicolumn{1}{c|}{$-$} & [48, 49, 50]\\
  \hline
\end{tabular}  
  \begin{tablenotes}
      \item[$\dagger$] The table does not include the 118 galaxies ($4.5\simless z \simless 6$) selected by the {\sc\small ALPINE} project. The information of the {\sc\small ALPINE} galaxies can be downloaded from the official webpage of the project: \url{https://cesam.lam.fr/a2c2s/data_release.php}. The {\sc\small ALPINE} galaxies are unlensed.
     \item[$\S$] $\rm SFR_{\rm UV}$ is converted from $L_{\rm UV}$ via ${\rm SFR_{UV}}\,(M_\odot\,{\rm yr^{-1})}{}={}1.58\times10^{-10}\,L_{\rm UV}\,(L_\odot)$ following \citet{Hao_2011} (see Table 3) for the \citet{Kroupa_2002} IMF.    
     \item[$\ddagger$] The number in the brackets indicates the specific {\sc\small ALMA} band at which dust continuum is measured.
     \item[$\P$] For the galaxies having no detection of dust thermal continuum ($\rm [C_{II}]$ emission), we show the $3\sigma$ upper confidence limit of $S$ ($L_{\rm [C_{II}]}$). 
     \item[$\#$] For the gravitationally-lensed galaxies, $L_{\rm UV}$ (and hence $\rm SFR_{UV}$), $S$, $L_{\rm IR}$ and $L_{\rm [C_{II}]}$ are de-magnified by $\mu$. 
     \item[$\parallel$]  $L_{\rm IR}$ (or the upper limit of $L_{\rm IR}$ for the dust-undetected sources) is converted from $S$ (the $3\sigma$ upper limit of $S$) via the standard MBB function with  $T_{\rm eqv}$ calculated by equation~(\ref{eq.4}) (assuming $\beta_{\rm dust}{}={}2.0$ and $\delta_{\rm dzr}{}={}0.4$). 
      \item[$\dagger\dagger$] SFR is derived using ${\rm SFR}\,(M_\odot\,{\rm yr^{-1})}{}={}{\rm SFR_{UV}}+{\rm SFR_{IR}}{}={}1.58 \times 10^{-10}\,(L_{\rm UV} + 0.46 L_{\rm IR})\,(L_\odot)$ following \citet{Hao_2011} (see Table 3) for the \citet{Kroupa_2002} IMF.   
      \item[$\ddagger\ddagger$] We only list here the 13 galaxies of the {\sc\small REBELS} sample that have confirmed detection of both $\rm [C_{II}]$ and dust continuum. The information of the other 5 galaxies having $\rm [C_{II}]$ but no dust detection is not yet publicly available.  
      \item[$\ast$]  References: (1): \citet{Capak_2015}, [2]: \citet{Barisic_2017}, [3]: \citet{Faisst_2017}, [4]: \citet{Fujimoto_2019}, [5]: \citet{Fujimoto_2016}, [6]: \citet{Knudsen_2016}, [7]: \citet{Harikane_2020}, [8]: \citet{Carniani_2018a}, [9]: \citet{Willott_2015b}, [10]: \citet{Willott_2013b}, [11]: \citet{Fujimoto_2021}, [12]: \citet{Welch_2023}, [13]: \citet{Fudamoto_2023a}, [14]: \citet{Fujimoto_2023}, [15]: \citet{Matthee_2019}, [16]: \citet{Kanekar_2013}, [17]: \citet{Hu_2002}, [18]: \citet{Ouchi_2013}, [19]: \citet{Carniani_2018b}, [20]: \citet{Sobral_2015}, [21]: \citet{Matthee_2017}, [22]: \citet{Pentericci_2016}, [23]: \citet{Schouws_2023}, [24]: \citet{Schouws_2022}, [25]: \citet{Bradac_2017}, [26]: \citet{Smit_2018}, [27]: \citet{Molyneux_2022}, [28]: \citet{Maiolino_2015}, [29]: \citet{Ota_2014}, [30]: \citet{Carniani_2017}, [31]: \citet{Carniani_2020}, [32]: \citet{Watson_2015}, [33]: \citet{Knudsen_2017}, [34]: \citet{Wong_2022}, [35]: \citet{Hashimoto_2019a}, [36]: \citet{Bowler_2018}, [37]: \citet{Inoue_2016},  [38]: \citet{Schaerer_2015b}, [39]: \citet{Finkelstein_2013}, [40]: \citet{Tamura_2019}, [41]: \citet{Bakx_2020}, [42]: \citet{Kawamata_2016}, [43]: \citet{Laporte_2017}, [44]: \citet{Laporte_2019}, [45]: \citet{Fujimoto_2022}, [46]: \citet{Heintz_2023b}, [47]: \citet{Hashimoto_2018}, [48]: \citet{Ferrara_2022}, [49]: \citet{Sommovigo_2022}, [50]: \citet{Bouwens_2022}.      
  \end{tablenotes}
  \end{threeparttable}
\label{T4x}
\end{table*}

\begin{table*}
\centering
\caption{Comparison between the mean `equivalent dust temperature' (<$T_{\rm eqv}$>) assumed by the {\sc\small ALPINE} and {\sc\small REBELS} projects and by this work.}
\begin{threeparttable}
\begin{tabular}{ p{2 cm} p{2.5 cm} p{1.5 cm} p{1 cm} p{2 cm}  p{2 cm}  p{1.8 cm} }
 \hline
  \hline
\multicolumn{1}{c|}{Project name} &  \multicolumn{1}{c|}{Reference} & \multicolumn{1}{c|}{No. of galaxies} & \multicolumn{1}{c|}{<$z$>}  & \;\;\;\;<$T_{\rm eqv}/\rm K$>    & \; \;\;<$T_{\rm eqv}/\rm K$>\tnote{$\dagger$} &  $\Delta$<${\rm log}(\frac{L_{\rm [C_{II}]}}{\rm SFR}$)>\tnote{$\ddagger$}  \\
   &    &   & & \;\;\;\;(literature)  & \; \;\;(this work)  & \;\;\;\;\;\;(dex) \\
 \hline
 \multicolumn{1}{c|}{{\sc\small ALPINE}} & \citet{Schaerer_2020} &  \multicolumn{1}{c|}{118} &  \multicolumn{1}{c|}{4.58} & \multicolumn{1}{c|}{42} & \multicolumn{1}{c|}{47.9} & \multicolumn{1}{c|}{-0.21} \\
 \multicolumn{1}{c|}{{\sc\small REBELS}} & \citet{Ferrara_2022}  & \multicolumn{1}{c|}{40} &  \multicolumn{1}{c|}{7.08} & \multicolumn{1}{c|}{55} & \multicolumn{1}{c|}{57.4} &  \multicolumn{1}{c|}{-0.12} \\
 \hline
\end{tabular}
  \begin{tablenotes}
     \item[$\dagger$] Calculated using equation~(\ref{eq.7}) with $\delta_{\rm dzr}{}={}0.4$. Note that with a lower $\delta_{\rm dzr}$, $T_{\rm eqv}$ is higher than the listed value in this column. 
     \item[$\ddagger$] The resulting difference in the derived mean $L_{\rm [C_{II}]}{}/{}\rm SFR$ ratio (in dex) of the galaxy samples due to the difference in $T_{\rm eqv}$ used by the previous studies (\citealt{Schaerer_2020} and \citealt{Ferrara_2019}) and this work.
    \end{tablenotes}
  \end{threeparttable}
\label{T5}
\end{table*}
 
Using a sample of high-$z$ galaxies produced by the {\sc\small MassiveFIRE} suite \citep{Feldmann_2016, Feldmann_2017b}, \citet{Liang_2019} derived the best-fitting formula for $T_{\rm eqv}$ using redshift and dust-to-gas mass ratio ($\delta_{\rm dzr}$) as variables, \ie
\begin{equation}
T_{\rm eqv} = T_0\,(1+z)^\alpha (\delta_{\rm dzr}/0.4)^\gamma.\;\;\rm (L19)
\label{eq.7}
\end{equation}
For {\sc\small ALMA} band 7 (6) fluxes, the best-fitting parameter values are $T_0{}={}26.9$ (24.5) K, $\alpha{}={}0.31$ (0.36) and $\gamma{}={}-0.13$ ($-0.15$). The increase of $T_{\rm eqv}$ with redshift is related to the enhanced level of star formation activity in galaxies (\ie~higher specific SFR) \citep{Safarzadeh_2016, Ma_2019, Liang_2019, Sommovigo_2020}. The anti-correlation with $\delta_{\rm dzr}$, on the other hand, is due to the fact that an increase of $\delta_{\rm dzr}$ leads to a higher dust opacity, which in turn results in a `colder' dust SED shape of galaxies \citep{Scoville_2013, Faisst_2017, Liang_2019}. Observationally, $\delta_{\rm dzr}$ of high-$z$ galaxies has not yet been constrained. 

Often, it is easier to detect the $\rm [C_{II}]$ line than the dust continuum of galaxies at $z{}\simgreat{}5$. For example, 75 out of the 118 ($63.6\%$) galaxies in the {\sc\small ALPINE} sample have confirmed detection of $\rm [C_{II}]$ emission, whilst only 21 ($17.8\%$) of them have confirmed detection of dust continuum. Almost all dust-detected galaxies have detection of $\rm [C_{II}]$ line. The detection limit of $\rm [C_{II}]$ of the current {\sc\small ALMA} observations is about $10^8\,L_\odot$.

We convert the sub-mm broad-band flux density ($S_{\nu_{\rm 0}}$) of the dust-detected galaxies (or the $3\sigma$ upper limit of $S_{\nu_0}$ for the dust-undetected galaxies) to $L_{\rm IR}$ (the upper limit of $L_{\rm IR}$) consistently using $T_{\rm eqv}$ that follows equation~(\ref{eq.7}) (assuming $\delta_{\rm dzr}{}={}0.4$) to make a fair comparison between different observed samples and our theoretical predictions using {\sc\small FIRE} galaxies. We compute the SFR of the observed galaxies using their measured $L_{\rm UV}$ and the derived $L_{\rm IR}$ following \citet{Hao_2011}, \ie~${\rm SFR}\,(M_\odot\,{\rm yr^{-1})}{}={}1.58\times10^{-10}\,(L_{\rm UV} + 0.46 L_{\rm IR})\,(L_\odot)$, for the \citet{Kroupa_2002} IMF. For the dust-undetected galaxies, we estimate the lower and upper bounds of their SFR, where the former is converted from their $L_{\rm UV}$ assuming no dust emission (\ie~$L_{\rm IR}{}={}0$), whilst the latter accounts for the upper limit of $L_{\rm IR}$ converted from the $3\sigma$ upper limit of $S_{\nu_0}$. 

In Fig.~\ref{fig.8}, we show the observed $L_{\rm [C_{II}]}$-SFR relation of the rest-UV-selected galaxy samples at $z{}\simgreat{}5$ (see Table~\ref{T4} for the details) together with the result of the {\sc\small FIRE} galaxies at $z{}={}4$, $z{}={}6$ and $z{}={}8$ in the two panels. For the observed galaxies having no detection of dust, we show the relation between their $L_{\rm [C_{II}]}$ (for the $\rm [C_{II}]$-undetected galaxies, the $3\sigma$ upper limit of their $L_{\rm [C_{II}]}$) and the lower and upper bound of their SFR, respectively, in the {\it left} and {\it right} panels of the figure. For reference, we also show in Fig.~\ref{fig.8} the observed $L_{\rm [C_{II}]}$-SFR relation of the local star-forming galaxies by \citetalias{De_Looze_2011}, \citetalias{De_Looze_2014} and \citetalias{Herrera_Camus_2015}.  

It can be seen that the {\sc\small FIRE} galaxies at $z{}={}4-8$ lie systematically below the observed local $L_{\rm [C_{II}]}$-SFR relations (and thus also the {\sc\small FIRE} galaxies at $z{}={}0$) over the broad SFR range of $\approx{}0.1-10^3\,M_\odot\,\rm yr^{-1}$, showing a $\rm [C_{II}]$ deficit. This appears to be in agreement with the observational data.

At $\rm SFR{}\simgreat{}100\,M_\odot\,\rm yr^{-1}$, most of the observed galaxies at $z{}\simgreat{}5$ have both $\rm [C_{II}]$ and dust detections and thus their (dust-obscured) SFR is more reliably constrained. The mean $L_{\rm [C_{II}]}{}/{}\rm SFR$ ratio of these galaxies is lower than the \citetalias{De_Looze_2011} relation (solid green line) by 0.22 dex, which is close to the $1\sigma$ scatter of the \citetalias{De_Looze_2011} relation (see Table~\ref{T2}). The {\sc\small FIRE} galaxies at $z\ge4$ are about $2\sigma$ below the \citetalias{De_Looze_2011} relation in the same SFR range, which seem to show a slightly more prominent `deficit' than the observed samples.

At $\rm SFR{}\simless{}100\,M_\odot\,\rm yr^{-1}$, most of the $z{}\simgreat{}5$ galaxies do not have confirmed dust detection with the current {\sc\small ALMA} observations, and a large fraction of them do not have confirmed $\rm [C_{II}]$ detections neither (marked by downward arrows). The uncertainty in the SFR estimate of these dust-undetected galaxies can be as large as a factor of $\sim5$ ($\approx{}20-100\,M_\odot\,\rm yr^{-1}$, see Fig.~\ref{fig.8}). Such a large uncertainty is due to the high $T_{\rm eqv}$ of galaxies at $z{}\simgreat{}5$ ($T_{\rm eqv}{}\simgreat{}45$ K for $\delta_{\rm dzr}{}={}0.4$, see equation~(\ref{eq.7})), so that even a low noise level (typically $\sigma\sim10\mu$Jy, see Table~\ref{T4}) of the {\sc\small ALMA} observations is converted to a relatively high upper bound of $L_{\rm IR}$ (and hence $\rm SFR_{IR}$). From Fig.~\ref{fig.8}, it can be seen that the predicted $L_{\rm [C_{II}]}$-SFR relation of the {\sc\small FIRE} galaxies does not conflict with the observational constraints over ${\rm SFR}{}\approx{}10-100\,M_\odot\,\rm yr^{-1}$. In particular, for the $\rm [C_{II}]$-undetected galaxies, the $3\sigma$ upper limit of their $L_{\rm [C_{II}]}$ (marked by downward arrows) appears to be above the data points of the {\sc\small FIRE} galaxies at similar SFR when their dust emission is insignificant, namely, $\rm SFR{}\approx{}SFR_{UV}$ (see the {\it left} panel of Fig.~\ref{fig.8}). 

At ${\rm SFR}{}\simless{}10\,M_\odot\,\rm yr^{-1}$, we lack enough observational data for a reliable constraint on the $L_{\rm [C_{II}]}$-SFR relation at $z{}\simgreat{}5$ because galaxies having such low SFR are intrinsically faint. The galaxy having the lowest SFR (${\rm SFR}{}\approx{}1\,M_\odot\,\rm yr^{-1}$) that has had $\rm [C_{II}]$ measurement to date at $z{}\simgreat{}5$ is MS0451-H \citep{Knudsen_2016}, a strongly lensed galaxy at $z{}={}6.7$ with an estimated magnification factor of $\mu=100\pm20$. MS0451-H has no confirmed $\rm [C_{II}]$ detection yet. The upper bound of its $L_{\rm [C_{II}]}{}/{}\rm SFR$ ratio is more than 1.5 dex below the \citetalias{De_Looze_2011} relation (even with the most conservative, UV-based SFR, see the {\it left} panel of Fig.~\ref{fig.8}), showing a strong $\rm [C_{II}]$ deficit. This appears to be in agreement with the {\sc\small FIRE} sample. It can be seen from the figure that the $\rm [C_{II}]$ deficit of the {\sc\small FIRE} galaxies extends to ${\rm SFR}{}\simless{}10\,M_\odot\,\rm yr^{-1}$ at $z{}\simgreat{}5$, which is even slightly more prominent than at higher SFR. Encouragingly, some of the {\sc\small FIRE} galaxies at $z{}\ge{}4$ show similarly low $L_{\rm [C_{II}]}{}/{}\rm SFR$ ratio as MS0451-H. 

The $L_{\rm [C_{II}]}$-SFR relation of the observed galaxies at $z{}\simgreat{}5$ reported in this work seems to have lower normalization than a number of the recent observational studies, including \eg~\citet{Schaerer_2020} ({\sc\small ALPINE} paper), \citet{Ferrara_2022} ({\sc\small REBELS} paper), \citet{Matthee_2017, Matthee_2019}, \citet{Carniani_2018a}, \citet{Harikane_2020} and \citet{Fujimoto_2021}. This is due to the fact that these studies have assumed a lower $T_{\rm eqv}$ than what we use for this study as derived using equation~(\ref{eq.7}). As has been mentioned in some of these studies, the largest uncertainty of the derived galaxy $L_{\rm [C_{II}]}$-SFR relation at $z{}\simgreat{}5$ is the assumed $T_{\rm eqv}$. In Table~\ref{T5}, we explicitly show the difference in the mean $T_{\rm eqv}$ adopted by the {\sc\small ALPINE}/{\sc\small REBELS} projects and this work (for $\delta_{\rm dzr}{}={}0.4$), as well as the resulting difference in the derived mean $L_{\rm [C_{II}]}{}/{}\rm SFR$ ratio (<$L_{\rm [C_{II}]}{}/{}\rm SFR$>) of the galaxies. Note that \citet{Ferrara_2022} has used very similar $T_{\rm eqv}$ compared to what is used in our work as fiducial (with $\delta_{\rm dzr}{}={}0.4$), whereas \citet{Schaerer_2020} has used significantly lower $T_{\rm eqv}$ ($<T_{\rm eqv}>{}={}42$ K) for the {\sc\small ALPINE} galaxies than us ($<T_{\rm eqv}>{}={}52.1$ K). Our estimate of the $L_{\rm [C_{II}]}$-SFR relation of the {\sc\small ALPINE} galaxies is therefore about 0.3 dex below the originally reported result. 
 \begin{figure}
 \includegraphics[width=85mm]{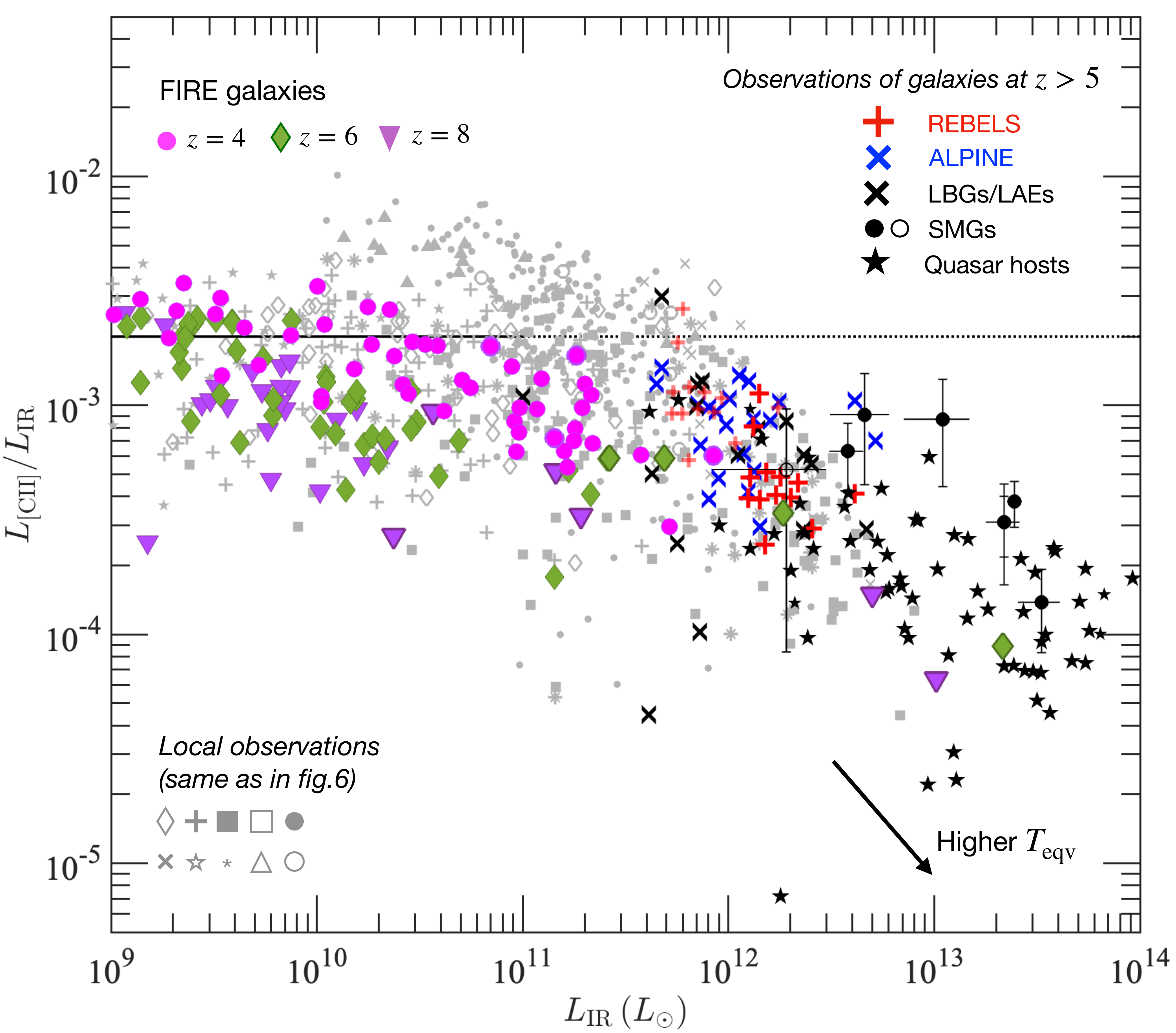}
 \caption{$L_{\rm [C_{II}]}{}/{}L_{\rm IR}$ vs. $L_{\rm IR}$ relation of galaxies at high redshifts. Filled coloured symbols indicate the data of the {\sc\small FIRE} galaxies (magenta circles for $z{}={}4$, green diamonds for $z{}={}6$ and purple downward triangles for $z{}={}8$). Red {crosses} and blue {`X's} represent the observational data of the {\sc\small REBELS} (<$z$>${}\approx{}7$) and {\sc\small ALPINE} (<$z$>${}\approx{}4.5$) galaxy samples, respectively. Black symbols represent the observational data of the other galaxy samples at $z{}\simgreat{}5$. Specifically, black {`X's}, black circles (filled and unfilled) and black stars correspond to the UV-selected galaxies, SMGs and quasar hosts, respectively. For the galaxies whose dust continuum is measured at only single {\sc\small ALMA} band, $L_{\rm IR}$ is derived using $T_{\rm eqv}$ that follows equation~(\ref{eq.7}) assuming $\delta_{\rm dzr}{}={}0.4$ except for the {\sc\small REBELS} galaxies, for which we show two different sets of data that are produced by using $\delta_{\rm dzr}{}={}0.4$ (semi-transparent red {crosses}) and $\delta_{\rm dzr}{}={}0.1$ (non-transparent red {crosses}). The lower $\delta_{\rm dzr}$ yields higher $T_{\rm eqv}$ (and hence $L_{\rm IR}$) estimates for the galaxies.  The black arrow indicates the direction along which the data points of these galaxies move on the diagram with increasing $T_{\rm eqv}$. For the SMGs, filled circles indicate the galaxies that are either confirmed as un-lensed or have observationally determined lensing magnification factor $\mu$, whereas unfilled circles indicate the lensed {\sc\small SPT} galaxies having no determined $\mu$ yet (see Section~\ref{Sec:4b}). Grey symbols in the background represent the observational data of the local $z{}={}0$ galaxy samples, as is shown in Fig.~\ref{fig.6} ({\it right} panel). Black horizontal line indicates the median $L_{\rm [C_{II}]}{}/{}L_{\rm IR}$ ratio (<$L_{\rm [C_{II}]}{}/{}L_{\rm IR}$>${}={}0.002$) of the local galaxies at $L_{\rm IR}{}<{}10^{11}\,L_\odot$. {\bf Galaxies at $z{}>{}5$ show a trend of declining $L_{\rm [C_{II}]}{}/{}L_{\rm IR}$ ratio with $L_{\rm IR}$ at $L_{\rm IR}{}\simgreat{}10^{11}\,L_\odot$ similar to the local samples. The {\sc\small FIRE} simulations successfully reproduced the observed $\rm [C_{II}]$ deficit at high $L_{\rm IR}$ at $z{}>{}5$}. }
\label{fig.9}
\end{figure}

\paragraph*{$L_{\rm [C_{II}]}{}/{}L_{\rm IR}$ of IR-luminous galaxies \\}

In addition to the LBGs/LAEs having moderate SFRs, there have been studies probing the more extreme systems at $z{}\simgreat{}5$, in particular, the {quasar hosts}. These systems are gas/dust-rich and very IR-luminous ($L_{\rm IR}{}\simgreat{}10^{12}\,L_\odot$). They typically are also bright $\rm [C_{II}]$ emitters, having $L_{\rm [C_{II}]}$ that spans across the range of $\approx{}10^8-10^{10}\,L_\odot$. We summarize the properties of the quasar hosts at $z{}\simgreat{}5$ having had $\rm [C_{II}]$ line detections to date in Table~\ref{T6} ($>{}65$ galaxies in total). Observations targeting the quasar hosts have a high successful detection rate for $\rm [C_{II}]$ line \citep[\eg][]{Decarli_2017, Venemans_2020}. 

Like most of the LBGs/LAEs at this epoch, the selected quasar hosts typically have one or two data points in their dust continuum (measured with {\sc\small ALMA} band 6 or 7) and their $L_{\rm IR}$ is converted from a single broad-band sub-mm flux density in the literature using the standard MBB function with an assumed $T_{\rm eqv}$. $L_{\rm IR}$ has generally been considered as a crude estimate of their SFR by the observational studies assuming that {these quasar hosts are gas and dust-rich and the stellar radiation of these galaxies is significantly dust-obscured}. It is, however, unknown to what degree the radiation from the accreting supermassive black hole affects the shape of the IR SED and the total IR luminosity of these early galaxies. Observations of galaxies at lower redshifts ($z{}\approx{}0-3$) demonstrate that the IR SED shape of galaxies becomes `warmer' (indicating higher $T_{\rm eqv}$) with increasing AGN power \citep{Kirkpatrick_2015}. {A similar conclusion was reached in the early study by \citet{Younger_2009} with hydrodynamic simulations of galaxy mergers that include AGN modelling. Note, however, that some recent studies \citep[\eg][]{Symeonidis_2016, McKinney_2021b} also suggest that AGN radiation may even dominate the cold-dust emission of the host galaxies at high redshifts. } 

In Fig.~\ref{fig.9}, we show the {$L_{\rm [C_{II}]}{}/{}L_{\rm IR}$ vs. $L_{\rm IR}$} relation of the quasar hosts, along with other galaxy populations at $z{}\simgreat{}5$, including the few SMGs (listed in Table~\ref{T3}), the {\sc\small ALPINE} and {\sc\small REBELS} galaxies and other rest-UV-selected galaxies at $z{}\simgreat{}5$ (we only show the galaxies having confirmed dust detection, which have more reliable constraints on $L_{\rm IR}$ than the dust-undetected galaxies). We convert the reported single-band sub-mm flux density of all the quasar hosts to $L_{\rm IR}$ using the standard MBB function and $T_{\rm eqv}$ that follows equation~(\ref{eq.7}) with the best-fit parameters derived by \citet{Liang_2019}. We note that for the quasar hosts, this is likely to be an underestimate because the best-fit parameters of \citet{Liang_2019} are derived using {\sc\small FIRE} simulations which do not include AGN feedback. Having a higher $T_{\rm eqv}$, the data points of the quasar hosts (black stars) will shift in the diagonal direction toward the bottom-right corner of the diagram (marked by the black arrow in Fig.~\ref{fig.9}). 

Looking at the observational data, we can see a clear trend of declining $L_{\rm [C_{II}]}{}/{}L_{\rm IR}$ ($\sim L_{\rm [C_{II}]}{}/{}\rm SFR$) ratio of the galaxies with $L_{\rm IR}$ ($\rm [C_{II}]$ deficit) at $L_{\rm IR}{}\simgreat{}10^{11.5}\,L_\odot$ at $z{}\simgreat{}5$, similar to the trend seen at lower redshifts. The $L_{\rm [C_{II}]}{}/{}L_{\rm IR}$-$L_{\rm IR}$ relation of these early galaxies appears to consistent with the local samples (grey symbols) in the overlapping $L_{\rm IR}$ regime and show similarly large scatter. 
 
We also show in Fig.~\ref{fig.9} the $L_{\rm [C_{II}]}{}/{}L_{\rm IR}$-$L_{\rm IR}$ relation of the {\sc\small FIRE} galaxies at $z{}={}4-8$. The result of the {\sc\small FIRE} galaxies is in good agreement with the observational data in overlapping $L_{\rm IR}$ range, except for the {\sc\small REBELS} sample (<$z$>${}\approx{}7$, indicated by red {`X's} in Fig.~\ref{fig.9}). Using $\delta_{\rm dzr}{}={}0.4$, the {\sc\small REBELS} galaxies (semi-transparent red {crosses}) show systematically higher $L_{\rm [C_{II}]}{}/{}L_{\rm IR}$ than the rest of the observed galaxy samples (blue and black {`X's}) as well as the {\sc\small FIRE} galaxies at similar $L_{\rm IR}$ ($\approx{}10^{12}\,L_\odot$) by $\sim0.5$ dex. Using $\delta_{\rm dzr}{}={}0.1$ instead, the expected mean $T_{\rm eqv}$ of the {\sc\small REBELS} sample increases by $\approx{}20\%$ (from 57 K to 71 K), and the derived mean $L_{\rm IR}$ ($L_{\rm [C_{II}]}{}/{}L_{\rm IR}$ ratio) of the galaxies increases (decreases) by a factor of $\sim{}3$. The data of the {\sc\small REBELS} sample for $\delta_{\rm dzr}{}={}0.1$ (non-transparent red {crosses}) appears to be consistent with the other observed samples as well as the {\sc\small FIRE} galaxies. 

The {\sc\small FIRE} galaxies at $z{}\ge{}4$ show a trend of declining $L_{\rm [C_{II}]}{}/{}L_{\rm IR}$ ratio with $L_{\rm IR}$, which agrees with the observational data. It is also clear to see that the $L_{\rm [C_{II}]}{}/{}L_{\rm IR}$ ratio of the {\sc\small FIRE} galaxies decreases with redshift at fixed $L_{\rm IR}$ at $z{}\ge{}4$. The trend of decreasing $L_{\rm [C_{II}]}{}/{}L_{\rm IR}$ ratio with both redshift and $L_{\rm IR}$ persists up to $z{}={}8$ in the {\sc\small FIRE} simulations.

Finally, we note that it is unclear whether AGN activity is directly related to the $\rm [C_{II}]$ deficit at high $L_{\rm IR}$ based on the current data, despite the large number of quasar hosts at $z{}\simgreat{}5$ showing strong $\rm [C_{II}]$ deficit. This is because most of the selected SMGs in the literature ($2{}\simless{}z{}\simless{}7$), having similar $L_{\rm IR}$ to the quasar hosts, have no identified AGN feature (see Table~\ref{T3}) but show similarly strong $\rm [C_{II}]$ deficit as the quasar hosts. In addition, the {\sc\small FIRE} simulations, which do not include AGN physics, have also successfully reproduced similarly low $L_{\rm [C_{II}]}{}/{}L_{\rm IR}$ ratio at high $L_{\rm IR}$. 

\begin{table*}
\centering
\caption{Characteristics of the high-$z$ quasar host galaxies. }
\begin{threeparttable}
\begin{tabular}{ p{3.2 cm} p{1.2 cm} p{2.3 cm}  p{2.5 cm}  p{3 cm} p{1.8 cm}}
 \hline
  \hline
 \multicolumn{1}{c|}{Name}  &  \multicolumn{1}{c|}{$z$}  &  \multicolumn{1}{c|}{$S_\nu$ ($\rm mJy$)\tnote{$\parallel$}}  &  \multicolumn{1}{c|}{${\rm log}\,(L_{\rm IR}{}/{}L_\odot)$\tnote{$\mathsection$}} & ${\rm log}\,(L_{\rm [C_{II}]}{}/{}L_\odot)$\tnote{$\parallel$} & References\tnote{$\ast$} \\
 \hline
SDSS J1015+0020 & 4.407 & 0.60 ({\sc\small ALMA} 7) & \multicolumn{1}{c|}{12.3} & 8.46 ({\sc\small ALMA} 7) & [1] \\ 
BRI 1335-0417 & 4.41 & 9.03 ({\sc\small ALMA} 6)  & \multicolumn{1}{c|}{14.0} & 10.21 ({\sc\small APEX/FLASH}) & [2, 3] \\  
BR 1202-0725 N & 4.691 & 18.8 ({\sc\small ALMA} 7) & \multicolumn{1}{c|}{13.8} & 10.00 ({\sc\small ALMA} 7) & [4, 5] \\  
BR 1202-0725 S & 4.694 & 18.0 ({\sc\small ALMA} 7) & \multicolumn{1}{c|}{13.8} & 9.81 ({\sc\small ALMA} 7) & [4, 5] \\  
SDSS J0338+0021 & 5.027 & 2.98 ({\sc\small ALMA} 6) & \multicolumn{1}{c|}{13.5} & 9.76 ({\sc\small ALMA} 6) & [6] \\
SDSS J0129-0035 & 5.779 & 2.61 ({\sc\small ALMA} 6) & \multicolumn{1}{c|}{13.5} & 9.28 ({\sc\small ALMA} 6) &  [7, 8, 9] \\  
SDSS J1044-0125 & 5.785 & 3.00 ({\sc\small ALMA} 6) & \multicolumn{1}{c|}{13.5} & 9.21 ({\sc\small ALMA} 6) &  [7, 8, 9] \\  
PSO J004+17 & 5.817 & 0.88 ({\sc\small ALMA} 6) &   \multicolumn{1}{c|}{13.0} & 8.31 ({\sc\small ALMA} 6) &  [10] \\  
PSO J352-15 & 5.832 & 0.34 ({\sc\small ALMA} 7) & \multicolumn{1}{c|}{12.1} & 9.09 ({\sc\small ALMA} 7) & [11] \\
HSC J1202-0057 & 5.929 & 0.25 ({\sc\small ALMA} 6) &  \multicolumn{1}{c|}{12.4} & 8.79 ({\sc\small ALMA} 7) & [12] \\
PSO J056+16 & 5.967 & 0.17 ({\sc\small ALMA} 6) &  \multicolumn{1}{c|}{12.3} & 7.11 ({\sc\small ALMA} 6) &  [10] \\  
PSO J007+04& 6.001 & 2.07 ({\sc\small ALMA} 6) & \multicolumn{1}{c|}{13.4} & 9.20 ({\sc\small ALMA} 6) & [9, 13] \\
SDSS J2310+1855\tnote{$\dagger$} & 6.003 & \multicolumn{1}{c|}{$-$} & \multicolumn{1}{c|}{13.2} & 9.94 ({\sc\small ALMA} 6) &  [7, 14] \\  
PSO J009-10 & 6.004 & 3.66 ({\sc\small ALMA} 6) & \multicolumn{1}{c|}{13.6} & 9.95 ({\sc\small ALMA} 6) & [9, 13] \\ 
CFHQS J0055+0146 & 6.006 & 0.21 ({\sc\small ALMA} 6) & \multicolumn{1}{c|}{12.4} & 8.92 ({\sc\small ALMA} 6) & [15] \\	
CFHQS J0216-0455 & 6.01 & $<0.04$ ({\sc\small ALMA} 6) & \multicolumn{1}{c|}{$<11.6$}  & $<7.85$ ({\sc\small ALMA} 6) & [16] \\
PSO J265+41 & 6.026 & 3.61 ({\sc\small ALMA} 6) & \multicolumn{1}{c|}{13.6}  & 9.96 ({\sc\small ALMA} 6) &  [10] \\  
SDSS J1306+0356 & 6.033 & 0.74 ({\sc\small ALMA} 6) & \multicolumn{1}{c|}{12.9} & 9.05 ({\sc\small ALMA} 6) & [9, 13] \\
ULAS J1207+0630 & 6.037 & 0.50 ({\sc\small ALMA} 6) & \multicolumn{1}{c|}{12.7} & 9.13 ({\sc\small ALMA} 6) & [13] \\
SDSS J2054-0005 & 6.039 & 3.15 ({\sc\small ALMA} 6) & \multicolumn{1}{c|}{13.5} & 9.49 ({\sc\small ALMA} 6)  &  [7, 9] \\  
VDESJ0454-4448 & 6.058 & 0.71 ({\sc\small ALMA} 6) & \multicolumn{1}{c|}{12.9} & 8.86 ({\sc\small ALMA} 6) & [13] \\
PSO J158+14 & 6.068 & 3.46 ({\sc\small ALMA} 6) & \multicolumn{1}{c|}{13.6} & 9.22 ({\sc\small ALMA} 6) &  [10] \\  
SDSS J0842+1218 & 6.075 & 0.68 ({\sc\small ALMA} 6) & \multicolumn{1}{c|}{12.9} & 8.88 ({\sc\small ALMA} 6) & [9, 13, 17]\\
HSC J2228+0152 & 6.081 & $<0.05$ ({\sc\small ALMA} 6) &  \multicolumn{1}{c|}{$<11.7$} & 8.39 ({\sc\small ALMA} 6) &  [18] \\  
CFHQS J2100-1715 & 6.081 & 0.56 ({\sc\small ALMA} 6) & \multicolumn{1}{c|}{12.8} & 9.12 ({\sc\small ALMA} 6)  &  [9, 13, 17, 19] \\  
HSC J2216-0016 & 6.096 & 0.14 ({\sc\small ALMA} 6) & \multicolumn{1}{c|}{12.2} & 9.01 ({\sc\small ALMA} 6) &  [12] \\  
PSO J239+07 & 6.110 & 0.23 ({\sc\small ALMA} 6) &  \multicolumn{1}{c|}{12.4} & 8.37 ({\sc\small ALMA} 6) &  [10] \\  
HSC J1208-0200 & 6.117 & 0.09 ({\sc\small ALMA} 6) & \multicolumn{1}{c|}{12.0} & 8.43 ({\sc\small ALMA} 6) &  [18] \\  
CFHQS J1509-1749	& 6.123 & 1.72 ({\sc\small ALMA} 6) &  \multicolumn{1}{c|}{13.3} & 9.37 ({\sc\small ALMA} 6) & [13] \\
PSO J065-19 & 6.125 & 0.46 ({\sc\small ALMA} 6) & \multicolumn{1}{c|}{12.7} & 8.97 ({\sc\small ALMA} 6) & [13] \\	
CFHQS J0221-0802 & 6.13 & 0.25 ({\sc\small ALMA} 6) &  \multicolumn{1}{c|}{12.4}  & $<8.08$ ({\sc\small ALMA} 6) & [16] \\
ULAS J1319+0950 & 6.135 & 5.13 ({\sc\small ALMA} 6) &  \multicolumn{1}{c|}{13.8} & 9.61 ({\sc\small ALMA} 6)  &  [7, 9, 20] \\   
VIK J2318-3029 & 6.146  & 3.11 ({\sc\small ALMA} 6) &  \multicolumn{1}{c|}{13.5} & 9.35 ({\sc\small ALMA} 6) & [9, 13] \\
VIMOS2911 & 6.149 & 0.77 ({\sc\small ALMA} 6) &  \multicolumn{1}{c|}{12.9} & 9.41 ({\sc\small ALMA} 6) & [16] \\
PSO J217-16 & 6.150 & 0.37 ({\sc\small ALMA} 6) &  \multicolumn{1}{c|}{12.6}  & 9.00 ({\sc\small ALMA} 6) & [13] \\	
CFHQS J2229+1457 & 6.152 & 0.05 ({\sc\small ALMA} 6) &  \multicolumn{1}{c|}{11.8} & 8.78 ({\sc\small ALMA} 6) & [15] \\	
PSO J359-06 & 6.172 & 0.79 ({\sc\small ALMA} 6) &  \multicolumn{1}{c|}{12.9} &  9.42 ({\sc\small ALMA} 6) & [9, 10, 13] \\
PSO J065-26 & 6.187 & 1.37 ({\sc\small ALMA} 6) &  \multicolumn{1}{c|}{13.2} & 9.23 ({\sc\small ALMA} 6) & [9, 13] \\
PSO J308-21 & 6.236 & 1.18 ({\sc\small ALMA} 6) &  \multicolumn{1}{c|}{13.1} & 9.53 ({\sc\small ALMA} 6) & [9, 13, 17] \\
HSC J2239+0207 & 6.250 & 1.11 ({\sc\small ALMA} 6) &  \multicolumn{1}{c|}{13.1} & 8.98 ({\sc\small ALMA} 6) &  [18] \\  
SDSS J0100+2802 & 6.327 & 1.37 ({\sc\small ALMA} 6) & \multicolumn{1}{c|}{13.2} & 9.58 ({\sc\small ALMA} 6) & [21, 22] \\
ATLAS J025-33 & 6.338 & 2.49 ({\sc\small ALMA} 6) &  \multicolumn{1}{c|}{13.4} & 9.75 ({\sc\small ALMA} 6) & [9, 13] \\	
VIK J2211-3206 & 6.339 & 0.57 ({\sc\small ALMA} 6) &  \multicolumn{1}{c|}{12.8} & 8.98 ({\sc\small ALMA} 6) & [13] \\
PSO J083+11 & 6.340 & 5.10 ({\sc\small ALMA} 6) &  \multicolumn{1}{c|}{13.8} & 10.02 ({\sc\small ALMA} 6) & [23] \\
VIK J1152+0055 & 6.364 & 0.22 ({\sc\small ALMA} 6) &  \multicolumn{1}{c|}{12.4} & 8.81 ({\sc\small ALMA} 6) & [12, 13] \\
PSO J159-02 & 6.381 & 0.65 ({\sc\small ALMA} 6) & \multicolumn{1}{c|}{12.9} & 9.05 ({\sc\small ALMA} 6) & [13] \\	
HSC J0859+0022 & 6.390 & 0.16 ({\sc\small ALMA} 6) &  \multicolumn{1}{c|}{12.2} & 8.66 ({\sc\small ALMA} 6) &  [12] \\  
J2329-0301& 6.417	& 0.04 ({\sc\small ALMA} 6) &  \multicolumn{1}{c|}{11.6} & 8.59 ({\sc\small ALMA} 6) & [16] \\ 	
SDSS J1148+5251\tnote{$\dagger$} & 6.42 &  \multicolumn{1}{c|}{$-$} &  \multicolumn{1}{c|}{13.3} & 9.64 ({\sc\small NOEMA}) &  [22, 24, 25, 26] \\  
CFHQS J0210-0456 & 6.432 & 0.12 ({\sc\small ALMA} 6) &  \multicolumn{1}{c|}{12.1} & 8.48 ({\sc\small ALMA} 6) & [27] \\
PSO J183+05 & 6.439 & 4.79 ({\sc\small ALMA} 6) &  \multicolumn{1}{c|}{13.7} & 9.85 ({\sc\small ALMA} 6) & [9, 13] \\
VIK J2318-3113 & 6.443 & 0.36 ({\sc\small ALMA} 6) &  \multicolumn{1}{c|}{12.6} & 9.20 ({\sc\small ALMA} 6) & [9, 13] \\
PSO J011+09 & 6.469 & 1.20 ({\sc\small ALMA} 6) &   \multicolumn{1}{c|}{13.1} & 8.47 ({\sc\small ALMA} 6) &  [10] \\  
PSO J167-13 & 6.514 & 0.89 ({\sc\small ALMA} 6) &  \multicolumn{1}{c|}{13.0} & 9.75 ({\sc\small ALMA} 6) & [9, 13, 16] \\
J043947+163415 (lensed$\ddagger$) & 6.519 & 3.27 ({\sc\small ALMA} 6) & \multicolumn{1}{c|}{13.6} & 9.54 ({\sc\small ALMA} 6) & [28, 29] \\
PSO J036+03 & 6.542 & 2.55 ({\sc\small ALMA} 6) &  \multicolumn{1}{c|}{13.5} & 9.53 ({\sc\small ALMA} 6) & [9, 30] \\	
PSO J231-20 & 6.587 & 4.37 ({\sc\small ALMA} 6) &  \multicolumn{1}{c|}{13.7} & 9.55 ({\sc\small ALMA} 6) & [9, 13, 17] \\
PSO J323+12 & 6.587 & 0.23 ({\sc\small ALMA} 6) &  \multicolumn{1}{c|}{12.4} & 9.16 ({\sc\small ALMA} 6) & [9, 31] \\		
PSO J006+39	& 6.610 & 0.55 ({\sc\small NOEMA}) &  \multicolumn{1}{c|}{12.8} & 8.95 ({\sc\small NOEMA}) & [32] \\	
VIK J030516-315056 & 6.614 & 5.34 ({\sc\small ALMA} 6) &  \multicolumn{1}{c|}{13.8} & 9.77 ({\sc\small ALMA} 6) & [9, 32, 33] \\	
PSO J338+29	& 6.658 & 0.97 ({\sc\small NOEMA}) &  \multicolumn{1}{c|}{13.0} & 9.30 ({\sc\small NOEMA}) & [31] \\			
VIK J1048-0109 & 6.676 & 2.84 ({\sc\small ALMA} 6) &  \multicolumn{1}{c|}{13.5} & 9.32 ({\sc\small ALMA} 6) & [9, 13] \\
 \hline
 \end{tabular}
 \;\;\;\;\;(Continue on next page)
  \end{threeparttable}
  \label{T6}
\end{table*}

\begin{table*}
\contcaption{}
\begin{threeparttable}
\begin{tabular}{ p{3.2 cm} p{1.2 cm} p{2.3 cm}  p{2.5 cm}  p{3 cm} p{1.8 cm}}
 \hline
  \hline
 \multicolumn{1}{c|}{Name}  &  \multicolumn{1}{c|}{$z$}  &  \multicolumn{1}{c|}{$S_\nu$ ($\rm mJy$)}  &  \multicolumn{1}{c|}{${\rm log}\,(L_{\rm IR}{}/{}L_\odot)$\tnote{$\mathsection$}} & ${\rm log}\,(L_{\rm [C_{II}]}{}/{}L_\odot)$ & References\tnote{$\ast$} \\
  \hline
HSC J1205-0000 & 6.723 & 1.17 ({\sc\small ALMA} 6) &  \multicolumn{1}{c|}{13.1} & 8.58 ({\sc\small ALMA} 6) & [34] \\
VIK J0109-3047 & 6.791 & 0.52 ({\sc\small ALMA} 6) &  \multicolumn{1}{c|}{12.8} & 9.38 ({\sc\small ALMA} 6) & [9, 33] \\
VIK J2348-3054 & 6.901 & 2.28 ({\sc\small ALMA} 6) &  \multicolumn{1}{c|}{13.4} & 9.25 ({\sc\small ALMA} 6) & [9, 33] \\	
HSC J1243+0100 & 7.075 &  1.52 ({\sc\small ALMA} 6) &  \multicolumn{1}{c|}{13.2} & 9.40 ({\sc\small ALMA} 6) & [35] \\	
ULAS J1120+0641 & 7.085 & 0.64 ({\sc\small ALMA} 6) &  \multicolumn{1}{c|}{12.9} & 9.08 ({\sc\small ALMA} 6)  & [9, 36] \\		
ULAS J1342+0928 & 7.541 & 0.34 ({\sc\small ALMA} 6) &  \multicolumn{1}{c|}{12.6} & 9.12 ({\sc\small ALMA} 6) & [9, 37] \\	
 \hline
  \end{tabular}
  \begin{tablenotes}
  \item[$\dagger$]  $L_{\rm IR}$ of SDSS J2310+1855 and SDSS J1148+5251 are derived by SED fitting \citep[\eg][]{Casey_2012, Casey_2014} to multiple data points at both Wien and Rayleigh-Jeans sides of the dust IR SED.   
  \item[$\ddagger$] J043947+163415 has been confirmed to be gravitationally-lensed, and its luminosities have been de-magnified by $\mu{}={}4.6\pm2.0$, estimated based on the lensing configuration from {\sc\small HST} imaging by \citet{Fan_2019}. 
   \item[$\parallel$] {\sc\small NOEMA}: NOrthern Extended Millimeter Array\\
   (Website: \url{https://www.iram-institute.org/EN/content-page-235-3-235-0-0-0.html}).  
    \item[$\mathsection$]  $L_{\rm IR}$ (or its upper $3\sigma$ limit) is converted from $S$ (its $3\sigma$ upper limit) using the standard MBB function and with $T_{\rm eqv}$ that follows equation~(\ref{eq.7}) (assuming $\beta_{\rm dust}{}={}2.0$ and $\delta_{\rm dzr}{}={}0.4$), except for SDSS J2310+1855 and SDSS J1148+5251. 
    \item[$\ast$]  References: [1]: \citet{Bischetti_2018}, [2]: \citet{Wagg_2010}, [3]: \citet{Lu_2018}, [4]: \citet{Wagg_2012}, [5]: \citet{Iono_2006},  [6]: \citet{Leipski_2014}, [7]: \citet{Wang_2013}, [8]: \citet{Wang_2019}, [9]: \citet{Venemans_2020}, [10]: \citet{Eilers_2020}: [11]: \citet{Rojas_Ruiz_2021}, [12]: \citet{Izumi_2018}, [13]: \citet{Decarli_2018}, [14]: \citet{Shao_2019}, [15]: \citet{Willott_2015a}, [16]: \citet{Willott_2017}, [17]: \citet{Decarli_2017}, [18]: \citet{Izumi_2019}, [19]: \citet{Walter_2018}, [20]: \citet{Shao_2017}, [21]: \citet{Wang_2016}, [22]: \citet{Leipski_2013}, [23]: \citet{Andika_2020}, [24]: \citet{Walter_2009}, [25]: \citet{Maiolino_2005}, [26]: \citet{Meyer_2022}, [27]: \citet{Willott_2013a}, [28]: \citet{YangY_2019}, [29]: \citet{Yue_2021}, [30]: \citet{Banados_2015}, [31]: \citet{Mazzucchelli_2017}, [32]: \citet{Venemans_2019} [33]: \citet{Venemans_2016}, [34]: \citet{Izumi_2021a}, [35]: \citet{Izumi_2021b}, [36]: \citet{Venemans_2012}, [37]: \citet{Venemans_2017}. 
  \end{tablenotes}
  \end{threeparttable}
  \label{T6x}
\end{table*}

\section{The physics of the $L_{\rm [C_{II}]}$-SFR scaling relation of galaxies}
\label{Sec:5}
 
In the previous section, we have shown that the {$L_{\rm [C_{II}]}$-SFR relation of the {\sc\small FIRE} galaxies predicted using our model is in good agreement with} the observational data of local and high-$z$ galaxies. In particular, our model reproduces the observed $\rm [C_{II}]$ deficit of galaxies at high $L_{\rm IR}$ and high redshifts. In this section, we explore the origin(s) of the $\rm [C_{II}]$ deficit of galaxies using the {\sc\small FIRE} galaxy sample.

In Section~\ref{Sec:5a}, we present the analytic solution of $\rm [C_{II}]$ line flux emerging from a plane-parallel gas slab. The toy model provides useful insights for understanding the $\rm [C_{II}]$ emission of galaxies. In Section~\ref{Sec:5b}, we derive an important scaling relation of galaxies between their $L_{\rm [C_{II}]}{}/{}\rm SFR$ ratio and other physical properties. Based on this scaling relation, we investigate the cause of the $\rm [C_{II}]$ deficit of galaxies in Section~\ref{Sec:5c}. Finally, in Section~\ref{Sec:5d}, we show the presence of two distinct physical regimes where the main reason for the $\rm [C_{II}]$ deficit of galaxies is different. 

\subsection{Insights from the plane parallel slab model}
\label{Sec:5a}

The $\rm [C_{II}]$ line flux emerging from a plane-parallel slab that is irradiated by an external radiation field has recently been studied by \citet[][hereafter \citetalias{Ferrara_2019}]{Ferrara_2019}. In this section, we summarize the key points of the \citetalias{Ferrara_2019} model. We refer interested readers to \citetalias{Ferrara_2019} for the details. 

The plane-parallel slab can be characterized by three distinct zones based on the ionization structures of gas, as has been discussed in Section~\ref{Sec:3a}. Right beneath the surface of the slab, ionizing radiation ($E_\gamma{}>{}13.6$ eV) creates a $\rm H^+$ region extending to a gas column density $N_{\rm s}$ (Zone I), where both hydrogen and carbon are ionized. Beyond $N_{\rm s}$, hydrogen becomes neutral but LW ($11.2<E_\gamma<13.6$ eV) photons maintain carbon in the singly ionized state (Zone II). The LW photons become fully absorbed by dust and $\rm H_2$ at a column density $N_{\rm F}$, beyond which hydrogen turns into $\rm H_2$ and carbon {becomes} neutral (Zone III). We have shown in Fig.~\ref{fig.2} the ionization structures of a plane-parallel slab calculated by {\sc\small CLOUDY} as an example (see also Fig. 1 of \citetalias{Ferrara_2019} for a schematic plot). 

$N_{\rm s}$ can be estimated by equating the photo-ionization rate to the recombination rate of hydrogen inside the $\rm H^+$ region (Zone I) assuming that dust extinction is negligible, which can be expressed as (see Appendix~\ref{Sec:Ap3} for the details)
\begin{equation}
N_{\rm s}=n_{\rm H}l_{\rm s}= \frac{Uc}{\alpha_{\rm B}} \approx 10^{23} U\;\rm cm^{-2},
\label{eq.8}
\end{equation}
where $l_{\rm s}$ is the distance from the surface of the slab to the end of Zone I, $U$ parameter represents the ionizing photon-to-gas density ratio, \ie
\begin{equation}
U=\frac{n_\gamma}{n_{\rm H}}, 
\label{eq.9}
\end{equation}

\noindent $c$ represents the speed of light, and $\alpha_{\rm B}{}={}2.6\times10^{-13}\, \rm cm^3\,s^{-1}$ is the Case-B recombination coefficient at gas temperature $T{}\approx{}10^4\,\rm K$ \citep{Ferland_1992}. For a slab with density $n_{\rm H}{}={}50\,\rm cm^{-3}$ that is exposed to a radiation field having $G{}={}200\,G_0$, we obtain $U{}={}n_\gamma/n_{\rm H}{}\approx{}1.3\times10^{-3}$ at and near the surface of the slab. Using equation~(\ref{eq.8}), we obtain $N_{\rm s}{}\approx{}1.3\times10^{20}\,\rm cm^{-2}$. We can see from Fig.~\ref{fig.2} that this estimated $N_{\rm s}$ is in good agreement with the result computed by {\sc \small CLOUDY}, in particular, for the metal-poor model (with $Z_{\rm gas}{}={}0.1\,Z_\odot$; {\it right} panels of Fig.~\ref{fig.2}), where dust extinction in the $\rm H^+$ (Zone I) region is negligible. $N_{\rm s}$ of the metal-rich model (with $Z_{\rm gas}{}={}Z_\odot$; {\it left} panels of Fig.~\ref{fig.2}) is smaller by about 1/4 due to higher absorption of ionizing photons by dust.

$N_{\rm F}$ can be estimated using
\begin{equation}
N_{\rm F}=n_{\rm H}l_{\rm F}= \bar{\sigma}^{-1}_{\rm d} \ln(1+10^5\omega U), \\
\label{eq.10}
\end{equation} 

\noindent which is obtained by performing a RT calculation \citep{Sternberg_2014} that accounts for the absorption of LW photons by dust grains and $\rm H_2$ as light propagates through the slab. In equation~(\ref{eq.10}), $l_{\rm F}$ represents the distance between the surface of the slab and the end of Zone II,
\begin{equation}
\bar{\sigma}_{\rm d} = 5.9\times10^{-22} \left(\frac{\delta_{\rm dgr}}{\delta_{\rm dgr,\,MW}}\right)\,\rm cm^2
\label{eq.11}
\end{equation}

\noindent represents the flux-weighted dust extinction cross section per H-atom, and 
\begin{equation}
\omega = \frac{1}{1+0.9(\delta_{\rm dgr}/\delta_{\rm dgr,\,MW})^{1/2}},
\label{eq.12}
\end{equation}

\noindent where $\delta_{\rm dgr, \,MW}=10^{-2}$ represents the Galactic dust-to-gas ratio \citep[see \eg][]{Gilmore_1989, Sodroski_1997, Zubko_2004, Remy_Ruyer_2014, McKinnon_2016, Li_2019}. For the two models where $Z_{\rm gas}{}={}Z_\odot$ and $Z_{\rm gas}{}={}0.1Z_\odot$, $N_{\rm F}$ is expected to be $\sim10^{21}\,\rm cm^{-2}$ and $\sim10^{22}\,\rm cm^{-2}$ (according to equation~\ref{eq.10}), respectively. This result is again in good agreement with the prediction of {\sc\small CLOUDY} as shown in Fig.~\ref{fig.2}. 

Now we can derive the $\rm [C_{\rm II}]$ line flux ($F_{\rm [C_{II}]}$) emerging from a plane-parallel slab following the three-zone model. $F_{\rm [C_{II}]}$ can be calculated using 
\begin{equation}
F_{\rm [C_{II}]} =\, \Lambda^{\rm (1)}_{\rm [C_{II}]}\, l_{\rm s} + \Lambda^{\rm (2)}_{\rm [C_{II}]} \, (l_{\rm F}-l_{\rm s}),
\label{eq.13}
\end{equation}
\noindent where the first and second terms correspond to the contribution of $\rm [C_{II}]$ line flux by Zone I and Zone II, respectively. $\Lambda^{\rm (1)}_{\rm [C_{II}]}$ ($\Lambda^{\rm (2)}_{\rm [C_{II}]}$) in the above equation represents the $\rm [C_{II}]$ cooling rate ($\rm erg\,s^{-1}\,cm^{-3}$) of gas in Zone I (II).  In the above and the following equations, the superscript ``(1)" (``(2)") indicates the properties of gas in Zone I (II).  {We neglect the $\rm [C_{II}]$ emission from the $\rm H_2$ region (Zone III)}.

Equation~(\ref{eq.13}) can be rewritten as (see Appendix~\ref{Sec:Ap4} for the details)
\begin{align}
F_{\rm [C_{II}]} \approx &\, h_{\rm P} \nu_{\rm [C_{II}]} \left (\frac{g_{\rm u}}{g_{\rm l}}\right ) R^{\rm e^-}_{\rm ul} (T^{(1)}) n^{(1)}_{\rm C^+} n^{(1)}_{\rm e^-} l_{\rm s} \nonumber \\ 
& + \frac{2}{5} h_{\rm P} \nu_{\rm [C_{II}]} \left (\frac{g_{\rm u}}{g_{\rm l}}\right ) R^{\rm H_I}_{\rm ul} (T^{(2)}) n^{(2)}_{\rm C^+} n^{(2)}_{\rm H_I} (l_{\rm F}-l_{\rm s}),
\label{eq.14}
\end{align}
\noindent where $h_{\rm P}$ is the Planck constant, $\nu_{\rm [C_{II}]}{}={}1900.5\,\rm GHz$ is the rest-frame frequency of the $\rm [C_{II}]$ line, $g_{\rm u}{}={}4$ ($g_{\rm l}{}={}2$) is the statistical weight of the ${}^{2}P_{3/2}$ (${}^{2}P_{1/2}$) state, $R^{\rm e^-}_{\rm ul}$ ($R^{\rm H_I}_{\rm ul}$) is the downward rate coefficient ($\rm s^{-1}$) for {${\rm C^+} + e^-$ ($\rm C^+ + H_I$)} collision, and {$n^{(1)}_{\rm C^+}$ (and $n^{(2)}_{\rm C^+}$)}, $n^{(1)}_{\rm e^-}$ and $n^{(2)}_{\rm H_I}$ represent the number density of {$\rm C^+$} ion, electron and $\rm H$ atom, respectively. Equation~(\ref{eq.14}) implies that in Zone I (II), the main collision partner of {$\rm C^+$} ion is electron ($\rm H$ atom). Knowing that $n^{(1)}_{\rm e^-}\approx n_{\rm H}$ and $n^{(2)}_{\rm H_I}\approx n_{\rm H}$ (see the {\it upper} panels of Fig.~\ref{fig.2}), we can rewrite equation~(\ref{eq.14}) to be
\begin{equation}
F_{\rm [C_{II}]} = h_{\rm P} \nu_{\rm [C_{II}]} \left ( \frac{g_{\rm u}}{g_{\rm l}} \right ) \left [ R^{\rm e^-}_{\rm ul} n^{(1)}_{\rm C^+} N_{\rm s} +  \frac{2}{5}R^{\rm H_I}_{\rm ul}  n^{(2)}_{\rm C^+} (N_{\rm F} - N_{\rm s}) \right ],
\label{eq.15}
\end{equation}
\noindent where $N_{\rm F}{}={}n_{\rm H}\,l_{\rm F}$ and $N_{\rm s}{}={}n_{\rm H}\,l_{\rm s}$. Furthermore, {$n^{(1)}_{\rm C^+}$ and $n^{(2)}_{\rm C^+}$} in the above equation can be rewritten as
\begin{equation}
n^{(1)}_{\rm C^+} = n_{\rm H} x^{(1)}_{\rm C^+} \mathcal{A}_{\rm C} \; \;\; {\rm and} \;\;\; n^{(2)}_{\rm C^+} = n_{\rm H} x^{(2)}_{\rm C^+} \mathcal{A}_{\rm C},
\label{eq.16}
\end{equation}
\noindent where
\begin{equation}
\mathcal{A}_{\rm C}=2.5\times10^{-4} \left(\frac{Z_{\rm gas}}{Z_\odot}\right)
\label{eq.17}
\end{equation} 
\noindent represents the abundance of carbon. The numerical factor $2.5\times10^{-4}$ in equation~(\ref{eq.17}) is the abundance of carbon in the solar photosphere \citep{Asplund_2009}. {$x^{(1)}_{\rm C^+}$ ($x^{(2)}_{\rm C^+}$)} in equation~(\ref{eq.16}) represents the fraction of carbon in {$\rm C^+$} form in Zone I (II). {$x^{(1)}_{\rm C^+}$} is roughly inversely proportional to $U$ (see Appendix~\ref{Sec:Ap5}), whereas {$x^{(2)}_{\rm C^+}{}\approx{}1$} (see the {\it middle} panels of Fig.~\ref{fig.2}). By inputting equation~(\ref{eq.16}) to equation~(\ref{eq.15}), we get
\begin{align}
F_{\rm [C_{II}]} = \, & n_{\rm H}  \mathcal{A}_{\rm C} N_{\rm F} h_{\rm P} \nu_{\rm [C_{II}]} \left ( \frac{g_{\rm u}}{g_{\rm l}} \right ) \left [ R^{\rm e^-}_{\rm ul}  x^{(1)}_{\rm C^+} \left ( \frac{N_{\rm s}}{N_{\rm F}} \right ) +  \frac{2}{5}R^{\rm H_I}_{\rm ul} \left ( \frac{N_{\rm F} - N_{\rm s}}{N_{\rm F}} \right ) \right ] \nonumber \\
= \,& n_{\rm H}  \mathcal{A}_{\rm C} N_{\rm F} \bar{\epsilon}_{\rm [C_{II}],\,slab}, 
\label{eq.18}
\end{align}

\noindent where we define
\begin{align}
\bar{\epsilon}_{\rm [C_{II}],\,slab} &= h_{\rm P} \nu_{\rm [C_{II}]} \left ( \frac{g_{\rm u}}{g_{\rm l}} \right ) \left [ R^{\rm e^-}_{\rm ul} x^{(1)}_{\rm C^+} \left ( \frac{N_{\rm s}}{N_{\rm F}} \right ) +  \frac{2}{5}R^{\rm H_I}_{\rm ul}  \left ( \frac{N_{\rm F} - N_{\rm s}}{N_{\rm F}} \right ) \right ]  \nonumber \\
&\equiv \alpha \, x^{(1)}_{\rm C^+} \left ( \frac{N_{\rm s}}{N_{\rm F}} \right ) + \gamma \left ( \frac{N_{\rm F} - N_{\rm s}}{N_{\rm F}} \right )
\label{eq.19}
\end{align}

\noindent as the \textit{{specific $\rm [C_{II}]$ cooling rate}} of the slab ($\rm erg\,s^{-1}\,cm^3$). It can be shown that (see Appendix~\ref{Sec:Ap4} for the details) 
\begin{align}
\alpha &\equiv h_{\rm P} \nu_{\rm [C_{II}]} \left ( \frac{g_{\rm u}}{g_{\rm l}} \right ) R^{\rm e^-}_{\rm ul} (T^{(1)}) \nonumber \\ 
 & \approx 10^{-21}\, {\rm erg\,s^{-1}\,cm^3} \; (T^{(1)}\approx10^4\,\rm K)
\label{eq.20}
\end{align}

\noindent and 
\begin{align}
\gamma & \equiv \frac{2}{5} h_{\rm P} \nu_{\rm [C_{II}]} \left ( \frac{g_{\rm u}}{g_{\rm l}} \right ) R^{\rm H_I}_{\rm ul} (T^{(2)}) \nonumber \\
& \approx 10^{-23} \,{\rm erg\,s^{-1}\,cm^3} \; (T^{(2)}\approx10^2\,\rm K)
\label{eq.21}
\end{align}
\noindent From equation~(\ref{eq.19}), we see that $\bar{\epsilon}_{\rm [C_{II}],\,slab}$ depends on {$x^{(1)}_{\rm C^+}$}, $N_{\rm s}$ and $N_{\rm F}$, and varies typically within the range $10^{-23}-10^{-21} \;{\rm erg\,s^{-1}\,cm^3}$. 

Likewise, we can derive the $\rm [C_{II}]$ luminosity of a spherical uniform gas cloud ($L_{\rm [C_{II}],\,cl}$). $L_{\rm [C_{II}],\,cl}$ can be expressed as
\begin{equation}
 L_{\rm [C_{II}],\,cl}=  
    \begin{cases}
     4\pi  \bigints_{\,0}^{R_{\rm cl}} \Lambda^{(1)}_{\rm [C_{II}]} r^2 {\rm d}r \;\;\;\;({\rm if}\; l_{\rm s}\ge R_{\rm cl})\\
     \\
     4\pi \left [ \bigints_{R_{\rm cl}-l_{\rm s}}^{R_{\rm cl}}  \Lambda^{(1)}_{\rm [C_{II}]} r^2 {\rm d}r + \bigints_{R_{\rm cl}-{\rm min} (l_{\rm F}, R_{\rm cl})}^{R_{\rm cl}-l_{\rm s}}  \Lambda^{(2)}_{\rm [C_{II}]} r^2 {\rm d}r \right ]. \\
        \;\; ({\rm if} \;l_{\rm s}< R_{\rm cl}) 
    \end{cases}       
\label{eq.22}
\end{equation}

\noindent The first condition of equation~(\ref{eq.22}) (\ie~$l_{\rm s}{}\ge{}R_{\rm cl}$) corresponds to when the cloud is fully ionized, while the second condition (\ie~$l_{\rm s}{}<{}R_{\rm cl}$) corresponds to when neutral hydrogen region (Zone II) forms in the cloud. Through simple re-arrangement, $L_{\rm [C_{II}],\,cl}$ can be expressed as 
\begin{equation}
L_{\rm [C_{II}],\,cl} =  f_{\rm [C_{II}],\, cl} \left ( \frac{M_{\rm cl}}{\mu_{\rm H} m_{\rm H}} \right ) n_{\rm H}\mathcal{A}_{\rm C} \bar{\epsilon}_{\rm [C_{II}],\,cl},
\label{eq.23}
\end{equation}

\noindent where $f_{\rm [C_{II}],\,cl}$ represents the fraction of the gas mass that is in $\rm H^+$ or $\rm H_I$ phases (Zone I and Zone II), $M_{\rm cl}$ indicates the mass of the gas cloud, $\mu_{\rm H}$ is the mean molecular weight of the gas and $m_{\rm H}$ {represents the proton mass}. By definition, $f_{\rm [C_{II}],\,cl}{}={}1$ when $l_{\rm F}{}>{}R_{\rm cl}$ and the cloud becomes $\rm H_2$-free. $\bar{\epsilon}_{\rm [C_{II}],\,cl}$ in equation~(\ref{eq.22}) represents the {\em {specific $\rm [C_{II}]$ cooling rate}} of the spherical uniform cloud, which accounts for the relative contribution of the $\rm [C_{II}]$ emission from $\rm H^+$ and $\rm H_I$ regions ($10^{-23}{}\simless{}\bar{\epsilon}_{\rm [C_{II}],\,cl}{}\simless{}10^{-21}\,\rm erg\,s^{-1}\,cm^3$). Like $\bar{\epsilon}_{\rm [C_{II}],\,slab}$ for the plane-parallel slab (equation~\ref{eq.19}), $\bar{\epsilon}_{\rm [C_{II}],\,cl}$ depends on {$x^{(1)}_{\rm C^+}$}, $N_{\rm s}$ and $N_{\rm F}$ but have different functional relation with these parameters due to the difference in geometry. We refer the readers to Appendix~\ref{Sec:Ap6}, where we present the derivation for $\bar{\epsilon}_{\rm [C_{II}],\,cl}$.

Note that we do not take into account the effects of the CMB background on the $\rm [C_{II}]$ cooling rate of gas in the analytic solution for the toy models presented in this section. While the CMB sets a floor for the excitation (or spin) temperature of gas and boosts the upper level (${}^{2}P_{3/2}$) population of the $\rm [C_{II}]$ transition (`CMB heating'), it acts as a background against which the $\rm [C_{II}]$ line is measured (`CMB attenuation'). The CMB effects (both heating and attenuation) can be important for the $\rm [C_{II}]$ emission from the low-density and low-temperature gas in galaxies at high redshifts ($z{}\simgreat{}6$) (see Appendix~\ref{Sec:Ap4}). We find, however, that the total $\rm [C_{II}]$ luminosity of the {\sc\small FIRE} sample is not significantly affected by the CMB (in agreement with \citealt{Lagache_2018}). This is due to the fact that the bulk of the $\rm [C_{II}]$ luminosity of the high-$z$ ($z{}\ge{}6$) galaxies in our sample originates from the gas of densities in excess of the densities where the CMB effects become important. 

\subsection{A scaling relation for the $L_{\rm [C_{II}]}$/SFR ratio of galaxies}
\label{Sec:5b}

\begin{figure}
 \includegraphics[width=85mm]{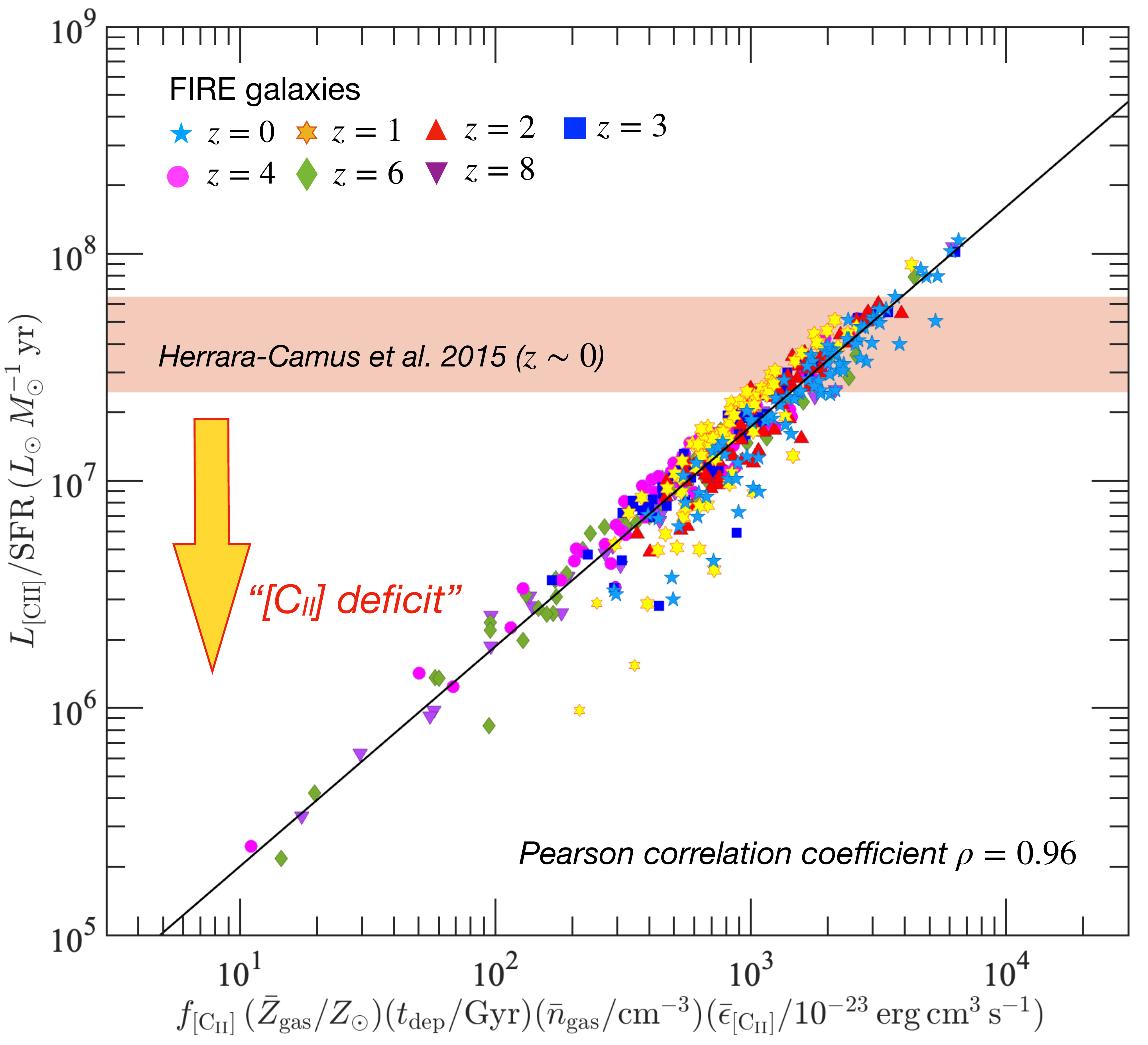}
 \caption{The relation between the $L_{\rm [C_{II}]}{}/{}\rm SFR$ ratio and $f_{\rm [C_{II}]}\,\bar{Z}_{\rm gas}\,t_{\rm dep}\bar{n}_{\rm gas}\bar{\epsilon}_{\rm [C_{II}]}$ of the {\sc \small FIRE} galaxies at different redshifts (cyan stars for $z{}={}0$, yellow hexagons for $z{}={}1$, red triangles for $z{}={}2$, blue squares for $z{}={}3$, magenta circles for $z{}={}4$, green diamonds for $z{}={}6$ and purple downward triangles for $z{}={}8$).  {The orange shaded band indicates the mean $L_{\rm [C_{II}]}{}/{}\rm SFR$ ratio of the local star-forming galaxy sample measured by \citetalias{Herrera_Camus_2015}. The width of the band indicates the $\pm1\sigma$ scatter.} The solid black line shows the best linear fit to the data of the {\sc\small FIRE} galaxies. The {\sc\small FIRE} galaxies show a strong linear correlation (Pearson correlation coefficient $\rho{}=0.96$) between $L_{\rm [C_{II}]}{}/{}\rm SFR$ and $f_{\rm [C_{II}]}\,\bar{Z}_{\rm gas}\,t_{\rm dep}\bar{n}_{\rm gas}\bar{\epsilon}_{\rm [C_{II}]}$. {A large number of the {\sc\small FIRE} galaxies in our sample are below the mean $L_{\rm [C_{II}]}{}/{}\rm SFR$ ratio of the \citetalias{Herrera_Camus_2015} sample (and those of the local \citetalias{De_Looze_2011, De_Looze_2014} samples, which are not shown in the figure), showing a $\rm [C_{II}]$ deficit. }}
    \label{fig.10}
\end{figure}

We have summarized the key points of the \citetalias{Ferrara_2019} model for the structures of a plane-parallel gas slab that is exposed to an external radiation field. We then derive the $\rm [C_{II}]$ luminosity of a uniform spherical gas cloud (equation~\ref{eq.23}). Following the results of the toy models, we now present a scaling relation for the $\rm [C_{II}]$ luminosity of galaxies, based on which we will explore the origins of the $\rm [C_{II}]$ deficit of galaxies.  

From equation~(\ref{eq.23}), one would expect that the $\rm [C_{II}]$ luminosity ($L_{\rm [C_{II}]}$) of galaxy has a similar expression, \ie 
\begin{equation}
L_{\rm [C_{II}]} \sim  f_{\rm [C_{II}]} \left (\frac{M_{\rm gas}}{\mu m_{\rm H}} \right) \bar{n}_{\rm gas}\bar{\mathcal{A}}_{\rm C} \bar{\epsilon}_{\rm [C_{II}]},
\label{eq.24}
\end{equation}

\noindent where we have replaced $M_{\rm cl}$ in equation~(\ref{eq.23}) by $M_{\rm gas}$, \ie~the gas mass of galaxy\footnote{{We calculate the gas mass of galaxy using the gas particles within $0.1R_{\rm vir}$ around the DM halo centre having $T{}<{}10^5$ K.}}. $f_{\rm [C_{II}]}$ ($={}1-f_{\rm H_2}$) in the above equation represents the fraction of the total gas mass in ionized or neutral atomic hydrogen forms (Zone I and Zone II), and $\bar{n}_{\rm gas}$, $\bar{\mathcal{A}}_{\rm C}$ and $\bar{\epsilon}_{\rm [C_{II}]}$ represent the {\em statistical average} of gas density, carbon abundance and specific $\rm [C_{II}]$ cooling rate of the galaxy, respectively. We can then divide the two sides of equation~(\ref{eq.24}) by galaxy SFR, and obtain
\begin{equation}
\frac{L_{\rm [C_{II}]}}{\rm SFR} \sim f_{\rm [C_{II}]}   t_{\rm dep} \bar{n}_{\rm gas}\bar{\mathcal{A}}_{\rm C} \bar{\epsilon}_{\rm [C_{II}]}   (\mu m_{\rm H})^{-1}
 \label{eq.25}
\end{equation}
where 
\begin{equation}
t_{\rm dep}\equiv \frac{M_{\rm gas}}{\rm SFR}
\label{eq.26}
\end{equation}
\noindent is the {\em gas depletion time} of {the} galaxy \citep[\eg,][]{Genzel_2015, Tacchella_2016, Semenov_2017, Scoville_2017, Tacconi_2018, Feldmann_2020}. Through further re-arrangement, equation~(\ref{eq.25}) can be expressed as
\begin{align}
\frac{L_{\rm [C_{II}]}/L_\odot}{{\rm SFR}/(M_\odot\,\rm yr^{-1})}{}\sim{}&4\times10^5 f_{\rm [C_{II}]}  \left( \frac{\bar{Z}_{\rm gas}}{Z_\odot} \right) \nonumber \\  
 & \times \left ( \frac{t_{\rm dep}}{\rm Gyr} \right ) \left(\frac{\bar{n}_{\rm gas}}{\rm cm^{-3}}\right)  \left ( \frac{\bar{\epsilon}_{\rm [C_{II}]}}{10^{-23}\,\rm erg\,s^{-1}\,cm^3} \right) \label{eq.27} \\
 \propto &  f_{\rm [C_{II}]}\, \bar{Z}_{\rm gas} \, t_{\rm dep} \,\bar{n}_{\rm gas}\, \bar{\epsilon}_{\rm [C_{II}]},
 \label{eq.28}
\end{align}
\noindent where we have replaced the carbon abundance $\bar{\mathcal{A}}_{\rm C}$ in equation~(\ref{eq.25}) by metallicity $\bar{Z}_{\rm gas}$ using equation~(\ref{eq.17}). 

\begin{figure}
 \includegraphics[width=86mm]{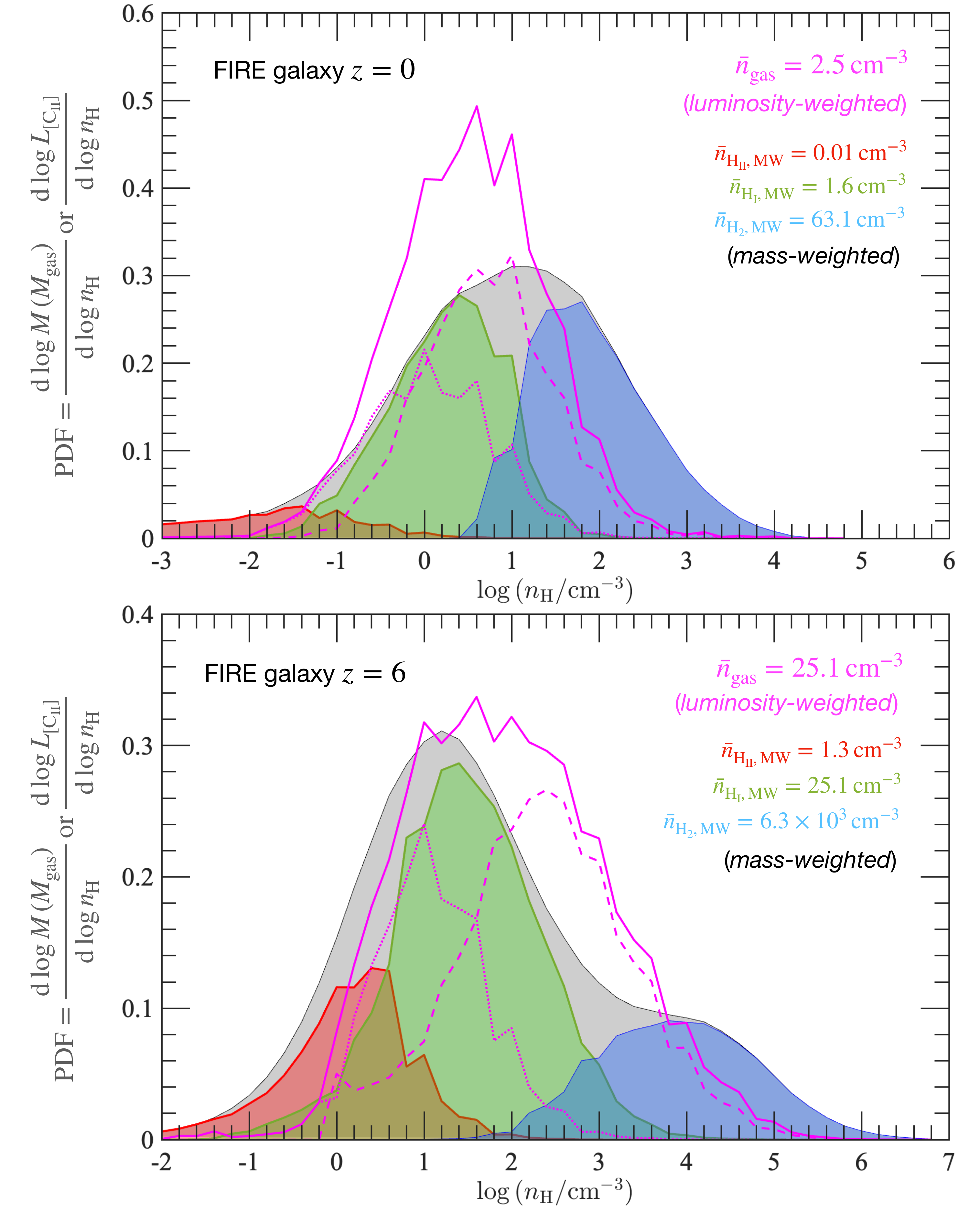}
 \includegraphics[width=86mm]{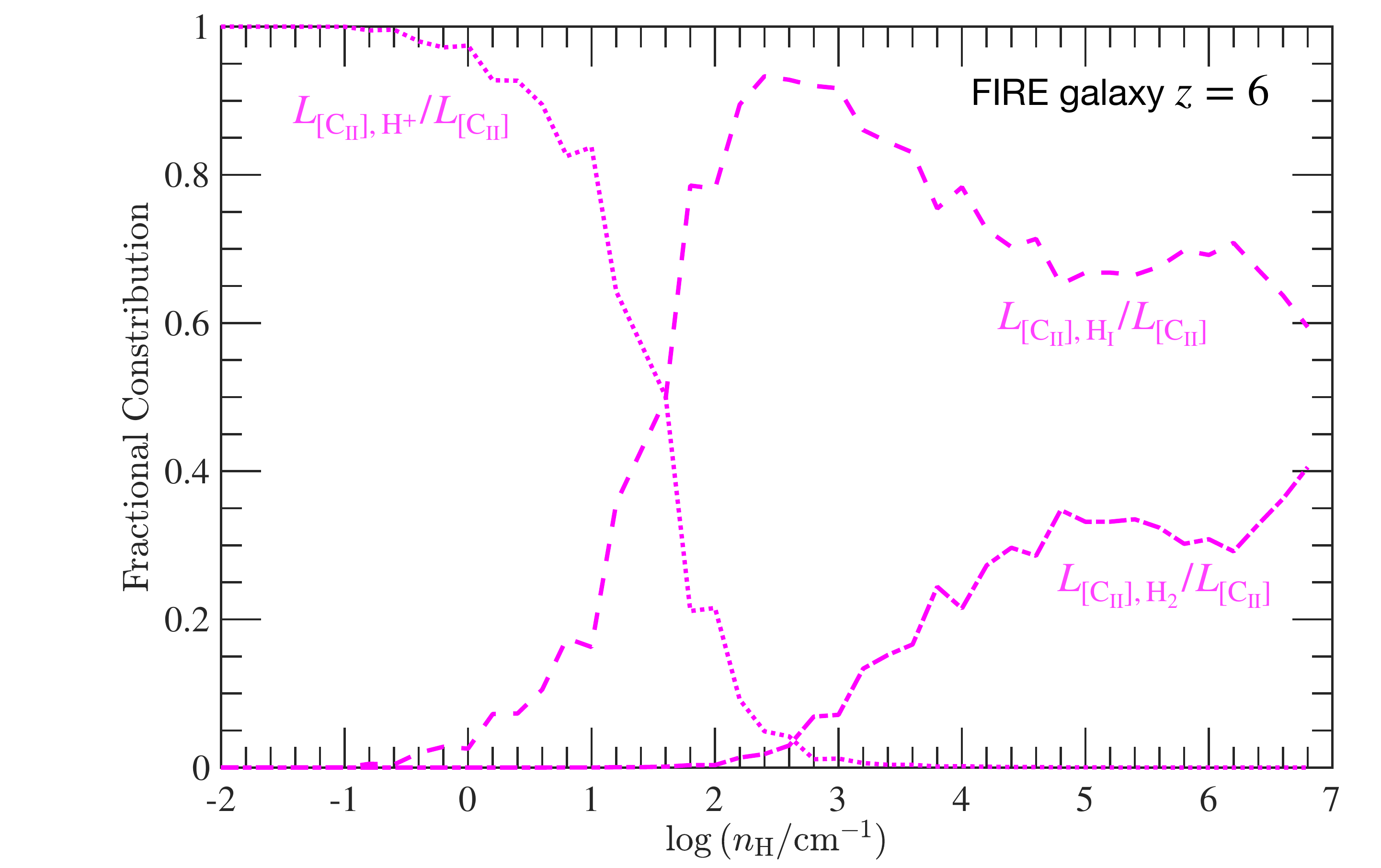}
 \caption{In the {\it top} and {\it middle} panels, we present the gas density Probability Density Functions (PDFs) for two selected \textsc{\small FIRE} galaxies at $z{}={}0$ and $z{}={}6$, respectively. The $z{}={}6$ galaxy exhibits a relatively denser ISM. Magenta lines in both panels indicate the {\em luminosity-weighted} PDFs. Specifically, solid, dotted, and dashed lines represent the results for the total gas, $\rm H^+$ gas (Zone I), and $\rm H_I$ gas (Zone II) in the ISM. The shaded areas in both panels depict the {\em mass-weighted} gas density PDFs. Grey, red, green, and blue areas represent the results for the total gas, $\rm H^+$ gas (Zone I), $\rm H_I$ gas (Zone II), and $\rm H_2$ gas (Zone III), respectively. {In the {\it bottom} panel, dotted, dashed, and dash-dotted lines show the fraction of the $\rm [C_{II}]$ emission from the selected $z{}={}6$ galaxy originating from the $\rm H^+$, $\rm H_I$, and $\rm H_2$ gas, respectively. }}
    \label{fig.11}
\end{figure}

\begin{figure}
 \includegraphics[width=86mm]{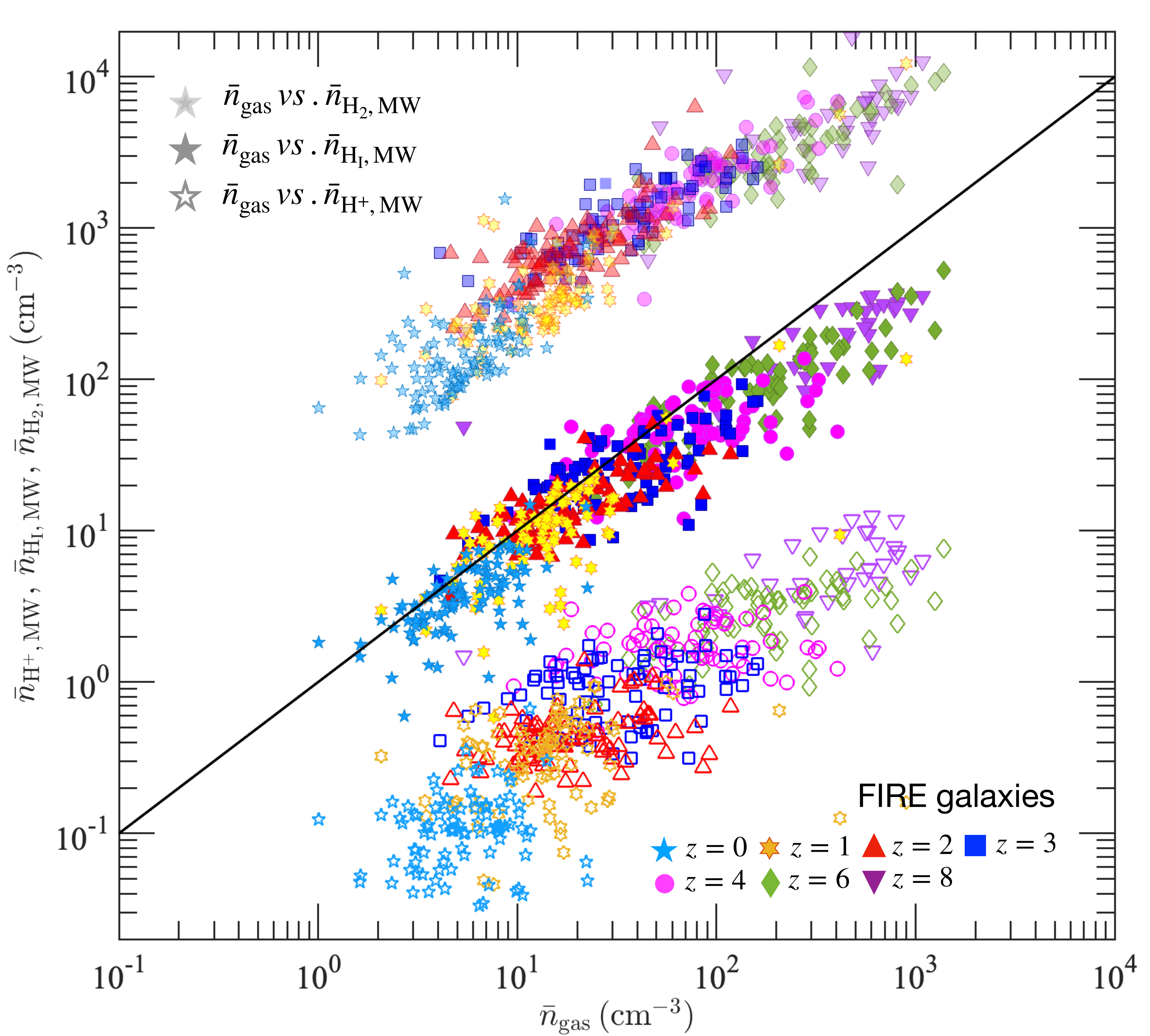}
 \caption{The relation between the $\rm [C_{II}]$ luminosity-weighted gas density ($\bar{n}_{\rm gas}$) and the mass-weighted density of the $\rm H^+$ ($\bar{n}_{\rm H^+,\,MW}$), $\rm H_I$ ($\bar{n}_{\rm H_{I},\,MW}$) and $\rm H_2$ gas ($\bar{n}_{\rm H_2,\,MW}$) of the {\sc\small FIRE} galaxies at $z{}={}0-8$. Filled, empty and semi-transparent symbols correspond to the $\bar{n}_{\rm H_{I},\,MW}$ vs. $\bar{n}_{\rm gas}$, the $\bar{n}_{\rm H^+,\,MW}$ vs. $\bar{n}_{\rm gas}$ and the $\bar{n}_{\rm H_2,\,MW}$ vs. $\bar{n}_{\rm gas}$ relations, respectively. The diagonal line indicates the one-to-one relationship. It can be seen that $\bar{n}_{\rm gas}$ appears to be close to $\bar{n}_{\rm H_{I},\,MW}$, both being systematically lower (higher) than $\bar{n}_{\rm H_2,\,MW}$ ($\bar{n}_{\rm H^+,\,MW}$).}
    \label{fig.12}
\end{figure}

\begin{figure}
 \includegraphics[width=84mm]{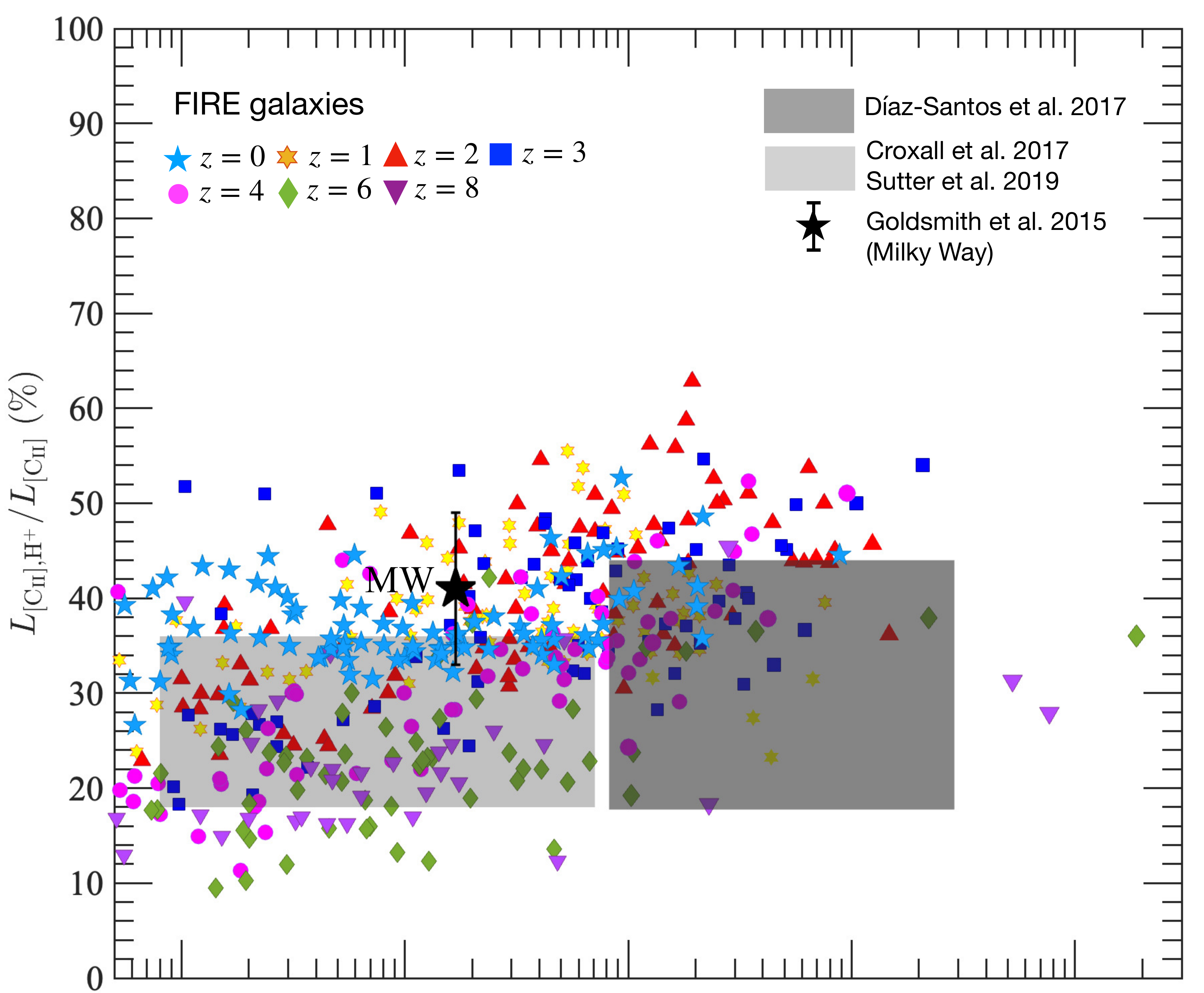}
  \includegraphics[width=84mm]{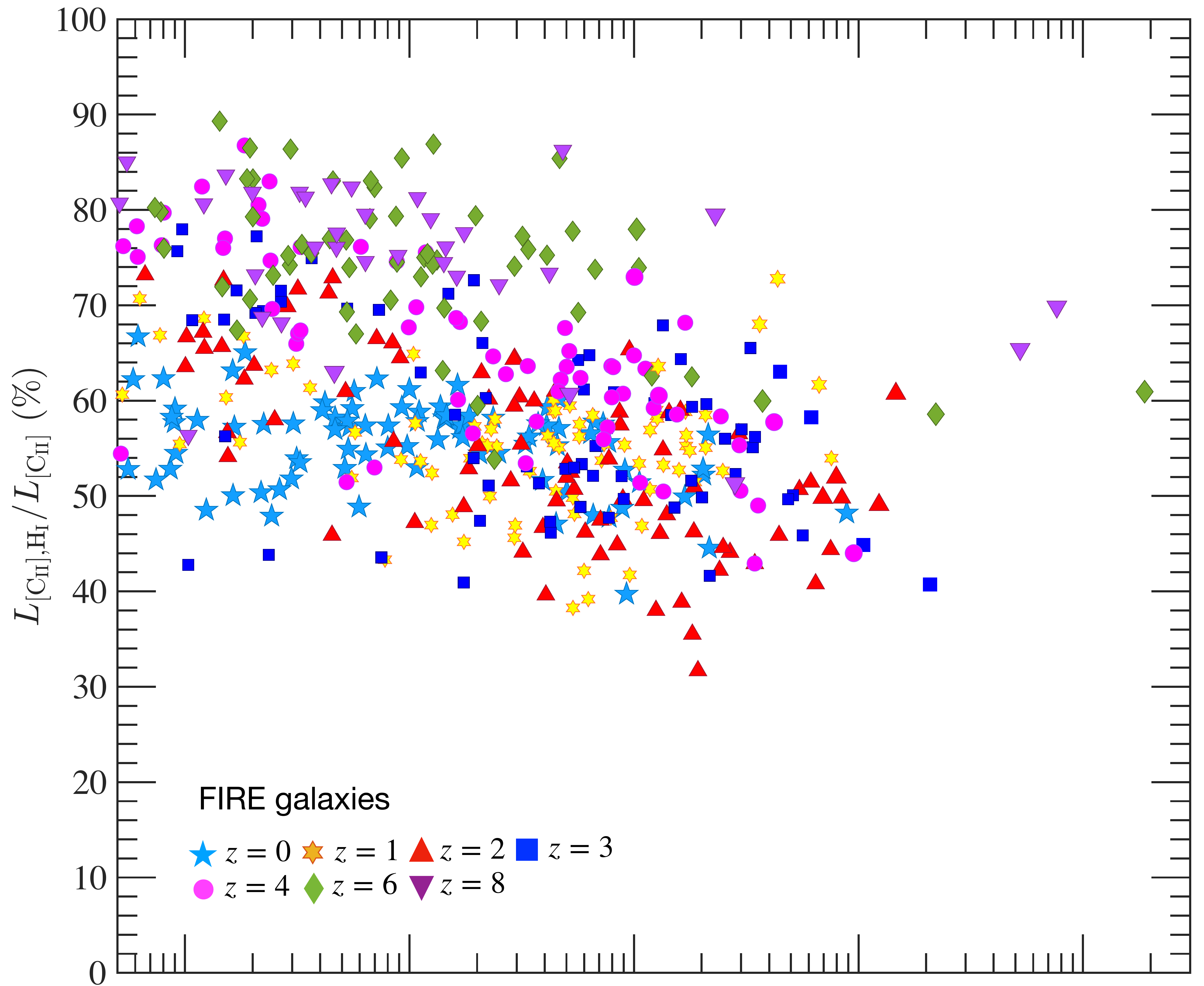}
  \includegraphics[width=84mm]{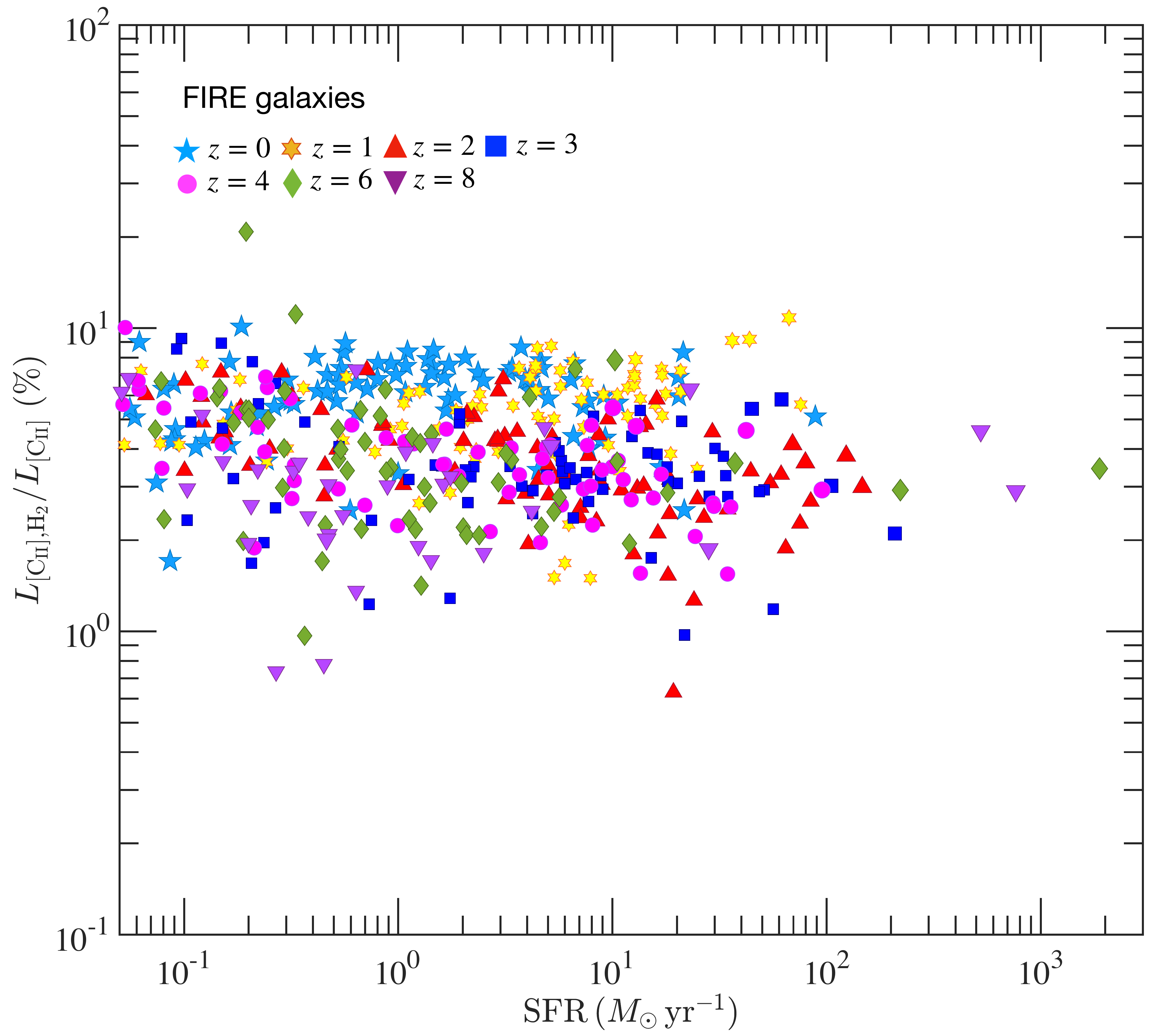}
  \caption{The fraction of the total $\rm [C_{II}]$ luminosity of the {\sc\small FIRE} galaxy sample that originates from the $\rm H^+$ (\textit{upper} panel), $\rm H_I$ ({\it middle} panel) and $\rm H_2$ gas phases ({\it lower} panel) as a function of their SFR. In the \textit{upper} panel, the dark (light) shaded area indicates the observational result ($\pm1\sigma$) of the local $z{}={}0$ samples by \citet{Sutter_2019}. {The black hexagon, along with the error bar, represents the constraint on the Galactic Plane measured by \citet{Goldsmith_2015}.}}
   \label{fig.13}
\end{figure}

Equation~(\ref{eq.27}) indicates that the $L_{\rm [C_{II}]}{}/{}\rm SFR$ ratio of galaxy is determined by five physical parameters, $f_{\rm [C_{II}]}$, $\bar{Z}_{\rm gas}$, $t_{\rm dep}$, $\bar{n}_{\rm gas}$ and $\bar{\epsilon}_{\rm [C_{II}]}$. Whilst $f_{\rm [C_{II}]}$ and $t_{\rm dep}$ are {\it global} properties of galaxy, which are well defined, the other three parameters are the {\em statistical average} of the corresponding physical properties of all different `gas clouds' in the ISM. This contrasts with the toy models (uniform plane-parallel slab or spherical cloud), where each of these properties (gas density, gas metallicity, and the specific $\rm [C_{II}]$ cooling rate) has a single, definite value. 

In Fig.~\ref{fig.10}, we show the relation between the $L_{\rm [C_{II}]}{}/{}\rm SFR$ ratio of the {\sc\small FIRE} sample at $z{}={}0-8$ and their $f_{\rm [C_{II}]}\, \bar{Z}_{\rm gas}\, t_{\rm dep}\, \bar{n}_{\rm gas} \bar{\epsilon}_{\rm [C_{II}]}$, where $\bar{n}_{\rm gas}$, $\bar{Z}_{\rm gas}$ and $\bar{\epsilon}_{\rm [C_{II}]}$ are the {\em luminosity-weighted} gas density \footnote{Note that we use the `luminosity-weighted median gas density', \ie~the gas density at the $50^{\rm th}$ percentile of $\rm [C_{II}]$ luminosity, instead of the `luminosity-weighted mean gas density'. This is because the gas density PDF of galaxy resembles a lognormal function, exhibiting an elongated tail at the high density end. Under certain circumstances, the `mean gas density' can be strongly biased by the $\rm [C_{II}]$-emitting gas at the highest density ({$n_{\rm H}\simgreat10^3\,\rm cm^{-3}$}, see the {\it lower} panel of Fig.~\ref{fig.11}), and hence is not statistically representative for the part of the gas that contributes the bulk of the $\rm [C_{II}]$ emission of galaxy. Throughout this paper, we use the term `luminosity-weighted' for simplicity when we refer to `luminosity-weighted median'. Similarly, `mass-weighted' in this paper refers to `mass-weighted median', \ie~value at the $50^{\rm th}$ percentile of mass. In Appendix~\ref{Sec:Ap7}, we show explicitly the difference between the `luminosity-weighted median gas density' and the `luminosity-weighted mean gas density' of the {\sc\small FIRE} galaxy sample. The former is higher by a factor of $\sim{}5$ on average.}, gas metallicity\footnote{Unlike the gas densities, the luminosity-weighted mean and the luminosity-weighted median gas metallicity are similar. Both are higher than the mass-weighted gas metallicity (see Appendix~\ref{Sec:Ap8}).} and specific $\rm [C_{II}]$ cooling rate of the galaxies, respectively. Our {\sc\small FIRE} sample follows a clear linear scaling relation on the diagram (Pearson correlation coefficient $\rho{}={}0.96$), which is in agreement with equation~(\ref{eq.27}). 

In the same figure, we explicitly show the mean $L_{\rm [C_{II}]}{}/{}\rm SFR$ ratio of the $z{}={}0$ star-forming galaxy sample of \citetalias{Herrera_Camus_2015} (shaded orange band). The \citetalias{Herrera_Camus_2015} sample demonstrates an almost linear correlation between $L_{\rm [C_{II}]}$ and SFR. As a result, the $L_{\rm [C_{II}]}{}/{}\rm SFR$ ratio remains nearly independent of SFR across the range of ${\rm SFR}{}\approx{}10^{-3}-10\,M_\odot\,\rm yr^{-1}$ (see Table~\ref{T2}). We therefore can use the mean $L_{\rm [C_{II}]}{}/{}\rm SFR$ ratio from the H15 sample as a reference point. Galaxies with significantly lower $L_{\rm [C_{II}]}{}/{}\rm SFR$ ratios than this reference point are considered to exhibit a $\rm [C_{II}]$ deficit. It is evident from the figure that a substantial number of the {\sc\small FIRE} galaxies in our sample, particularly the early galaxies, display a $\rm [C_{II}]$ deficit.

One crucial question is identifying the primary contributor to the $\rm [C_{II}]$ emission in a galaxy's ISM. The ISM exhibits a wide density range spanning several orders of magnitude, with denser regions dominated by $\rm H_2$ and diffuse regions by $\rm H^+$ gas. In Fig.~\ref{fig.11}, we depict the $\rm [C_{II}]$ luminosity-weighted (magenta lines) and gas mass-weighted (grey and coloured shaded areas) probability density functions (PDFs) for $n_{\rm H}$ in two selected \textsc{\small FIRE} galaxies at $z{}={}0$ ({\it top} panel) and $z{}={}6$ ({\it middle} panel). The figure illustrates that $\rm [C_{II}]$ emission in \textsc{\small FIRE} galaxies originates from gas spanning a wide density range across several orders of magnitude. Interestingly, we observe that the luminosity-weighted gas density ($\bar{n}_{\rm gas}$) of \textsc{\small FIRE} galaxies closely aligns with the mass-weighted density of $\rm H_I$ gas ($\bar{n}_{\rm H_I,\,MW}$) in the ISM. Both are notably higher (lower) than the mass-weighted density of $\rm H^+$ ($\rm H_2$) gas. This relationship is more evident in Fig.~\ref{fig.12}, where we depict the correlation between $\bar{n}_{\rm gas}$ and the mass-weighted gas density of $\rm H^+$, $\rm H_I$, and $\rm H_2$ gas for the \textsc{\small FIRE} sample at $z{}={}0-8$.

This observation can be explained by the inefficiency of the bulk of the diffuse, ionized $\rm H^+$ gas in producing $\rm [C_{II}]$ emission due to its low gas density ($L_{\rm [C_{II}],\,cl}{}/{}M_{\rm cl}{}\propto{}n_{\rm H}$, see equation~\ref{eq.23}). Conversely, in the densest ISM regions where gas is primarily in molecular hydrogen form (Zone III), there is not much $\rm [C_{II}]$ emission due to the scarcity of ionized carbon (predominantly in Zone I and Zone II) in those areas. Consequently, the majority of the $\rm [C_{II}]$ luminosity in \textsc{\small FIRE} galaxies at $z{}={}0-8$ originates from gas within the intermediate density range.

We present in Fig.\ref{fig.13} the fractional contribution of $\rm [C_{II}]$ emission from different gas phases ($\rm H^+$, $\rm H_I$ and $\rm H_2$) in the \textsc{\small FIRE} galaxies. Notably, $50\%-80\%$ of the total $\rm [C_{II}]$ emission originates from $\rm H_I$ gas regions, with the majority of the remaining emission attributed to $\rm H^+$ gas. {Inside the galaxy, the contribution of $\rm H_I$ gas dominates in intermediate and high-density regions, while $\rm H^+$ gas dominates in the diffuse regions in the ISM. This trend is illustrated in the {\it bottom} panel of Fig.\ref{fig.11}.}

{The predicted fractional contribution of the $\rm H^+$ gas aligns closely with the upper limits of the observational data reported by \citet{Diaz_Santos_2017}, who investigated the LIRGs in the {\sc\small GOALS} \citep{Armus_2009} sample, as well as \citet{Sutter_2019}, who studied a sample of normal star-forming galaxies from the {\sc\small KINGFISH} \citep{Kennicutt_2011} catalogue. Additionally, our results demonstrate strong agreement with the findings of \citet{Goldsmith_2015}, who conducted measurements along the Galactic Plane.}

Finally, we find that only $<{}10\%$ of the $\rm [C_{II}]$ emission originates from $\rm H_2$ gas in our sample (see Section~\ref{Sec:6a} for further discussions).

\begin{figure*}
 \includegraphics[width=170mm]{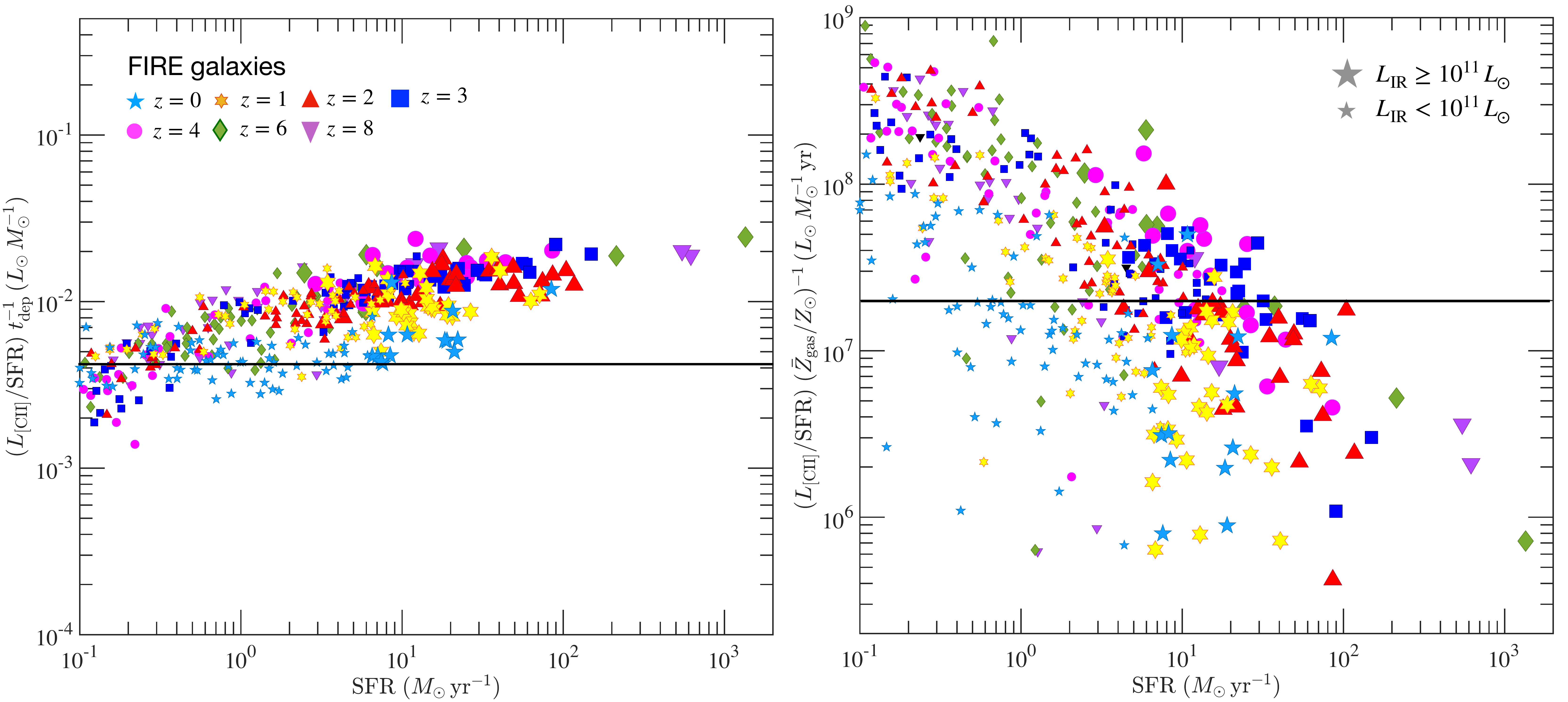}
  \includegraphics[width=175mm]{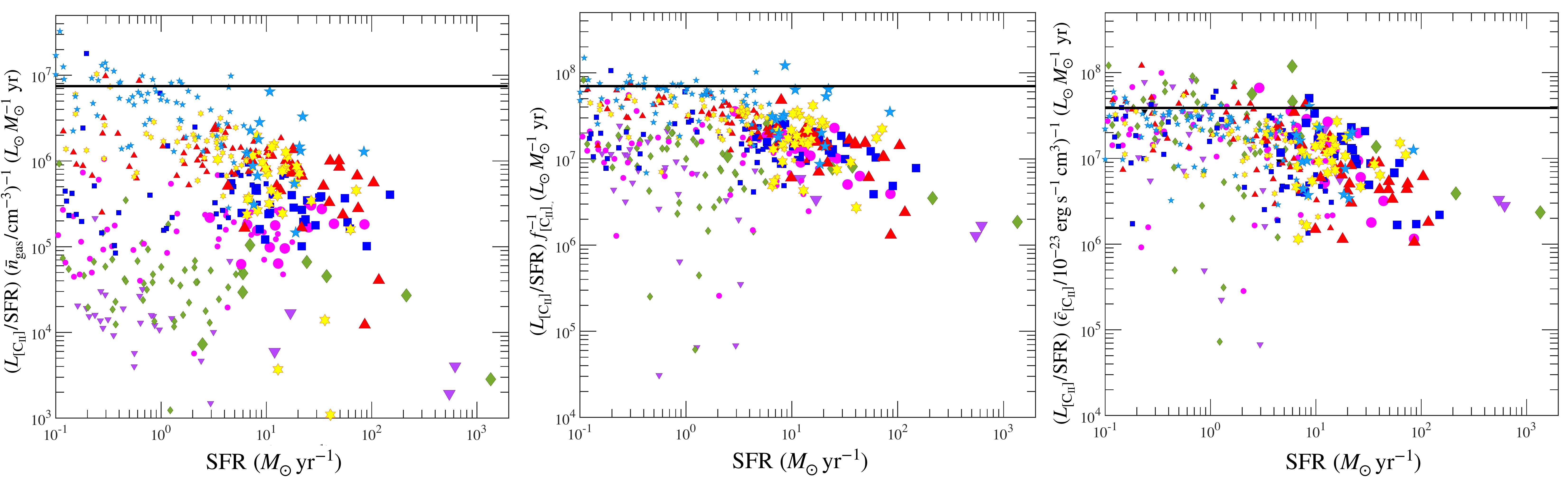}
\caption{ {The relation between $L_{\rm [C_{II}]}{}/{}{\rm SFR}\,t^{-1}_{\rm dep}$ ({\it upper left}), $L_{\rm [C_{II}]}{}/{}{\rm SFR}\,\bar{Z}^{-1}_{\rm gas}$ ({\it upper right}), $L_{\rm [C_{II}]}{}/{}{\rm SFR}\,\bar{n}^{-1}_{\rm gas}$ ({\it lower left}), $L_{\rm [C_{II}]}{}/{}{\rm SFR}\,f^{-1}_{\rm [C_{II}]}$ ({\it lower middle}) and $L_{\rm [C_{II}]}{}/{}{\rm SFR}\,\bar{\epsilon}^{-1}_{\rm [C_{II}]}$  ({\it lower right}) against SFR} of the {\sc\small FIRE} galaxies at different redshifts (cyan stars for $z{}={}0$, yellow hexagons for $z{}={}1$, red triangles for $z{}={}2$, blue squares for $z{}={}3$, magenta circles for $z{}={}4$, green diamonds for $z{}={}6$ and purple downward triangles for $z{}={}8$). {In each panel, large symbols denote galaxies with $L_{\rm IR}{}\ge{}10^{11}\,L_\odot$, while small symbols denote galaxies with $L_{\rm IR}{}<{}10^{11}\,L_\odot$. The solid black line indicates the mean value of the normal SFGs at $z{}={}0$. The figure reveals that the reduced $L_{\rm [C_{II}]}{}/{}\rm SFR$ ratio of the galaxies with high SFR (at high $z$) is primarily due to a relatively low $t_{\rm dep}$ (gas metallicity). (see Section~\ref{Sec:5c} for the details).}}
    \label{fig.14}
\end{figure*}

\subsection{The physical origins of $\rm [C_{II}]$ deficit of galaxies}
\label{Sec:5c}

\begin{figure*}
 \includegraphics[width=165mm]{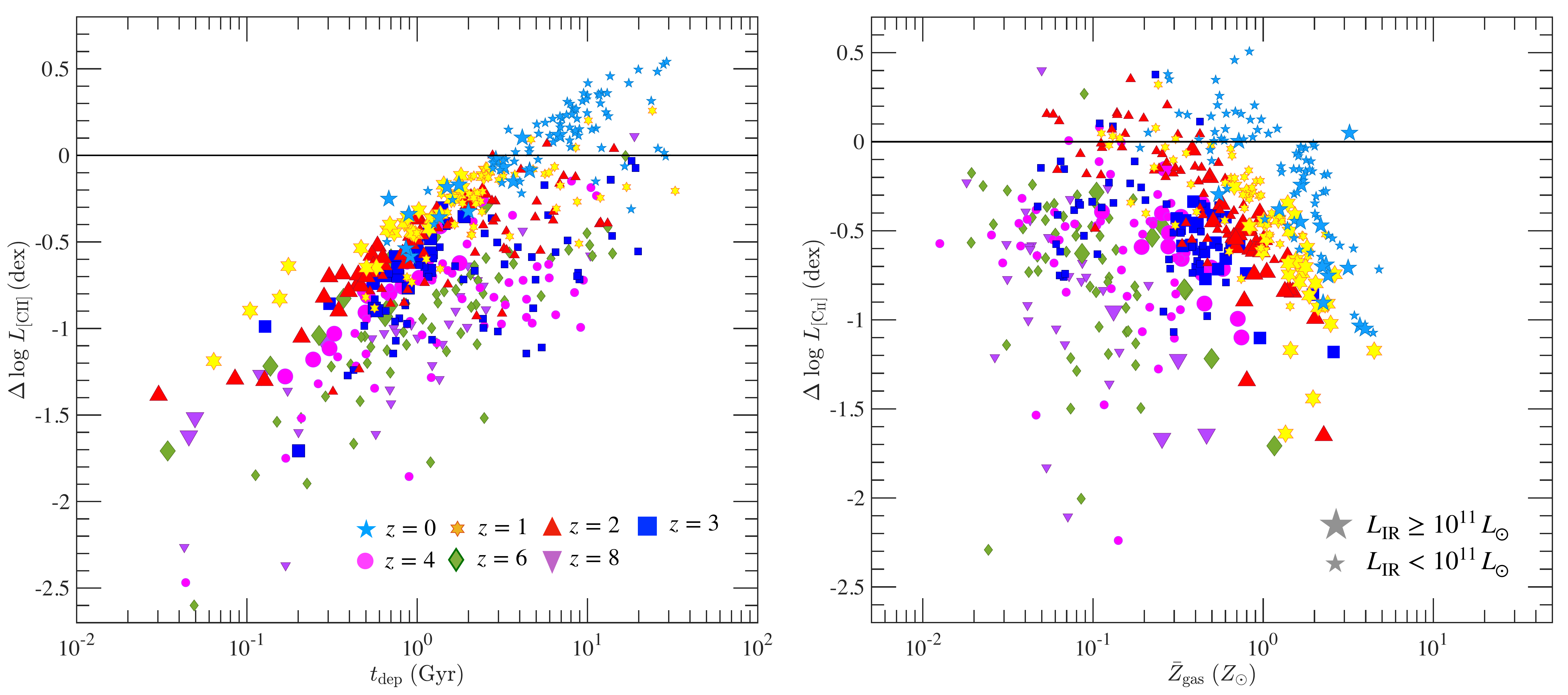}
  \includegraphics[width=175mm]{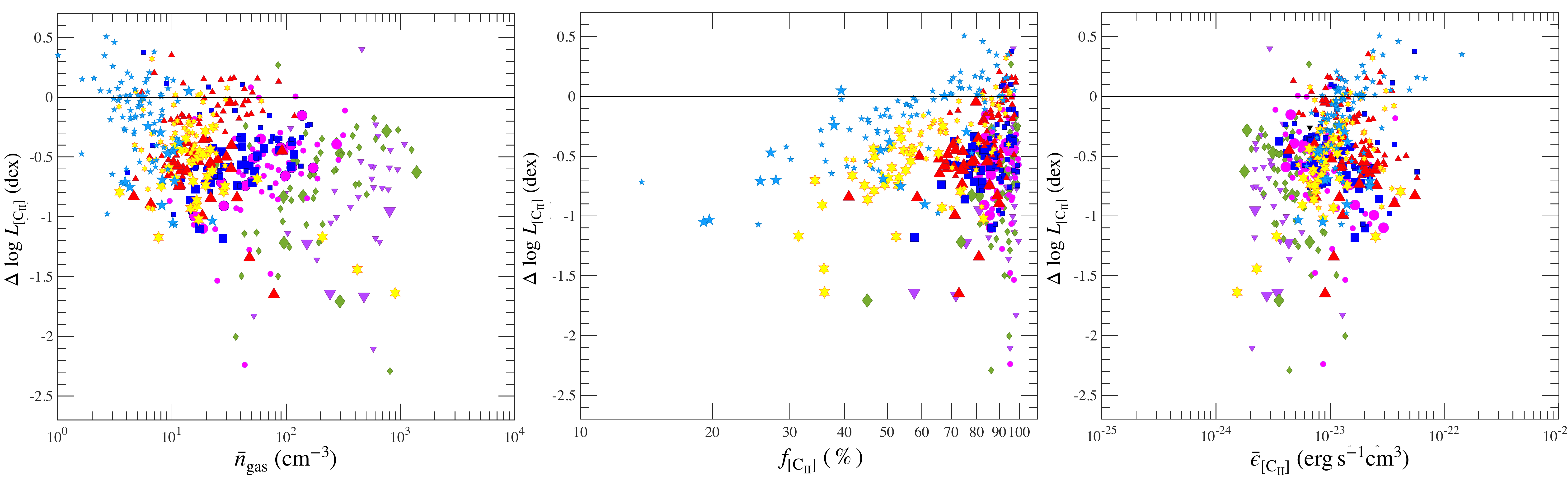}
  \caption{{$\Delta({\rm log}\,L_{\rm [C_{II}]})$ as a function of $t_{\rm dep}$ ({\it upper left}), $\bar{Z}_{\rm gas}$ ({\it upper right}), $\bar{n}_{\rm gas}$ ({\it lower left}), $f_{\rm [C_{II}]}$ ({\it lower middle}) and $\bar{\epsilon}_{\rm [C_{II}]}$ ({\it lower right})} of the {\sc\small FIRE} galaxies at different redshifts, where $\Delta({\rm log}\,L_{\rm [C_{II}]})$ represents the offset between the $L_{\rm [C_{II}]}{}/{}\rm SFR$ ratio of the galaxies and the observed mean value of the local star-forming sample of \citetalias{Herrera_Camus_2015} ($4.3\times10^7\,L_\odot\,M^{-1}_\odot\,\rm yr$). In each panel, large (small) symbols correspond to the {\sc\small FIRE} galaxies having $L_{\rm IR}{}\ge{}10^{11}\,L_\odot$ ($L_{\rm IR}{}<{}10^{11}\,L_\odot$). }
   \label{fig.15}
\end{figure*}

In the previous section, we have presented a simple analytic expression for the $L_{\rm [C_{II}]}{}/{}\rm SFR$ ratio of galaxies (equation~\ref{eq.27}) found with the {\sc\small FIRE} galaxy sample. Based on this result, we will probe in this section the origins of the observed $\rm [C_{II}]$ deficit of galaxies. 

Equation~(\ref{eq.27}) indicates that the $L_{\rm [C_{II}]}{}/{}\rm SFR$ ratio of the galaxies depends on five parameters: the fraction of gas in the $\rm [C_{II}]$-emitting regions (Zone I and Zone II), the depletion time (\ie~gas mass per unit SFR), gas density, gas metallicity and the specific $\rm [C_{II}]$ cooling rate. {\em Hence, the $\rm [C_{II}]$ deficit of the galaxies can, in principle, be due to a strong deficit of one or few of the five parameters with respect to the observed local star-forming samples} (\eg,~\citetalias{De_Looze_2011}, \citetalias{De_Looze_2014} and \citetalias{Herrera_Camus_2015}). It should be noted that the observed $\rm [C_{II}]$ deficit in the two regimes, high redshifts and high $L_{\rm IR}$, may not be due to the same reason. We will separately discuss the origin of the $\rm [C_{II}]$ deficit in these two regimes in this section. 

To investigate the factors influencing the $\rm [C_{II}]$ deficit in the FIRE sample, we analyze the $L_{\rm [C_{II}]}{}/{}\rm SFR$ ratio in relation to a range of parameters. We assess whether the `$\rm [C_{II}]$ deficit'\footnote{{In Section~\ref{Sec:5c}, the concept of the `$\rm [C_{II}]$ deficit' extends beyond comparing a galaxy's $L_{\rm [C_{II}]}{}/{}\rm SFR$ ratio to that of $z{}={}0$ normal SFGs; it also encompasses the consideration of the five new parameters. We establish the mean values of these new parameters for local SFGs as the new reference points. Galaxies with significantly lower values for any of the new parameters compared to the $z{}={}0$ SFGs are categorized as {\em having a `$\rm [C_{II}]$ deficit' in that particular parameter space}. For those galaxies that exhibit a `$\rm [C_{II}]$ deficit' in the $L_{\rm [C_{II}]}{}/{}\rm SFR$ ratio but possess similar or higher values for one of the new parameters than the $z{}={}0$ SFGs, we consider their `$\rm [C_{II}]$ deficit' as {\em `disappearing' within the new parameter space}. }} diminishes or disappears in new parameter spaces, including ($L_{\rm [C_{II}]}{}{\rm SFR}^{-1}){}f^{-1}_{\rm [C_{II}]}$, $(L_{\rm [C_{II}]}{}{\rm SFR}^{-1}){}\bar{Z}^{-1}_{\rm gas}$,  $(L_{\rm [C_{II}]}{}{\rm SFR}^{-1}){}\bar{n}^{-1}_{\rm gas}$,  $(L_{\rm [C_{II}]}{}{\rm SFR}^{-1}){}t^{-1}_{\rm dep}$ and $(L_{\rm [C_{II}]}{}{\rm SFR}^{-1}){}\bar{\epsilon}^{-1}_{\rm [C_{II}]}$. 

In Fig.~\ref{fig.14}, we illustrate these new parameters as a function of SFR for the galaxies in our sample at different redshifts. {Note that previous observations have indicated a $\rm [C_{II}]$ deficit at high $L_{\rm IR}$ ($L_{\rm IR}{}\simgreat{}10^{11}\,L_\odot$), where SFR and $L_{\rm IR}$ are closely correlated (Fig.~\ref{fig.5}). By graphing these new parameters as a function of SFR, it becomes clearer which parameters contribute to the $\rm [C_{II}]$ deficit at high SFR ($\sim{}L_{\rm IR}$).} 

In Fig.~\ref{fig.15}, we also demonstrate how $L_{\rm [C_{II}]}{}/{}\rm SFR$ in the FIRE sample depends on $f_{\rm [C_{II}]}$, $\bar{Z}_{\rm gas}$, $\bar{n}_{\rm gas}$, $t_{\rm dep}$ and $\bar{\epsilon}_{\rm [C_{II}]}$, each presented in separate panels. For reference, readers can find the mean values of $f_{\rm [C_{II}]}$, $\bar{Z}_{\rm gas}$, $\bar{n}_{\rm gas}$, $t_{\rm dep}$ and $\bar{\epsilon}_{\rm [C_{II}]}$, as well as the values of the five new parameters specific to the FIRE sample at each redshift, in Table~\ref{T8} and Table~\ref{T7}, respectively.

\begin{table*}
\centering
\caption{{The difference between the mean values of $L_{\rm [C_{II}]}{}/{}\rm SFR$, $L_{\rm [C_{II}]}{}/{}{\rm SFR}\,t^{-1}_{\rm dep}$, $L_{\rm [C_{II}]}{}/{}{\rm SFR}\,\bar{Z}^{-1}_{\rm gas}$, $L_{\rm [C_{II}]}{}/{}{\rm SFR}\,\bar{n}^{-1}_{\rm gas}$, $L_{\rm [C_{II}]}{}/{}{\rm SFR}\,f^{-1}_{\rm [C_{II}]}$ and $L_{\rm [C_{II}]}{}/{}{\rm SFR}\,\bar{\epsilon}^{-1}_{\rm [C_{II}]}$ for the {\sc\small FIRE} galaxies at redshift $z$ and the values of $z{}={}0$ normal star-forming galaxies (SFGs) in the sample.} }
\begin{threeparttable}
\begin{tabular}{ p{1.8 cm} p{1.7 cm} p{1.7 cm}  p{1.7 cm}  p{1.7 cm} p{1.7 cm} p{1.7 cm} }
 \hline
  \hline
 \multicolumn{1}{c|}{$z$}  &  \multicolumn{1}{c|}{$\Delta\,{\rm log}\,\left (\frac{L_{\rm [C_{II}]}}{\rm SFR} \right)$ }  &  \multicolumn{1}{c|}{$\Delta\,{\rm log}\,\left (\frac{L_{\rm [C_{II}]}}{\rm SFR} t^{-1}_{\rm dep} \right)$}  & \multicolumn{1}{c|}{$\Delta\,{\rm log}\, \left (\frac{L_{\rm [C_{II}]}}{\rm SFR} \bar{Z}^{-1}_{\rm gas} \right)$} & \multicolumn{1}{c|}{$\Delta\,{\rm log}\,\left (\frac{L_{\rm [C_{II}]}}{\rm SFR} \bar{n}^{-1}_{\rm gas} \right)$} & \multicolumn{1}{c|}{$\Delta\,{\rm log}\,\left (\frac{L_{\rm [C_{II}]}}{\rm SFR} f^{-1}_{\rm [C_{II}]} \right)$} & \multicolumn{1}{c|}{$\Delta\,{\rm log}\,\left (\frac{L_{\rm [C_{II}]}}{\rm SFR} \bar{\epsilon}^{-1}_{\rm [C_{II}]} \right)$} \\
  \multicolumn{1}{c|}{}  &  \multicolumn{1}{c|}{(dex)}  &  \multicolumn{1}{c|}{(dex)}  &  \multicolumn{1}{c|}{(dex)} & \multicolumn{1}{c|}{(dex)} & \multicolumn{1}{c|}{(dex)} & \multicolumn{1}{c|}{(dex)}  \\
 \hline
 \multicolumn{1}{c|}{1}  &  \multicolumn{1}{c|}{$-0.38$}  &  \multicolumn{1}{c|}{\bf 0.39}  &  \multicolumn{1}{c|}{$-0.11$} &  \multicolumn{1}{c|}{$-0.67$} &  \multicolumn{1}{c|}{$-0.25$} &  \multicolumn{1}{c|}{$-0.23$} \\
 \multicolumn{1}{c|}{2}  &  \multicolumn{1}{c|}{$-0.32$}  &  \multicolumn{1}{c|}{\bf 0.36}  &  \multicolumn{1}{c|}{0.31} &  \multicolumn{1}{c|}{$-0.71$} &  \multicolumn{1}{c|}{$-0.34$}  &  \multicolumn{1}{c|}{$-0.30$} \\
 \multicolumn{1}{c|}{3}  &  \multicolumn{1}{c|}{$-0.50$}  &  \multicolumn{1}{c|}{\bf 0.41}  &  \multicolumn{1}{c|}{0.27} &  \multicolumn{1}{c|}{$-1.15$} &  \multicolumn{1}{c|}{$-0.51$} &  \multicolumn{1}{c|}{$-0.30$} \\
 \multicolumn{1}{c|}{4}  &  \multicolumn{1}{c|}{$-0.58$}  &  \multicolumn{1}{c|}{0.37}  &  \multicolumn{1}{c|}{\bf 0.53} &  \multicolumn{1}{c|}{$-1.54$} &  \multicolumn{1}{c|}{$-0.60$} & \multicolumn{1}{c|}{$-0.22$} \\
  \multicolumn{1}{c|}{6}  &  \multicolumn{1}{c|}{$-0.70$}  &  \multicolumn{1}{c|}{0.21}  &  \multicolumn{1}{c|}{\bf 0.78} &  \multicolumn{1}{c|}{$-2.08$} &  \multicolumn{1}{c|}{$-0.67$} & \multicolumn{1}{c|}{$-0.09$} \\
 \multicolumn{1}{c|}{8}  &  \multicolumn{1}{c|}{$-0.81$}  &  \multicolumn{1}{c|}{0.22}  &  \multicolumn{1}{c|}{\bf 0.67} &  \multicolumn{1}{c|}{$-2.58$} &  \multicolumn{1}{c|}{$-0.86$} & \multicolumn{1}{c|}{$-0.11$} \\
 \hline
 \end{tabular}
  \end{threeparttable}
  \label{T7}
\end{table*}

\begin{table}
\caption{The mean of $t_{\rm dep}$, $\bar{Z}_{\rm gas}$, $\bar{n}^{-1}_{\rm gas}$, $f_{\rm [C_{II}]}$ and $\bar{\epsilon}_{\rm [C_{II}]}$ of the {\sc\small FIRE} galaxy sample at different redshifts. }
\begin{threeparttable}
\begin{tabular}{ p{0.3 cm} p{0.4 cm}  p{0.4 cm}  p{0.4 cm} p{0.4 cm} p{0.4 cm}}
 \hline
  \hline
 \multicolumn{1}{c|}{$z$}   &  \multicolumn{1}{c|}{<$\frac{t_{\rm dep}}{\rm Gyr}$>}  &  \multicolumn{1}{c|}{<$\frac{\bar{Z}_{\rm gas}}{Z_\odot}$>} & \multicolumn{1}{c|}{<$\frac{\bar{n}_{\rm gas}}{\rm cm^{-3}}$>} & \multicolumn{1}{c|}{<$f_{\rm [C_{II}]}$>} &  \multicolumn{1}{c|}{<$\frac{\bar{\epsilon}_{\rm [C_{II}]}}{10^{-23}\,{\rm erg\,s^{-1}\,cm^3}}$>}  \\
  \hline 
  & & &  \multicolumn{1}{c|}{Total} & & \\
 \hline 
 \multicolumn{1}{c|}{0}  &  \multicolumn{1}{c|}{6.30}  &  \multicolumn{1}{c|}{1.69} &  \multicolumn{1}{c|}{5.2} & \multicolumn{1}{c|}{0.57} & \multicolumn{1}{c|}{1.2} \\
 \multicolumn{1}{c|}{1}  &  \multicolumn{1}{c|}{2.02}  &  \multicolumn{1}{c|}{1.08} &  \multicolumn{1}{c|}{14.6}  & \multicolumn{1}{c|}{0.63} & \multicolumn{1}{c|}{1.1} \\
 \multicolumn{1}{c|}{2}  &  \multicolumn{1}{c|}{1.02}  &  \multicolumn{1}{c|}{0.56} &  \multicolumn{1}{c|}{17.3} & \multicolumn{1}{c|}{0.85} & \multicolumn{1}{c|}{1.6} \\
 \multicolumn{1}{c|}{3}  &  \multicolumn{1}{c|}{1.10}  &  \multicolumn{1}{c|}{0.43} &  \multicolumn{1}{c|}{30.3}  & \multicolumn{1}{c|}{0.88} & \multicolumn{1}{c|}{1.2} \\
 \multicolumn{1}{c|}{4}  &  \multicolumn{1}{c|}{1.14}  &  \multicolumn{1}{c|}{0.24} &  \multicolumn{1}{c|}{63.6} & \multicolumn{1}{c|}{0.92} & \multicolumn{1}{c|}{0.8} \\
 \multicolumn{1}{c|}{6}  &  \multicolumn{1}{c|}{0.86}  &  \multicolumn{1}{c|}{0.12} &  \multicolumn{1}{c|}{180.9} & \multicolumn{1}{c|}{0.95} & \multicolumn{1}{c|}{0.5} \\
 \multicolumn{1}{c|}{8}  &  \multicolumn{1}{c|}{0.73}  &  \multicolumn{1}{c|}{0.09} &  \multicolumn{1}{c|}{468.8} & \multicolumn{1}{c|}{0.97} & \multicolumn{1}{c|}{0.3} \\
 \hline
  & &  \multicolumn{3}{c|}{$L_{\rm IR}{}\ge{}10^{11}\,L_\odot$}  & \\
  \hline
 \multicolumn{1}{c|}{0}  &  \multicolumn{1}{c|}{1.88}  &  \multicolumn{1}{c|}{2.40} &  \multicolumn{1}{c|}{8.7} & \multicolumn{1}{c|}{0.43} & \multicolumn{1}{c|}{1.2} \\
 \multicolumn{1}{c|}{1}  &  \multicolumn{1}{c|}{1.32}  &  \multicolumn{1}{c|}{1.45} &  \multicolumn{1}{c|}{17.3}  & \multicolumn{1}{c|}{0.52} & \multicolumn{1}{c|}{0.8} \\
 \multicolumn{1}{c|}{2}  &  \multicolumn{1}{c|}{0.52}  &  \multicolumn{1}{c|}{1.07} &  \multicolumn{1}{c|}{17.8} & \multicolumn{1}{c|}{0.74} & \multicolumn{1}{c|}{1.6} \\
 \multicolumn{1}{c|}{3}  &  \multicolumn{1}{c|}{0.83}  &  \multicolumn{1}{c|}{0.54} &  \multicolumn{1}{c|}{40.6}  & \multicolumn{1}{c|}{0.78} & \multicolumn{1}{c|}{0.9} \\
 \multicolumn{1}{c|}{4}  &  \multicolumn{1}{c|}{0.69}  &  \multicolumn{1}{c|}{0.36} &  \multicolumn{1}{c|}{59.9} & \multicolumn{1}{c|}{0.85} & \multicolumn{1}{c|}{0.9} \\
 \multicolumn{1}{c|}{6}  &  \multicolumn{1}{c|}{0.51}  &  \multicolumn{1}{c|}{0.32} &  \multicolumn{1}{c|}{217.7} & \multicolumn{1}{c|}{0.81} & \multicolumn{1}{c|}{0.4} \\
 \multicolumn{1}{c|}{8}  &  \multicolumn{1}{c|}{0.09}  &  \multicolumn{1}{c|}{0.59} &  \multicolumn{1}{c|}{360.0} & \multicolumn{1}{c|}{0.74} & \multicolumn{1}{c|}{0.3} \\
 \hline
 \end{tabular}
  \end{threeparttable}
  \label{T8}
\end{table}

\paragraph*{$\rm [C_{II}]$ deficit at high redshifts\\}

The normalization of the $L_{\rm [C_{II}]}$-SFR relation for the {\sc\small FIRE} sample consistently decreases with increasing redshift. The mean $L_{\rm [C_{II}]}{}/{}\rm SFR$ ratio of the galaxies reduces by 0.8 dex (approximately a factor of 6) from $z{}={}0$ to $z{}={}8$ (as shown in column 2 of Table~\ref{T7}).

Table~\ref{T7}, as well as Fig.~\ref{fig.14}, demonstrates that the evolution of the $L_{\rm [C_{II}]}{}/{}\rm SFR$ ratio in galaxies is primarily influenced by $\bar{Z}_{\rm gas}$ and $t_{\rm dep}$, as the $\rm [C_{II}]$ deficit diminishes at almost all redshifts in the parameter spaces of $(L_{\rm [C_{II}]}{}/{}{\rm SFR}){}t^{-1}_{\rm dep}$ and  $(L_{\rm [C_{II}]}{}/{}{\rm SFR}){}\bar{Z}^{-1}_{\rm gas}$. This suggests that the $\rm [C_{II}]$ deficit in high-redshift galaxies is attributed to either low gas metallicity or a deficiency of gas capable of producing $\rm [C_{II}]$ emission per unit SFR.

A closer look at Table~\ref{T7} reveals that $t_{\rm dep}$ is the key parameter driving the evolution of the $L_{\rm [C_{II}]}$-SFR relation at $z{}\le{}3$, while $\bar{Z}_{\rm gas}$ plays a more critical role at $z{}\ge{}4$. This shift is due to $t_{\rm dep}$ decreasing more significantly from $z{}={}0$ to 3 (from 6.3 to 1.1 Gyr, by a factor of $\sim{}6$) compared to the change from $z{}={}3$ to 8 (from 1.1 to 0.73 Gyr, by only $\sim{}30\%$) as outlined in Table~\ref{T8}. In contrast, $\bar{Z}_{\rm gas}$ for the FIRE sample decreases sharply with redshift at $z{}=3-8$ (from $0.43\,Z_\odot$ to $0.09\,Z_\odot$, by a factor of $\sim{}5$), exerting a more pronounced impact on the evolution of $L_{\rm [C_{II}]}{}/{}\rm SFR$ than $t_{\rm dep}$. 

Unlike $t_{\rm dep}$ and $\bar{Z}_{\rm gas}$, $\bar{\epsilon}_{\rm [C_{II}]}$ has a relatively modest effect on the redshift evolution of $L_{\rm [C_{II}]}{}/{}\rm SFR$. From $z{}={}0$ to $z{}={}8$, the mean $\bar{\epsilon}_{\rm [C_{II}]}$ of the FIRE sample experiences a slight decrease with redshift (by a factor of 4, as seen in Table~\ref{T8}). The $\rm [C_{II}]$ deficit persists at high redshifts in the parameter space of$(L_{\rm [C_{II}]}{}/{}{\rm SFR})\,\bar{\epsilon}^{-1}_{\rm [C_{II}]}$ (Table~\ref{T7}). 

The other two parameters, $\bar{n}_{\rm gas}$ and $f_{\rm [C_{II}]}$, have {\em completely} no contribution to the $\rm [C_{II}]$ deficit at high redshifts. Both of these parameters increase with redshift, with higher $\bar{n}_{\rm gas}$ indicating a more compact ISM in earlier galaxies. While it may seem that an increase in gas density should lead to higher $L_{\rm [C_{II}]}{}/{}\rm SFR$ (according to the relationship $L_{\rm [C_{II}],\,cl}{}/{}M_{\rm cl}{}\propto{}n_{\rm H}$, equation~\ref{eq.23}), this effect is overshadowed by the combined impact of $t_{\rm dep}$ and $\bar{Z}_{\rm gas}$ on $L_{\rm [C_{II}]}{}/{}\rm SFR$.

The increase in $f_{\rm [C_{II}]}$ with redshift suggests that our sample includes more $\rm H_2$ gas-poor galaxies at higher redshifts, where a larger fraction of carbon in the ISM gas becomes ionized. Nevertheless, the influence of $f_{\rm [C_{II}]}$ on the evolution of $L_{\rm [C_{II}]}{}/{}\rm SFR$ is insignificant, as the mean $f_{\rm [C_{II}]}$ of the galaxies in our sample increases by no more than a factor of 2 from $z{}={}0$ to $z{}={}8$ (from $57\%$ to $97\%$, as shown in Table~\ref{fig.8}).

In summary, the decrease in $L_{\rm [C_{II}]}/{\rm SFR}$ for the FIRE sample with redshift is primarily driven by a reduction in $t_{\rm dep}$ and gas metallicity. While $t_{\rm dep}$ plays a more significant role at $z{}\le{}3$, gas metallicity becomes the key parameter driving the $\rm [C_{II}]$ deficit in galaxies at higher redshifts. The redshift evolution of $\bar{n}_{\rm gas}$, $f_{\rm [C_{II}]}$, and $\bar{\epsilon}_{\rm [C_{II}]}$ has either no or limited impact.

 \paragraph*{$\rm [C_{II}]$ deficit at high $L_{\rm IR}$\\}
  
The {\sc \small FIRE} sample shows a consistent trend of decreasing $L_{\rm [C_{II}]}{}/{}L_{\rm IR}$ ratio with $L_{\rm IR}$ at each redshift. To identify the primary driver of the $\rm [C_{II}]$ deficit in IR-luminous galaxies, we examined how each of the five physical parameters ($f_{\rm [C_{II}]}$, $t_{\rm dep}$, $\bar{n}_{\rm gas}$, $\bar{Z}_{\rm gas}$, and $\bar{\epsilon}_{\rm [C_{II}]}$) depends on $L_{\rm IR}$. 

In Table~\ref{T8}, we present the mean values of $f_{\rm [C_{II}]}$, $t_{\rm dep}$, $\bar{n}_{\rm gas}$, $\bar{Z}_{\rm gas}$, and $\bar{\epsilon}_{\rm [C_{II}]}$ for galaxies with $L_{\rm IR}{}\ge{}10^{11}\,L_\odot$ (where galaxies are observed to exhibit a $\rm [C_{II}]$ deficit) at different redshifts. We also include the mean values for the entire sample, which includes fainter galaxies. The table, as well as Fig.~\ref{fig.15}, reveals that IR-luminous galaxies ($L_{\rm IR}{}\ge{}10^{11}\,L_\odot$) typically have lower $t_{\rm dep}$ and $f_{\rm [C_{II}]}$, but higher $\bar{n}_{\rm gas}$ and $\bar{Z}_{\rm gas}$ compared to the rest of the sample at a given redshift. This suggests that IR-luminous galaxies are richer in metals and $\rm H_2$ gas, have more compact ISM, and shorter gas depletion time. The mean $\bar{\epsilon}_{\rm [C_{II}]}$ of these galaxies shows no significant dependence on $L_{\rm IR}$.

Therefore, the reduced $L_{\rm [C_{II}]}{}/{}\rm SFR$ ratio in IR-bright galaxies can be attributed to their lower $t_{\rm dep}$ (\ie, gas mass per SFR) and $f_{\rm [C_{II}]}$. Fig.~\ref{fig.14} indicates that $t_{\rm dep}$ plays a more significant role than $f_{\rm [C_{II}]}$. While these galaxies still exhibit a `$\rm [C_{II}]$ deficit' in the space of ($L_{\rm [C_{II}]}{}/{}{\rm SFR}){}f^{-1}_{\rm [C_{II}]}$ ({\it lower middle} panel), their ($L_{\rm [C_{II}]}{}/{}{\rm SFR}){}t^{-1}_{\rm dep}$ ({\it upper left} panel) is higher than that of local star-forming galaxies. Hence, the primary reason for the reduced $L_{\rm [C_{II}]}{}/{}{\rm SFR}$ in these galaxies compared to local star-forming galaxies is their lower $t_{\rm dep}$.

{It's worth noting that ($L_{\rm [C_{II}]}{}/{}{\rm SFR}){}t^{-1}_{\rm dep}$ can be re-written as $L_{\rm [C_{II}]}{}/{}M_{\rm gas}$, following equation~\ref{eq.26}. Therefore, an alternative interpretation of the upper left panel of Fig.~\ref{fig.14} is that the ISM of IR-luminous galaxies produces more $\rm [C_{II}]$ emission per unit gas mass than fainter ones (due to higher gas metallicity and density). If $t_{\rm dep}$ were a constant, meaning that $M_{\rm gas}$ is proportional to SFR, IR-luminous galaxies should exhibit an excess in $L_{\rm [C_{II}]}{}/{}{\rm SFR}$ rather than a deficit. The fact that they exhibit a reduced $L_{\rm [C_{II}]}{}/{}{\rm SFR}$ ratio compared to local star-forming galaxies is due to their low gas mass relative to their SFR.}

\subsection{The two regimes of $\rm [C_{II}]$ emission of galaxies} 
\label{Sec:5d}

In the previous section, we have shown with the {\sc\small FIRE} sample that the main driver of the $\rm [C_{II}]$ deficit at high redshifts and high $L_{\rm IR}$ is different. The observed $\rm [C_{II}]$ deficit of the galaxies at $z{}\simgreat{}4$ (at $L_{\rm IR}{}\simgreat{}10^{11}\,L_\odot$) may be due to their low gas metallicity (gas depletion time). In this section, we explore the fundamental reason for galaxies having different origin of $\rm [C_{II}]$ deficit in the two regimes.

We at first discuss the {$L_{\rm [C_{II}]} {} / {} \rm SFR$ vs. $t_{\rm dep}$} relation of the {\sc\small FIRE} galaxies (Section~\ref{Sec:5d1}). We subsequently explore the reason for galaxies showing two distinct regimes on the {$\Delta\,({\rm log}\,L_{\rm [C_{II}]})$ vs. $t_{\rm dep}$} diagram (Section~\ref{Sec:5d2}). Finally, we discuss how this is related to the distinct origin of $\rm [C_{II}]$ deficit at high redshifts and high $L_{\rm IR}$ (Section~\ref{Sec:5d3}).

\begin{figure}
 \includegraphics[width=86mm]{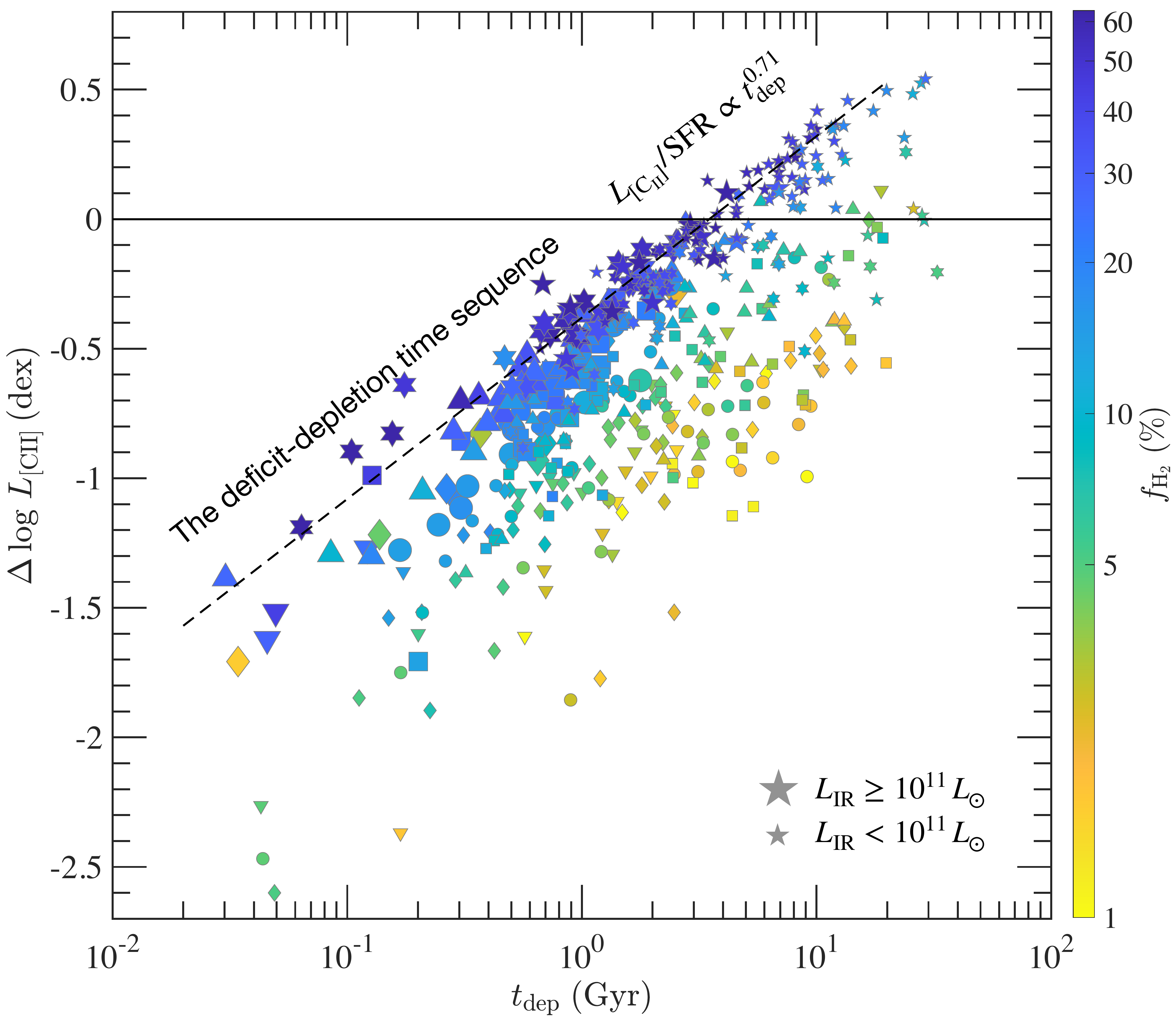}
 \caption{The relation between $t_{\rm dep}$ and $\Delta\,({\rm log}\,L_{\rm [C_{II}]})$ of the {\sc\small FIRE} sample at $z{}={}0-8$ (same as the {\it upper left} panel of Fig.~\ref{fig.15} except for the colour). The data points are coloured-coded by $f_{\rm H_2}$ of the galaxies. The large (small) symbols represent the galaxies having $L_{\rm IR}{}\ge{}10^{11}\,L_\odot$ ($L_{\rm IR}{}<{}10^{11}\,L_\odot$). The $\rm H_2$ gas-rich galaxies ({$f_{\rm H_2}{}\simgreat{}50\%$}) exhibit a linear correlation between ${\rm log}\,(t_{\rm dep}{}/{}\rm Gyr)$ and $\Delta\,({\rm log}\,L_{\rm [C_{II}]})$ (indicated by the black dashed line), which can be converted to a power-law relation $L_{\rm [C_{II}]}{}/{}{\rm SFR}{}\propto{}t^{0.71}_{\rm dep}$ (equation~\ref{eq.30}).}
   \label{fig.16}
\end{figure}
\begin{figure}
 \includegraphics[width=84mm]{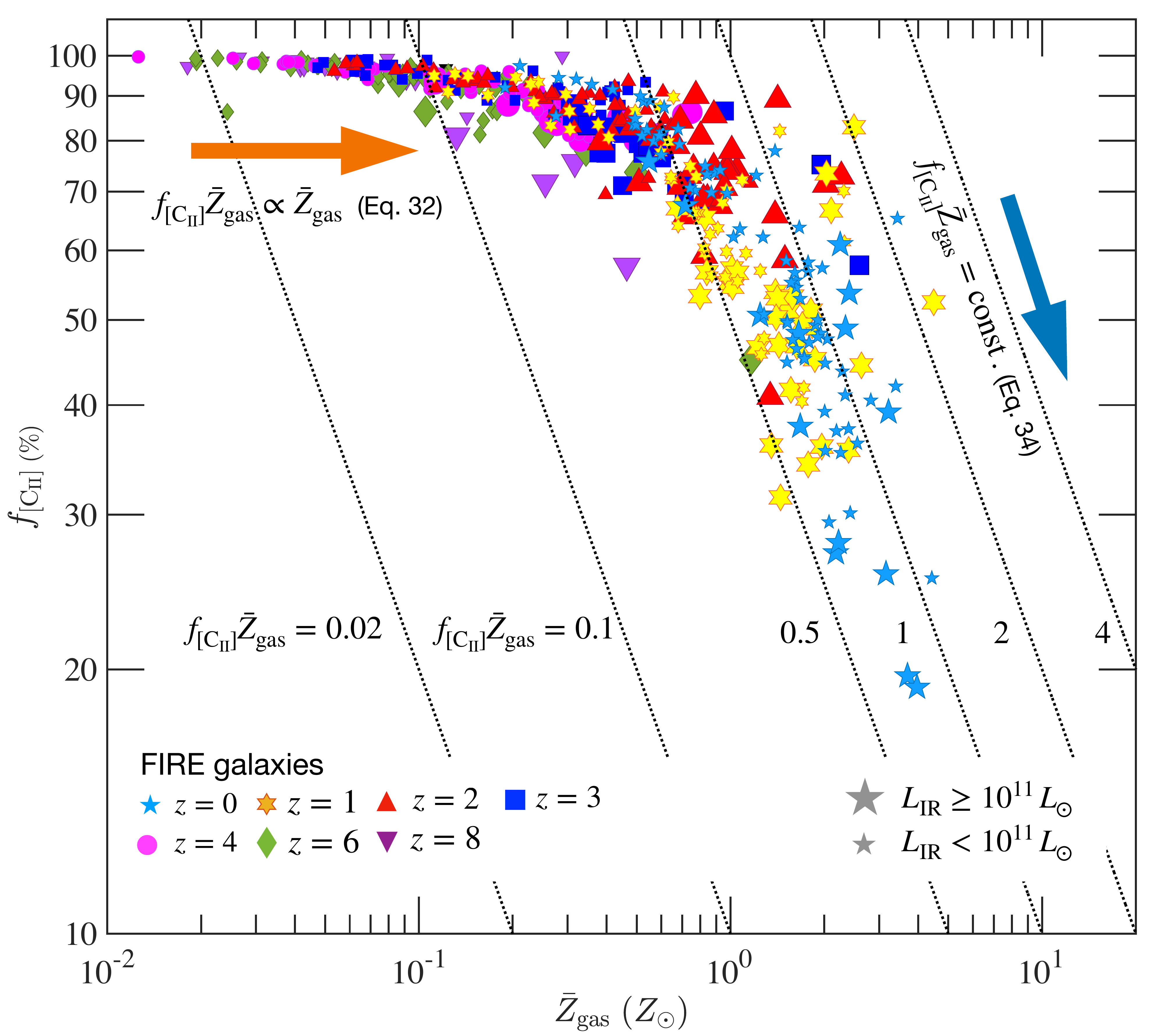}
 \caption{The relation between $\bar{Z}_{\rm gas}$ and $f_{\rm [C_{II}]}$ of the {\sc\small FIRE} sample at different redshifts. The large (small) symbols represent the galaxies having $L_{\rm IR}{}\ge{}10^{11}\,L_\odot$ ($L_{\rm IR}{}<{}10^{11}\,L_\odot$). The black dotted lines indicate the relation of $f_{\rm [C_{II}]}\bar{Z}_{\rm gas}{}={}0.02$, 0.1, 0.5, 1, 2 and 4 (from left to right). At $\bar{Z}_{\rm gas}{}\simless{}Z_\odot$, where galaxies are $\rm H_2$ gas-poor, $f_{\rm [C_{II}]}{}\approx{}1$ and $f_{\rm [C_{II}]}\bar{Z}_{\rm gas}{}\approx{}\bar{Z}_{\rm gas}$ (\cf~equation~\ref{eq.32}). At larger $\bar{Z}_{\rm gas}$, $f_{\rm [C_{II}]}$ scales roughly inversely with $\bar{Z}_{\rm gas}$ and hence $f_{\rm [C_{II}]}\bar{Z}_{\rm gas}{}\approx{}$constant (\cf~equation~\ref{eq.34}). }
   \label{fig.17}
\end{figure}

\begin{figure*}
 \includegraphics[width=180mm]{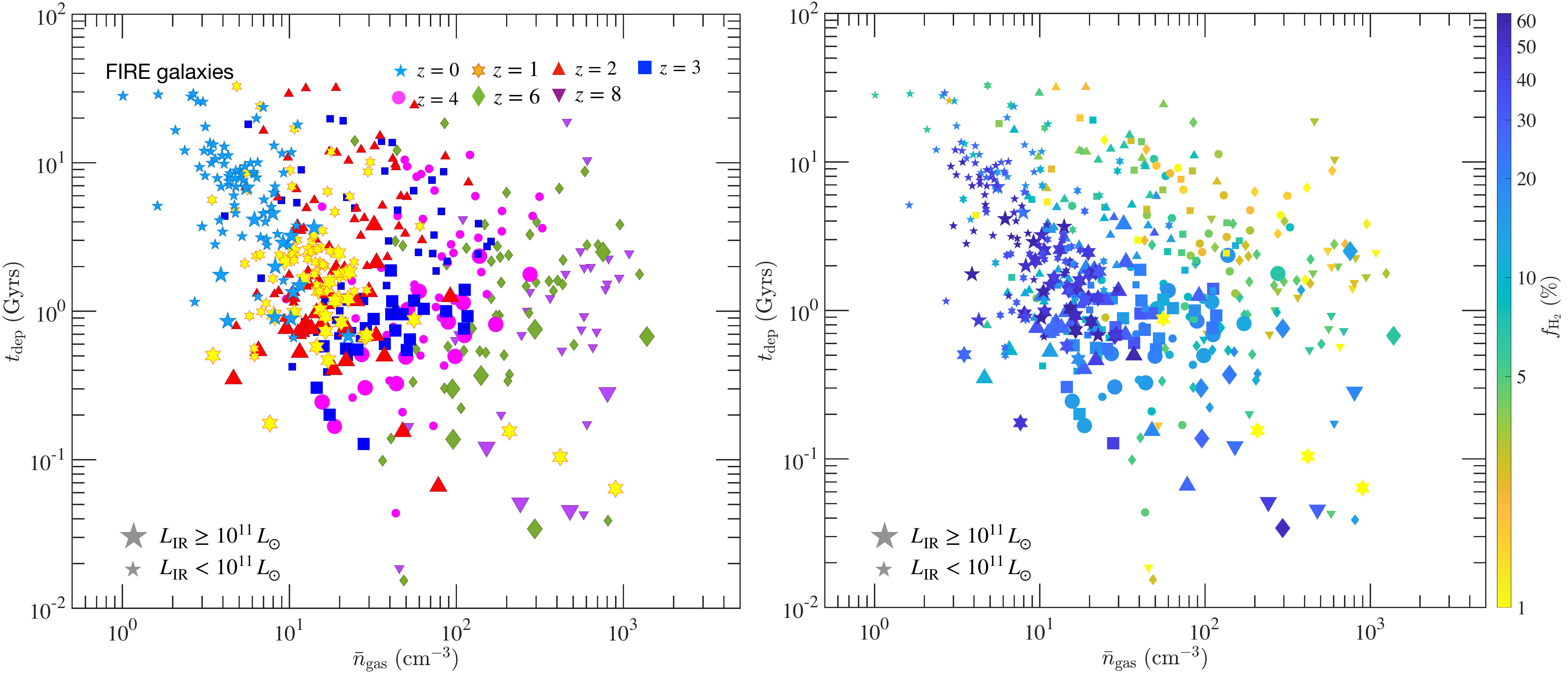}
 \caption{The relation between $\bar{n}_{\rm gas}$ and $t_{\rm dep}$ of the {\sc\small FIRE} sample at different redshifts. The data points in the {\it left} ({\it right}) panel are colour-coded by the redshift ($f_{\rm H_2}$) of the galaxies. Large (small) symbols represent the galaxies having $L_{\rm IR}\ge10^{11}\,L_\odot$ ($L_{\rm IR}<10^{11}\,L_\odot$). The {\sc\small FIRE} galaxies show a clear anti-correlation between $t_{\rm dep}$ and $\bar{n}_{\rm gas}$, in particular, the $\rm H_2$ gas-rich galaxies in the sample.}
   \label{fig.18}
\end{figure*}

\subsubsection{The {$L_{\rm [C_{II}]} {} / {} \rm SFR$ vs. $t_{\rm dep}$} relation}
\label{Sec:5d1}

The {\sc\small FIRE} galaxies exhibit two distinct regimes on the {$\Delta\,({\rm log}\,L_{\rm [C_{II}]})$ vs. $t_{\rm dep}$} diagram. While a considerable number of the galaxies show a tight linear correlation between their ${\rm log}\,(t_{\rm dep}{}/{}\rm Gyr)$ and $\Delta\,({\rm log}\,L_{\rm [C_{II}]})$, exhibiting a linear sequence {(we hereafter refer to it as the `deficit-depletion time sequence', or DDS)}, others show larger scatter on the diagram and fall systematically below the DDS.  

The galaxies on the DDS  appear to be more $\rm H_2$ gas-rich. In Fig.~\ref{fig.16}, we show the same {$L_{\rm [C_{II}]} {} / {} \rm SFR$ vs. $t_{\rm dep}$} relation of the {\sc\small FIRE} sample as in Fig.~\ref{fig.15} ({\it upper left} panel), but colour-code the data points by the $\rm H_2$ gas mass fraction, $f_{\rm H_2}$, of the galaxies instead of their redshift. It can be seen from Fig.~\ref{fig.16} that the galaxies along the DDS tend to be more $\rm H_2$ gas-rich, having {$f_{\rm H_2}{}\simgreat{}50\%$} (equivalent to {$f_{\rm [C_{II}]}{}\simless{}50\%$}). Besides, we see from the two figures that the majority of the low-redshift ($z{}={}0-2$, {shown by cyan stars, yellow hexagons and red triangles in Fig.~\ref{fig.15}}) and IR-luminous ($L_{\rm IR}{}\simgreat{}10^{11}\,L_\odot$, {indicated by large symbols in Fig.~\ref{fig.15} and Fig.~\ref{fig.16}}) galaxies locate on or close to the DDS.

We derive the best-fit linear scaling relation between ${\rm log}\,(t_{\rm dep}{}/{}\rm Gyr)$ and $\Delta\,( {\rm log}\,L_{\rm [C_{II}]})$ for the $\rm H_2$ gas-rich galaxies in our sample having {$f_{\rm H_2}{}\simgreat{}50\%$}, \ie~
\begin{equation}
\Delta({\rm log}\,L_{\rm [C_{II}]})  = (-0.38\pm0.01) + (0.71\pm0.03) \;{\rm log}\left ( \frac{t_{\rm dep}}{\rm Gyr}\right),
\label{eq.29}
\end{equation}
\noindent which can be rewritten as
\begin{equation}
\frac{L_{\rm [C_{II}]}/L_\odot}{{\rm SFR}/(M_\odot\,\rm yr^{-1})} = 1.78\times10^7 \left ( \frac{t_{\rm dep}}{\rm Gyr}\right)^{0.71}.
\label{eq.30}
\end{equation}
The coefficient of determination is $R^2{}={}0.936$.

\subsubsection{The two regimes on the {$\Delta\,({\rm log}\,L_{\rm [C_{II}]})$ vs. $t_{\rm dep}$} diagram}
\label{Sec:5d2}

The reasons for the $\rm H_2$ gas-rich galaxies ({$f_{\rm H_2}{}\simgreat{}50\%$}) showing a linear sequence on the {$\Delta\,({\rm log}\,L_{\rm [C_{II}]})$ vs. $t_{\rm dep}$} diagram are threefolds: i) Their $f_{\rm [C_{II}]}\bar{Z}_{\rm gas}$ `saturates', meaning that it becomes almost like a constant and hence $L_{\rm [C_{II}]}{}/{}{\rm SFR}$ of the galaxies simply scales to $t_{\rm dep} \bar{n}_{\rm gas}$, ii) their $t_{\rm dep}$ and $\bar{n}_{\rm gas}$ anti-correlate with each other, and iii) $\bar{\epsilon}_{\rm [C_{II}]}$ has relatively small variation among different galaxies.

Let us at first understand the $f_{\rm [C_{II}]}$ vs. $\bar{Z}_{\rm gas}$ relation. In Fig.~\ref{fig.17}, we show the $f_{\rm [C_{II}]}$ vs. $\bar{Z}_{\rm gas}$ relation for the {\sc\small FIRE} sample. It can be seen that at $\bar{Z}_{\rm gas}{}\simless{}Z_\odot$, $f_{\rm [C_{II}]}$ barely declines from unity ($f_{\rm [C_{II}]}{}\approx{}1$) with increasing $\bar{Z}_{\rm gas}$, whereas at higher $\bar{Z}_{\rm gas}$, $f_{\rm [C_{II}]}$ declines sharply and $f_{\rm [C_{II}]}\bar{Z}_{\rm gas}$ becomes approximately a constant (`saturates') with increasing $\bar{Z}_{\rm gas}$ (or decreasing $f_{\rm [C_{II}]}$). 

The shape of the $f_{\rm [C_{II}]}$ vs. $\bar{Z}_{\rm gas}$ relation of the {\sc\small FIRE} galaxies can be understood as follows. Consider a spherical gas cloud having a radius $R_{\rm cl}$ and a surface-to-centre column density $N_{\rm cl}$ ($=n_{\rm H}R_{\rm cl}$). When the cloud is metal and dust-poor (having very low $Z_{\rm gas}$ and $\delta_{\rm dgr}$), the LW photons from the radiation field can penetrate the entire cloud (\ie~$l_{\rm F}{}>{}R_{\rm cl}$) and dissociate all the molecular hydrogen ($\rm H_2$) and neutral carbon ($\rm C_I$ and $\rm CO$) in the cloud. In such a low-metallicity (or $\delta_{\rm dgr}$) regime, we have

\begin{equation}
 f_{\rm [C_{II}],\,cl}\approx 1
 \label{eq.31}
\end{equation}
\noindent and 
\begin{equation}
f_{\rm [C_{II}],\,cl}\,Z_{\rm gas}\propto Z_{\rm gas}.
 \label{eq.32}
\end{equation}

\noindent Since $N_{\rm F}{}\propto{}l_{\rm F}{}\propto{} \delta^{-1}_{\rm dgr}\propto Z^{-1}_{\rm gas}$ (equation~\ref{eq.10} and~\ref{eq.11}), indicating stronger dust absorption of UV photons with increasing gas metallicity, $l_{\rm F}$ decreases with $Z_{\rm gas}$ and will become equal or less than $R_{\rm cl}$ when $Z_{\rm gas}$ becomes sufficiently large. Through simple mathematics, it can be derived that for a spherical geometry, $f_{\rm [C_{II}]}Z_{\rm gas}$ increases {\it sub-linearly} with $Z_{\rm gas}$ until when $l_{\rm F}\ll R_{\rm cl}$, we have
\begin{equation}
  f_{\rm [C_{II}],\,cl} \propto \frac{N_{\rm F}}{N_{\rm cl}} \propto (Z_{\rm gas}N_{\rm cl})^{-1}
 \label{eq.33}
\end{equation}
\noindent or
\begin{equation}
f_{\rm [C_{II}],\,cl}\,Z_{\rm gas}=\rm constant. 
 \label{eq.34}
\end{equation}
\noindent It is not surprising to find similar scaling relations with the {\sc\small FIRE} galaxies, $f_{\rm [C_{II}]}\,\bar{Z}_{\rm gas}{}\approx{}\bar{Z}_{\rm gas}$ at low $\bar{Z}_{\rm gas}$ and $f_{\rm [C_{II}]}\,\bar{Z}_{\rm gas}{}\approx{}$const. at high $\bar{Z}_{\rm gas}$ (as shown in Fig.~\ref{fig.17}), given that the ISM of the galaxies can be viewed as being made up of numerous such idealized gas `clouds'. The `saturation' of $f_{\rm [C_{II}]}\,\bar{Z}_{\rm gas}$ at high $\bar{Z}_{\rm gas}$ indicates that the $\rm [C_{II}]$ cooling rate of the galaxies does not increase much with gas metallicity due to the shrinking of the size of the $\rm [C_{II}]$-emitting region (Zone I + Zone II).
 
Another important reason for the $\rm H_2$ gas-rich galaxies showing a clear sequence on the {$\Delta\,({\rm log}\,L_{\rm [C_{II}]})$ vs. $t_{\rm dep}$} diagram is that their $t_{\rm dep}$ and $\bar{n}_{\rm gas}$ have clear anti-correlation. In Fig.~\ref{fig.18}, we show the {$t_{\rm dep}$ vs. $\bar{n}_{\rm gas}$} relation of the {\sc\small FIRE} sample. This anti-correlation is due to the fact that the local free-fall timescale of star-forming clouds decreases with gas density ($t_{\rm ff}{}\propto{} \rho^{-1/2}$), and hence gas is converted into stars more rapidly in the galaxies having denser ISM. It also accounts for the sub-linearity (power law index $n{}={}0.71$) of the {$L_{\rm [C_{II}]}/\rm SFR$ vs. $t_{\rm dep}$} scaling relation of the $\rm H_2$ gas-rich galaxies on the DDS (equation~\ref{eq.30}). 

For the $\rm H_2$ gas-poor galaxies, the fact that they lie below the DDS on the {$\Delta\,({\rm log}\,L_{\rm [C_{II}]})$ vs. $t_{\rm dep}$} diagram (Fig.~\ref{fig.16}) is because of their low gas metallicity (and hence low $f_{\rm [C_{II}]}\bar{Z}_{\rm gas}$). From equation~(\ref{eq.27}), we see that at fixed $L_{\rm [C_{II}]}{}/{}{\rm SFR}$ (equivalently, at fixed $\Delta\,{\rm log}\,L_{\rm [C_{II}]}$), their $t_{\rm dep}$ has to be higher than that of the galaxies on the DDS so as to compensate for their having lower $f_{\rm [C_{II}]}\bar{Z}_{\rm gas}$. Besides, the fact that the $\rm H_2$ gas-poor galaxies show a larger scatter of $t_{\rm dep}$ at given $\Delta ({\rm log}\,L_{\rm [C_{II}]})$ (Fig.~\ref{fig.16}) than the $\rm H_2$ gas-rich galaxies is due to the non-trivial scatter of $f_{\rm [C_{II}]}\bar{Z}_{\rm gas}$ among these galaxies, as opposed to $f_{\rm [C_{II}]}\bar{Z}_{\rm gas}$ being like a constant for the $\rm H_2$ gas-rich galaxies (Fig.~\ref{fig.17}). 

\subsubsection{The physical origins of $\rm [C_{II}]$ deficit of galaxies (a revisit)}
\label{Sec:5d3}

The important consequence of $f_{\rm [C_{II}]}\,\bar{Z}_{\rm gas}$ being `saturated' for the $\rm H_2$ gas-rich galaxies is that the overall $L_{\rm [C_{II}]}{}/{}\rm SFR$ ratio of the galaxies shows a tight and steep dependence on $t_{\rm dep}$ (equation~\ref{eq.30}). As a result, $t_{\rm dep}$ becomes the dominating parameter that determines the $L_{\rm [C_{II}]}{}/{}\rm SFR$ ratio of these galaxies. Their $L_{\rm [C_{II}]}{}/{}\rm SFR$, in contrast, does not shows a clear correlation with any of the other four parameters ($f_{\rm [C_{II}]}$, $\bar{Z}_{\rm gas}$, $\bar{n}_{\rm gas}$ or $\bar{\epsilon}_{\rm [C_{II}]}$).

Now we should be able to understand the fundamental reason for $t_{\rm dep}$ being the main driver of the $\rm [C_{II}]$ deficit at high $L_{\rm IR}$. The IR-luminous galaxies are $\rm H_2$ gas-rich (due both to their being
dust-rich and having high gas column density). Hence, they are in the regime where the $L_{\rm [C_{II}]}{}/{}\rm SFR$ ratio of galaxies is determined primarily by $t_{\rm dep}$ (\ie~they lie on the DDS in the {$\Delta\,({\rm log}\,L_{\rm [C_{II}]})$ vs. $t_{\rm dep}$} diagram) and their $\rm [C_{II}]$ deficit is due to their low $t_{\rm dep}$.

Besides, we can now understand the redshift evolution of the $L_{\rm [C_{II}]}{}/{}\rm SFR$ ratio of the {\sc\small FIRE} sample at $z{}={}0-2$. At these low redshifts, our sample includes more galaxies that are $\rm H_2$ gas-rich as a result of their being more metal and dust-rich than the galaxies at higher redshifts. The $L_{\rm [C_{II}]}{}/{}\rm SFR$ ratio of these low-$z$ galaxies therefore depends more sensitively on $t_{\rm dep}$. 

At higher redshifts, in contrast, our sample includes a large fraction of metal and dust-poor galaxies that are also $\rm H_2$ gas-poor. They are off the DDS in the {$\Delta\,({\rm log}\,L_{\rm [C_{II}]})$ vs. $t_{\rm dep}$} diagram. For these galaxies, $f_{\rm [C_{II}]}\,\bar{Z}_{\rm gas}{}\approx{}\bar{Z}_{\rm gas}$ (Fig.~\ref{fig.17}) and hence $L_{\rm [C_{II}]}{}/{}\rm SFR$ of the galaxies depends more sensitively on $\bar{Z}_{\rm gas}$. As a result, gas metallicity becomes the main driver of the $\rm [C_{II}]$ deficit of the high-$z$ galaxies in our sample.

\section{Discussions}
\label{Sec:6}

\subsection{The origins of $\rm [C_{II}]$ emission in galaxies}
\label{Sec:6a}

In Section~\ref{Sec:5b}, we presented the fractional contributions of the $\rm [C_{II}]$ emission from various gas phases in the {\sc \small FIRE} galaxies (fig.~\ref{fig.13}). Here in this section, we will compare our findings with the observational results in more details.

\paragraph*{Observational results\\}
Observational studies on the origins of $\rm [C_{II}]$ emission in galaxies have been limited to the Milky Way and local galaxies.

Estimating the fraction of $\rm [C_{II}]$ emission that originates from the $\rm H^+$ gas ($L_{\rm [C_{II}],\,H^+}{}/L_{\rm [C_{II}]}$) is relatively straightforward. The common approach is by using the $\rm [N_{II}]$ $\rm 205\,\mu m$ fine-structure line. This line has a critical density ($\rm \sim{}32\,cm^{-3}$) that is similar to that of the $\rm [C_{II}]$ $158\,\rm \mu m$ line ($\sim{}45\,\rm cm^{-3}$) in ionized gas, resulting in a negligible dependence of the $\rm [\rm C_{II}]_{\rm 158\,\mu m}{}/{} [N_{II}]_{\rm 205\,\mu m}$ ratio on gas density in the $\rm H^+$ regions \citep{Oberst_2006, Croxall_2012}. {\citet{Goldsmith_2015} conducted the first large-scale Galactic survey of the $\rm [N_{II}]_{\rm 205\,\mu m}$ line, comprising 149 positions in the Galactic Plane. They showed that $1/3-1/2$ of the $\rm [C_{II}]$ emission originates from the $\rm H^+$ gas in those regions. Using the {\sc\small GOALS} sample, \citet{Diaz_Santos_2017} found that $\rm H^+$ gas contributes to $18\%-35\%$ ($\pm1\sigma$) of the total $\rm [C_{II}]$ emission of the LIRGs. A similar result has been reported by \citet{Croxall_2017} using the {\sc\small KINGFISH} sample, which incorporates more moderately star-forming galaxies (see also the updated result by \citealt{Sutter_2019} using the same sample). Studies probing small-scale regions in other nearby galaxies \citep[\eg][]{Okada_2015, Jameson_2018, Tarantino_2021} have also indicated a lower $L_{\rm [C_{II}],\,H^+}{}/L_{\rm [C_{II}]}$ ratio.} Overall, ionized gas does not appear to be the dominant source of $\rm [C_{II}]$ emission in galaxies based on the local observations.

The $\rm [C_{II}]$ emission that originates from the $\rm H_2$ gas regions has been used as a tracer of `CO-dark' $\rm H_2$ gas \citep{Grenier_2005, Langer_2010, Langer_2014, Wolfire_2010}. To disentangle this component from the others, the common method is to compare the velocity profile of $\rm [C_{II}]$ to those of CO and $\rm H_I$ 21 cm, typically considered tracers of `CO-bright' and $\rm H_I$ gas, respectively. The remaining $\rm [C_{II}]$ emission attributed to `CO-dark' $\rm H_2$ gas. Using this method, \citet{Pineda_2013, Pineda_2014} find that $\sim25\%$ of the total $\rm [C_{II}]$ luminosity of the Milky Way is associated with the `CO-dark' $\rm H_2$ gas. Similar analyses have been conducted for the Magellanic Clouds \citep{Requena_Torres_2016, Pineda_2017, Lebouteiller_2019, Tarantino_2021} and nearby low-metallicity dwarf galaxies \citep{Fahrion_2017, Madden_2020}. The reported fractional contributions of $\rm H_2$ gas to the total $\rm [C_{II}]$ emission exhibit a significant scatter among different studies, ranging from $\sim{}20\%$ to over $50\%$. However, it's important to note that these studies typically probe individual star-forming regions rather than providing a complete mapping of emissions across entire galaxies. Consequently, their results may be biased toward the densest regions within the ISM. Additionally, the findings are constrained by small sample sizes and may be influenced by the sensitivity limits of the observations.
 
\paragraph*{Simulated vs. observational results \\}

In Section~\ref{Sec:5b}, we demonstrated that the primary source of $\rm [C_{II}]$ emission in the \textsc{\small FIRE} galaxies is the $\rm H_I$ gas phase, constituting $50\%$ to $80\%$ of the total luminosity. The majority of the remaining emission originates from the $\rm H^+$ gas phase, while the $\rm H_2$ gas phase contributes only around $10\%$. The fractional contributions of $\rm [C_{II}]$ emission from these phases do not strongly depend on the galaxy's SFR.

{We find that the $L_{\rm [C_{II}],\,H^+}{}/{}L_{\rm [C_{II}]}$ ratio at $z{}={}0$ broadly aligns with the constraints from observations by \citet{Goldsmith_2015}, \citet{Diaz_Santos_2017}, \citet{Croxall_2017}, and \citet{Sutter_2019} over a wide range of overlapping SFR values (SFR${}\approx{}0.1-100\,M_\odot\,\rm yr^{-1}$), except that our simulations do not produce any system at $z{}={}0$ that shows a very small ($\simless{}20\%$) contribution of $\rm H^+$ gas as some of the local observations have found \citep[\eg][]{Okada_2015, Jameson_2018, Sutter_2019, Tarantino_2021}. This may suggest that our simulations over-predict the amount of diffuse gas in the ISM, where the contribution by the $\rm H^+$ gas is more significant (see the {\it bottom} panel of Fig.~\ref{fig.11}).}

The predicted $L_{\rm [C_{II}],\,H_2}{}/{}L_{\rm [C_{II}]}$ ratio ($\simless{}10\%$) at $z{}={}0$ appears to be lower than what {recent observational studies \citep[\eg][]{Pineda_2013, Pineda_2014, Tarantino_2021} have reported.} The disparity between the simulated and observed $L_{\rm [C_{II}],\,H_2}{}/{}L_{\rm [C_{II}]}$ ratio may suggest that the ISM of the $z{}={}0$ \textsc{\small FIRE} galaxies, especially the low-metallicity dwarf systems, has lower gas column densities than the observed samples in the star-forming regions that observations have mainly probed. Studies have shown that self-shielding of $\rm H_2$ from Lyman-Werner radiation can become significant at high column densities \citep[\eg][]{Draine_1996, Madden_1997, Madden_2020, Wolfire_2010}. Consequently, a significant amount of $\rm C^+$ can be found within the envelope of the $\rm H_2$ regions, and the contribution of $\rm H_2$ gas to the $\rm [C_{II}]$ emission can be non-trivial. However, it's worth noting that the reported high $L_{\rm [C_{II}],,H_2}{}/{}L_{\rm [C_{II}]}$ ratio for local galaxies may be largely influenced by several systematic factors, as mentioned earlier.

\subsection{Comparison with the previous studies}
\label{Sec:6b}

Here we discuss the relation between the findings of the previous studies to this from this work. Specifically, we will discuss the conclusions regarding the origin of the $\rm [C_{II}]$ deficit at high $L_{\rm IR}$ in Section~\ref{Sec:6b1}, whereas in Section~\ref{Sec:6b2}, we will compare the predictions of the $L_{\rm [C_{II}]}$-SFR relation of galaxies at redshift $z{}\simgreat{}5$ from the recent studies with ours.

\subsubsection{The $\rm [C_{II}]$ deficit at high $L_{\rm IR}$}
\label{Sec:6b1}

\paragraph*{$\rm [C_{II}]$ deficit due to a strong ISRF\\}

A number of studies suggest that the observed $\rm [C_{II}]$ deficit at high $L_{\rm IR}$ is due to a strong ISRF in IR-luminous galaxies. This can result in large positive grain charges,  leading to inefficient heating of gas through photo-electric processes in the neutral galactic medium \citep{Tielens_1985, Kaufman_1999, Malhotra_2001, Croxall_2012, Mckinney_2021a}. Consequently, the rate of gas cooling via $\rm [C_{II}]$ line drops. Additionally, a strong ISRF (and hence high $U$) may also give rise to ``dust-bounded" $\rm H^+$ regions near the newly formed young stars \citep{Bottorff_1998, Abel_2009}, where $N_{\rm s}{}\approx{}N_{\rm F}$ (note: $N_{\rm s}$ increases about linearly with $U$ until $N_{\rm s}{}\approx{}N_{\rm F}$). In this scenario, gas cooling through $\rm [C_{II}]$ can become inefficient due to a lack of {$\rm C^+$} ions in the $\rm H^+$ regions --- when $U$ is large, a significant fraction of carbon can be ionized further into {$\rm C^{2+}$} ions (In Zone I, {$x^{(1)}_{\rm C^+}{}\approx{}1-x^{(1)}_{\rm C^{2+}}{}\propto{}U^{-1}$}, see Appendix~\ref{Sec:Ap5}). Both mechanisms can ultimately lead to a reduced $\bar{\epsilon}_{\rm [C_{II}]}$ in galaxies.

Examining the {\sc\small FIRE} sample, we do not find that the IR-luminous galaxies in our sample exhibit significantly lower $\bar{\epsilon}_{\rm [C_{II}]}$ compared to the fainter galaxies at each given redshift, as indicated in Table~\ref{T8}. It is important to note that {\sc\small CLOUDY} (version 17.01) incorporates grain charging physics \citep{Baldwin_1991, van_Hoof_2004, Abel_2005}, and our approach of conducting dust Radiative Transfer (RT) calculations with {\sc\small SKIRT} provides a more accurate estimate of the of the ISRF (and hence $U$) distribution within galaxies compared to previous studies. Our findings suggest that the $\rm [C_{II}]$ deficit at high $L_{\rm IR}$ is not primarily caused by a high $U$ in these galaxies. 

However, we do observe that the mean $\bar{\epsilon}_{\rm [C_{II}]}$ decreases with redshift from $z{}={}0$ to $z{}={}8$ by a factor of $\sim{}4$, which is associated with an increasing value of $U$ with redshift. We will delve into this effect in more detail in a follow-up study.  

\paragraph*{$\rm [C_{II}]$ deficit due to a high gas density\\}

It has also been suggested that the $\rm [C_{II}]$ deficit in IR-luminous galaxies can be driven by the high density of the star-forming gas in these galaxies \citep[\eg][]{Narayanan_2017}. With increasing density, ISM gas becomes more shielded from ionizing radiation of massive young stars and therefore more carbon in the ISM gas becomes neutral (in $\rm CO$ or $\rm C_I$). The $\rm [C_{II}]$ deficit is thus due to a lack of {$\rm C^+$} ions in the ISM gas in this scenario (\ie~due to a low $f_{\rm [C_{II}]}$).

This, however, does not seem to be exactly like what we find with the {\sc\small FIRE} simulations. The ISM of the {\sc\small FIRE} galaxies spans a very wide range of density (see Fig.~\ref{fig.11}), and even for the most massive starburst galaxies in our sample, much of their $\rm [C_{II}]$ luminosity originates from the gas having intermediate density ($\bar{n}_{\rm gas}{}\approx{}\bar{n}_{\rm H_{I},\,MW}$, see Fig.~\ref{fig.11}). Overall, the luminosity-weighted gas density ($\bar{n}_{\rm gas}$) of the IR-luminous galaxies ($L_{\rm IR}{}\ge{}10^{11}\,L_\odot$) is not much higher than that of the IR-faint galaxies in our sample at any given redshift (see Table~\ref{T8}), and the difference is not as strong as that in $t_{\rm dep}$. Therefore, the $\rm [C_{II}]$ deficit of the IR-luminous galaxies in {\sc\small FIRE} simulations does not appear to be mainly driven by their having too dense ISM gas. 

\subsubsection{The $L_{\rm [C_{II}]}$-SFR relation at redshift $z{}\simgreat{}5$}
\label{Sec:6b2}

\begin{figure}
 \includegraphics[width=87mm]{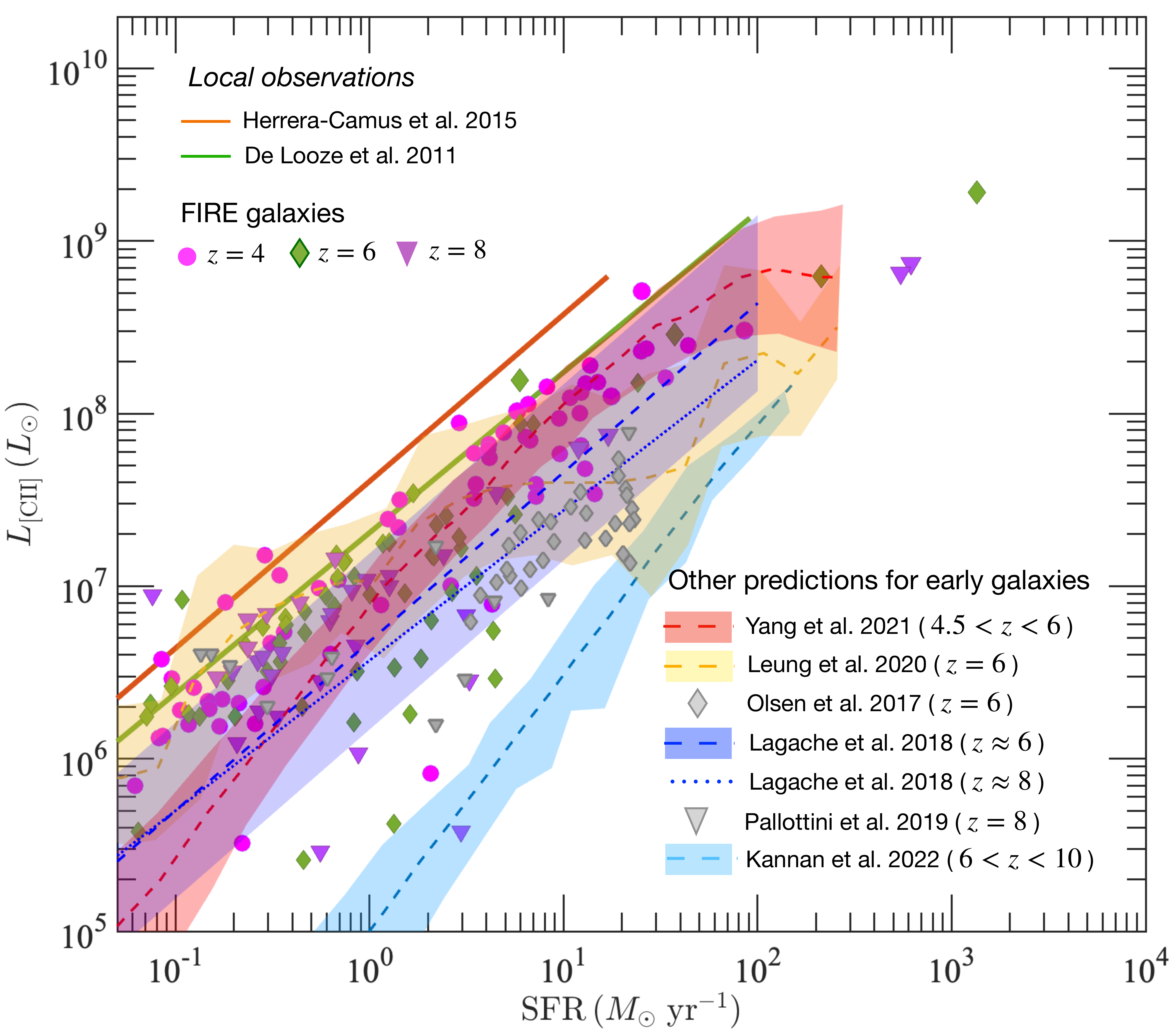}
 \caption{{The $L_{\rm [C_{II}]}$-SFR relation at $z{}\simgreat{}5$ predicted by different simulation groups. Red, yellow, blue and cyan lines indicate the mean result of \citet{Yang_2021} ($4.5<{}z{}<6$), \citet{Leung_2020} ($z{}={}6$), \citet{Lagache_2018} (dashed blue line for $z{}\approx{}6$ and dotted blue line for $z{}\approx{}8$), and \citet{kannan_2022b} ($6{}<z{}<10$). These studies use statistically significant samples. The corresponding coloured shaded areas represent the $1\sigma$ dispersion of the data around the mean relation of each sample. In addition, we also show the data of individual galaxies of the \citet{Olsen_2017} ($z{}={}6$) and \citet{Pallottini_2019} ($z{}={}8$) samples by grey diamonds and grey downward triangles, respectively. For reference, we show the observed $L_{\rm [C_{II}]}$-SFR relation of the local star-forming samples of \citetalias{Herrera_Camus_2015} (solid orange line) and \citetalias{De_Looze_2011} (solid green line) as well as the the data of the {\sc\small FIRE} sample at $z{}={}4$ (magenta circles), $z{}={}6$ (green diamonds) and $z{}={}8$ (purple downward triangles). {\bf A $\rm [C_{II}]$ deficit at $z{}\simgreat{}5$ is generally predicted by various simulation groups.} }}
    \label{fig.19}
\end{figure}

As mentioned in the Introduction, several planned ground-based $\rm [C_{II}]$ line intensity mapping (LIM) experiments will target the emitting sources at redshift $z{}\simgreat{}5$ \citep{Kovetz_2017}, including {\sc\small CCAT-prime}, {\sc\small CONCERTO} and {\sc\small TIME}. Predicting the $L_{\rm [C_{II}]}$-SFR relation of galaxies at this early epoch has thus become extremely important for interpreting the upcoming data of these experiments \citep[see \eg][]{Visbal_2011, Gong_2012, Serra_2016, Fonseca_2017, Padmanabhan_2019, Yue_2019, Chung_2020, Padmanabhan_2022, Karoumpis_2022, Sun_2023, Murmu_2023, Horlaville_2023}.

In Fig.~\ref{fig.19}, we present the results from a number of recent studies. These include the ones using SAMs \citep{Lagache_2018, Yang_2021} as well as those using hydrodynamic simulations \citep{Olsen_2017, Pallottini_2019, Leung_2020, kannan_2022b}. It can be seen that different studies have generally predicted a clear $\rm [C_{II}]$ deficit at $z{}\simgreat{}5$ with respect to the local samples of \citetalias{De_Looze_2011} and \citetalias{Herrera_Camus_2015}, similar to this work using the {\sc\small FIRE} simulations. Some have also predicted a mild trend of growing deficit with increasing redshift \citep[\eg][]{Lagache_2018, kannan_2022b}. The predicted $1\sigma$ scatter of the $L_{\rm [C_{II}]}$-SFR relation at a given redshift of these studies is typically as large as $0.3-0.5$ dex (except \citealt{kannan_2022b}, which shows noticeably smaller scatter than the others).

There is, however, a clear difference in the normalization and slope of the $L_{\rm [C_{II}]}$-SFR relation predicted by the different groups. In particular, \citet{Yang_2021} (\citealt{kannan_2022b}) produce the highest (lowest) normalization among all different groups at ${\rm SFR}{}\approx{}1-100\,M_\odot\,\rm yr^{-1}$. Both also produce a considerably steeper power-law slope ($\approx{}1.5$) than the others. 

The difference in the $L_{\rm [C_{II}]}$-SFR relation indicates that the predicted ISM properties (\eg~$\bar{Z}_{\rm gas}$, $t_{\rm dep}$) of the galaxies at $z{}\simgreat{}5$ are not well converged between the current simulations. We highlight that the data of the upcoming LIM experiments may provide useful constraints on the ISM properties of the galaxies in this early epoch, given that direct measurement of these properties is very challenging using the current techniques.

\section{Summary and conclusions}
\label{Sec:7}

The $\rm 158\,\mu m$ fine structure line of singly ionized carbon ($\rm [C_{II}]$) has been considered as a SFR indicator since observations of nearby star-forming galaxies found a linear correlation between their $L_{\rm [C_{II}]}$ and SFR. There is, however, evidence showing that IR-bright ($L_{\rm IR}{}\simgreat{}10^{11}\,L_\odot$), starburst galaxies as well as early galaxies at $z{}\simgreat{}5$ have reduced $L_{\rm [C_{II}]}{}/{}\rm SFR$ with respect to the local star-forming samples (so-called `$\rm [C_{II}]$ deficit' problem). Different models have been posited to explain the origin of the $\rm [C_{II}]$ deficit of galaxies at high $L_{\rm IR}$ or at high redshifts and yet no consensus has been reached at both regimes.

In this work, we present a comprehensive analysis on the $L_{\rm [C_{II}]}$-SFR relation of galaxies using a galaxy sample at $z{}={}0-8$ ($M_*{}={}10^7-5\times10^{11}\,M_\odot$) extracted from the cosmological hydrodynamic simulations, which are part of the {\sc\small FIRE} project \citep{Hopkins_2014, Hopkins_2018, Hopkins_2023}, coupled with {\sc\small CLOUDY} \citep{Ferland_1998, Ferland_2017} models. The sample consists mainly of galaxies ($N_{\rm gal}{}\sim{}500$) from {\sc\small FIREbox} \citep{Feldmann_2023}, a high-resolution cosmological-volume hydrodynamic simulation run with {{\sc\small FIRE}-2} physics, and is supplemented with a few dozen of high-$z$ massive galaxies from the cosmological `zoom-in' simulations of the {\sc\small MassiveFIRE} suite \citep{Feldmann_2016, Feldmann_2017b, Angles_Alcazar_2017}. The sample covers {an} unprecedentedly broad dynamic range among all studies on $\rm [C_{II}]$, including normal star-forming galaxies, (U)LIRG and SMG candidates as well as UV-bright galaxies at EoR, which can be used to study the full range of the observational data on $\rm [C_{II}]$ currently available. 

The predicted $L_{\rm [C_{II}]}$-SFR relation of the {\sc\small FIRE} sample {agrees well} with the observational data. In particular, we successfully reproduce the observed linear correlation of the local star-forming samples over the SFR range $\approx{}0.1-10\,M_\odot\,\rm yr^{-1}$ (Fig.~\ref{fig.4} and Fig.~\ref{fig.6}). Apart from that, we also reproduce the sharp decline of $L_{\rm [C_{II}]}{}/{}\rm SFR$ with $L_{\rm IR}$ ($\sim{}\rm SFR$) at $L_{\rm IR}{}\simgreat{}10^{11}\,L_\odot$ at low and high redshifts, which is consistent with the data of the (U)LIRGs and SMGs in this $L_{\rm IR}$ regime (Fig.~\ref{fig.7} and Fig.~\ref{fig.9}).

Our sample shows a general decline of $L_{\rm [C_{II}]}{}/{}\rm SFR$ with redshift, in particular, at low SFR (Fig.~\ref{fig.8}). The mean $L_{\rm [C_{II}]}{}/{}\rm SFR$ ratio of the early EoR galaxies at $z{}>{}5$ in our sample is about one order of magnitude below the local galaxies, showing a clear $\rm [C_{II}]$ deficit, similar to what has been previously found with other simulations {(Section~\ref{Sec:6b2})}. Observations of galaxies at EoR have drawn divergent conclusions on their $L_{\rm [C_{II}]}$-SFR relation, which is largely due to the uncertainty in the dust SED shape (or `dust temperature') of the galaxies at these high redshifts. We analyze the sub-mm data of all the observed EoR galaxies and derive their dust-obscured $\rm SFR$ using the `dust temperature' estimated from the SED templates of the {\sc\small FIRE} samples self-consistently. We conclude that the $L_{\rm [C_{II}]}$-SFR relation of the {\sc\small FIRE} galaxies at $z{}>{}5$ is in no conflict with the current observational constraints, including those placed by the recent {\sc\small ALPINE} and {\sc\small REBELS} surveys.

The $L_{\rm [C_{II}]}{}/{}{\rm SFR}$ ratio of the {\sc\small FIRE} sample roughly follows a simple linear scaling relationship (equation~\ref{eq.27})
\begin{equation}
\frac{L_{\rm [C_{II}]}}{\rm SFR} \propto f_{\rm [C_{II}]}\bar{Z}_{\rm gas}t_{\rm dep}\bar{n}_{\rm gas}, \nonumber
\label{eq.35}
\end{equation} 
where $f_{\rm [C_{II}]}$ is the mass fraction of ionized or neutral atomic hydrogen gas in the ISM, $t_{\rm dep}$ is the gas depletion time ($={}M_{\rm gas}{}/{}\rm SFR$), and $\bar{Z}_{\rm gas}$ and $\bar{n}_{\rm gas}$ indicate the gas metallicity and gas density that are weighted by $\rm [C_{II}]$ luminosity. Following this scaling relationship, we find that the key driver of the $\rm [C_{II}]$ deficit is different at high $L_{\rm IR}$ and high redshifts (Section~\ref{Sec:5c}). At high $L_{\rm IR}$, the $\rm [C_{II}]$ deficit is mainly due to the low $t_{\rm dep}$ of galaxies, indicating that IR-luminous, starburst galaxies have less amount of gas that is able to produce $\rm [C_{II}]$ emission per unit SFR than the normal star-forming galaxies with moderate SFR. The $\rm [C_{II}]$ deficit at $z{}\simgreat{}5$, in contrast, is mainly driven by the low gas metallicity of galaxies at this epoch.

The underlying reason for $\rm [C_{II}]$ deficit being driven by different physical parameters at high $L_{\rm IR}$ and high redshifts is as follows. In the low-metallicity regime (corresponding to high-$z$ galaxies), $L_{\rm [C_{II}]}$ of galaxies depends sensitively on metallicity because line emissivity scales linearly with metallicity. In the high-metallicity regime (corresponding to low-$z$, massive and starburst galaxies), however, such dependence can become weak. This is because dust-to-gas ratio ($\delta_{\rm dgr}$) in the ISM increases with metallicity, which leads to the shrinking of the size of $\rm [C_{II}]$-emitting region (Section~\ref{Sec:5d}). The shrinking of its size almost cancels out the effect of increasing emissivity with metallicity (in this case, $f_{\rm [C_{II}]}\bar{Z}_{\rm gas}{}\approx{}$constant). As a result, $L_{\rm [C_{II}]}{}/{}\rm SFR$ of galaxies does not depend much on metallicity --- but instead, on $t_{\rm dep}{}={}M_{\rm gas}{}/{}\rm SFR$, see equation~(\ref{eq.30}) --- for massive, metal (dust) and $\rm H_2$ gas-rich starburst galaxies at low redshifts.

{In summary, the {\sc\small FIRE} simulations have predicted a reduced $L_{\rm [C_{II}]}{}/{}\rm SFR$ ratio in early high-redshift galaxies, as well as in IR-luminous galaxies, compared to local normal star-forming galaxies, which aligns with what observations have indicated.} The results suggest that {the} `$\rm [C_{II}]$ deficit' may be a common phenomenon among galaxies. This finding has significant implications for the interpretation of data from several major upcoming $\rm [C_{II}]$ line intensity mapping experiments, such as {\sc\small EXCLAIM} \citep{Ade_2020}, {\sc\small TIME} \citep{Sun_2021}, {\sc\small CCAT-prime} \citep{CCAT_2021} and {\sc\small CONCERTO} \citep{CONCERTO_2020, Gkogkou_2023}. Our results further imply that utilizing a constant linear $L_{\rm [C_{II}]}$-SFR relation derived from nearby star-forming galaxies \citep[\eg][]{De_Looze_2011, De_Looze_2014, Herrera_Camus_2015} may lead to a systematic overestimation of the cosmic star formation rate density in the high-redshift Universe.

\section*{Acknowledgements}

We thank the anonymous referee for useful comments which have helped improved the quality of this manuscript. LL acknowledges financial support from the Swiss National Science Foundation (hereafter SNSF) (grant no P2ZHP2\_199729) and the University of Toronto Faculty of Arts and Science. RF acknowledges financial support from the SNSF (grant no PP00P2\_194814, 200021\_188552). NM was supported by the Natural Sciences and Engineering Research Council of Canada (grant no RGPIN-2023-04901). DN acknowledges funding from the NSF via AST-1909153. DAA acknowledges support by NSF grants AST-2009687 and AST-2108944, CXO grant TM2-23006X, and Simons Foundation award CCA-1018464. LB acknowledge financial support from the SNSF (grant no PP00P2\_194814). CAFG was supported by NSF through grants AST-1715216, AST-2108230, and CAREER award AST-1652522; by NASA through grants 17-ATP17-0067 and 21-ATP21-0036; by STScI through grants HST-AR-16124.001-A and HST-GO-16730.016-A; by CXO through grant TM2-23005X; and by the Research Corporation for Science Advancement (RCSA) through a Cottrell Scholar Award. DTC is supported by a CITA/Dunlap Institute postdoctoral fellowship. The Dunlap Institute is funded through an endowment established by the David Dunlap family and the University of Toronto. DTC also acknowledges support through the Vincent and Beatrice Tremaine Postdoctoral Fellowship at CITA during the preparation and review of this work. JYHC acknowledges support from a CITA postdoctoral fellowship. DK were supported by NSF grant AST-1715101 and the Cottrell Scholar Award from the RCSA. Support for PFH was provided by NSF Research Grants 1911233, 20009234, 2108318, NSF CAREER grant 1455342, NASA grants 80NSSC18K0562, HST-AR-15800. This work was performed in part at the Aspen Center for Physics, which is supported by National Science Foundation grant PHY-2210452. The Flatiron Institute is supported by the Simons Foundation. 

We acknowledge PRACE for awarding us access to MareNostrum at the Barcelona Supercomputing Center (BSC), Spain. This research was partly carried out via the Frontera computing project at the Texas Advanced Computing Center. Frontera is made possible by National Science Foundation award OAC-1818253. Computations were performed on the Niagara supercomputer at the SciNet HPC Consortium. SciNet is funded by Innovation, Science and Economic Development Canada; the Digital Research Alliance of Canada; the Ontario Research Fund: Research Excellence; and the University of Toronto. This work was supported in part by a grant from the Swiss National Supercomputing Centre (CSCS) under project IDs s697 and s698. We acknowledge access to Piz Daint at the Swiss National Supercomputing Centre, Switzerland under the University of Zurich’s share with the project ID uzh18. This work made use of infrastructure services provided by S3IT (\url{www.s3it.uzh.ch}), the Service and Support for Science IT team at the University of Zurich. 

The authors would also like to acknowledge the use of AI language model ChatGPT for assistance in refining the writing and language of this manuscript.

\section*{data availability statement}

The data underlying this article will be shared on reasonable request to the corresponding author.

\bibliographystyle{mnras}
\bibliography{CII}

\appendix

\section{The radiative cooling rate of gas from the $\rm [C_{II}]$ fine structure transition --- I. The general case}
\label{Sec:Ap1}

The {$\rm C^+$} ion has two fine structure levels in the ground electronic state. The radiative cooling rate of gas from the $\rm [C_{II}]$ transition can therefore be calculated by solving a classical two-level problem \citep{Goldsmith_2012}.

The cooling rate in $\rm erg\,s^{-1}\,cm^{-3}$ can be written as
\begin{equation}
\Lambda_{\rm [C_{II}]} = [A_{\rm ul} n_{\rm u} + B_{\rm ul} n_{\rm u} U (T^{\rm b})  -  B_{\rm lu} n_{\rm l} U (T^{\rm b}) ]  E_{\rm ul}
\label{eq.a1}
\end{equation}
\noindent where $n_{\rm u}$ and $n_{\rm l}$ represent the densities of the upper (${}^{2}P_{3/2}$) and lower level (${}^{2}P_{1/2}$) {$\rm C^+$} ions ($\rm cm^{-3}$) that result from the combination of collisional and radiative processes. $A_{\rm ul}$, $B_{\rm ul}$ and $B_{\rm lu}$ in the above equation represent the Einstein coefficients for spontaneous emission ($\rm s^{-1}$), stimulated emission ({$\rm erg^{-1}\,s^{-2}\,cm^3$}) and stimulated absorption ({$\rm erg^{-1}\,s^{-2}\,cm^3$}), respectively. {$E_{\rm ul}$ ($\equiv{}h_{\rm P} \nu_{\rm [C_{II}]}$, where $\nu_{\rm [C_{II}]}=1900.5\,\rm GHz$) represents the transition energy of the $\rm [C_{II}]$ line.} $U(T^{\rm b})$ indicates the radiative energy density at $\nu_{\rm [C_{II}]}$ and $T^{\rm b}$ is the brightness temperature of the background radiation field. The source of the background radiation may be the CMB {and/or} the thermal emission of warm dust.   

$\Lambda_{\rm [C_{II}]}$ can be rewritten as a function of the excitation (or spin) temperature for the transition ($T^{\rm ex}$) and the temperature of the background radiation field ($T^{\rm b}$). The excitation temperature is defined by the relative populations of the upper and lower levels through
\begin{equation}
\frac{n_{\rm u}}{n_{\rm l}} \equiv \frac{g_{\rm u}}{g_{\rm l}} e^{-T^*/T^{\rm ex}},
\label{eq.a2}
\end{equation}
\noindent where $T^*{}={}h_{\rm P} \nu_{\rm [C_{II}]}/ k_{\rm B}{}={}91.8\,\rm K$ is the equivalent temperature of the $\rm [C_{II}]$ transition, and $g_{\rm u}{}={}4$ ($g_{\rm l}{}={}2$) is the statistical weight of the upper (lower)-level state. Given the relationships between the Einstein coefficients, \ie
\begin{equation}
B_{\rm lu} = (g_{\rm u}/g_{\rm l})B_{\rm ul}
\label{eq.a3}
\end{equation}
\noindent and
\begin{equation}
\frac{A_{\rm ul}}{B_{\rm ul}} = \frac{8\pi h_{\rm P} \nu_{\rm [C_{II}]}^3}{c^3},
\label{eq.a4}
\end{equation}
\noindent and substituting equation~(\ref{eq.a2}) into equation~(\ref{eq.a1}), we obtain
\begin{equation}
\Lambda_{\rm [C_{II}]} = n_{\rm u} A_{\rm ul}  h_{\rm P} \nu_{\rm [C_{II}]}  \left [1 -  \frac{{\rm e}^{(T^*/T^{\rm ex})}-1}{{\rm e}^{(T^*/T^{\rm b})}-1} \right ].
\label{eq.a5}
\end{equation}
\noindent Neglecting background radiation (\ie~$T^{\rm b}{}\simeq{}0$), we get
\begin{equation}
\Lambda_{\rm [C_{II}]} =n_{\rm u} A_{\rm ul}  h_{\rm P} \nu_{\rm [C_{II}]}, 
\label{eq.a6}
\end{equation}
\noindent which is the usual expression for the cooling rate. The term in the square brackets in equation~(\ref{eq.a5}) is the background correction term for attenuation (see \citealt{daCunha_2013} for the details). From equation~(\ref{eq.a2}), we have
\begin{equation}
n_{\rm u}  =  n_{\rm C^+} \left [1+\left (\frac{g_{\rm l}}{g_{\rm u}}\right ) {\rm e}^{T^*/T^{\rm ex}} \right ]^{-1}.
\label{eq.a7}
\end{equation}
\noindent By substituting equation~(\ref{eq.a7}) into equation~(\ref{eq.a5}), we then obtain the analytic expression for the $\rm [C_{II}]$ cooling rate when a background is included, 
\begin{equation}
\Lambda_{\rm [C_{II}]} = n_{\rm C^+} A_{\rm ul}  h_{\rm P} \nu_{\rm [C_{II}]} \Psi (T^{\rm ex}, T^{\rm b}),
\label{eq.a8}
\end{equation}
\noindent where 
\begin{equation}
 \Psi (T^{\rm ex}, T^{\rm b}) =  \left [1 -  \frac{{\rm e}^{(T^*/T^{\rm ex})}-1}{{\rm e}^{(T^*/T^{\rm b})}-1} \right ] \left[ 1+\left (\frac{g_{\rm l}}{g_{\rm u}}\right ) {\rm e}^{T^*/T^{\rm ex}}\right ]^{-1}. 
\label{eq.a9}
\end{equation}
{Equations}~(\ref{eq.a8})-(\ref{eq.a9}) indicates that one can derive $\Lambda_{\rm [C_{II}]}$ by solving for $T^{\rm ex}$. 

\section{Excitation temperature for the $\rm [C_{II}]$ transition}
\label{Sec:Ap2}

Here we present the analytic expression for the excitation temperature ($T^{\rm ex}$) for the $\rm [C_{II}]$ transition. 

The rate equation that determines the upper and lower level {$\rm C^+$} densities, $n_{\rm u}$ and $n_{\rm l}$, includes both collisional and radiative processes, and is 
\begin{equation}
n_{\rm u} [A_{\rm ul} + B_{\rm ul} U (T^{\rm b}) + C_{\rm ul}] = n_{\rm l} [ B_{\rm lu} U (T^{\rm b}) + C_{\rm lu} ],
\label{eq.b1}
\end{equation}
\noindent where $C_{\rm ul}$ ($C_{\rm lu}$) represents the collisional de-excitation (excitation) rate ($\rm s^{-1}$). The Einstein coefficients, $A_{\rm ul}$, $B_{\rm ul}$ and $B_{\rm lu}$, are related by {equations}~(\ref{eq.a3}) and (\ref{eq.a4}). For a single collision partner, the collision rates are equal to the rate coefficients ($\rm cm^3\,s^{-1}$) times the density $n_{\rm X}$ of that collision partner (${\rm X} {}= {}e^-$, $\rm H_I$ or $\rm H_2$), \ie
\begin{equation}
C_{\rm ul} = R^{\rm X}_{\rm ul} \, n_{\rm X} \;\;\;{\rm and}\;\;\;C_{\rm lu}= R^{\rm X}_{\rm lu} \, n_{\rm X},
\label{eq.b2}
\end{equation}
\noindent where $R^{\rm X}_{\rm ul}$ ($R^{\rm X}_{\rm lu}$) is the downward (upward) rate coefficient for collision partner $\rm X$. The two rate coefficients are related by detailed balance
\begin{equation}
R^{\rm X}_{\rm lu}/R^{\rm X}_{\rm ul} = (g_{\rm u}/g_{\rm l}){\rm e}^{-T^*/T},
\label{eq.b3}
\end{equation}
\noindent where $T$ is the kinetic temperature of gas. By substituting {equations}~(\ref{eq.a2})-(\ref{eq.a4}), (\ref{eq.b1})-(\ref{eq.b3}) into equation~(\ref{eq.b1}) and through rearrangement, we obtain the analytic expression for the excitation temperature 
\begin{equation}
{\rm e}^{T^*/T^{\rm ex}} = \frac{(1+G)A_{\rm ul} + n_{\rm X}\,R^{\rm X}_{\rm ul}}{G A_{\rm ul} + n_{\rm X}\,R^{\rm X}_{\rm ul} {\rm e}^{-T^*/T}} \label{eq.b4}
\end{equation}
\noindent where we define
\begin{equation}
G = \frac{1}{{\rm e}^{T^*/T^{\rm b}}-1}
\label{eq.b5}
\end{equation}
\noindent following \citet{Goldsmith_2012}. For the $\rm [C_{II}]$ transition, we have \citep[see \eg][]{Suginohara_1999, Goldsmith_2012}
\begin{align}
&A_{\rm ul} = 2.36\times10^{-6}\,\rm s^{-1}, \label{eq.b6} \\
& R^{\rm e^-}_{\rm ul} (T) = 8.7\times 10^{-8}  (T/2000)^{-0.37} \,\rm cm^3\,s^{-1}, \label{eq.b7} \\
& R^{\rm H_I}_{\rm ul} (T) = 4.0\times10^{-11}  (16+0.35T^{0.5} + 48 T^{-1}) \,\rm cm^3\,s^{-1},
\label{eq.b8}
\end{align}
\noindent and 
\begin{equation}
R^{\rm H_2}_{\rm ul} (T) = 3.8\times10^{-10}  (T/100)^{0.14} \,\rm cm^3\,s^{-1}.
\label{eq.b9}
\end{equation}
We can see from equations~(\ref{eq.b4}) and (\ref{eq.b5}) that for no background radiation (\ie~$T^{\rm b}{}\simeq{}0$) and high gas density (\ie~$n_{\rm X}{}\gg{}A_{\rm ul}{}/{}R^{\rm X}_{\rm ul}$), $G{}\rightarrow{}0$ and $T^{\rm ex}{}\rightarrow{}T$. In this case, $T^{\rm ex}$ (and hence the {$\rm C^+$} level populations) is set totally by the kinetic temperature of gas. The impact of background radiation on $T^{\rm ex}$ can be important in low-density environments (\ie~$n_{\rm X}{}\ll{}A_{\rm ul}{}/{}R^{\rm X}_{\rm ul}$).

\section{The Strömgren depth of a plane-parallel slab}
\label{Sec:Ap3}

The Strömgren depth ($l_{\rm s}$) can be derived by equating the ionizing photon rate ($\dot{N}_{\rm ion}$) to the hydrogen recombination rate ($\dot{N}_{\rm rec}$) in the $\rm H^+$ region. $\dot{N}_{\rm ion}$ can be expressed as
\begin{equation}
\dot{N}_{\rm ion} = F_{\rm ion} A,
\label{eq.c1}
\end{equation} where 
\begin{equation}
F_{\rm ion} = \int^\infty_{\rm \nu_{\rm L}} \frac{F_\nu}{h_{\rm P} \nu} {\rm d}\nu, 
\label{eq.c2}
\end{equation}
is the ionizing photon flux ($\rm cm^{-2}\,s^{-1}$) and $A$ is the surface area of the slab. $F_\nu$ indicates the specific energy flux ($\rm cm^{-2}\,s^{-1}\,Hz^{-1}$) at frequency $\nu$ and $\nu_{\rm L}{}={}3.2\times10^6\,\rm GHz$ is the frequency corresponding to the ionization energy of hydrogen, \ie~$h_{\rm P} \nu_{\rm L}=13.6\,\rm eV$. $\dot{N}_{\rm rec}$ can be expressed as
\begin{equation}
\dot{N}_{\rm rec} = n_{\rm e^-} n_{\rm p} \alpha_{\rm B}l_{\rm s}{\rm d}A \approx n^2_{\rm H}  \alpha_{\rm B}l_{\rm s} A, 
\label{eq.c3}
\end{equation}
\noindent where $\alpha_{\rm B}{}={}2.6\times10^{-13}\,\rm cm^3\,s^{-1}$ is the Case-B recombination coefficient at temperature $T{}\approx{}10^4\,\rm K$. Combining equation~(\ref{eq.c1}) and equation~(\ref{eq.c3}), we have 
\begin{equation}
l_{\rm s} = \frac{F_{\rm ion}}{n^2_{\rm H} \alpha_{\rm B}}.
\label{eq.c4}
\end{equation}
\noindent Hence, the gas column density at the Strömgren depth is
\begin{equation}
N_{\rm s} = n_{\rm H} l_{\rm s} = \frac{F_{\rm ion} }{n_{\rm H}\alpha_{\rm B}}= \frac{Uc}{\alpha_{\rm B}} \approx 10^{23} U \,\rm cm^{-2},
\label{eq.c5}
\end{equation}
\noindent where
\begin{equation}
U =  \frac{F_{\rm ion}}{n_{\rm H} c} = \frac{n_{\gamma}}{n_{\rm H}} 
\label{eq.c6}
\end{equation}
\noindent is the ionizing photon-to-gas density ratio. 

\section{The radiative cooling rate of gas from the $\rm [C_{II}]$ fine structure transition --- II. The plane-parallel slab model}
\label{Sec:Ap4}

Following Appendix~\ref{Sec:Ap1}, we present specifically here an analytic expression for the gas cooling rate via $\rm [C_{II}]$ line in the $\rm H^+$ (Zone I) and $\rm H_I$ regions (Zone II) of a plane-parallel slab. The superscript ``(1)" and ``(2)" in the following equations indicate the properties of gas in Zone I and II, respectively.

\paragraph*{$\rm H^+$ region \\ }
For $\rm H^+$ region (Zone I), where $T^{(1)}{}\approx{}10^4\,\rm K$ (hence ${\rm e}^{-T*/T^{(1)}}{}\approx{}1$) and the main collision partner of {$\rm C^+$} ions is $e^-$, we can rewrite equation~(\ref{eq.b4}) to be
\begin{equation}
{\rm e}^{T^*/T^{\rm ex}} =  \frac{A_{\rm ul} + n^{(1)}_{\rm e^-} R^{\rm e^-}_{\rm ul} (T^{(1)}) }{ n^{(1)}_{\rm e^-} R^{\rm e^-}_{\rm ul} (T^{(1)}) },
\label{eq.d1}
\end{equation}
where we neglect the effect of background field. For densities below the critical one (\ie~$n^{(1)}_{\rm e^-}{}\simless{}A_{\rm ul}/R^{\rm e^-}_{\rm ul}$),
\begin{equation}
{\rm e}^{T^*/T^{\rm ex}} \approx  \frac{A_{\rm ul}}{n^{(1)}_{\rm e^-} R^{\rm e^-}_{\rm ul} (T^{(1)}) }.
\label{eq.d2}
\end{equation}
Given $A_{\rm ul}{}={}2.36\times10^{-6}\,\rm s^{-1}$ and $R^{\rm e^-}_{\rm ul} (T^{(1)}) \approx 5\times10^{-8} \,\rm cm^3\,s^{-1}$ (equation~(\ref{eq.b7})), equation~(\ref{eq.d2}) can be rewritten as
\begin{equation}
{\rm e}^{T^*/T^{\rm ex}} \approx  \frac{50}{n^{(1)}_{\rm e^-} }.
\label{eq.d3}
\end{equation}
\noindent Substituting equation~(\ref{eq.d3}) into equation~(\ref{eq.a9}) gives
\begin{equation}
\Psi^{(1)} \approx \left [1+ \left (\frac{g_{\rm l}}{g_{\rm u}} \right ) {\rm e}^{T^*/T^{\rm ex}} \right ]^{-1} \approx \frac{n^{(1)}_{\rm e^-}}{25}.
\label{eq.d4}
\end{equation}
\noindent Finally, by substituting equation~(\ref{eq.d4}) into equation~(\ref{eq.a8}), we obtain the expression for the $\rm [C_{II}]$ cooling rate in $\rm H^+$ region
\begin{align}
\Lambda^{(1)}_{\rm [C_{II}]} &= n^{(1)}_{\rm C^+} A_{\rm ul}  h_{\rm P} \nu_{\rm [C_{II}]} \Psi^{(1)}  \nonumber \\
&= \left [ A_{\rm ul}  h_{\rm P} \nu_{\rm [C_{II}]}\, \left ( \frac{g_{\rm u}}{g_{\rm l}}\right ) {\rm e}^{-T^*/T^{\rm ex}} \right ] n^{(1)}_{\rm C^+}  \nonumber \\
&\approx 10^{-21} \,n^{(1)}_{\rm C^+} n^{(1)}_{\rm e^-}\;\;\rm erg\,s^{-1}\,cm^{-3}. 
\label{eq.d5}
\end{align}

\paragraph*{$\rm H_{\rm I}$ region \\ }

Now consider the $\rm [C_{II}]$ cooling rate in $\rm H_{I}$ region (Zone II), where $T^{(2)}{}\approx{}100\,\rm K$ (hence ${\rm e}^{-T*/T^{(2)}}{}\approx{}\frac{2}{5}$) and the main collision partner of {$\rm C^+$} ions is $\rm H_I$. In this case, equation~(\ref{eq.b5}) can be rewritten as
\begin{align}
{\rm e}^{T^*/T^{\rm ex}} &= \frac{(1+G)A_{\rm ul} + n^{(2)}_{\rm H_I} R^{\rm H_I}_{\rm ul} }{G A_{\rm ul} +  n^{(2)}_{\rm H_I} R^{\rm H_I}_{\rm ul} \,{\rm e}^{-T^*/T^{(2)}} } \nonumber \\
& \approx \frac{1}{G + \frac{2}{5} n^{(2)}_{\rm H_I} (R^{\rm H_I}_{\rm ul}/A_{\rm ul})}. 
\label{eq.d6}
\end{align} 
Given $R^{\rm H_I}_{\rm ul} (T^{(2)}) \approx 8\times10^{-10} \,\rm cm^3\,s^{-1}$ (equation~(\ref{eq.b8})), we have
\begin{equation}
{\rm e}^{T^*/T^{\rm ex}} \approx \frac{1}{ G + n^{(2)}_{\rm H_I}/7400}.
\label{eq.d7}
\end{equation}
For the case when background radiation is unimportant (\eg~low-$z$ CMB), $T^{\rm b}{}\rightarrow{}0$ and thus $G{}\rightarrow{}0$, we get
\begin{equation}
{\rm e}^{T^*/T^{\rm ex}} \approx 7400/n^{(2)}_{\rm H_I}.
\label{eq.d8}
\end{equation}
Substituting equation~(\ref{eq.d8}) into equation~(\ref{eq.a9}) and equation~(\ref{eq.a8}) gives
\begin{equation}
\Psi^{(2)} (T^{\rm b}=0) \approx\left [1+ \left (\frac{g_{\rm l}}{g_{\rm u}} \right ) {\rm e}^{T^*/T^{\rm ex}} \right ]^{-1} \approx 2.7\times10^{-4}\;n^{(2)}_{\rm H_I} 
\label{eq.d9}
\end{equation}
\noindent and 
\begin{align}
\Lambda^{(2)}_{\rm [C_{II}]} (T^{\rm b}=0) &= n^{(2)}_{\rm C^+} A_{\rm ul}  h_{\rm P} \nu_{\rm [C_{II}]}\, \Psi^{(2)} (T^{\rm b}=0)   \nonumber \\
&= \left [ A_{\rm ul}  h_{\rm P} \nu_{\rm [C_{II}]}\, \left ( \frac{g_{\rm u}}{g_{\rm l}}\right ) {\rm e}^{-T^*/T^{\rm ex}} \right ] n^{(2)}_{\rm C^+}  \nonumber \\
&\approx 10^{-23} \,n^{(2)}_{\rm C^+} n^{(2)}_{\rm H_I}\;\;\rm erg\,s^{-1}\,cm^{-3}.
\label{eq.d10}
\end{align}
\noindent Equation~(\ref{eq.d10}) is the expression for the $\rm [C_{II}]$ cooling rate in $\rm H_I$ region when background radiation is neglected.

\begin{figure}
 \includegraphics[width=83mm]{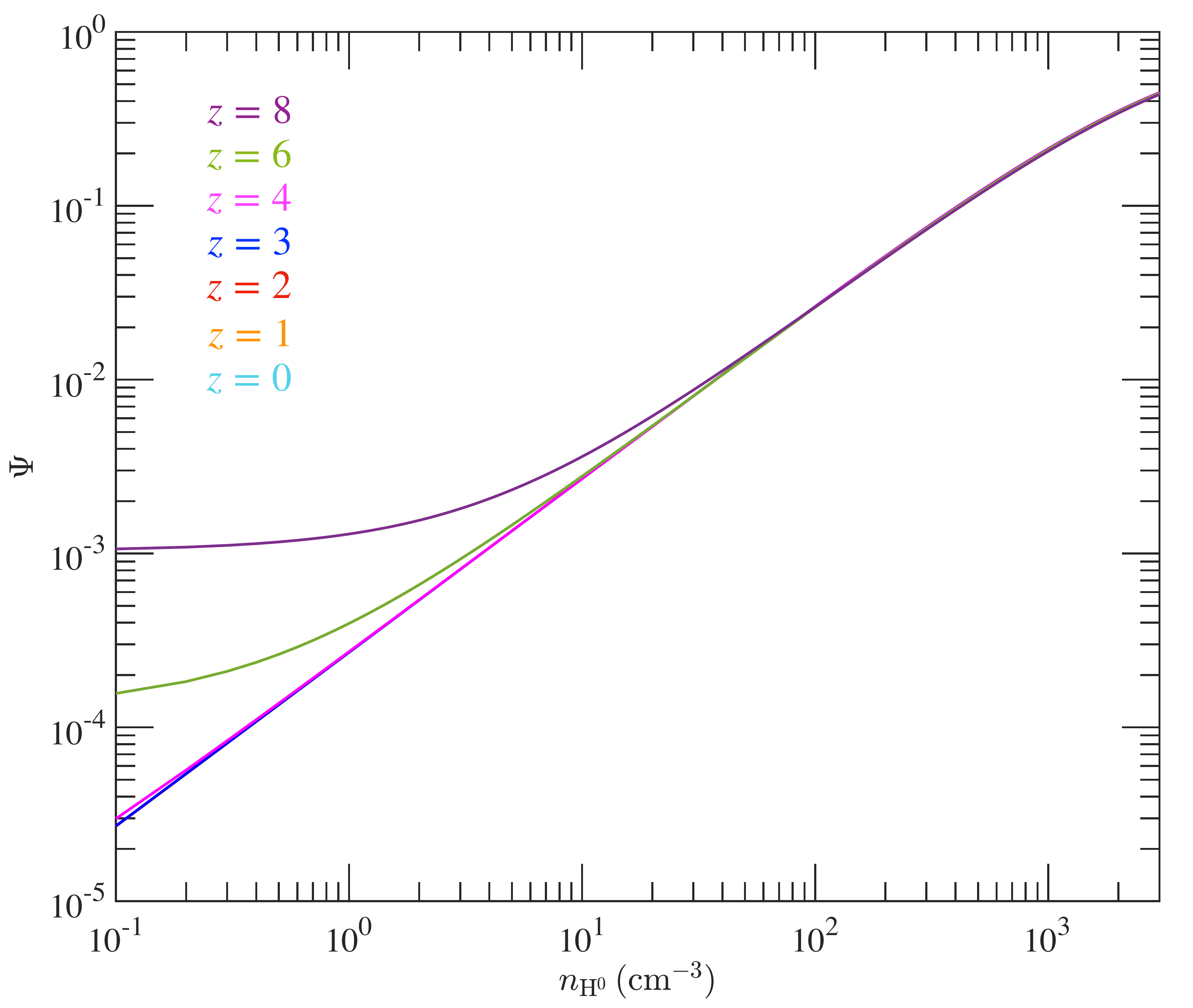}
 \caption{ The relation between $\Psi$ (equation~\ref{eq.d12}) and gas density for $\rm H_I$ gas ($T{}={}100\,\rm K$) at different redshifts. $\Psi$ is unaffected by the CMB at redshift $0\le{}z{}\le4$. At $z{}={}6-8$, $\Psi$ (and hence the $\rm [C_{II}]$ cooling rate) can be much affected by the CMB in low-density gas. }
 \label{fig.d1}
\end{figure}

\begin{figure}
 \includegraphics[width=83mm]{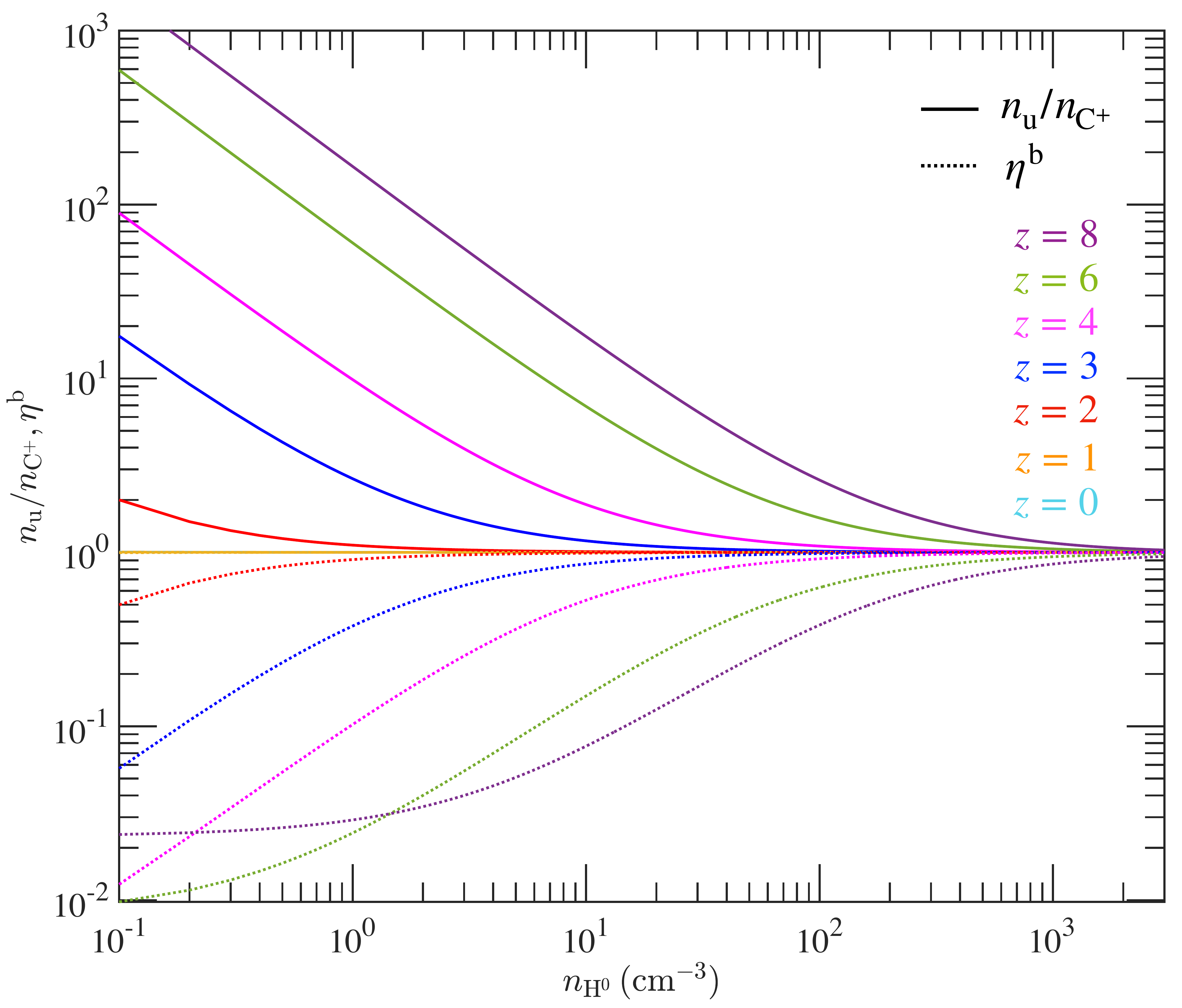}
 \caption{ Solid (dotted) lines indicate the relation between $\eta^{\rm b}$ ($n_{\rm u}{}/{}n_{\rm C^+}$) and gas density for $\rm H_I$ gas ($T{}={}100$ K) at different redshifts. At a given redshift, both the effects of CMB heating and attenuation increases with decreasing gas density. }
 \label{fig.d2}
\end{figure}

Taking into account background radiation, equation~(\ref{eq.a9}) can be expressed as
\begin{equation}
\Psi^{(2)} = \eta^{b} \, (n_{\rm u} / n_{\rm C^+}), 
\label{eq.d11}
\end{equation}
where 
\begin{equation}
 \eta^{b} \equiv  1 -  \frac{{\rm e}^{(T^*/T^{\rm ex})}-1}{{\rm e}^{(T^*/T^{\rm b})}-1} \approx \frac{G+n^{(2)}_{\rm H_I} / (7400\,G)}{1+ n^{(2)}_{\rm H_I} / (7400\,G)} 
\label{eq.d12}
\end{equation}
is the background attenuation term and 
\begin{equation}
\frac{n_{\rm u}}{n_{\rm C^+}} =  \left[ 1+\left (\frac{g_{\rm l}}{g_{\rm u}}\right ) {\rm e}^{T^*/T^{\rm ex}}\right ]^{-1} \approx \left [ 1+ \frac{1}{2\,(G+ n^{(2)}_{\rm H_I} / 7400)} \right ]^{-1}.
\label{eq.d13}
\end{equation}
Equation~(\ref{eq.d13}) indicates that background radiation (\eg~the CMB) leads to increased upper level (${}^{2}P_{3/2}$) population of the $\rm [C_{II}]$ transition (`background heating'). Using the above equations, we obtain the level of change of the $\rm [C_{II}]$ cooling rate by the CMB at redshift $z$, 
\begin{align}
\mathcal{R} &\equiv \frac{\Lambda^{(2)}_{\rm [C_{II}]} (T^{\rm CMB}(z)) }{\Lambda^{(2)}_{\rm [C_{II}]} (T^{\rm b} = 0)} = \frac{\Psi^{(2)} (T^{\rm CMB}(z)) }{\Psi^{(2)} (T^{\rm b} = 0)} \nonumber \\
&\approx  \left [\frac{G+n^{(2)}_{\rm H_I} / (7400\,G)}{1+ n^{(2)}_{\rm H_I} / (7400\,G)} \right]  \left [ \frac{2}{7400} n^{(2)}_{\rm H_I} + \frac{1}{ 7400\,G/n^{(2)}_{\rm H_I}+ 1} \right ]^{-1}. 
\label{eq.d14}
\end{align}

We show in Fig.~\ref{fig.d1} the relation between $\Psi^{(2)}$ (equation~\ref{eq.d11}) and gas density for $\rm H_I$ gas ($T^{(2)}{}\approx{}100\rm \,K$) at different redshifts ($z{}={}0-8$), where we account for the effects of the CMB background. It can be seen that $\Psi^{(2)}$ shows almost no redshift evolution at $z{}={}0-4$ over the wide density range being considered. At higher redshifts, $\Psi^{(2)}$ (and hence $\Lambda^{(2)}_{\rm [C_{II}]}$) is raised by the CMB in low-density gas. At $z{}={}6$ ($z{}={}8$), for example, $\Psi^{(2)}$ appears to be much higher than that of the lower redshifts at densities below $\sim{}1\,\rm cm^{-3}$ ($\sim{}10\,\rm cm^{-3}$). 

It should be noted, however, that although the net effect of CMB heating and attenuation on the $\rm [C_{II}]$ cooling rate is negligible except for the low-density gas at $z{}\simgreat{}6$, their own effect can be prominent at various densities and at lower redshifts. This can be seen from Fig.~\ref{fig.d2}, where we explicitly show how $n_{\rm u}{}/{}n_{\rm C^+}$ (indicating heating) and $\eta^{\rm b}$ (indicating attenuation) depend on gas density for $\rm H_I$ gas ($T^{(2)}{}\approx{}100\,\rm K$) at different redshifts \citep[\cf~][]{Kohandel_2019}. Both the effects of CMB heating and attenuation becomes stronger with decreasing gas density, but they almost cancel out each other at above $0.1\,\rm cm^{-3}$ at $z{}={}0-4$ (and at higher densities at $z{}={}6-8$). As a result, the $\rm [C_{II}]$ cooling rate becomes almost unaffected by the CMB in that regime.

\section{Carbon ionization in the $\rm H^+$ region}
\label{Sec:Ap5}
Here we present the analytic expression for the abundance of {$\rm C^+$} ions in the $\rm H^+$ region. Consider the carbon ionization equilibrium equation:
\begin{equation}
\Gamma_{\rm C} n_{\rm C^+} = \alpha_{\rm C} n_{\rm C^{2+}} n_{\rm e^-},
\label{eq.e1}
\end{equation}
\noindent where we only account for the {$\rm C^+{}\Leftrightarrow{}C^{2+}$} equilibrium. $\Gamma_{\rm C}$ is the optically thin carbon photo-ionization rate ($\rm s^{-1}$) and $\alpha_{\rm C}{}={}6.02\times10^{-12}\,\rm cm^3\,s^{-1}$ is the recombination coefficient \citep{Nahar_1997}. Given {$n_{\rm C^+}{}={}x_{\rm C^+} n_{\rm C}$} and {$n_{\rm C^{2+}}{}={}(1-x_{\rm C^+}) n_{\rm C}$}, we can rewrite equation~(\ref{eq.e1}) to be
\begin{equation}
x_{\rm C^+} = \left (1+\frac{\Gamma_{\rm C}}{n_{\rm e^-} \alpha_{\rm C}} \right)^{-1} \approx \frac{n_{\rm e^-}\alpha_{\rm C}}{\Gamma_{\rm C}}. 
\label{eq.e2}
\end{equation}
Following \citet{Ferrara_2019}, we have 
\begin{equation}
\Gamma_{\rm C} = F_{\rm ion} \bar{\sigma}_{\rm C} = U n_{\rm H} c \bar{\sigma}_{\rm C},
\label{eq.e3}
\end{equation}
\noindent where $\bar{\sigma}_{\rm C}{}\approx{}4\times10^{-18}\,\rm cm^2$ is the flux-weighted carbon photo-ionization cross section \citep{Spitzer_1998}. Substituting equation~(\ref{eq.e3}) into equation~(\ref{eq.e2}) and given $n_{\rm e^-}{}\approx{}n_{\rm H}$ for the $\rm H^+$ region, we then get
\begin{equation}
x_{\rm C^+} \approx \frac{\alpha_{\rm C}}{U c \bar{\sigma}_{\rm C} } \propto U^{-1}.
\label{eq.e4}
\end{equation}
Hence, $\rm x_{\rm C^+}$ is inversely proportional to $U$. 

\section{$\rm [C_{II}]$ luminosity of a uniform spherical gas cloud}
\label{Sec:Ap6}
Here we derive the specific $\rm [C_{II}]$ cooling rate ($\rm erg\,cm^3\,s^{-1}$) for a spherical uniform cloud ($\bar{\epsilon}_{\rm [C_{II}],\,cl}$). For the case where the cloud is fully photo-ionized by the external UV radiation (\ie~$l_{\rm s}{}\ge{}R_{\rm cl}$), the luminosity of the cloud ($L_{\rm [C_{II}],\,cl}$) can be expressed as
\begin{equation}
L_{\rm [C_{II}],\,cl} = 4\pi \int_{\,0}^{R_{\rm cl}}  \Lambda^{(1)}_{\rm [C_{II}]} r^2 {\rm d}r. 
\label{eq.f1}
\end{equation}
\noindent Substituting equation~(\ref{eq.d5}) into the above equation, we get
\begin{equation}
L_{\rm [C_{II}],\,cl} = \left ( \frac{4\pi}{3} n_{\rm H} R^3_{\rm cl} \right )  n_{\rm H} \mathcal{A}_{\rm C} \left [ h_{\rm P} \nu_{\rm [C_{II}]} \left (\frac{g_{\rm u}}{g_{\rm l}} \right) R^{\rm e^-}_{\rm ul} (T^{(1)}) x^{(1)}_{\rm C^+} \right ]. 
\label{eq.f2}
\end{equation}
For the case where $\rm H_I$ region forms in the cloud (\ie~$l_{\rm s}{}<{}R_{\rm cl}$), $L_{\rm [C_{II}],\,cl}$ can be expressed as 
\begin{equation}
L_{\rm [C_{II}],\,cl} = 4\pi \left [ \int_{R_{\rm cl}-l_{\rm s}}^{R_{\rm cl}}  \Lambda^{(1)}_{\rm [C_{II}]} r^2 {\rm d}r + \int^{R_{\rm cl}-l_{\rm s}}_{{\rm max}(0,\,R_{\rm cl}-l_{\rm F})}  \Lambda^{(2)}_{\rm [C_{II}]} r^2 {\rm d}r \right ],
\label{eq.f3}
\end{equation}
\noindent where the first and second terms on the right-hand side of the equation correspond to the $\rm [C_{II}]$ emission from $\rm H^+$ (Zone I) and $\rm H_I$ regions (Zone II), respectively. By substituting equation~(\ref{eq.d5}) into the first term and equation~(\ref{eq.d13}) into the second term, we can rewrite the above equation to be
\begin{align}
\begin{split}
	&L_{\rm [C_{II}],\,cl} = f_{\rm [C_{II}],\,cl}  \left ( \frac{4\pi}{3} n_{\rm H} R^3_{\rm cl} \right )  n_{\rm H} \mathcal{A}_{\rm C} \\
	&\times h_{\rm P} \nu_{\rm [C_{II}]} \left ( \frac{g_{\rm u}}{g_{\rm l}} \right ) \frac{  \bigints_{R_{\rm cl}-l_{\rm s}}^{R_{\rm cl}}  x^{(1)}_{\rm C^+} R^{\rm e^-}_{\rm ul} r^2{\rm d}r+ \bigints_{{\rm max}(0,\,R_{\rm cl}-l_{\rm F})}^{R_{\rm cl}-l_{\rm s}} (2/5) R^{\rm H_I}_{\rm ul} r^2{\rm d}r }{\bigints_{{\rm max}(0,\,R_{\rm cl}-l_{\rm F})}^{R_{\rm cl}} r^2{\rm d}r},
\label{eq.f4}
     \end{split}
\end{align}
\noindent where $f_{\rm [C_{II}],\,cl}$ represents the total fraction of gas mass in $\rm H^+$ or $\rm H_I$ regions (Zone I + Zone II). Combining equation~(\ref{eq.f2}) and equation~(\ref{eq.f4}), and substituting $M_{\rm cl}=\frac{4}{3}\pi R^3_{\rm cl} (\mu_{\rm H} m_{\rm H}  n_{\rm H})$ into the equations, we obtain
\begin{equation}
L_{\rm [C_{II}],\,cl} =  f_{\rm C_{II},\, cl} \left ( \frac{M_{\rm cl}}{\mu_{\rm H} m_{\rm H}} \right ) n_{\rm H}\mathcal{A}_{\rm C} \bar{\epsilon}_{\rm [C_{II}],\,cl},
\label{eq.f5}
\end{equation}
\noindent where
\begin{equation}
f_{\rm [C_{II}],\,cl}=  
    \begin{dcases}
     1\; \;\;({\rm if}\; l_{\rm F}\ge R_{\rm cl})\\
      3 \int_{R_{\rm cl}-l_{\rm s}}^{R_{\rm cl}} (r/R_{\rm cl})^2{\rm d}(r/R_{\rm cl}) \;\;({\rm if} \;l_{\rm F}< R_{\rm cl})\\
    \end{dcases}       
\label{eq.f6}
\end{equation}
\noindent and 
\begin{align}
\begin{split}
& \bar{\epsilon}_{\rm [C_{II}],\,cl}=  h_{\rm P} \nu_{\rm [C_{II}]} \left (\frac{g_{\rm u}}{g_{\rm l}} \right) \\
 & \times  \begin{dcases}
     R^{\rm e^-}_{\rm ul} (T^{(1)}) x^{(1)}_{\rm C^+} \;\;\;\; ({\rm if}\; l_{\rm s}\ge R_{\rm cl})\\
     \frac{  \bigints_{R_{\rm cl}-l_{\rm s}}^{R_{\rm cl}}  x^{(1)}_{\rm C^+} R^{\rm e^-}_{\rm ul} r^2{\rm d}r + \bigints_{{\rm max}(0,\,R_{\rm cl}-l_{\rm F})}^{R_{\rm cl}-l_{\rm s}} \left (\frac{2}{5} \right) R^{\rm H_I}_{\rm ul} r^2{\rm d}r }{\bigints_{{\rm max}(0,\,R_{\rm cl}-l_{\rm F})}^{R_{\rm cl}} r^2{\rm d}r} \;\; ({\rm if}\; l_{\rm s}< R_{\rm cl}) 
    \end{dcases}       
\label{eq.f7}
\end{split}
\end{align}
Equation~(\ref{eq.f7}) is the analytic expression for the specific $\rm [C_{II}]$ cooling rate for a uniform spherical gas cloud.

\section{Luminosity-weighted gas density of galaxies}
\label{Sec:Ap7}

\begin{figure}
 \includegraphics[width=83mm]{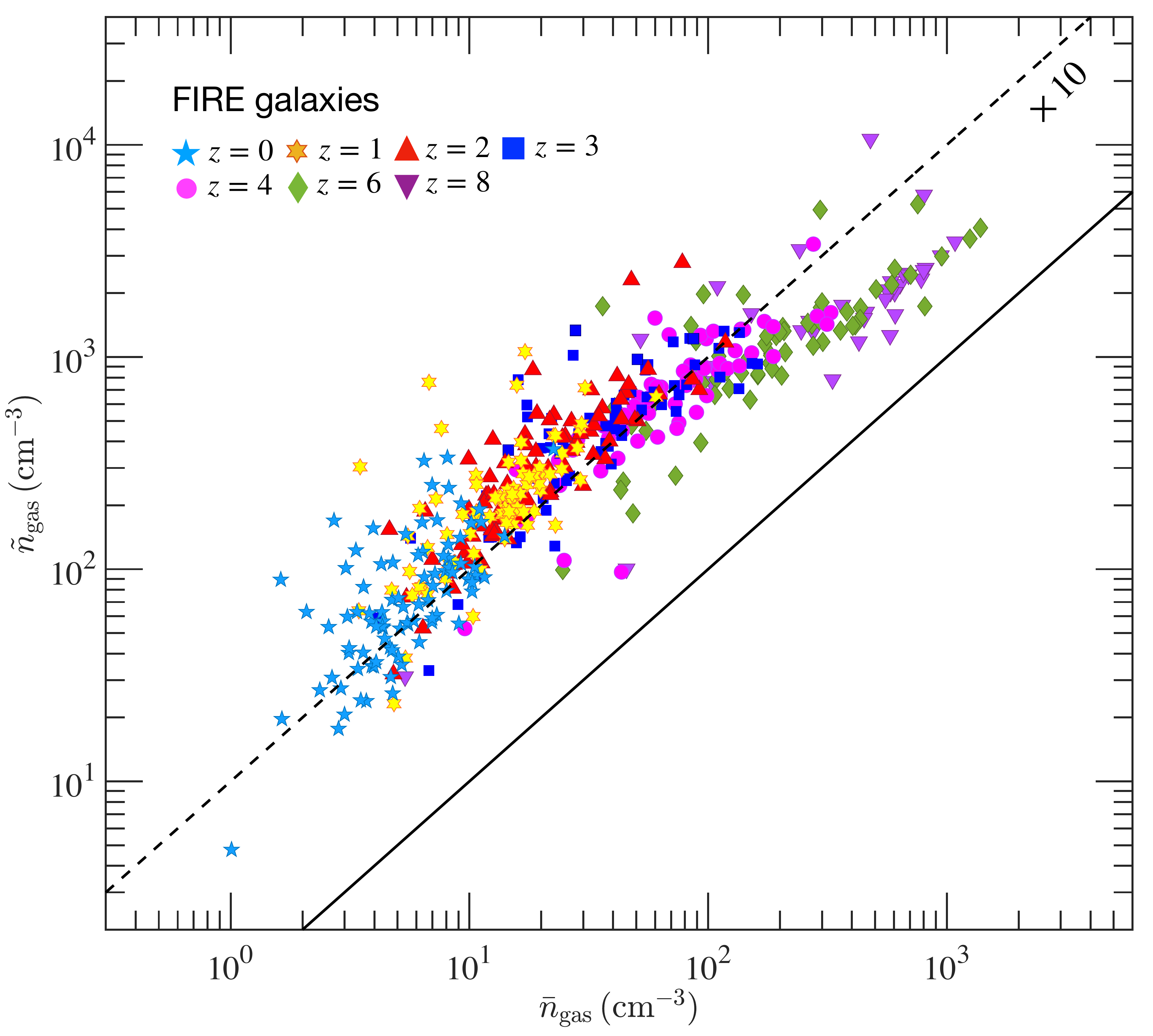}
 \caption{The relation between the $\rm [C_{II}]$ luminosity-weighted {\em median} gas density ($\bar{n}_{\rm gas}$) and the $\rm [C_{II}]$ luminosity-weighted {\em mean} gas density ($\tilde{n}_{\rm gas}$) of the {\sc\small FIRE} galaxy sample at $z{}={}0-8$. The solid black line indicates the one-to-one relationship, whilst the dashed black line indicates the relation $\tilde{n}_{\rm gas}{}={}10\bar{n}_{\rm gas}$. $\tilde{n}_{\rm gas}$ is systematically higher than $\bar{n}_{\rm gas}$.}
    \label{fig.g1}
\end{figure}

\begin{figure}
 \includegraphics[width=83mm]{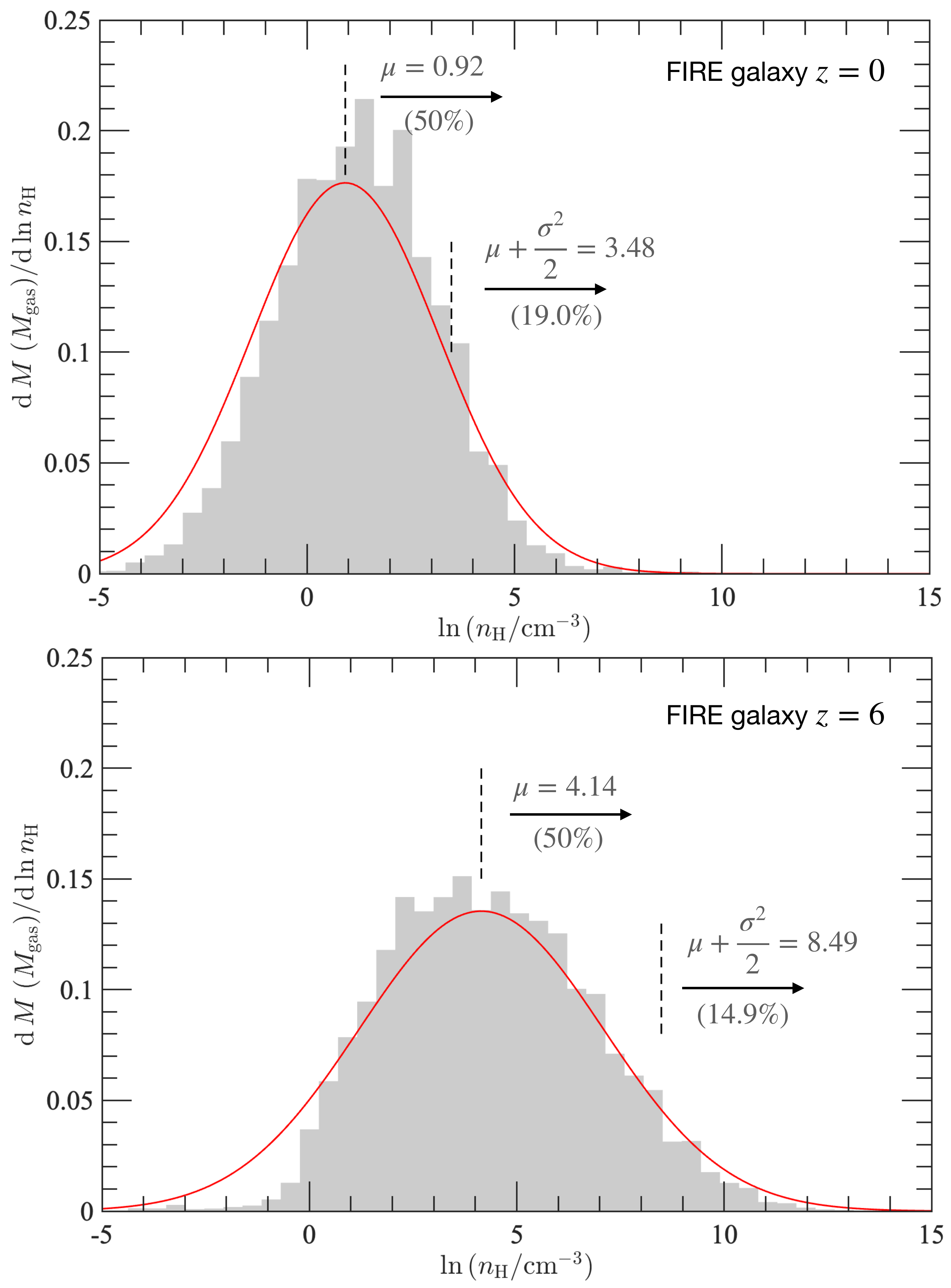}
 \caption{The $\rm [C_{II}]$-luminosity-weighted PDF of gas density of two selected {\sc\small FIRE} galaxies at $z{}={}0$ ({\it upper} panel) and $z{}={}6$ ({\it lower} panel) and the best-fit lognormal function (equation~\ref{eq.g1}) to the PDF.  In each panel, shaded grey area represents the original PDF whereas solid red line indicates the best-fit lognormal function. The luminosity-weighted mean gas density ($\bar{n}_{\rm H}$; marked by the vertical dashed line on the right) of the galaxies is higher than the luminosity-weighted median density ($\tilde{n}_{\rm H}$; marked by the vertical dashed line on the left).}
    \label{fig.g2}
\end{figure}

In Fig.~\ref{fig.g1}, we show the relation between the $\rm [C_{II}]$ luminosity-weighted {\em median} gas density ($\bar{n}_{\rm gas}$) and the $\rm [C_{II}]$ luminosity-weighted {\em mean} gas density ($\tilde{n}_{\rm gas}$) of the {\sc\small FIRE} sample at different redshifts ($z{}={}0-8$). It can be seen from the figure that the latter is systematically higher. 

The reason for this result is that the $\rm [C_{II}]$ luminosity-weighted PDF of gas density ($n_{\rm H}$) of the galaxies resembles a lognormal function (see Fig.~\ref{fig.g2} for an example), showing an elongated tail at high density end. Consider a lognormal function with two parameters $\mu$ and $\sigma$, \ie~
\begin{equation}
P(n_{\rm H}; \mu, \sigma) = \frac{1}{n_{\rm H}\sqrt{2\pi}\sigma} {\rm e}^{-\frac{({\rm ln}(n_{\rm H})-\mu)^2}{2\sigma^2}}.
\label{eq.g1}
\end{equation}
The cumulative distribution function (CDF) for a lognormal distribution is
\begin{align}
C(n_{\rm H}; \mu, \sigma) &\equiv \int^{n_{\rm H}}_{-\infty} P(x;\mu, \sigma) {\rm d}x \nonumber \\
&= \frac{1}{2}\left[ 1+{\rm erf} \left(\frac{{\rm ln}(n_{\rm H}) -\mu}{\sqrt{2} \sigma}\right)\right],
\label{eq.g2}
\end{align}
where {\small erf} is the error function. It is easy to show that the mean density ($\tilde{n}_{\rm H}$) of a lognormal distribution is 
 \begin{align}
 \begin{split}
\tilde{n}_{\rm H}=&\int^{\infty}_{-\infty} x P(x; \mu, \sigma) {\rm d} x = \int^{\infty}_{-\infty} \frac{1}{\sqrt{2\pi}\sigma} {\rm e}^{-\frac{({\rm ln}(x)-\mu)^2}{2\sigma^2}} {\rm d} x \\
=& {\rm e}^{\mu+\frac{\sigma^2}{2}},
 \end{split}
\label{eq.g3}
\end{align}
\noindent whereas the median density ($\bar{n}_{\rm H}$), \ie~the density at which $C(n_{\rm H}; \mu, \sigma){}={}\frac{1}{2}$, is
\begin{equation}
\bar{n}_{\rm H} = {\rm e}^\mu. 
\label{eq.g4}
\end{equation}
Hence, $\tilde{n}_{\rm H}$ is higher than $\bar{n}_{\rm H}$ by a factor of $\tilde{n}_{\rm H}/\bar{n}_{\rm H} = e^\frac{\sigma^2}{2}$. 

In Fig.~\ref{fig.g2}, we show the luminosity-weighted density PDF of two selected {\sc\small FIRE} galaxies at $z{}={}6$ ({\it lower} panel) and $z{}={}0$ ({\it upper} panel) as well as the best-fit lognormal function to their PDF (note: the same galaxies as for Fig.~\ref{fig.11}) as an example. The luminosity-weighted median gas density $\bar{n}_{\rm gas}$ of the $z{}={}0$ ($z{}={}6$) galaxy is $\rm 2.5\,cm^{-3}$ ($\rm 25.1\,cm^{-3}$), whereas its luminosity-weighted mean density $\tilde{n}_{\rm gas}$ is $\rm 4.2\,cm^{-3}$ (754.4 $\rm cm^{-3}$). For the $z{}={}6$ ($z{}={}0$) galaxy, only $19.0\%$ ($14.9\%$) of the total $\rm [C_{II}]$ luminosity originates from the gas at density above $\tilde{n}_{\rm gas}$. It is therefore not statistically representative for the bulk of the gas in galaxies emitting $\rm [C_{II}]$. 

\section{Luminosity-weighted gas metallicity of galaxies}
\label{Sec:Ap8}

\begin{figure}
 \includegraphics[width=83mm]{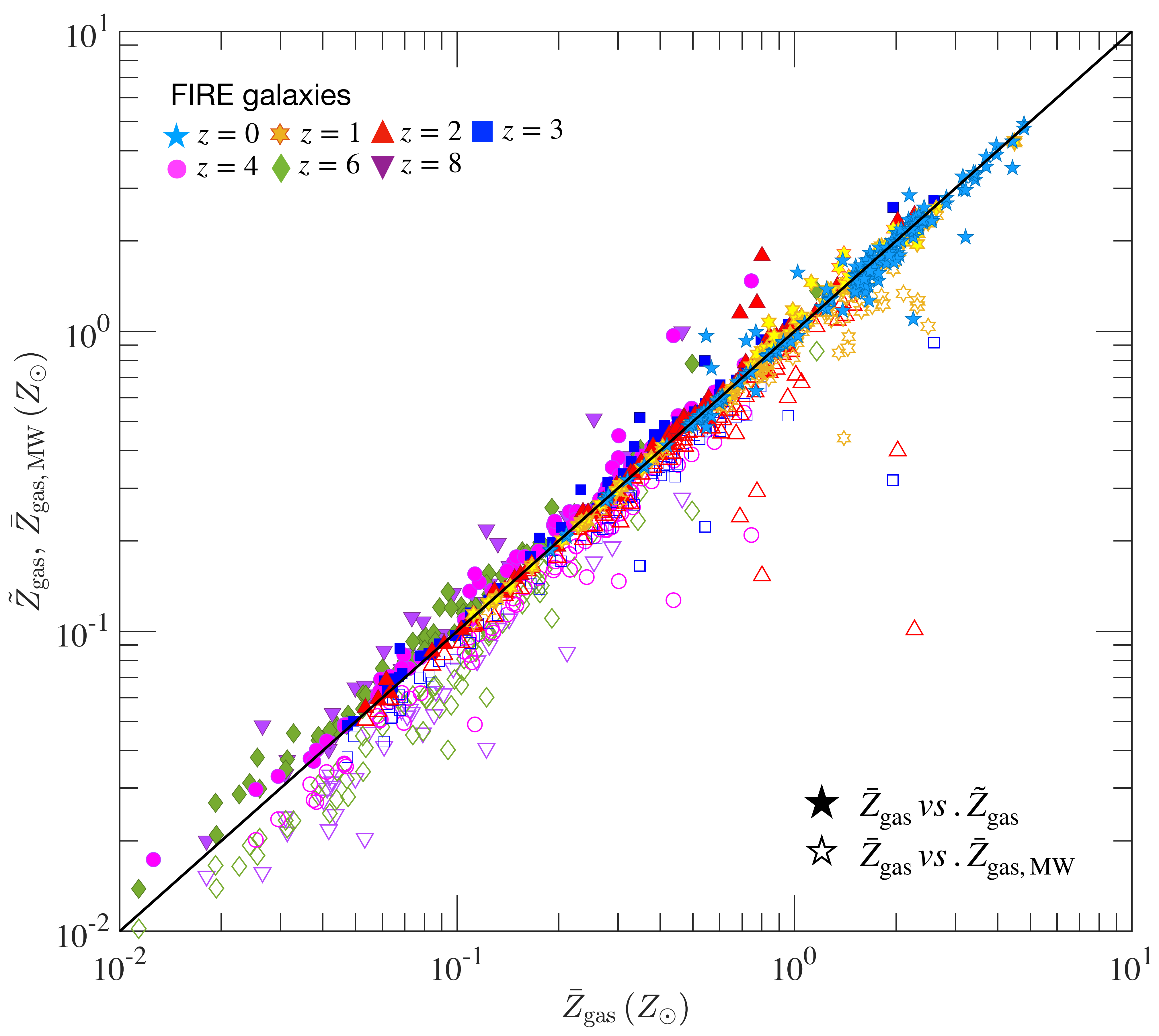}
 \caption{The $\bar{Z}_{\rm gas}$ vs. $\tilde{Z}_{\rm gas}$ relation and the $\bar{Z}_{\rm gas}$ (filled coloured symbols) vs. $\tilde{Z}_{\rm gas,\,MW}$ (empty symbols) relation of the {\sc\small FIRE} sample at $z{}={}0-8$, where $\bar{Z}_{\rm gas}$, $\tilde{Z}_{\rm gas}$ and $\tilde{Z}_{\rm gas,\,MW}$ represent the luminosity-weighted median, luminosity-weighted mean and mass-weighted median gas metallicity, respectively. The solid black line indicates the one-to-one relationship.  $\bar{Z}_{\rm gas}$, $\tilde{Z}_{\rm gas}$ and $\tilde{Z}_{\rm gas,\,MW}$ of the galaxies are very similar to each other.}
    \label{fig.h1}
\end{figure}

In Fig.~\ref{fig.h1}, we show the relation between the luminosity-weighted median ($\bar{Z}_{\rm gas}$) and the luminosity-weighted mean gas metallicity ($\tilde{Z}_{\rm gas}$) of the {\sc\small FIRE} galaxy sample at $z{}={}0-8$. $\tilde{Z}_{\rm gas}$ and $\bar{Z}_{\rm gas}$ are very close to each other. The former is higher by only 0.02 dex ($4\%$) on average.

Both $\tilde{Z}_{\rm gas}$ and $\bar{Z}_{\rm gas}$ of the galaxies are similar to their mass-weighted gas metallicity ($\bar{Z}_{\rm gas,\,MW}$). In the same figure, we show the $\bar{Z}_{\rm gas}$ vs. $\bar{Z}_{\rm gas,\,MW}$ relation for the {\sc\small FIRE} sample. $\bar{Z}_{\rm gas}$ is on average higher than $\bar{Z}_{\rm gas,\,MW}$ by 0.10 dex ($20\%$). 

\bsp
\label{lastpage}
\end{document}